\documentclass[12pt]{report}
\usepackage{nmtthes2000}

\usepackage{amsthm,amscd}
\usepackage{mathrsfs}
\usepackage{verbatim}
\usepackage{amsfonts,amsmath,latexsym,amssymb,txfonts,bm}
\usepackage[american]{babel}

\newcommand{\bra}[1]{\langle{#1}}
\newcommand{\ket}[1]{{#1}\rangle}

\def\II{{\mathbb I}}
\def\RR{{\mathbb R}}

\def\tr{\mathrm{ tr\,}}
\def\Tr{\mathrm{ Tr\,}}

\def\Det{\mathrm{ Det\,}}

\def\Res{\mathrm{ Res\,}}

\def\vol{\mathrm{ vol\,}}

\def\be{\begin{equation}}
\def\ee{\end{equation}}
\def\bea{\begin{eqnarray}}
\def\eea{\end{eqnarray}}
\def\bed{\begin{definition}{\ }}
\def\eed{\end{definition}}
\def\bd{\begin{description}}
\def\ed{\end{description}}
\def\bc{\begin{center}}
\def\ec{\end{center}}

\newtheorem{lemma}{Lemma}

\newtheorem{definition}{Definition}

\dissertation

\author{Guglielmo Fucci}

\title{NON-PERTURBATIVE ASPECTS OF QUANTUM ELECTRODYNAMICS ON CURVED SPACE\newheadline AND INVESTIGATIONS IN MATRIX GRAVITY}

\degree{Doctor of Philosophy in Physics \\ with
Dissertation in Mathematical Physics}

\graduationdate{April, 2009}

\supervisor{Ivan G. Avramidi}

\committeesize{5}

\begin{document}



	%
	%

\titlepage			

\begin{abstract}

This Dissertation is devoted to the detailed study of two major subjects. In the
first part, we study non-perturbative aspects of
quantum electrodynamics on Riemannian manifolds by using
heat kernel asymptotic expansion techniques. Here, we established the existence of a new
non-perturbative heat kernel asymptotic expansion for a Laplace type operator
on homogeneous Abelian bundles with parallel curvature,
and we evaluated explicitly the first three coefficients of the expansion.
As an application of this important result, we computed the imaginary part
of the non-perturbative effective action in quantum electrodynamics and
derived a generalization of the classical Schwinger's result for the creation
of scalar and spinor particles in an electromagnetic field induced by the
gravitational field. We also discovered new infrared divergences due to
the gravitational corrections, which represents a completely new physical
effect.

In the second part of the Dissertation, we studied some aspects of a
newly developed non-commutative theory of the gravitational field called Matrix Gravity.
There are two versions of Matrix Gravity, in the first one, called Matrix General Relativity, the action
functional is obtained by generalizing the Hilbert-Einstein functional to
matrix-valued quantities. In the second one, called Spectral Matrix Gravity,
the action is constructed form the first two spectral invariants of a
non-Laplace type second order partial differential operator.
For the first version, we found the dynamical equations of the
theory, while, for the second version, we computed the first non-commutative
corrections to Einstein equations in the weak deformation limit.
For Spectral Matrix Gravity we analyzed the spectrum of the theory on
DeSitter space and found that the dynamical degrees of freedom are represented
by a number of massive spin-2 and massive scalar particles. Furthermore,
we developed the kinematics of test particles in Matrix Gravity
and found the first and second order corrections to the usual
Riemannian geodesic flow. We evaluated the anomalous non-geodesic
acceleration in a particular case of static spherically symmetric background.
We applied this result to study the problem of the
Pioneer anomaly.

\end{abstract}

\begin{acknowledgments}

There are many people who deserve to be thanked for my personal and professional growth here at New Mexico
Tech.

I would like to thank my advisor Prof. Ivan G. Avramidi for all he has taught me during
these years and for being such an inspiring researcher. Thank you, Ivan, for all the time spent
with me talking about Mathematics and Physics
and for always suggesting new and interesting directions of research. You have been a great
mentor and collaborator helping me to become a better scientist and to be prepared for my future career.

I would like to thank Dr. Giampiero Esposito for introducing me to the exciting
field of Mathematical Physics and for his genuine support over the years. Thank you,
Giampiero, for always believing that I can succeed.

I would like to thank Prof. Klaus Kirsten for working with me on previous papers
and for his valuable suggestions on how to improve my techniques. I am sure that
my time at Baylor working with you will be a productive one.

I thank the members of my Ph.D. committee for valuable suggestions on how to improve
the present manuscript.

A special thank you goes to Megan, her love and support over the years that we
have been together helped me to overcome difficult times and to focus on my studies.
Thank you, Megan, for being such a wonderful and loving person.

A well deserved thank you goes to my family and friends back in Italy. I thank them
for believing in me and for respecting the choice that I made to leave the country.
I greatly appreciated their encouragement to pursue the path I choose.

I would also like to thank the many friends here at New Mexico Tech for making life
easy and pleasant in Socorro and for all the ''good times``. I thank Gina for
her help with my bureaucratic issues and for the many laughs.

I would like to conclude by citing some verses form Dante Alighieri's \emph{Divine Comedy}, \emph{Canto XXVI};
They inspired my life.

\begin{verse}

``\emph{Consider well the seed that gave you birth: \newline 	
 you were not made to live your lives as brutes,\newline
  but to be followers of worth and knowledge.}''
\end{verse}

\end{acknowledgments}

\tableofcontents	


\chapter{INTRODUCTION}

One of the most important achievements in theoretical and mathematical physics in the
$20^{\textrm{th}}$ century was the development of a quantized theory of fields. Today,
this area of research is still thriving and presenting researchers with complex mathematical and
experimental challenges. Quantum field theory begun when physicists tried to unify the newly
discovered special theory of relativity with quantum mechanics. The seminal paper by P. A. M Dirac
in 1927 \cite{dirac27}, \emph{The Quantum Theory of the Emission and Absorption of Radiation},
represents the first attempt to create a quantized theory of the electromagnetic field and
is generally recognized as the beginning of quantum field theory.

After the work of Dirac an extraordinary amount of work in this field, both theoretical
and experimental, was carried out by many physicists leading to striking discoveries.
Amongst the most important are the prediction of the existence of antiparticles, relativistic
effects on the spectra of atoms like Lamb shift, the discovery of a number of elementary particles, the
formulation of the scattering matrix which describes the interaction and decay of particles, and
the development of gauge theories which lead to a unified formulation of three of the
fundamental forces of Nature.

Perhaps one of the most fundamental contributions to the development of the quantum
theory of fields was given by Feynman in his dissertation \cite{feynman42},
\emph{The Principle of Least Action in Quantum Mechanics}, where he developed
the formalism that is known as path integral. This method is based on the
description of quantum systems by using the Lagrangian formalism instead of the
Hamiltonian one. This method leads to a covariant formulation of quantum
mechanics and quantum field theory, namely a formulation in which the form
of the fundamental relations does not depend on the system of reference used.
The extreme importance of this formalism is promptly recognized when one attempts
to unify quantum mechanics with Relativity which is, essentially, a covariant theory.

The path integral gives the transition amplitudes of quantum processes as an
integral (sum) over all the possible paths in the configuration space
joining the initial and final states of the process.  Since the integral, in
Euclidean formulation, contains the exponential of the (negative) classical action,
it is straightforward to realize that only the paths that are close to the one
that extremises the classical action contributes the most to the sum.
The paths that are far from the classical path become exponentially small and are
negligibly small in the sum. Unfortunately, in many cases of physical interest,
the path integral cannot be evaluated exactly, and therefore some approximations
need to be used. The features of the path integral make this formalism particularly
suitable for a semiclassical (or WKB) approximation.

It is well known that this is not the only case in which one uses approximation techniques.
Several approximation methods
have been the most important tools for studying phenomena in quantum field theory.
The reason for that is very simple: as any other field of study in Physics, quantum
field theory presents very complicated non-linear problems for which a closed analytic
solution cannot be found explicitly. Therefore various approximation schemes have been developed
in order to obtain observable predictions at different regimes.

One of the most important approximation schemes, especially for its
manifestly covariant form, is the \emph{background field method} which
was developed in major part by DeWitt \cite{dewitt67a,dewitt67b}. This method
is, actually, a generalization of the method of generating functionals developed by
Schwinger \cite{schwinger51,schwinger54}. The most important object in
the background field method is the so called \emph{effective action}.
The effective action is a functional of the background field and, in principle,
contains all the information about the quantum theory.
In order to yield observable predictions, quantum field theory needs to give the
probability amplitudes of a variety of scattering processes in which the initial interacting
states are known and the final products can be measured by using particle detectors.
Theoretically, this interaction is studied by means of the so called $S$-matrix
(or scattering matrix). In the Feynman diagrammatic technique the $S$-matrix
is described in terms of the propagator and vertex functions.
The effective action determines the propagator and the vertex
functions with regard to all quantum corrections. Of course, once these basic
ingredients are known, the complete $S$-matrix is determined and, hence, measurable
predictions can be made \cite{dewitt65}. Moreover the effective action yields,
upon variation, the effective equations of motion which describe the back-reaction
of quantum processes on the classical background field. Another important feature
of the effective action is its low-energy limit (also called effective potential), which
is the natural tool for studying the vacuum structure of the quantum theory under
consideration.

The effective action is one of the most powerful tools in quantum field theory
and quantum gravity (see
\cite{schwinger51,dewitt03,avramidi00,avramidi02,avramidi08c}). The effective
action is a functional of the background fields that encodes, in principle, all
the information of quantum field theory. It determines the full one-point
propagator and the full vertex functions and, hence, the whole $S$-matrix.
Moreover, the variation of the effective action gives the effective equations
for the background fields, which makes it possible to study the back-reaction
of quantum processes on the classical background. In particular, the low energy
effective action (or the effective potential) is the most appropriate tool for
investigating the structure of the physical vacuum in quantum field theory.

The effective action is expressed in terms of the propagators and the vertex
functions. One of the most powerful methods to study the propagators is the
heat kernel method, which was originally proposed by Fock \cite{fock37} and
later generalized by Schwinger \cite{schwinger51} who also applied it to the
calculation of the one-loop  effective action in quantum electrodynamics.
Finally, DeWitt reformulated it in the geometrical language and applied it to
the case of gravitational field (see his latest book \cite{dewitt03}).

Unfortunately, in most interesting physical cases the effective action cannot be computed exactly, and therefore
approximation methods need to be developed. Since the effective action can be written in
terms of a path integral, one of the most effective approximation methods is the semiclassical
perturbative expansion of the path integral in the number of loops, also known as the \emph{loop expansion}.
The basic idea is the following: all the fields are decomposed in a classical
background part $\phi$ and a quantum disturbance $h$, like $\varphi=\phi+\sqrt{\hbar}\,h$, where $\hbar$ is
the Planck constant. By substituting this decomposition in the
classical action, one can expand the action in the quantum fields. The quadratic part
of the expansion in the quantum fields gives the propagator and the higher order
terms give the various vertex functions. This information is basically all we need in order to
obtain the effective action, because, as we mentioned before, it is constructed in terms
of the propagator and the vertex functions \cite{dewitt65,dewitt03}.
The number of loops in the perturbative expansion corresponds to the power of the
Planck's constant: $\hbar$ corresponds to one-loop expansion, $\hbar^{2}$
represents two-loop expansion, and so on.
One of the most effective mathematical tools to study the propagators in quantum
field theory is the \emph{proper time method} or \emph{heat kernel method} which
was developed by Schwinger in \cite{schwinger51,schwinger54} and then generalized
to include curved spacetime by DeWitt in \cite{dewitt67a,dewitt75}. Some nice
reviews on this subject can be found in \cite{barvin85,vassile03}.

It is clear that the various approximation methods only give correct results
within their own specific regime. In other words each approximation method has
its limits of validity. There are, mainly, three types of regimes that have been
extensively studied in the literature (a good review can be found in \cite{avramidi94}),

\begin{itemize}

\item{\emph{Semiclassical Approximation}: This approximation is used in the case in which the fields of interest
have a large mass. The main idea of the method is to expand all the relevant quantities (like effective action)
in a series of inverse powers of the mass. In this way, since the mass is large, the higher order terms become
smaller and smaller and can be treated as perturbations. Of course this approximation fails completely if one considers
small or vanishing masses because the higher order terms, in the expansion in inverse powers of the mass, would became
larger and larger posing problems for the convergence of the expansion.}

\item{\emph{High Energy Approximation}: This approximation is particularly useful in the case in which
one is interested in weak, rapidly varying background fields. This approximation analyzes the short-wave,
and hence the high energy, part of the spectrum of the background field. The idea is to construct an
expansion in powers of the field strength. Obviously this expansion will not produce correct results
when the field strength becomes large.}

\item{\emph{Low Energy Approximation}: This approximation is the opposite of the previous one, and
it is used when one is interested in strong and slowly varying background fields. This approximation probes
the long-wave, and therefore the low energy, part of the spectrum of the background field. In this approximation
the idea is to construct an expansion in the derivatives of the fields. Exactly as in the other cases
described above, this approximation fails if the field strength becomes small and the derivatives become large.}

\end{itemize}

In the first part of this Dissertation we will mainly study quantum electrodynamics on
curved spacetime. However, before describing in details the work done here, it is important
to briefly review the past literature and results about this subject. One of the first attempts
to utilize the heat kernel method for studying the effective action in quantum
electrodynamics was carried out by Schwinger in \cite{schwinger51}. In his paper
he was able to successfully derive the effective action for quantum electrodynamics
and evaluate its imaginary part which, in turn, gives the probability amplitude
for the creation of pairs of particles in an electromagnetic field. This result,
although of fundamental importance, was obtained on a flat (Minkowski) spacetime.
The generalization of the results obtained by Schwinger in curved spacetime
represent a much more complicated task.
As mentioned before, in general cases the heat kernel and, therefore, the effective action
cannot be computed exactly: the curved spacetime case is one of them.
For this reason the research was focused on trying to obtain some results in
particular regimes by using suitable approximation methods.

In order to study such case one of the oldest methods used was the
Mina-ckshisundaram-Pleijel short-time asymptotic expansion of the heat
kernel. This method is essentially perturbative, in fact in this approach
the expansion is written in terms of powers of the curvature of the spacetime
and its derivatives. Of course this method is inadequate when the curvature
becomes large (strongly curved space-times).
If some of the geometric invariants of the curvature are large, the method of
partial summation has been developed in which one sums the terms in curvature that
are large. This is still a perturbative approach because the expansion is written
in terms of powers of the geometric invariants of the curvature that are not large.

In the low-energy approximation, when the curvature, but not their derivatives, is large,
an effective manifestly covariant method for the evaluation of the heat
kernel asymptotics has been developed \cite{avramidi93,avramidi94,avramidi95a}.
In the case the derivatives of the curvature are small, this method effectively
sums the contribution from the curvature of the spacetime in the heat kernel expansion.
However this study does not contain the electromagnetic field.
A recent paper \cite{avramidi08a} overcomes this difficulty by treating covariantly
constant gauge fields over symmetric spaces which are manifolds with covariantly constant
Riemann tensor.

The first part of this Dissertation deals with the evaluation of the heat kernel asymptotic
expansion for covariantly constant electromagnetic field on arbitrary Riemannian
manifolds. We obtain, for the first time in literature, an expansion, for the heat kernel,
in powers of the geometric invariants of the curvature of the spacetime but to
\emph{all orders of the electromagnetic field}. This represents a completely
new, non-perturbative, heat kernel asymptotic expansion.
Before this important result, the heat kernel expansion in
presence of an electromagnetic field in curved spacetime, and hence the effective action,
were only known as power series in the electromagnetic field strength and its derivatives.
Our result, instead, presents analytic functions of the electromagnetic field which means
that we have effectively summed all the infinite contributions coming from the
electromagnetic field to the effective action in curved spacetime. In this sense
our results are \emph{non-perturbative} in the electromagnetic field.
In physical language this means that we have a tool, which was missing before,
for studying the behavior of matter in a gravitational field under the influence
of a very strong electromagnetic field. One can promptly realize that these results can
be applied, for instance, to the study of matter close to any astrophysical
object which possesses a strong electromagnetic field (like magnetars, pulsars, etc.).
Moreover our results can be used as a tool in order to study charged black holes, in
particular the important subject of creation of pairs of particles near a black hole
possessing a strong electromagnetic field. In other words, we can apply our
non-perturbative method to any physical system under the influence of the gravitational
field and a strong electromagnetic field. Since this result is completely new,
it is possible that new, non-perturbative, physical phenomena, unpredictable if one
uses perturbative methods, might be found in the near future. The results obtained
in the first part of this Dissertation have a profound impact also in mathematics especially
in the area of study of K\"{a}hler manifolds which are, basically, complex manifolds
with some specific algebraic conditions imposed on them.
The complex structure on K\"{a}hler manifolds is a covariantly constant
antisymmetric two-tensor which plays the role of our covariantly constant electromagnetic
field. Our result, applied to this case, would give a new heat kernel asymptotic
expansion of K\"{a}hler manifolds which can be used to obtain specific geometric information
about these manifolds. This is an important topic that would be interesting to
study in the near future.

As an application of the new non-perturbative heat kernel asymptotic expansion
analyzed in this Dissertation, we studied the effective action for scalar and spinor fields
under the influence of the gravitational and the electromagnetic field.
In particular we considered the imaginary part of the effective action.
The imaginary part actually measures the probability amplitude for the creation of
pairs of particles, in particular if it vanishes no particles are created.
In this Dissertation we applied the heat kernel asymptotic expansion in order
to find the imaginary part of the effective action for particles in a
gravitational field under the influence of an electromagnetic field. This
important study generalizes the result obtained by Schwinger in \cite{schwinger51}
to curved spacetime and effectively yields the probability amplitude
of creation of pairs in an electromagnetic field induced by the gravitational field.
Our result is of particular importance because, in this effect, we take into account,
in a non-perturbative fashion, the effect of the electromagnetic field in particle creation
in curved spacetime. This is a completely new result that was absent in the literature.

\bigskip

Undeniably the Standard Model is the most successful achievement of quantum field theory,
unifying strong, weak and electromagnetic interactions. However a quantized
theory of the remaining fundamental force of Nature, namely gravitational interaction,
remains still elusive. Researchers have put a great deal of effort in order to
find a consistent theory of quantum gravity. This work, over the years, produced a
number of different theories trying to reconcile quantum theory and gravitation but
none of them has been proved to be the correct one yet. A review of the present status
of quantum gravity can be found in \cite{dewitt08} and references therein.

For this, and other reasons, recent research has been focused on the development
of alternative theories of gravity. The aim is to find a new theory which would
address the problem of quantization of the gravitational field and also some recent
discrepancies between the predictions of General Relativity and the observations in
particular physical systems.

In the second part of this Dissertation we will present various results concerning
a newly developed theory of the gravitational field called \emph{Matrix Gravity}.
Gravity is a universal physical phenomenon. It is this universality
that leads to a successful geometric interpretation of gravity in terms of Riemannian
geometry in General Relativity. General Relativity is widely accepted as a good
approximation to the physical reality at large range of scales.

However some important open issues in General Relativity are still under debate.
First of all, the experimental evidence points to the fact that all matter exhibits quantum behavior at microscopic scales. Thus, it is
generally believed
that the classical
general relativistic description of gravity is inadequate at short distances
due to quantum fluctuations.
However, despite the enormous efforts to unify
gravity and quantum mechanics during the last several
decades we still do not have a consistent theory of quantum gravity.
There are, of course, some promising approaches, like string theory,
loop gravity and non-commutative geometry. But, at the time, none of them
provides a complete consistent theory that can be verified by existing or
realistic future experiments.

Secondly, in the last decade or so it became more and more evident that there
might be a few problems  in
the {\it classical
domain} as well. In addition to the old problem of gravitational singularities
in General Relativity these {\it gravitational anomalies} include such effects
as dark matter, dark energy, Pioneer anomaly, flyby anomaly, and others \cite{laemmerzahl06}. They
might signal to {\it new physics} not only at the Planckian scales but
at very large (galactic) scales as well.

This suggests that General Relativity, that works perfectly well at macroscopic
scales, should be { modified (or deformed)
both at microscopic and at galactic (or cosmological)
 scales} (or, in the language of high energy physics, both in the ultraviolet
and the infrared). It is very intriguing to imagine that these effects
(that is, the quantum origin of gravity and gravitational anomalies at large
scales) could be somehow related.
Of course, this modification should be done in such a way that at the usual
distances the usual General Relativity is recovered. This condition puts some
constraints (experimental bounds) on the deformation parameters; in the case of non-commutative field theory such bounds on the non-commutativity parameter were
obtained in \cite{carroll01}.

The main ideas of General Relativity are closely related to the geometric
interpretation of linear second-order partial differential operators which describe
the propagation of waves, in particular light, in the spacetime \cite{synge60}.
In fact, in Einstein's General Relativity, light is used to measure distances
and to synchronize clocks in different points of the space. At the time of discovery of
General Relativity the electromagnetic radiation (light) was the only known field that
could serve the purpose. Today it is known that there exist different kinds of fields that
can transmit information in the spacetime in particular fields with some internal structure
(like gauge fields). Matrix Gravity is based on the idea that the structure of spacetime
can be analyzed with fields possessing an internal structure rather than light. The role of the
electromagnetic field in General Relativity is played, in Matrix Gravity, by some other gauge field
(e.g. gluons or other vector bosons) \cite{avramidi04}.

This generalization to gauge fields completely changes the structure of the
spacetime. In General Relativity one is concerned with propagation of
light which is described by a hyperbolic partial differential operator. The
matrix of the second derivatives is smooth symmetric and non-degenerate, moreover
it transforms like a contravariant two-tensor of type $(2,0)$. These properties
allow us to interpret the matrix of the second derivatives as a Riemannian metric.
Obviously, once the metric on the manifold is known, one can construct all the
relevant geometric quantities that are used in General Relativity, like the Christoffel
symbols, Riemann tensor and the invariants coming from it. In Matrix Gravity the
picture is different. The propagation of gauge fields is determined by a system of
hyperbolic partial differential operators. In this case the matrix of the
second derivatives becomes endomorphism-valued, or, in other words, it becomes
a ``matrix of matrices". The latter object does not describe a Riemannian
metric but rather a more general collection of Finsler metrics. This metric is
a generalization of the Riemann metric in which the distance between neighboring points
is an homogeneous function of the point and the tangent vectors. The spacetime
manifold is, now, equipped with a matrix-valued metric which describes a collection of
Finsler metrics. At this point the construction of all the geometric quantities
we need is the same as in General Relativity, however, in Matrix Gravity, the
abovementioned quantities are matrix-valued.

Once the geometric framework is set, we need to focus on the dynamics of the gravitational
field. General Relativity is nothing but the dynamical theory of the metric tensor defined
on the spacetime manifold. Analogously, Matrix Gravity is the dynamical theory
of the matrix-valued metric. However, because of the matrix-valued nature of the geometrical
quantities, namely their non-commutative nature, in this theory, the definition of an action which yields the dynamics is not unique.
If fact, if we try to generalize the Hilbert-Einstein action to matrix-valued quantities, we are
soon faced with the problem of ordering the matrix-valued measure on the spacetime manifold
with the matrix-valued scalar curvature.
In order to avoid this problem we can define the action for Matrix Gravity via spectral
invariants of the partial differential operator which describes the dynamics of the theory.
In general Relativity the action can be written as a specific combination of the
first two global spectral invariants of a Laplace type operator (which in Euclidean formulation
describes the propagation of light). In Matrix Gravity we can write the action as same combination
of the first two global spectral invariants of a more general non-Laplace type partial differential
operator. In this way the action is uniquely defined and does not depend on the order in
which we write the geometric quantities.

The evaluation of the dynamical equations of the theory is an important step because
we can use them in order to study some particular case of physical interest.
In this Dissertation we analyzed the kinematics of test particles in the ambit of Matrix Gravity.
We found that the motion of test particles in the spacetime is quite different from the
predictions of General Relativity. Since Matrix Gravity is basically a dynamical
theory of a collection of Finsler metrics, a particle is described by $N$ different
mass parameters instead of one mass parameter $m$: For each Finsler metric we have a mass parameter.
Every single Finsler metric in the collection determines a particular
geodesic which is followed by the corresponding mass parameter. The sum of all the different mass
parameters is the usual mass. The idea here is similar to the concept of colors in Quantum Chromodynamics (QCD):
In Matrix Gravity the mass consists of different mass parameters as in QCD; the proton, for
example, consists of three quarks of different color.
The different mass parameters describe the tendency for a particle to move along
a particular geodesic. We would like to stress, at this point, that in this picture the
trajectory of a particle ``splits" in a system of trajectories (Finsler geodesics) close
to the Riemannian geodesic. Obviously when everything commutes the bundle of trajectories
collapses in one trajectory which coincides with the one predicted by General Relativity.
As a result, the test particles exhibit a \emph{non-geodesic motion} in the sense that they
do not follow any geodesic derived from any Riemannian metric. This non-geodesic motion
can be interpreted in terms of an anomalous acceleration affecting the test particles.
Driven by this interesting result, we applied the kinematics of test particle
in Matrix Gravity to the Pioneer spacecrafts which present an unexplained (in the
ambit of General Relativity) acceleration.


\bigskip

The outline of the Dissertation is as follows.
In the next chapter we will review some basic technical material and concepts that have been
used throughout the Dissertation. In the third Chapter we will derive the first three heat kernel
asymptotic coefficients for a covariant Laplace type operator in powers of the Riemannian curvature
but to \emph{all orders} of the electromagnetic field. In the fourth Chapter we will use the
results obtained in Chapter three in order to study the effective action in non-perturbative
quantum electrodynamics and compute, in particular, its imaginary part which gives the probability
amplitude for the creation of pairs of particles in the electromagnetic field induced by the
gravitational field. In the fifth Chapter we will be mainly concerned with the computation
of the non-commutative Einstein equations derived from the action in Matrix Gravity constructed
by generalizing the usual Hilbert-Einstein action to matrix-valued quantities. In the sixth Chapter
we will focus on the action for Matrix Gravity derived from spectral invariants of a non-Laplace type
operator. We utilize the heat kernel asymptotic expansion technique to compute the spectral invariants
that form the action. In the second part of the Chapter we use the spectral action to find the non-commutative
corrections to Einstein equations in the low-energy limit, and discuss the spectrum of the theory.
In the seventh Chapter we analyze the kinematics of test particles in Matrix Gravity, especially
the new phenomenon of non-geodesic motion which is related to some anomalous acceleration. In order to
obtain a more specific expression we study the motion of test particles in a static and spherically symmetric
spacetime by using an algebra of $2\times 2$ commuting matrices.
As an application of Matrix Gravity, in the eighth Chapter, we describe the Pioneer anomaly in the ambit
of the anomalous acceleration of test particles in Matrix Gravity. At the end of the Chapter we give
an estimate of the free parameters of the theory to match the value of the observed anomalous acceleration
of the Pioneer spacecrafts.
We conclude, then, the Dissertation with a summary of the most important results obtained in this work and
some ideas for future directions of research.

\bigskip

We would like, at this point, to fix the units which will be used throughout this Dissertation.
In quantum field theory it is convention to set
\begin{displaymath}
\hbar=c=e=1\;.
\end{displaymath}
According to these units, all the relevant quantities can be expressed in terms of length $l$.
More precisely we obtain
\begin{displaymath}
[x]=[x^{0}]=l\;, \quad [m]=[{\rm Energy}]=l^{-1}\;, \quad [F]=[R]=l^{-2}\;,
\end{displaymath}
where $m$ is the mass, $F$ is the electromagnetic field strength and $R$ is the scalar curvature.
Moreover, for the gravitational constant $G$ and the cosmological constant $\Lambda$ we have
\begin{displaymath}
[G]=l^{2}\;,\quad [\Lambda]=l^{-2}\;.
\end{displaymath}




\chapter{MATHEMATICAL BACKGROUND IN QUANTUM FIELD THEORY AND GRAVITY}

In this chapter we review the basic ideas and techniques that have been used over the
years in the literature in order to develop a covariant formalism in quantum field theory.
We will present the main ideas of the formal development of covariant methods in quantum
field theory and the derivation and use of the effective action.
Moreover, we will discuss the heat kernel method as a way (mostly used in this Dissertation) to deal
with the effective action calculations in quantum field theory.
In the second part of this chapter we will describe the basic mathematical tools
that are used in Matrix Gravity.

\section{Introduction to Quantum Field Theory}

Classical mechanics is one of the milestones of human understanding of the physical
world. Newton's dynamical equations describe, by means of a set of second order differential equations,
the motion of a system of point masses. Given the initial position and the initial velocity
for each mass, the subsequent motion is completely determined. This description
of nature was the paradigm for many years. With the discovery of electric
and magnetic phenomena, Maxwell realized that these entities were better described by utilizing
the concept of field.

The solution of Newton's equations is a dynamical trajectory, which is an object that
associates to every instant in time a vector in the space.
A field, instead, is a relation that associates to every point in space and every instant in time
 an object in a particular space. Maxwell described the electric and magnetic fields
as vector fields. One of the most important achievements of Maxwell was the formulation
of a dynamical theory in which the electric and magnetic fields were described as two
manifestations of one single entity: the electromagnetic field.

One of the most striking prediction of the theory of electromagnetism was the finiteness
of the speed of propagation of the electromagnetic radiation in any inertial system of reference.
This was a completely new feature which was not present in classical mechanics
in which information could propagate instantly. Moreover, the constancy of the speed of light
in any reference frame was in direct contrast with the well established classical Galilean
transformations.
In order to overcome these difficulties, Einstein developed a theory, special relativity,
in which the constancy of speed of light in any inertial reference frame
was the starting point. In this theory, time and space are treated on an equal footing
forming a single $4$-dimensional entity called spacetime. In special relativity the
electromagnetic field is described by a $4\times 4$ antisymmetric matrix $F_{\mu\nu}$
which transforms as a tensor (a $2$-form) under Lorentz transformations.
It is soon realized, in the framework of
special relativity, that the usual electric and magnetic fields do not have
an absolute meaning. In fact, in different reference frames, the electric field
can become a magnetic field and vice versa. This means that observers
in different inertial systems of reference would not agree on what to call electric and
magnetic field. In these circumstances, only certain invariant combinations of the electric
and magnetic fields are physical observables. For this reason it is
often convenient to utilize the spectral decomposition of the $2$-form $F_{\mu\nu}$.
In this way the electromagnetic field is represented in terms of invariants and projection on invariant subspaces.
This means that there is a reference frame in which $F_{\mu\nu}$ can be written as a block diagonal matrix where the
entries are the field invariants.

In the period in which Einstein developed the theory of special relativity, another
important theory, which would change the vision of the world, was just at its early stages.
Quantum mechanics was developed in order to explain certain experimental observations,
especially in atomic physics, which could not be predicted in the ambit of classical mechanics and
electromagnetism. A few years later, quantum mechanics and special relativity became accepted
in the scientific community as the theories best suited to describe the physical world.
However, it was soon realized that the two theories were not completely compatible
with each other. The reason is the following: in special relativity, space and time play the same role and
are treated equally in the formulation of the fundamental equations. In quantum mechanics,
instead, time still plays a privileged role, as one can see by just analyzing the Schr\"{o}dinger equation.

Quantum field theory was developed in order to reconcile quantum mechanics and special relativity in a quantum theory that would be relativistically invariant.
In this theory, the microscopic world is described in terms of fields which transform, in
a certain specific way, under the Poincar\'{e} group. The Poincar\'{e} group contains two subgroups,
the Abelian group of translations along the four coordinates and the non-Abelian Lorentz group
of rotations in Minkowski space. The fields are characterized by their transformation
properties under the Lorentz group. More specifically, the Lorentz group has, in general,
two types of representations in terms of matrices of the corresponding algebra. The first type representations
are called \emph{single-valued} and the second type representations are \emph{double-valued}.
Fields that transform under the single-valued representations of the Lorentz group
are tensor fields, such as scalar and vector fields. The fields that transform under
the double-valued representation of the Lorentz group are called spinor fields.
These representations are also called spinor representations of the
Lorentz group. The more familiar concept of spin of a field is related to the particular dimension
of the representation of the Lorentz group under which the field transforms.

The dynamics of a field is described by an action functional. From the action
one derives the dynamical equations for the fields by utilizing the stationary action
principle. For instance,
the dynamics of scalar fields is described by the Klein-Gordon equation and
the dynamics of fields of spin $1/2$ is described by the Dirac equation. In general
the fields satisfy hyperbolic second order partial differential equations
together with some suitable boundary conditions which are usually determined from
the particular physical situation. The canonical quantization procedure
for field theory is the following: one solves the dynamical equations
by using the Fourier transform method. In this way the field will be described
by a Fourier integral containing a combination of positive and negative frequencies.
The field is quantized in the canonical quantization scheme by treating the
field as an operator and by imposing specific equal time commutation relations.
These relations induce similar commutation relations to the coefficients
of the negative and positive frequencies in the solution for the field.
These coefficients, which are now operators as well, are interpreted as
creation and annihilation operators.

Let us, now, describe in more detail the Klein-Gordon and the Dirac equations in
the light of their use in Chapter 4. It is well known that the Klein-Gordon
equation can be derived by varying the following action functional with respect to $\varphi$
\begin{equation}\label{newwilly12a}
S_{KG}=\int_{M}dx\,g^{1/2}\,(g^{\mu\nu}\nabla_{\mu}\varphi^{\ast}\nabla_{\nu}\varphi+m^{2}\varphi^{\ast}\varphi)\;,
\end{equation}
where $\varphi$ represents a complex field, $m$ its mass, $M$ the spacetime manifold, $dx\,g^{1/2}$ the invariant measure on $M$ and $\nabla_{\mu}$ is the covariant derivative.
Since in this Dissertation we are primarily interested in curved
spacetimes, we consider the following generalization to curved spacetime of the above action functional
\begin{equation}\label{newwilly12}
S_{KG}=\int_{M}dx\,g^{1/2}\,(g^{\mu\nu}\nabla_{\mu}\varphi^{\ast}\nabla_{\nu}\varphi+\xi R\varphi^{\ast}\varphi+m^{2}\varphi^{\ast}\varphi)\;,
\end{equation}
where $R$ is the scalar curvature, and $\xi$ is a dimensionless coupling constant which
represents the interaction of $\varphi$ with the gravitational field.
In general, one tries to find the simplest generalization from flat to curved spacetime.
However, more complicated choices can be made if there are important, physical or
technical, reasons to do so \cite{fulling89}.

We would like to stress, at this point, that the action (\ref{newwilly12})
describes the dynamics of a free scalar field in curved space. In order to
describe a self-interacting scalar field, one can add a self-interaction term to
the action (\ref{newwilly12}); The most common term is $\lambda(\varphi^{\ast}\varphi)^{2}$, where $\lambda$ is the coupling constant for self-interaction.
In this case, by varying the action functional containing $\lambda(\varphi^{\ast}\varphi)^{2}$,
one would get the following dynamical equation
\begin{equation}
L_{\rm scalar}\varphi=0\;,
\end{equation}
where
\begin{equation}
L_{\rm scalar}=-\Delta+\xi R+m^{2}+Q_{\rm scalar}\;,
\end{equation}
where $\Delta$ is the Laplacian containing covariant
derivatives and $Q_{\rm scalar}=2\lambda\varphi^{\ast}(\varphi^{\ast}\varphi)$. In Chapter 4 we consider
charged free scalar fields. In this case no self-interaction is present and the term $Q_{\rm scalar}$
is set to zero.

The dynamics of fermions, such as electrons, in flat spacetime is described by the following action
\begin{equation}
S_{D}=\int_{M}dx\,g^{1/2}\,\bar{\psi}(i\gamma^{\mu}\partial_{\mu}+m)\psi\;,
\end{equation}
where $\psi$ represents the spinor field, $\bar{\psi}$ is the Dirac conjugate and $\gamma^{\mu}$ are the gamma matrices. The variation
of this action with respect to the independent field leads to the Dirac equation.
The generalization of the above action to curved spacetime is obtained by replacing the
ordinary derivative with covariant derivative. No scalar analytic term proportional to the
curvature of the spacetime can be added to the above action because any invariant
of the spacetime curvature does not have the right dimensions. In order to
obtain a wave equation for spinors, one considers the square of the Dirac operator $D=i\gamma^{\mu}\nabla_{\mu}+m$.
More specifically, one has
\begin{equation}
L_{\rm spinor}=\bar{D}D=(-i\gamma^{\mu}\nabla_{\mu}+m)(i\gamma^{\mu}\nabla_{\mu}+m)=\gamma^{\mu}\gamma^{\nu}\nabla_{\mu}\nabla_{\nu}+m^{2}\;.
\end{equation}
By writing the coefficient of the second derivatives as sum of the commutator and anticommutator of the gamma matrices
and by taking into account that the commutator of the covariant derivatives introduces the Riemann curvature tensor $R^{ab}{}_{\mu\nu}$ as follows
\begin{equation}
[\nabla_{\mu},\nabla_{\nu}]\psi=\frac{1}{4}R^{ab}{}_{\mu\nu}\gamma_{[a}\gamma_{b]}\,\psi\;,
\end{equation}
it is not difficult to prove that one obtains
\begin{equation}\label{newwilly13}
L_{\rm spinor}=-\Delta+\frac{1}{4}R+m^{2}\;.
\end{equation}
As we can notice, from the last expression, there is a very specific value, $\xi=1/4$,
of the coupling constant for spinors in the gravitational field which is the same in any dimension.

Let us consider, now, the case in which a background electromagnetic field $F_{\mu\nu}$ is present,
which will be studied in Chapters 3 and 4. As it is explained in detail in the next section,
introducing a background electromagnetic field in the formalism is equivalent to replacing the covariant
derivatives in the action $\nabla_{\mu}$ with $\mathscr{D}_{\mu}=\nabla_{\mu}+iA_{\mu}$, where $A_{\mu}$
represents the vector potential.
For charged scalar fields, the introduction of the electromagnetic field is treated in
detail in the next section. It is interesting to consider spinor fields in curved spacetime
under the influence of a background electromagnetic field.
By using the exact same argument that lead us to equation (\ref{newwilly13}),
and by noticing that, in this case,
\begin{equation}
[\nabla_{\mu},\nabla_{\nu}]\psi=\left(\frac{1}{4}R^{ab}{}_{\mu\nu}\gamma_{[a}\gamma_{b]}+iF_{\mu\nu}\right)\psi\;,
\end{equation}
one gets
\begin{equation}
L_{\rm spinor}=-\Delta+\frac{1}{4}R+m^{2}+Q_{\rm spinor}\;,
\end{equation}
where in this case it is not difficult to show that
\begin{equation}
Q_{\rm spinor}=-\frac{i}{2}F_{\mu\nu}\gamma^{[\mu}\gamma^{\nu]}\;.
\end{equation}
Obviously, in absence of the underlying electromagnetic field, the term $Q_{\rm spinor}$ vanishes.
The presence of the term $Q_{\rm spinor}$ in the previous equation represents the reason why
spinors and scalars behave differently in an electromagnetic field. It is the spin,
represented by the antisymmetric product of gamma matrices, that directly couples with the
electromagnetic field.

In fact, in the semiclassical approximation, charged scalar fields are described by just their
electric charge. Spinor fields, instead, are described by the electric charge and the so called
magnetic moment
\boldmath
\begin{equation}
\mu\mbox{\unboldmath$=\frac{1}{m}$}s\mbox{\unboldmath$\;,$}
\end{equation}
\unboldmath
where $\mathbf{s}$ represents the spin of the particle. For instance, the intrinsic magnetic moment
of the electron is the negative of the Bohr magneton $\mu_{B}$ which, in the usual units, has the value
\begin{equation}
\mu_{B}=\frac{e\hbar}{2m_{e}}=9.27\cdot 10^{-24}\,J\, T^{-1}\;.
\end{equation}
The magnetic moment couples to the electromagnetic field
making spinors behave differently from scalars in an electromagnetic field.
This different behavior is the reason why we obtain different results for the creation of scalars and spinors
in curved spacetime under the influence of a strong electromagnetic field
in Chapter 4.

An important difference between quantum mechanics and quantum field theory is
in the description of particles. In quantum mechanics, a certain wave function
is an element of the Hilbert space and describes
a system with a fixed number of quantum particles. Moreover it is well known that the number of
particles is conserved. In quantum field theory a field is an element of a more general
Fock space which is a direct sum of Hilbert spaces; this field describes quantum states
with variable number of particles. This means that in quantum field theory the number
of particles is not constant leading to the interesting phenomenon of creation of particles.
In this framework, the operators introduced in the fields above are interpreted as
creation and annihilation of particles.

Quadratic actions containing single fields describe the propagation of free fields without interactions.
However, interesting physical processes arise from the interaction of two or more fields.
These processes are described by adding a Lagrangian
of interaction to the free Lagrangian of the field. In particle physics one tries to predict the outcome of
a process of interaction of two or more fields in a finite region of the spacetime.
In particular, one is interested in the probability that specific ``in" quantum states become,
after interaction, some ``out" quantum states, in other words scattering of particles.
This probability amplitude is given in terms of the so called $S$-matrix.
As explained later, these fundamental quantities are expressed in terms
of the so called effective action.

One of the most interesting features of quantum field theory is the structure of the
vacuum. The vacuum is defined as the quantum state with no particles. However, in quantum field theory
the vacuum is not really ``empty". In fact, there are continuous processes
of creation and annihilation of pairs of particles. This means that the vacuum is a very dynamical
entity. In the mathematical framework of the effective action, these processes of creation
and annihilation of pairs in the vacuum are described by the imaginary part of the effective action.
One interesting process related to the properties of the vacuum is the \emph{Schwinger
mechanism}. Suppose that we introduce a constant electric field in a region where the are no particles (vacuum).
As we already mentioned above, pairs of particles and antiparticles are created and annihilated.
Now if the electric field is not strong enough (of energy less than the rest mass of the pairs created),
all the created pairs will behave like small electric dipoles and will align with the field generating a ``dielectric effect".
This dielectric effect of the vacuum is also known as vacuum polarization.
Now let us suppose that the energy density of the electric field is much greater than the rest mass density
of the pairs produced in the vacuum. At this point once a pair (of opposite charges) is created
the positive one tends to move parallel to the electric field lines and the other, charged negatively,
tends to move in the opposite direction. In the assumption of strong electric field, the two particles
that are created drift apart to the point in which they cannot annihilate each other.
This means that particles \emph{have been created} because of the presence of a strong electric field.
Schwinger found that the rate of particle production of mass $m$ and charge $e$ per unit volume and per unit time in a
constant electric field $E$ is
given by \cite{schwinger51}
\begin{equation}
R=\frac{1}{16\pi^{4}}E^{2}\sum_{n=1}^{\infty}n^{-2}\exp\left(-\frac{n\pi m^{2}}{E}\right)\;.
\end{equation}
Formally, the presence of the electric field leads to an imaginary contribution
in the effective action which is interpreted as probability of creation of particles.
This process of particle creation in an electric field was studied by Schwinger
in a Minkowski (flat) space and for strong constant electric fields. In this Dissertation, we generalize his result
to Riemannian (curved) spaces and strong covariantly constant electric fields.

In all the discussion above there is an important element missing; namely
the above theories do not take into account General Relativity (GR). GR is a theory
of the gravitational field in which the spacetime is described by a manifold
and the gravitational field is described by a symmetric non-degenerate $2$-tensor
field $g^{\mu\nu}$. GR successfully describes a wide range of physical phenomena at
large scales. Despite this, it is believed that at a more fundamental level
any theory describing the physical world should be quantized. Great efforts
have been made in order to find a consistent theory of quantum gravity and yet
this is still an open problem. Many theories have been conceived in order to
unify quantum mechanics and General Relativity, but none of them seems to
be correct. One of the most difficult problems to overcome
when trying to construct a quantum theory of gravity is that General Relativity
falls in a class of theories that are non-renormalizable. This means that
the infinities that appear in the quantized theory cannot be consistently removed,
unlike renormalizable theories in which this procedure is well defined.
One of the most important achievements in the direction of a full quantized theory
of gravity, was the development of quantum field theory on curved space.
In this theory, the gravitational background is treated classically and the fields
defined in the spacetime are quantized. Even though quantum field theory in curved space
is just an effective theory (it should be the result of a certain limiting case of the full and still unknown
quantum theory of gravity), it has predicted some important phenomena. The most famous
one is the Hawking radiation. Hawking discovered that a black hole emits a thermal radiation
with a black body spectrum due to quantum effects. The Hawking radiation process reduces
the mass of the black hole and is also known as black hole evaporation.
Besides the problem of quantization, it has been recently discovered
that the predictions of General Relativity are not in full agreement
with the observations.
Because of the open issues in General Relativity and its problems with quantization,
it is generally believed that a new theory of gravity needs to be found.
For this reason, in this Dissertation we study a modified theory of gravity
called Matrix Gravity in which the gravitational field is described by a matrix-valued
metric tensor. The idea is to understand whether or not this modified theory
is able to address and solve the open issues that are present in General Relativity.

\section{The Electromagnetic Field as $U(1)$ Gauge Theory}

In this section we will explicitly show that the electromagnetic field can be described
as the gauge theory of the $U(1)$ group following the discussion in \cite{ryder96}. We will see that the electromagnetic field arises
naturally by requiring the invariance under local $U(1)$ transformations of the Lagrangian for a
charged scalar field. 
 A charged scalar field is described by two real functions or,
equivalently, by one complex function. In what follows we will restrict ourselves to the case of a
(flat) Minkowski spacetime.
The Lagrangian for a complex field $\phi$ can be written as follows
\be\label{tut0}
{\cal L}=(\partial_{\mu}\phi^{\ast})(\partial^{\mu}\phi)-m^{2}\phi^{\ast}\phi\;.
\ee
Obviously, this Lagrangian is invariant under internal rotations of the field; namely under the following
transformation
\be
\phi\to e^{-i\Lambda}\phi\;,
\ee
with $\Lambda$ being an arbitrary constant. This transformation implies that the field
is rotated at every point in the space at the same time. This is in contrast with the principles
of special relativity in which information propagates at finite speed. For this reason a
more appropriate transformation (which respects the principles of Relativity), is the following
\be\label{tut1}
\phi\to e^{-i\Lambda(x)}\phi\;, \quad \phi^{\ast}\to e^{i\Lambda(x)}\phi^{\ast}\;.
\ee
These transformations are called local $U(1)$ transformations. At this point one can easily see that
the original Lagrangian is no longer invariant under the transformations (\ref{tut1}).
More precisely, under the transformations (\ref{tut1}) the field and its derivative vary as follows
\be
\delta\phi=-i\Lambda(x)\phi\;,\quad \delta(\partial_{\mu}\phi)=-i\Lambda(x)(\partial_{\mu}\phi)-i[\partial_{\mu}\Lambda(x)]\phi\;.
\ee
By using the above variations, one can show that the variation of the Lagrangian (\ref{tut0})
is not vanishing, but acquires an additional term
\be
\delta{\cal L}=j^{\mu}\partial_{\mu}\Lambda(x)\;,
\ee
where
\be
j_{\mu}=i\left[-\phi\partial_{\mu}\phi^{\ast}+\phi^{\ast}\partial_{\mu}\phi\right]\;.
\ee
Now, in order to make the original Lagrangian invariant under the transformations (\ref{tut1}),
we need to introduce an additional term for which the variation cancels exactly the variation
of the original Lagrangian. This means that we add to (\ref{tut0}) the following term
\be
{\cal L}_{1}=-j^{\mu}A_{\mu}\;,
\ee
where $A_{\mu}$ is a vector field which varies, under the transformations (\ref{tut1}) as follows
\be
A_{\mu}\to A_{\mu}+\partial_{\mu}\Lambda(x)\;.
\ee

We now proceed with the variation of ${\cal L}+{\cal L}_{1}$, and check if this combination
is invariant under local $U(1)$ transformations. By taking the variation of both terms we obtain
\be
\delta{\cal L}+\delta{\cal L}_{1}=-(\delta j^{\mu})A_{\mu}=-2\phi^{\ast}\phi A_{\mu}\,\partial^{\mu}\Lambda(x)\;.
\ee
As we can see, the last combination is not yet invariant. In order to cancel the above
term, we introduce, in the same spirit as earlier, the following term
\be
{\cal L}_{2}=A_{\mu}A^{\mu}\phi^{\ast}\phi\;.
\ee
Now, by taking the variation of the combination ${\cal L}+{\cal L}_{1}+{\cal L}_{2}$, we finally obtain
\be
\delta{\cal L}+\delta{\cal L}_{1}+\delta{\cal L}_{2}=0\;.
\ee
This means that the correct Lagrangian for a charged field which is invariant
under local $U(1)$ transformations is the following
\be
{\cal L}_{\rm CF}=(\partial_{\mu}\phi^{\ast})(\partial^{\mu}\phi)-m^{2}\phi^{\ast}\phi-i\left[-\phi\partial^{\mu}\phi^{\ast}+\phi^{\ast}\partial^{\mu}\phi\right]A_{\mu}+A_{\mu}A^{\mu}\phi^{\ast}\phi\;.
\ee

At this point we see that the vector field $A_{\mu}$ enters the Lagrangian and interacts with the field $\phi$.
One, therefore, expects that this field would be dynamical. Let us then, write a Lagrangian
that describes the dynamics of the new field $A_{\mu}$. The Lagrangian for $A_{\mu}$, in general, must be a scalar
quadratic in the derivatives of the field and invariant under local gauge transformations.
The simplest choice would be ${\cal L}_{\rm A}=(\partial_{\mu}A_{\nu})(\partial^{\mu}A^{\nu})$.
However, this choice is not gauge invariant. A gauge invariant quantity which
can be constructed from the derivatives of $A_{\mu}$ is the following $2$-form
\be
F_{\mu\nu}=\partial_{\mu}A_{\nu}-\partial_{\nu}A_{\mu}\;.
\ee
Therefore, the Lagrangian describing the dynamics of the field $A_{\mu}$ invariant
under the local gauge transformations (\ref{tut1}) is
\be\label{tut2}
{\cal L}_{\rm A}=-\frac{1}{4}F_{\mu\nu}F^{\mu\nu}\;.
\ee
The dynamics of the complex scalar field $\phi$ coupled with the $A_{\mu}$ field is described by
the Lagrangian
\be\label{tut3}
{\cal L}_{\rm Total}=(\partial_{\mu}\phi^{\ast})(\partial^{\mu}\phi)-m^{2}\phi^{\ast}\phi-i\left[-\phi\partial^{\mu}\phi^{\ast}+\phi^{\ast}\partial^{\mu}\phi\right]A_{\mu}+A_{\mu}A^{\mu}\phi^{\ast}\phi
-\frac{1}{4}F_{\mu\nu}F^{\mu\nu}\;.
\ee

Let us, now, see what the dynamical equations for the field $A_{\mu}$ are.
It is not difficult to prove, by varying the action constructed from (\ref{tut2}), that the
dynamical equations for $A_{\mu}$ are
\be
\partial_{\mu}F^{\mu\nu}=-\mathscr{J}^{\nu}\;,
\ee
where
\begin{equation}
\mathscr{J}_{\nu}=i(\phi^{\ast}\partial_{\nu}\phi-\phi\partial_{\nu}\phi^{\ast})-2A_{\nu}\phi^{\ast}\phi\;.
\end{equation}
These are nothing but the covariant version of Maxwell's equations for the electromagnetic field, where $A_{\mu}$
is the vector potential.


We would like to mention, here, that this discussion has a nice geometrical interpretation.
The final Lagrangian (\ref{tut3}) can be written, by rearranging the terms, as follows
\be
{\cal L}_{\rm CF}=(\nabla_{\mu}\phi^{\ast})(\nabla^{\mu}\phi)-m^{2}\phi^{\ast}\phi-\frac{1}{4}F_{\mu\nu}F^{\mu\nu}\;,
\ee
where we have defined a \emph{covariant derivative} as
\be
\nabla_{\mu}=\partial_{\mu}+iA_{\mu}\;.
\ee
This derivative is nothing but the covariant derivative defined on the $U(1)$ bundle
which transforms covariantly under $U(1)$ transformations. In this interpretation,
the vector potential $A_{\mu}$ is nothing but the connection coefficient.
Since, in general, the covariant derivatives do not commute, their commutator is
called curvature. The commutator of the covariant derivative on the $U(1)$ bundle is
\be
\left[\nabla_{\mu},\nabla_{\nu}\right]\phi=iF_{\mu\nu}\,\phi\;.
\ee
Therefore, the electromagnetic $2$-form $F_{\mu\nu}$ is nothing but the curvature of the
$U(1)$ bundle. This discussion shows that there exists a nice geometrical interpretation
of the origin of the electromagnetic field as a gauge theory.

\section{Effective Action in Quantum Field Theory}

In this section we will describe the role of the effective action within the framework
of quantum field theory. We will be mainly concerned, here, with the description of the
quantization of non-gauge field theories (for a detailed description of the quantization
of gauge field theories see \cite{dewitt03}).

The basic object of any physical theory is the spacetime which is described as a $n$-dimensional
manifold, say $M$, with the following topological structure
\begin{equation}
M=I\times \Sigma\;,
\end{equation}
where $I$ is a one-dimensional manifold diffeomorphic either to part of or to the whole real line, and
$\Sigma$ is an $(n-1)$-dimensional manifold. The manifold $M$ is assumed to be globally hyperbolic
and equipped with a pseudo-Riemannian metric. These conditions are sufficient for saying that
the spacetime manifold $M$ possesses a foliation into spacelike sections diffeomorphic to
$\Sigma$. This topology is necessary in order to have the correct causal structure of the spacetime.
The additional structure of a vector bundle $\mathcal{V}$ can be defined over the spacetime manifold $M$, where
each fiber is isomorphic to a vector space $V$. The sections of the vector bundle $\mathcal{V}$ over
the manifold $M$, which we can denote by $\varphi^{i}$, are the fields. The fields describe different kinds of particles (depending
on the structure of the vector bundle) in quantum field theory \cite{avramidi08c}.
In what follows we will consider bosonic fields.
The label ``i" attached to the field represents a compact notation introduced by DeWitt and
denotes not only its components but also the point in spacetime where the field is defined.
This label can be considered as a set of two labels $i\equiv(A,x)$ where $A$ is a discrete index taking values from $1$ to some $D$
associated with the field, and $x$ is the spacetime point \cite{avramidi01a}, for instance we can write
\begin{equation}
\varphi^{i}\equiv\varphi^{A}(x)\qquad\textrm{and}\qquad \varphi^{i^{\prime}}\equiv\varphi^{A}(x^{\prime})\;.
\end{equation}
The set of all possible fields $\varphi^{i}$ on every point of the spacetime manifold
is a manifold itself, $\mathcal{M}$, and is called configuration space.

The first and most important assumption in quantum field theory is that every
isolated dynamical system is describable in terms of a characteristic action functional, $S$,
defined on the configuration space $\mathcal{M}$ with values on the real line,
\begin{equation}\label{intro7}
S:\mathcal{M}\longrightarrow \mathbb{R}\;.
\end{equation}
The dynamics of the isolated system is described by the least action principle;
In other words, the dynamics is determined by setting the first functional derivative of
the action with respect to the independent fields to zero,
\begin{equation}
\frac{\delta S}{\delta\varphi^{i}}=0\;.
\end{equation}
The fields that satisfy the above equation, and suitable boundary and initial conditions,
are called dynamical fields. They form a subspace of the configuration space $\mathcal{M}$
called dynamical subspace which is often called, in quantum field theory, the \emph{mass shell}
\cite{avramidi01a}.

Most of the problems in quantum field theory deal with processes of
scattering of particles. In more details, in the remote past we have well defined
measurable field states (or particles) which are described by the linearized
equations of motion. As the system evolves in time,
the field states interact in a specific finite region in the spacetime. The equations
describing this interaction are highly non-linear and cannot be solved exactly.
After the interaction, in the remote future, we have again well defined measurable
field states which, in general, are different from the initial states. We will call
the initial state $|\ket{\textrm{in}}$ and the final states $|\ket{\textrm{out}}$ which
are defined, respectively, in the remote past and in the remote future.
The scattering process is, then, essentially described by the transition
amplitude $\bra{\textrm{in}}|\ket{\textrm{out}}$.

A powerful method for studying the transition amplitudes is given by the
Schwinger variational principle which gives a relation between the variation of the
transition amplitude $\bra{\textrm{in}}|\ket{\textrm{out}}$ and the variation of the
action describing the dynamical system in the region of interaction. In more detail,
the principle states that
\begin{equation}
\delta\bra{\textrm{in}}|\ket{\textrm{out}}=\frac{i}{\hbar}\bra{\textrm{in}}|\delta S|\ket{\textrm{out}}\;.
\end{equation}
This principle is generally recognized as the principle of quantization because
all the information about the quantum system can be derived from the above equation
\cite{avramidi01a,dewitt65,dewitt03}. In general the action is replaced with a
functional obtained by adding, to the previous action,
a linear interaction with an external classical source $J$,
namely
\begin{equation}
S(\varphi)\rightarrow S(\varphi)+J_{i}\varphi^{i}\;,
\end{equation}
where repeated indices mean a summation over the discrete labels and an integration
over the continuous ones. Under this variation of the action functional the transition amplitude
$\bra{\textrm{in}}|\ket{\textrm{out}}$ becomes a functional of the external source $J$, in other words
\begin{equation}
Z(J)=\bra{\textrm{in}}|\ket{\textrm{out}}|_{S(\varphi)\rightarrow S(\varphi)+J_{i}\varphi^{i}}\;.
\end{equation}
Some of the most important objects in quantum field theory are the
chronological mean values of the quantum fields defined as
\begin{equation}
\frac{\bra{\textrm{in}}|T(\varphi^{i_{n}}\cdots\varphi^{i_{1}})|\ket{\textrm{out}}}{\bra{\textrm{in}}|\ket{\textrm{out}}}\;,
\end{equation}
where $T$ represents the chronological ordering operator which orders the noncommuting
quantum fields with respect to their time label from right to left.

By considering the following specific variation of the action
\begin{equation}
\delta S=\delta J_{i}\varphi^{i}\;,
\end{equation}
it can be shown \cite{avramidi01a,avramidi08c,dewitt65,dewitt03} that $Z(J)$ is
the generating functional for all the chronological amplitudes, namely
\begin{equation}
Z(J+\eta)=\sum_{n\geq 0}\frac{i^{n}}{n!}\eta_{i_{1}}\cdots\eta_{i_{n}}\bra{\textrm{in}}|T(\varphi^{i_{n}}\cdots\varphi^{i_{1}})|\ket{\textrm{out}}\;.
\end{equation}
The chronological amplitudes are easily obtained by repeatedly differentiating the above functional with
respect to the auxiliary fields $\eta$ and then setting $\eta$ equal to zero.

From the functional $Z(J)$ one can construct a new functional $W(J)$ defined as follows
\begin{equation}
Z(J)=e^{iW(J)}\;.
\end{equation}
The utility of this newly introduced functional $W(J)$ is soon recognized by taking its functional
derivatives with respect to the external sources; explicitly, one obtains \cite{avramidi01a,avramidi08c,dewitt65,dewitt03}
\begin{equation}
\bra{}\varphi^{i_{n}}\cdots\varphi^{i_{1}}\ket{}=(-i)^{n}e^{-iW(J)}\frac{\delta^{n}}{\delta J_{i_{n}}\cdots\delta J_{i_{n}}}e^{iW(J)}\;,
\end{equation}
where $\bra{}\varphi^{i_{n}}\cdots\varphi^{i_{1}}\ket{}$ is the mean value of the quantum fields.
In particular one has \cite{avramidi01a}
\begin{equation}
\bra{}\varphi^{i}\ket{}=\phi^{i}\;,
\end{equation}
\begin{equation}
\bra{}\varphi^{i}\varphi^{j}\ket{}=\phi^{i}\phi^{j}+\frac{1}{i}G^{ij}\;,
\end{equation}
\begin{equation}
\bra{}\varphi^{i}\varphi^{j}\varphi^{k}\ket{}=\phi^{i}\phi^{j}\phi^{k}+\frac{3}{i}\phi^{(i}G^{jk)}-G^{ijk}\;,
\end{equation}
where the parentheses $()$ denote symmetrization over the included indices, $\phi^{i}$ represents
the background (or mean) field, $G^{ij}$ is the one-point Green function (or propagator) and $G^{ijk}$ is called
multi-point Green function. To summarize, $Z(J)$ is the generating functional for the chronological
products and $W(J)$ is the generating functional for the Green functions.

It is clear that the mean field $\phi$ is a functional of the external source $J$. It is not difficult
to show that the functional derivative of the mean field with respect to the external source
is the propagator $G^{ij}$. Therefore, if the matrix (propagator) $G^{ij}$ is non-degenerate
we can write the external source $J$ as a functional of $\phi$. By using this property, it
can be shown \cite{dewitt65,dewitt03} that there exists a functional $\Gamma(\phi)$, the \emph{effective action}
which depends on the mean (background) field $\phi$ and is the functional Legendre transform of $W(J)$, namely
\begin{equation}
\Gamma(\phi)=W(J)-J_{i}\phi^{i}\;.
\end{equation}
In terms of the effective action the dynamical equations of the theory take the form
\begin{equation}
\frac{\delta \Gamma}{\delta\phi^{i}}=-J_{i}\;,
\end{equation}
\begin{equation}
\frac{\delta^{2}\Gamma}{\delta J_{i}\delta J_{k}}G^{km}=-\delta_{i}{}^{m}\;,
\end{equation}
where the first equation represents the effective equations of motion determining
the dynamics of the background field, and the second relation defines the full propagator
of the theory, namely the propagator of the background field with regards to all quantum corrections.
The higher functional derivatives of the effective action determine the full vertex functions.
We can say, then, that the effective action is the generating functional of the full vertex functions.
The vertex functions and the full propagator determine the full Green functions and, hence, the
chronological amplitudes which, in turn, give the complete matrix of scattering processes.
It is clear, at this point of the discussion, that the effective action is the most important
object in the theory because it encodes all the information about the quantum fields
\cite{avramidi01a,avramidi08c,dewitt65,dewitt03}. An advantage of basing the theory on the
effective action is that the external sources no longer appear. Moreover, as we will see later in this
section, the functional integral representation of the effective action has a particularly
suitable form for a perturbative analysis \cite{esposito97}.

A very useful representation of the effective action is via the Feynman path integral.
By integrating the Schwinger variational principle, one obtains the following expression
for the $\bra{\textrm{in}}|\ket{\textrm{out}}$ amplitudes \cite{avramidi08c}, namely
\begin{equation}\label{lala1}
\bra{\textrm{in}}|\ket{\textrm{out}}=\int_{\mathcal{M}}\;\mathcal{D}\varphi\exp\left\{\frac{i}{\hbar}[S(\varphi)+J_{k}\varphi^{k}]\right\}\;,
\end{equation}
where $\mathcal{D}\varphi$ represents the functional measure defined on the configuration space.
From this last expression it is possible to get a useful representation
of the effective action \cite{avramidi01a,avramidi08c}, more precisely
\begin{equation}\label{newwilly8}
\exp\left\{\frac{i}{\hbar}\Gamma(\phi)\right\}=\int_{\mathcal{M}}\mathcal{D}\varphi\exp\left\{\frac{i}{\hbar}\left[S(\varphi)-\frac{\delta\Gamma(\phi)}{\delta\phi^{k}}(\varphi^{k}-\phi^{k})\right]\right\}\;.
\end{equation}
Strictly speaking this expression is purely formal, however meaningful results
can be obtained in the framework of perturbation theory. For this reason it is
convenient to use the semi-classical approximation of the effective action. One decomposes the effective
action according to:
\begin{equation}\label{lala2}
\Gamma= S+\Sigma\;,
\end{equation}
where $\Sigma$ is called the
self-energy functional which describes all the radiative corrections to the classical theory \cite{avramidi01a}.
The self-energy functional is computed in terms of an asymptotic expansion in powers of $\hbar$ as follows
\begin{equation}\label{lala3}
\Sigma\sim\sum_{k\geq 1}\hbar^{k}\Gamma_{(k)}\;.
\end{equation}
The next step is to substitute the above expansion for $\Gamma$ in the functional integral representation (\ref{newwilly8}), and to
make a change of variables in the functional integral
\begin{equation}\label{lala4}
\varphi=\phi+\sqrt{\hbar}\;h\;,
\end{equation}
where $\phi$ is the background field and $h$ represents a small quantum perturbation.
Because of the above change of variables, the measure in the functional integral (\ref{newwilly8})
transforms as $\mathcal{D}\varphi=\mathcal{D} h$ and the classical action in the functional integral
is expanded in terms of $h$ as follows \cite{avramidi08c},
\begin{eqnarray}\label{lala5}
S(\phi+\sqrt{\hbar}\;h)&=&S(\phi)+\sqrt{\hbar}\;\frac{\delta S(\phi)}{\delta\phi^{i}}h^{i}-\frac{\hbar}{2}h^{i}\mathscr{L}_{ij}(\phi)h^{j}\nonumber\\
&+&\sum_{n=3}^{\infty}\frac{\hbar^{\frac{n}{2}}}{n!}\frac{\delta^{(n)}S(\phi)}{\delta\phi^{i_{1}}\cdots\delta\phi^{i_{n}}}h^{i_{n}}\cdots h^{i_{1}}\;,
\end{eqnarray}
where $\mathscr{L}$ is a partial differential operator, which is also called the operator of small disturbances,
defined as the second variation of the classical action
\begin{equation}
\mathscr{L}_{ij}=-\frac{\delta^{2}S}{\delta\varphi^{i}\delta\varphi^{j}}\;.
\end{equation}

We would like to stress that in a non-gauge field theory this operator
is non-degenerate, and therefore has a well defined Green function $G=\mathscr{L}^{-1}$ and
a well defined functional determinant. In what follows we will consider non-gauge field theories.
At this point all the needed quantities are expanded in powers of the small disturbances $h$. The final step
is to expand both sides of the functional integral relation (\ref{newwilly8}) in powers of $\hbar$ and equate the terms
of equal powers in $\hbar$. This expansion is called the \emph{loop expansion} where the number of loops is
given by the power of $\hbar$.
The terms in the final expansion in $\hbar$ are functional integrals as well.
However, since we previously expanded in powers of the small disturbances, the functional
integrals are of Gaussian form and can actually be computed. These Gaussian integrals
contain the quadratic form $h^{i}\mathscr{L}_{ij}(\phi)h^{j}$ and the result of the integration
is written in terms of the functional determinant of the operator $\mathscr{L}$, namely $\textrm{Det}\,\mathscr{L}$ and
also in terms of the bare propagator $G=\mathscr{L}^{-1}$. In general one finds the following integrals
\cite{avramidi01a,avramidi08c}
\begin{eqnarray}
\int_{\mathcal{M}}\mathcal{D}h\exp\left\{-\frac{i}{2}h^{i}\mathscr{L}_{ij}h^{j}\right\}&=&(\textrm{Det}\,\mathscr{L})^{-\frac{1}{2}}\;,\label{lala8}\\
\int_{\mathcal{M}}\mathcal{D}h\exp\left\{-\frac{i}{2}h^{i}\mathscr{L}_{ij}h^{j}\right\}h^{k_{1}}\cdots h^{k_{2n+1}}&=&0\;,\\
\int_{\mathcal{M}}\mathcal{D}h\exp\left\{-\frac{i}{2}h^{i}\mathscr{L}_{ij}h^{j}\right\}h^{k_{1}}\cdots h^{k_{2n}}&=&\frac{(2n)!}{2^{n}n!i^{n}}(\textrm{Det}\,\mathscr{L})^{-\frac{1}{2}}G^{(k_{1}k_{2}}\cdots G^{k_{2n-1}k_{2n})}\;.\;\;\;\;\;\;\;\;\;
\end{eqnarray}
The above expansion and the Gaussian integrals give a method to evaluate recursively
all the terms $\Gamma_{(k)}$ of the expansion of the effective action.
Since we are particularly interested in the one-loop effective action,
we will explicitly compute $\Gamma_{(1)}$.
By substituting the expansions (\ref{lala5}) and (\ref{lala3}) in the expression
(\ref{newwilly8}), one obtains
\bea
&\phantom{}&\exp\left\{i\Gamma_{(1)}(\phi)\right\}\exp\left\{i\sum_{k=2}^{\infty}\hbar^{k-1}\Gamma_{(k)}(\phi)\right\}=\nonumber\\
&\phantom{}&\int_{\mathcal{M}}\mathcal{D}h\exp\Bigg\{-\frac{i}{2}h^{i}\mathscr{L}_{ij}h^{j}\Bigg\}\exp\Bigg\{i\sum_{n=3}^{\infty}\frac{\hbar^{\frac{n}{2}-1}}{n!}\frac{\delta^{(n)}S(\phi)}{\delta\phi^{i_{1}}\cdots\delta\phi^{i_{n}}}h^{i_{n}}\cdots h^{i_{1}}\nonumber\\
&-&i\sum_{k=1}^{\infty}\hbar^{k-\frac{1}{2}}\frac{\delta\Gamma_{(k)}}{\delta\phi^{j}}h^{j}\Bigg\}\;.
\eea
By equating the same power of $\hbar$ on both sides of this equation we obtain,
in particular for the one-loop effective action, the following expression
\be
\exp\left\{i\Gamma_{(1)}(\phi)\right\}=\int_{\mathcal{M}}\mathcal{D}h\exp\Bigg\{-\frac{i}{2}h^{i}\mathscr{L}_{ij}h^{j}\Bigg\}\;.
\ee
By using the integral in (\ref{lala8}) it is easy to show that \cite{avramidi01a,dewitt03}
\begin{equation}\label{intro1}
\Gamma_{(1)}=\frac{i}{2}\log\textrm{Det}\,\mathscr{L}\;.
\end{equation}

It is clear now that the effective action is a fundamental object in quantum field theory.
All the information about quantum theory is encoded in the functional structure of the
effective action, its functional derivatives give the full vertex functions and therefore the full
propagators of the theory with are used to build the scattering matrix.
We would like to stress that neither the classical action nor the self-energy functional are physical
objects by themselves, only the effective action, $\Gamma=S+\Sigma$,
describes physical and measurable processes. If the self-energy functional has some divergent terms
one can add equal and opposite counter-terms to the classical action. The coupling constants of these terms
are the observable ones. The classical action with the addition of these counter-terms is called the
renormalized classical action. This is, in a nutshell, the main idea of renormalization theory \cite{avramidi01a}.

In a physical theory the effective action describes the
in-out vacuum transition amplitude via
\be
\left<{\rm out}|{\rm in}\right>=\exp[i\Gamma_{(1)}]\,.
\ee
The real part of the effective
action
describes the polarization of the vacuum of quantum fields by the background
fields and the imaginary part describes
the creation of particles. Namely,
the probability of production of particles (in the whole spacetime)
is given by
\be
P=1-\left|\left<{\rm out}|{\rm in}\right>\right|^2
=1-\exp\left[-2\,{\rm Im}\,\Gamma_{(1)}\right]\,.
\ee
Unitarity requires that the imaginary part of the effective
action should be non-negative
\be
{\rm Im}\,\Gamma_{(1)}\ge 0\,.
\ee
Notice that when the imaginary part of the effective action
is small, one has
\be
P\approx 2\,{\rm Im}\,\Gamma_{(1)}\,.
\ee
The one-loop effective Lagrangian is defined by
\be
\Gamma_{(1)}=\int\limits\limits_M dx\;g^{1/2}{\cal L}\,.
\ee
Therefore, the rate of particle production per unit volume per
unit time is given by the imaginary part of the effective Lagrangian
\be
R=\frac{P}{VT}
\approx 2\,{\rm Im}\,{\cal L}\,.
\ee

We would like to make a final remark here. In quantum field theory, the operator $\mathscr{L}$
describes the propagation of small disturbances in the spacetime and is a hyperbolic operator.
By performing a Wick rotation, $t\to it$, Minkowski space gets mapped into the Euclidean space.
In particular, the hyperbolic operator $\mathscr{L}$ becomes an elliptic operator. The Euclidean
formulation has some advantages: Elliptic operators have been intensively studied and
important information is known about their spectrum. Moreover, it is easier to study the convergence of the
integrals presented above resulting from the path integral formulation.
Of course once a solution is found in Euclidean formulation, one can rotate back to Minkowski
formulation and obtain the solution for Minkowski space.
After a Wick rotation one obtains the Euclidean effective action which
is defined by
\begin{equation}
\Gamma_{(1)}=\varrho\frac{1}{2}\log{\rm Det}\,\mathscr{L}\;,
\end{equation}
where $\varrho$ is the fermionic number of the field, $(+1)$ for bosons and $(-1)$ for fermions.
This particular form of the one-loop effective action will be used in Chapter 4.

In this section we exposed the very basic and main ideas of the effective action
approach to quantum field theory. A more detailed description of the subject (including gauge theories)
can be found in the references that have been cited throughout the section.

\section{The Heat Kernel Method}

In the previous section we saw that the one-loop effective action is written in terms
of the functional determinant of the operator of small disturbances which, in general,
is a second order partial differential operator. The main challenge, at this point, is
to find a formal way to deal with the determinant of an operator. The functional
determinant is defined as a formal expression, and therefore needs to be regularized.
In renormalizable field theories this procedure can be carried out in a consistent way.
However, many field theories of physical interest, including General Relativity,
are non-renormalizable. The short time asymptotic expansion of the trace
of the heat kernel is constructed in terms of spectral invariants which determine the
spectral asymptotics of the operator. Before starting to analyze the heat kernel methods,
we will introduce the Laplace type operator on manifolds and present its main features.

\subsection{Laplace Type Operators}

We consider an $n$-dimensional manifold $M$ which is smooth, compact and without boundary equipped with a
positive definite Riemannian metric $g$. The couple $(M,g)$ will denote a Riemannian manifold
with the properties described above. The coordinates on $M$ will be denoted by $x^{\mu}$
where $\mu$ ranges over $\{1,\cdots n\}$. For any point $p$ on the manifold we can locally define the
tangent space $T_{p}M$ to the manifold $M$ at the point $p$. The space $T_{p}M$ is the vector
space of tangent vectors to $M$ at the point $p$. Moreover, the space $T^{\ast}_{p}M$ is the cotangent space to the
manifold $M$ at the point $p$ and represents the space of the linear functionals, also called forms, acting on the tangent
vectors.

The tangent bundle $TM$ is the disjoint union of the
tangent spaces at each point of the manifold $M$. The notion of bundle
can be easily generalized, in particular, to vector spaces $V$. A vector bundle $\mathcal{V}$ is the disjoint union
of the vector spaces $V$ at each point of the manifold $M$, moreover the dual bundle $\mathcal{V}^{\ast}$ is the
vector space of all the linear functionals defined on $\mathcal{V}$. A section of the vector bundle
is a smooth map
\begin{equation}
\varphi:M\longrightarrow\mathcal{V}\;,
\end{equation}
such that at each point of the manifold $M$ it associates a vector in the vector bundle $\mathcal{V}$.
It is easily recognized that this map represents a vector field defined on the manifold $M$ which
we will denote by $\varphi^{A}$ where $A$ ranges over $\{1,\cdots,\textrm{dim}V \}$.
Of course this idea can be generalized to functions, tensors, etc. giving functions, tensor fields etc.
defined on a manifold. The vector bundle $\mathcal{V}$ itself has the structure of a manifold. We equip
the vector bundle with a non degenerate, Hermitian positive definite metric
\begin{equation}
E:\mathcal{V}\times\mathcal{V}\longrightarrow \mathbb{R}\;,
\end{equation}
which we will call the fiber metric on the vector bundle $\mathcal{V}$ and which can
be naturally identified with the map
\begin{equation}
E:\mathcal{V}\longrightarrow\mathcal{V}^{\ast}\;.
\end{equation}

Since we aim at introducing operators on manifolds, particularly the Laplacian, we need
to construct a suitable space where the operator can be defined, namely a functional space.
For $M$ we introduce the natural Riemannian volume element defined by the Riemannian
metric $g_{\mu\nu}$ on the manifold $M$ as follows: $d{\rm vol}(x)=dx\;g^{1/2}$,
where $g=|{\rm det}g_{\mu\nu}|$.
We introduce the set $C^{\infty}(\mathcal{V})$ of all smooth sections of the bundle $\mathcal{V}$,
which is a vector space. Let $\varphi$ and $\psi$ be in $C^{\infty}(\mathcal{V})$. By using the fiber metric,
in the above functional space we introduce the following inner product
\begin{equation}
(\varphi,\psi)=\int_{M}d{\rm vol}(x)\varphi^{A}(x)E_{AB}(x)\psi^{B}(x)\;.
\end{equation}
Therefore the space $C^{\infty}(\mathcal{V})$, equipped with this inner product,
is an inner product space. The inner product space can be made into a normed space by
defining the following norm
\begin{equation}
||\varphi||^{2}=(\varphi,\varphi)=\int_{M}d{\rm vol}(x)\varphi^{A}(x)E_{AB}(x)\varphi^{B}(x)\;.
\end{equation}
By completing the space $C^{\infty}(\mathcal{V})$ with respect to this norm
one obtains the Hilbert space $\mathscr{L}^{2}(\mathcal{V})$ of square integrable sections of
the vector bundle $\mathcal{V}$.

A connection on the vector bundle $\mathcal{V}$ is a linear map
\begin{equation}
\nabla:C^{\infty}(\mathcal{V})\longrightarrow C^{\infty}(T^{\ast}M\otimes\mathcal{V})\;,
\end{equation}
from the smooth sections of the bundle $\mathcal{V}$ to 1-form valued sections of the bundle $\mathcal{V}$
obeying the Leibnitz rule. We assume that the connection is compatible with the Hermitian metric
on the vector bundle $\mathcal{V}$. In a more explicit form we write
\begin{equation}
\nabla_{\mu}\varphi=(\mathbb{I}\,\partial_{\mu}+\mathscr{A}_{\mu})\varphi\;,
\end{equation}
where $\mathbb{I}$ represents the identity on the vector bundle $\mathcal{V}$ and $\mathscr{A}_{\mu}$ are the connection coefficients which bear a mixture of fiber and manifold
indices. We would like to stress here that the operator defined above is a derivative operator. By explicitly writing
all the indices we have
\begin{equation}
\nabla_{\mu}\varphi^{A}=\left(\delta^{A}{}_{B}\otimes\partial_{\mu}+\mathscr{A}_{\mu}{}^{A}{}_{B}\right)\varphi^{B}\;,
\end{equation}
 We can also define the
formal adjoint $\nabla^{\ast}$ of the derivative operator $\nabla$ with the help of the
Riemannian metric on $M$ and the Hermitian structure on $\mathcal{V}$. Finally, let $Q\in C^{\infty}(\textrm{End}(\mathcal{V}))$
be a smooth section of the bundle of the endomorphisms of the vector bundle $\mathcal{V}$.

A Laplace type operator is a linear partial differential operator
\begin{equation}
\mathscr{L}:C^{\infty}(\mathcal{V})\longrightarrow C^{\infty}(\mathcal{V})\;,
\end{equation}
 of the following form
\begin{equation}
\mathscr{L}=\nabla^{\ast}\nabla+Q=-g^{\mu\nu}\nabla_{\mu}\nabla_{\nu}+Q\;.
\end{equation}
In local coordinates the Laplacian operator can be written, in a manifestly self-adjoint form, as
\begin{equation}
g^{\mu\nu}\nabla_{\mu}\nabla_{\nu}=\Delta=g^{-1/2}(\partial_{\mu}+A_{\mu})g^{1/2}g^{\mu\nu}(\partial_{\mu}+A_{\mu})\;.
\end{equation}
One can write the above operator by explicitly separating the terms with second, first and zeroth order
in the derivatives, namely it is possible to show that
\begin{equation}
\mathscr{L}=-g^{\mu\nu}\partial_{\mu}\partial_{\nu}+b^{\mu}\partial_{\mu}+c\;,
\end{equation}
where
\begin{eqnarray}
b^{\mu}&=&-2g^{\mu\nu}A_{\mu}-g^{-1/2}\partial_{\nu}(g^{1/2}g^{\mu\nu})\;,\\
c&=&Q-g^{\mu\nu}A_{\mu}A_{\nu}-g^{-1/2}\partial_{\mu}(g^{1/2}g^{\mu\nu}A_{\nu})\;.
\end{eqnarray}

Associated to any partial differential operator there is a function $\sigma(x,\xi)$ called the symbol,
\begin{equation}
\sigma:M\times T^{\ast}_{x}M\longrightarrow C^{\infty}\left(\textrm{End}(\mathcal{V})\right)\;,
\end{equation}
which
is obtained from the operator by replacing the derivatives with a covector $i\xi_{\mu}$ (momentum).
For the operator $\mathscr{L}$ considered above, it reads
\begin{equation}
\sigma(x,\xi)=\mathbb{I}g^{\mu\nu}(x)\xi_{\mu}\xi_{\nu}+ib^{\mu}(x)\xi_{\mu}+c(x)\;.
\end{equation}
The leading symbol $\sigma_{L}$ of $\mathscr{L}$ is the part of the symbol with the highest power of
the covector $\xi$. In the case we are analyzing, it has the form
\begin{equation}
\sigma_{L}(x,\xi)=\mathbb{I}g^{\mu\nu}(x)\xi_{\mu}\xi_{\nu}\;.
\end{equation}

From the form of the Laplace type operator we can say that the second order part of $\mathscr{L}$ is determined
by the metric $g^{\mu\nu}$ on the manifold $M$, the first order part of $\mathscr{L}$ is determined by the connection
$\mathcal{A}_{\mu}$ on the vector bundle $\mathcal{V}$ and the zeroth order part of $\mathscr{L}$ is given
by the endomorphism $Q$. It is important to say that any second order partial differential operator with a scalar
leading symbol given by the metric is of Laplace type and can be put in the above form by
a suitable choice of the connection and the endomorphism.

Second order partial differential operators can be classified by utilizing their leading
symbol. For Laplace type operators this classification is equivalent to the following:
\begin{itemize}

\item{The operator $\mathscr{L}$ is \emph{elliptic} if the eigenvalues of $g^{\mu\nu}$ are all different from zero and
have the same sign.}

\item{The operator $\mathscr{L}$ is \emph{hyperbolic} if the eigenvalues of $g^{\mu\nu}$ are all non-vanishing and
all have the same sign except one that has the opposite sign.}

\item{The operator $\mathscr{L}$ is \emph{parabolic} if the eigenvalues of $g^{\mu\nu}$ have all the same sign except
one which is zero.}

\end{itemize}

It can be proved that the Laplacian, and therefore the operator $\mathscr{L}$, is an elliptic
and symmetric partial differential operator.
An important property that we will assume for the second order partial differential operators in the rest of this work is
for them to be self-adjoint.
An operator is called essentially self-adjoint if
for any $\varphi, \phi\in C^{\infty}(\mathcal{V})$,
\begin{equation}
(\mathscr{L}\varphi,\psi)=(\varphi,\mathscr{L}\psi)\;.
\end{equation}
It can be proved that we can always find a unique self-adjoint extension. Moreover,
we will assume, from now on, that the operators have a positive definite leading symbol.

It is worth mentioning, here, that for a Laplace type operator the leading symbol is scalar.
However, in Matrix Gravity we consider more general partial differential operators of non-Laplace type with
non-scalar leading symbol. In fact, the coefficient of the second derivative of the operator we
consider in Matrix Gravity bears two spacetime indexes and two fiber indexes making it a
matrix-valued symmetric, non-degenerate tensor of type $(0,2)$. A general second order partial differential
operator acting on a vector bundle has the general form
\begin{equation}\label{newwilly11}
L=-a^{\mu\nu}(x)\partial_{\mu}\partial_{\nu}+b^{\mu}(x)\partial_{\mu}+c(x)\;,
\end{equation}
with coefficients $(a^{\mu\nu}(x))_{A}{}^{B}$, $(b^{\mu}(x))_{A}{}^{B}$ and $(c(x))_{A}{}^{B}$.
The operator $L$ is called of non-Laplace type if it is self-adjoint and if the leading symbol \emph{is not} scalar, which means that it cannot
be written as product of the identity $\mathbb{I}$ on the bundle and a 2-times contravariant tensor.
More explicitly, the leading symbol of a non-Laplace type operator is
\begin{equation}
\sigma_{L}(x,\xi)=a^{\mu\nu}(x)\xi_{\mu}\xi_{\nu}\;.
\end{equation}
We will study non-Laplace type operators and their heat kernel asymptotic expansion in Chapter 6.

For an elliptic self-adjoint second order partial differential operator 
with positive definite leading symbol on a compact manifold, the following
properties are well known \cite{gilkey95}:
\begin{itemize}

\item{The spectrum, $\{\lambda_{n}\}_{n=1}^{\infty}$ is real discrete and bounded
from below
\begin{displaymath}
\lambda_{0}<\lambda_{1}<\cdots<\lambda_{n}<\cdots
\end{displaymath}
with some real constant $\lambda_{0}$.}

\item{The eigenvalues have the following asymptotic behavior $\lambda_{k}\sim Ck^{\frac{n}{2}}$ as
$k\rightarrow\infty$, where $n=\textrm{dim}(M)$.}

\item{ The eigenspaces are finite-dimensional.}

\item{The eigenvectors, $\{\varphi_{n}\}_{n=1}^{\infty}$, are smooth sections
of the vector bundle $\mathcal{V}$ and form a complete orthonormal basis for the functional space $\mathscr{L}^{2}(\mathcal{V})$.}

\end{itemize}


\subsection{Spectral Functions}

In this section we will present some basic material regarding spectral functions, in particular
the spectral zeta function. This particular object is of primary interest in quantum field theory
because it gives a way to define the regularized functional determinant of an operator, which
appears in the one-loop effective action.
The spectrum of an operator contains important information both from the
physical and mathematical points of view. That is why many efforts have been put in order to
find objects that would give knowledge about the spectrum. Of fundamental importance to the
study of the spectrum of an operator are particular spectral invariants called spectral functions.
Here, we will be mainly interested in two particular spectral functions, namely the
\emph{heat trace} and the \emph{spectral zeta function}. The heat trace is the function $\Theta$ defined as
\begin{equation}\label{intro9}
\Theta(t)=\sum_{n=1}^{\infty}e^{-t\lambda_{n}}\;,
\end{equation}
where $\lambda_{n}$ are the eigenvalues of the operator counted with their multiplicity.
The heat trace is related, as we will show later in this Chapter, to the spectral zeta function
and, therefore, to the one-loop effective action.
The heat trace is a well defined function for positive $t$, moreover it can be analytically
continued to the plane $\textrm{Re}\;t>0$. We would like to mention that if the spectral functions
are known exactly, then the spectrum is completely determined. Unfortunately, in general,
the spectral functions are not known exactly. However, their asymptotic expansions are known, and
they give important information about certain parts of the spectrum.

The spectral zeta function $\zeta$ is a generalization of the Riemann zeta function,
defined as follows:
\begin{equation}
\zeta(s)=\sum_{n=1,\lambda_{n}\neq 0}^{\infty}\lambda_{n}^{-s}\;,
\end{equation}
where $\lambda_{n}$ are the non-vanishing eigenvalues of the operator counted with their multiplicity.
The spectral zeta function can be analytically continued on the whole complex $s$-plane yielding
a meromorphic function with only simple poles and regular at the origin.
There is a nice representation of the spectral zeta function in terms of the heat trace.
One can utilize the well known integral representation of the Gamma function
\begin{equation}
\Gamma(s)=\int_{0}^{\infty}dt\;t^{s-1}e^{-t}\;.
\end{equation}
By performing the following change of variables $t\rightarrow\lambda_{n}t$, it is not difficult to
show that
\begin{equation}
\lambda_{n}^{-s}=\frac{1}{\Gamma(s)}\int_{0}^{\infty}dt\;t^{s-1}e^{-\lambda_{n}t}\;.
\end{equation}
By taking the sum of this last expression one obtains
\begin{equation}
\zeta(s)=\sum_{n=1,\lambda_{n}\neq 0}^{\infty}\lambda_{n}^{-s}=\frac{1}{\Gamma(s)}\int_{0}^{\infty}dt\;t^{s-1}\sum_{n=1,\lambda_{n}\neq 0}^{\infty}e^{-\lambda_{n}t}\;.
\end{equation}
Alternatively, by recalling the expression for the heat trace, one can write
\begin{equation}\label{intro2}
\zeta(s)=\frac{1}{\Gamma(s)}\int_{0}^{\infty}dt\;t^{s-1}\left[\Theta(t)-N\right]\;,
\end{equation}
where $N=\dim\ker\mathscr{L}$ is the number of zero eigenvalues of $\mathscr{L}$.
From this expression we clearly notice that the heat trace and the spectral zeta function are
related to one another by a Mellin transformation.

In particular the spectral zeta function is used in order to define complex powers of an
elliptic non-degenerate self-adjoint operator \cite{gilkey95}. One can use this general
result to evaluate the determinant of a suitable operator which appears in the semiclassical approximation
of the path integral in quantum mechanics and quantum field theory.
The relation between the spectral zeta function and the determinant of an operator
can be formally shown as follows. Let $\mathscr{L}$ be a Laplace type operator. Because of its
discrete spectrum one defines the logarithm of the determinant of $\mathscr{L}$ in complete
analogy to the finite dimensional case, namely
\begin{equation}
\log\textrm{Det}_{N}(\mathscr{L})=\log\prod_{n=1,\lambda_{n}\neq 0}^{N}\lambda_{n}=\sum_{n=1,\lambda_{n}\neq 0}^{N}\log(\lambda_{n})\;.
\end{equation}
Now, by recalling the expression for the spectral zeta function, and taking the derivative
with respect to the parameter $s$, one gets
\begin{equation}
\zeta^{\prime}_{N}(s)=-\sum_{n=1,\lambda_{n}\neq 0}^{N}\lambda_{n}^{-s}\log(\lambda_{n})\;.
\end{equation}
By setting $s=0$ in the last expression, one obtains a formula for $\log\textrm{Det}_{N}(\mathscr{L})$,
more explicitly
\begin{equation}
\log\textrm{Det}_{N}(\mathscr{L})=-\zeta^{\prime}_{N}(0)\;.
\end{equation}
In particular the determinant of the operator $\mathscr{L}$ can be obtained by taking the limit as $N\to\infty$ and by exponentiating the last expression,
more explicitly
\begin{equation}
\textrm{Det}(\mathscr{L})=e^{-\zeta^{\prime}(0)}\;.
\end{equation}
In order to relate the spectral zeta function to the one-loop effective action in quantum field theory,
we recall equation (\ref{intro1}). The operator of small disturbances on the right hand side
of equation (\ref{intro1}) is an elliptic, non-degenerate and self-adjoint second order
partial differential operator. By using the relation obtained above, one can
write an expression for the one-loop effective action in the zeta function regularization as follows
\begin{equation}
\Gamma_{(1)}=-\frac{i}{2}\zeta^{\prime}(0)\;.
\end{equation}

Since the spectral zeta function and the heat trace are related by the Mellin transform,
we can find an expression for the one-loop effective action and the heat trace. By taking
the derivative with respect to the parameter $s$ in (\ref{intro2}) it is not difficult
to show that
\begin{equation}
\Gamma_{(1)}=-\frac{i}{2}\int_{0}^{\infty}dt\;t^{-1}\Theta(t)\;.
\end{equation}
The one-loop effective action will be of primary interest in Chapter 5 where we will
evaluate its imaginary part.
This integral needs to be regularized because it diverges at $t=0$ (ultraviolet divergences),
moreover it could also diverge at $t=\infty$ (infrared divergence). The integral can be regularized
by means, for example, of a cutoff regularization.
It is clear, now, from the last expression, that the knowledge of the heat trace
is of fundamental importance in order to obtain information about the one-loop
effective action. Unfortunately the heat trace cannot be computed exactly in
many interesting cases. However, one can find suitable asymptotic expansions
to get some information on $\Gamma_{(1)}$. In the next section we will analyze
the heat kernel (and in particular the heat trace) and some methods for
the computation of its asymptotic expansion.

\subsection{The Heat Kernel}
Let $\mathscr{L}$ be the operator that we utilized so far. For $t>0$ the one-parameter
family of operators
\begin{equation}
U(t)=\exp\left(-t \mathscr{L}\right)\;,
\end{equation}
forms a semigroup of bounded operators on $\mathscr{L}^{2}(\mathcal{V})$ which is
called the \emph{heat semigroup}. Associated with the heat semigroup one can define the
heat kernel as follows \cite{gilkey95}
\begin{equation}\label{intro80}
U(t|x,x^{\prime})=\sum_{n=1}^{\infty}e^{-t\lambda_{n}}\varphi_{n}(x)\otimes\varphi^{\ast}_{n}(x^{\prime})\;,
\end{equation}
where $\lambda_{n}$ are the eigenvalues of the operator $\mathscr{L}$ counted
with their multiplicity, and $\varphi_{n}$ are
the corresponding eigenfunctions. The heat kernel satisfies the following partial
differential equation
\begin{equation}\label{intro4}
\left(\partial_{t}+\mathscr{L}\right)U(t|x,x^{\prime})=0\;,
\end{equation}
with the initial condition
\begin{equation}\label{intro5}
U(0|x,x^{\prime})=\delta(x,x^{\prime})\;,
\end{equation}
where $\delta(x,x^{\prime})$ represents the covariant Dirac delta function.
It can be shown \cite{gilkey95} that the heat semigroup is a trace-class operator,
namely its $\mathscr{L}^{2}$-trace is well defined
\begin{equation}
\textrm{Tr}\,\exp\left(-t \mathscr{L}\right)=\int_{M}d\textrm{vol}\;\textrm{tr}_{\mathcal{V}}U(t|x,x)\;,
\end{equation}
where $\textrm{tr}_{\mathcal{V}}$ represents the trace over the vector bundle indexes.
By using the definition of the $\mathscr{L}^{2}$-trace and the explicit expression (\ref{intro80}) for
$U(t|x,x')$, it is easy to realize that the heat trace, in (\ref{intro9}), is equal to the trace of
the heat semigroup
\begin{equation}
\Theta(t)=\textrm{Tr}\,\exp\left(-t \mathscr{L}\right)\;.
\end{equation}
It is completely clear, now, that the knowledge of the trace of the heat kernel is equivalent
to the knowledge of the one-loop effective action.

\subsection{Asymptotic Expansion of the Heat Kernel}

As we mentioned earlier, the heat kernel, and therefore its trace, cannot be evaluated
explicitly in many cases of interest. For this reason some approximation
schemes have been developed. 
It is important, at this point, to briefly introduce
some two-point geometric quantities that will be used throughout the Dissertation. A more complete
introduction to this subject can be found in \cite{dewitt03,synge60}.


Let us fix a point, say
$x^{\prime}$, on the manifold $M$ and consider a sufficiently small
neighborhood of $x^{\prime}$, say a geodesic ball with a radius smaller than
the injectivity radius of the manifold. Then, it can be proved that there exists a unique geodesic
that connects every point $x$ to the point $x^{\prime}$.
In what follows we will restrict ourselves to this neighborhood.
In order to avoid a cumbersome notation, we will denote by \emph{Latin letters} tensor indices
associated to the point $x$ and by \emph{Greek letters} tensor indices
associated to the point $x^{\prime}$. Of course, the indices associated with
the point $x$ (resp. $x^{\prime}$) are raised and lowered with the metric at
$x$ (resp. $x^{\prime}$). Also, we will denote by $\nabla_a$ (resp.
$\nabla'_\mu$) the covariant derivative with respect to $x$ (resp. $x'$). We will
use the standard notation of square brackets to denote the coincidence limit of two-point
functions, more precisely, for any functions of $x$ and $x^{\prime}$ we define
\begin{equation}
\left[f\right](x)\equiv\lim_{x\to x^{\prime}}f(x,x^{\prime})\;.
\end{equation}

The world function $\sigma(x,x^{\prime})$
is defined as one half of the square of the length of the geodesic between the
points $x$ and $x^{\prime}$. It satisfies the equation \cite{avramidi91,avramidi00,dewitt03,dewitt65}
\be\label{newwilly20}
\sigma=\frac{1}{2}u^a u_a=\frac{1}{2}u_\mu u^\mu\,,
\ee
where
\begin{equation}
\label{34}
u_{a}=\nabla_{a}\sigma,
\qquad
u_{\mu}=\nabla^{\prime}_{\mu}\sigma\;.
\end{equation}
The variables $u^\mu$ are nothing but the normal coordinates with the origin at the point $x'$.
Next, one defines the tensors which are the second derivative of the world function \cite{avramidi91,avramidi00}
\begin{equation}
\label{3c}
\eta^{\mu}{}_{b}=\nabla_{b}\nabla^{\prime\mu}\sigma\;,\qquad \xi^{a}{}_{b}=\nabla^{a}\nabla_{b}\sigma\;,
\end{equation}
and the tensor
\begin{equation}\label{newwilly1}
X^{\mu\nu}=\eta^{\mu}{}_{a}\eta^{\nu a}\;.
\end{equation}
In particular, one needs the tensor $\gamma^a{}_\mu$ inverse to $\eta^\mu{}_a$ defined by
\be
\gamma^a{}_\mu\eta^\mu{}_b=\delta^a_b\,,\qquad
\eta^\mu{}_b\gamma^b{}_\nu=\delta^\mu_\nu\,.
\ee
The two-point quantities defined above satisfy the following equations \cite{avramidi91,avramidi00,dewitt03,dewitt65}
\begin{equation}
\label{6b}
\xi^{a}{}_{b}u^{b}
=u^{a}\;,\qquad\eta^{\mu}{}_{a}u^{a}=u^{\mu}\;,
\qquad
\eta^{\mu}{}_{a}u_{\mu}=u_{a}\;,
\end{equation}
with the boundary conditions
\begin{equation}
\label{65}
[\sigma]=[u^{a}]=[u^{\mu}]=0\;,
\end{equation}
\begin{equation}
\label{6ca}
[\xi^{a}{}_{b}]
=\delta^{a}{}_{b}\;,
\qquad[\eta^{\mu}{}_{a}]=-\delta^{\mu}{}_{a}\;.
\end{equation}

Another useful two-point quantity is the Van Vleck-Morette determinant which is defined as
\begin{equation}
\label{3e}
\Delta(x,x^{\prime})
=g^{-1/2}(x)\det[-\nabla_{a}\nabla^{\prime}_{\nu}
\sigma(x,x^{\prime})]g^{-1/2}(x^{\prime})\;.
\end{equation}
This quantity should not be confused with the Laplacian
$\Delta=g^{ab}\nabla_a\nabla_b$. Usually, the meaning of $\Delta$ will
be clear from the context. Following \cite{avramidi91,avramidi00}, we find it convenient to parameterize it by
\begin{equation}\label{36b}
\Delta(x,x^{\prime})=\exp[{2\zeta(x,x')}]\;.
\end{equation}

Next, one defines new derivative operators by \cite{avramidi91,avramidi00}
\begin{equation}
\bar{\nabla}_{\mu}=\gamma^{a}{}_{\mu}\nabla_{a}\;.
\end{equation}
These operators commute when acting on objects that have been
parallel transported to the point $x^{\prime}$
(in other words the objects that do not have Latin indices).
In fact, when acting on such objects these operators are just partial
derivatives with respect to normal coordinate $u$
\be
\bar\nabla_\mu=\frac{\partial}{\partial u^\mu}\,.
\ee

Next, the parallel displacement operator $\mathcal{P}(x,x^{\prime})$ of sections
of the vector bundle ${\cal V}$ along the geodesic from the point
$x^{\prime}$ to the point $x$ is defined as the solution of the
equation \cite{avramidi91,avramidi00,dewitt03,dewitt65}
\begin{equation}\label{intro8}
u^{a}\nabla_{a}\mathcal{P}(x,x^{\prime})=0\;,
\end{equation}
with the initial condition
\begin{equation}\label{36c}
\left[\mathcal{P}\right]=\mathbb{I}\;.
\end{equation}
It is not difficult to show that the parallel displacement operator satisfies the equation
\begin{equation}
\label{6we}
[\nabla_{a},\nabla_{b}]\mathcal{P}
=\mathcal{R}_{ab}\mathcal{P}\;,
\end{equation}
where $\mathcal{R}_{ab}$ represents the curvature of the connection on the vector bundle ${\cal V}$.
Finally, one defines the two-point quantity which is the derivative of the parallel transport operator \cite{avramidi91,avramidi00}
\begin{equation}\label{310}
\mathscr{A}_{\mu}=\mathcal{P}^{-1}\bar{\nabla}_{\mu}\mathcal{P}\;.
\end{equation}

It is important, for future reference, to present here the coincidence limits of higher
derivatives of the two-point functions introduced above. They are expressed in terms of the
curvature and, in particular, the ones that we will need are found to be
\cite{avramidi00,avramidi91,dewitt03,dewitt65}
\bea
\label{627} [\zeta_{;\;a}]&=&0\\{}
[\zeta_{;\;bc}]&=&\frac{1}{6}R_{bc}\;,\\
\left[\eta^{\mu}{}_{a;\;b}\right]&=&0\;,\\
\left[\eta^{\mu}{}_{c;\;ab}\right]
&=&-\frac{1}{3}(R^{\mu}{}_{\nu\alpha\beta}
+R^{\mu}{}_{\alpha\nu\beta})\;,\\{}
[\mathscr{A}_{\nu}]&=&0\\{}
[\mathscr{A}_{\nu;\;b}]&=&-\frac{1}{2}\mathcal{R}_{\nu b}\;,
\eea
where $R^{\mu}_{\;\;\nu\rho\sigma}$ is the Riemann tensor and
$R_{\mu\nu}=R^\alpha{}_{\mu\alpha\nu}$ is the Ricci tensor.

We will briefly present, here, the role of the previous geometric quantities in the evaluation of
the heat kernel and, therefore, of its trace. We consider a second order partial differential operator
of Laplace type, $\mathscr{L}=-\Delta+Q$. The heat kernel cannot be computed exactly in this case,
however, one can find an asymptotic expansion for small time following \cite{avramidi91,avramidi00,dewitt65}. We know that the heat kernel $U(t|x,x^{\prime})$
satisfies the equation (\ref{intro4}) with the initial condition (\ref{intro5}). To start, we
consider the following ansatz,
\begin{equation}
U(t|x,x^{\prime})=A(t;x,x^{\prime})\exp\left\{-\frac{S(x,x^{\prime})}{t}\right\}\;,
\end{equation}
where $S$ represents the action (\ref{intro7}).
By substituting the ansatz in the heat equation it is not difficult to show that
\begin{equation}\label{intro6}
\partial_{t}A+\frac{1}{t^{2}}\left[S-g^{ab}(\nabla_{a}S)(\nabla_{b}S)\right]A+\frac{1}{t}\left[2g^{ab}(\nabla_{a}S)\nabla_{b}-\Delta S\right]A+\left(-\Delta+Q\right)A=0\;.
\end{equation}
Since we are considering the asymptotic expansion for small $t$, we set the coefficient of $t^{-2}$ in
the previous equation to zero, by doing so we obtain
\begin{equation}
g^{ab}(\nabla_{a}S)(\nabla_{b}S)=S\;.
\end{equation}
These are nothing but the Hamilton-Jacobi equations for the action $S$. By using (\ref{newwilly20}), it can be proved that the solution
of this equation is
\begin{equation}
S(x,x^{\prime})=\frac{\sigma(x,x^{\prime})}{2}\;.
\end{equation}
By substituting the explicit solution for $S$ in (\ref{intro6}), we obtain
\begin{equation}
\partial_{t}A+\frac{1}{t}\left[u^{a}\nabla_{a}+\frac{1}{2}\xi_{a}{}^{a}\right]A=-\left(-\Delta+Q\right)A\;.
\end{equation}

At this point we introduce an ansatz for the function $A(t;x,x^{\prime})$. By careful inspection of the previous
equation it is useful to write
\begin{equation}
A(t;x,x^{\prime})=(4\pi t)^{-n/2}\Delta^{\frac{1}{2}}(x,x^{\prime}){\cal P}(x,x^{\prime})\Omega(t;x,x^{\prime})\;,
\end{equation}
where ${\cal P}$ is the parallel transport operator defined in (\ref{intro8}) and the choice of the factor $(4\pi t)^{-n/2}$ will assure that the solution satisfies the
initial condition (\ref{intro5}).
By utilizing this last ansatz we obtain
\begin{equation}
\partial_{t}\Omega+\frac{1}{2t}\left[\Delta^{-1}u^{a}(\nabla_{a}\Delta)+\xi_{a}{}^{a}-n\right]\Omega+\frac{1}{t}u^{a}\nabla_{a}\Omega
=-\Delta^{-1/2}{\cal P}^{-1}(-\Delta+Q){\cal P}\Delta^{1/2}\Omega\;.
\end{equation}
At this point, it can be proved \cite{avramidi91,avramidi00,dewitt65} that the Van Vleck-Morette determinant (\ref{3e})
satisfies the equation
\begin{equation}
\Delta^{-1}u^{a}\nabla_{a}\Delta+\xi_{a}{}^{a}-n=0\;,
\end{equation}
By using this remark
the equation for $\Omega$ becomes
\begin{equation}
\left[\partial_{t}+\frac{1}{t}u^{a}\nabla_{a}+\Delta^{-1/2}{\cal P}^{-1}(-\Delta+Q){\cal P}\Delta^{1/2}\right]\Omega=0\;.
\end{equation}

Since we want an asymptotic expansion as $t\to 0$ we write $\Omega$ as asymptotic series
\begin{equation}
\Omega(t;x,x^{\prime})\sim\sum_{k=0}^{\infty}a_{k}(x,x^{\prime})t^{k}\;.
\end{equation}
By substituting this expression in the equation above we finally obtain an expression
for the asymptotic expansion of the heat kernel
\begin{equation}\label{newwilly7}
U(t|x,x^{\prime})\sim(4\pi t)^{-n/2}\Delta^{\frac{1}{2}}(x,x^{\prime}){\cal P}(x,x^{\prime})\exp\left\{-\frac{\sigma(x,x^{\prime})}{2t}\right\}\sum_{k=0}^{\infty}a_{k}(x,x^{\prime})t^{k}\;,
\end{equation}
where the heat kernel coefficients $a_{k}$ satisfy the DeWitt recurrence relation
\begin{equation}
[(k+1)+u^{a}\nabla_{a}]a_{k+1}(x,x^{\prime})=\Delta^{-1/2}{\cal P}^{-1}(-\Delta+Q){\cal P}\Delta^{1/2}a_{k}(x,x^{\prime})\;,
\end{equation}
with the initial condition
\begin{equation}
a_{0}(x,x^{\prime})=1\;.
\end{equation}
This expansion will be generalized in Chapter 3 in order to include, in a non-perturbative way,
the electromagnetic field.

The lower order diagonal heat kernel coefficients are well known and have the form
\cite{gilkey95,avramidi91,avramidi00}
\bea\label{3120igaw}
a^{\rm diag}_0&=&1\,,
\\[10pt]
a^{\rm diag}_1&=&\frac{1}{6}R\,,\label{3120igaw1}
\\[10pt]
a^{\rm diag}_2&=&
{1\over 30}\Delta R
+{1\over 72}R^2
-{1\over 180} R_{\mu\nu}R^{\mu\nu}
+{{1}\over {180}}R_{\alpha\beta\mu\nu}R^{\alpha\beta\mu\nu}
\nonumber\\
&&
+{1\over 12}{\cal R}_{\mu\nu}{\cal R}^{\mu\nu}
\,,
\label{3120iga}
\eea
where $a^{\rm diag}_{k}=a_{k}(x,x)$.
To avoid confusion we should stress that the normalization of the coefficients
$a_k$ differs from the papers \cite{avramidi91,avramidi99,avramidi00}.

It can be easily proven, by taking the trace of both sides of (\ref{newwilly7}), that the heat trace,
which is of primary interest here, has the following asymptotic expansion
\begin{equation}\label{newwilly2}
{\rm Tr}\,\exp\{-t\mathscr{L}\}\sim(4\pi t)^{-n/2}\sum_{k=0}^{\infty}t^{k}A_{k}\;,
\end{equation}
where $A_{k}$ are global heat kernel coefficients
\begin{equation}\label{newwilly3}
A_{k}=\int_{M}d{\rm vol}\;{\rm tr}_{{\cal V}}\;a^{\rm diag}_{k}\;.
\end{equation}
One should point out, here, that the heat trace for a non-Laplace type operator
has the same asymptotic expansion as $t\to 0$ \cite{gilkey95}.

In the next section we will describe two methods for the evaluation of the
heat kernel asymptotic expansion: the covariant Taylor expansion method which we use
in the computation of the non-perturbative heat kernel asymptotics on
homogeneous Abelian bundles in Chapters 3 and 4, and the covariant Fourier transform method that
we use for the evaluation of the action in Spectral Matrix Gravity in Chapter 6.

\subsection{Covariant Taylor Expansion and Covariant Fourier Transform}

In this section we will briefly discuss two methods for evaluating the coefficients of the
heat kernel asymptotic expansion which are different from the DeWitt method presented
in the previous section.
In the previous section we described some two-point geometric functions that are widely
used for heat kernel calculations on Riemannian manifolds. Since we ultimately
want to find an expansion for the heat kernel and its trace, we need to develop
an expansion for the two-point quantities. In the first part of this section we will briefly describe the Taylor
expansion of the two-point quantities following \cite{avramidi91,avramidi00}..
A more detailed discussion on this subject can be found in \cite{avramidi91,avramidi00}.

Let us consider, as before, two neighboring points $x$ and $x^{\prime}$ connected by a unique
geodesic. The derivatives, $u^{\mu}$, of the world function $\sigma(x,x^{\prime})$ form a set of
normal coordinates for the neighborhood, $\mathcal{U}$, under consideration. Our aim is to
find an expansion in terms of the normal coordinates for functions defined on $\mathcal{U}$.
Since the language that we use is covariant, we need an expansion which is independent on the
system of coordinates that we choose. It is evident, from the discussion above, that the
quantity $u^{\mu}$ is a vector at the point $x^{\prime}$ and a scalar at the point $x$.
Since scalars are the simplest invariant quantities, one can develop the Taylor expansion for a scalar function $f$
at the point $x$ by using the coordinate $u^{\mu}$. By parameterizing the geodesic
between $x$ and $x^{\prime}$ with an affine parameter one can show that for a
scalar function we obtain the expansion \cite{avramidi00,avramidi91}
\be
f(x)=\sum_{n=0}^{\infty}\frac{(-1)^{n}}{n!}u^{\mu_{1}}\cdots u^{\mu_{n}}\left[g_{\mu_{1}}{}^{a_{1}}\cdots g_{\mu_{n}}{}^{a_{n}}\nabla_{a_{1}}\cdots\nabla_{a_{n}}f\right]\;,
\ee
where $g_{\mu}{}^{a}$ is the parallel transport of covectors from the point $x$ to the point $x^{\prime}$.
By multiplying this expression by the parallel transport operators (\ref{intro8}), as many as needed, we can
find an expression for the covariant Taylor expansion of arbitrary tensors. In particular,
we can find the covariant Taylor expansion for the two-point quantities that we need.
It can be shown that \cite{avramidi00,avramidi91}
\bea\label{newwilly4}
\eta^{\mu}{}_{b}&=&g_{b}{}^{\nu}\Bigg[-\frac{1}{3}R^{\mu}{}_{\alpha\nu\beta}u^{\alpha}u^{\beta}-\frac{1}{2}\nabla_{\alpha}R^{\mu}{}_{\beta\nu\gamma}u^{\alpha}u^{\beta}u^{\gamma}
-\frac{3}{5}\nabla_{\alpha}\nabla_{\beta}R^{\mu}{}_{\gamma\nu\delta}u^{\alpha}u^{\beta}u^{\gamma}u^{\delta}
\nonumber\\
&-&\frac{7}{15}R^{\mu}{}_{\alpha\rho\beta}R^{\rho}{}_{\gamma\nu\delta}u^{\alpha}u^{\beta}u^{\gamma}u^{\delta}+O(u^{5})\Bigg]\;,
\eea
\begin{eqnarray}\label{38}
\zeta&=&
\frac{1}{12}R_{\alpha\beta}u^{\alpha}u^{\beta}
-\frac{1}{24}\nabla_{\alpha}R_{\beta\gamma}u^{\alpha}u^{\beta}u^{\gamma}
+\frac{1}{80}\nabla_{\alpha}\nabla_{\beta}
R_{\gamma\delta}u^{\alpha}u^{\beta}u^{\gamma}u^{\delta}
\nonumber\\
&+&
\frac{1}{360}R_{\mu\alpha\nu\beta}
R^{\mu}{}_{\gamma}{}^{\nu}{}_{\delta}u^{\alpha}u^{\beta}u^{\gamma}u^{\delta}
+O(u^{5})\;,
\end{eqnarray}
\begin{eqnarray}\label{38a}
\Delta^{1/2}&=&1
+\frac{1}{12}R_{\alpha\beta}u^{\alpha}u^{\beta}
-\frac{1}{24}\nabla_{\alpha}R_{\beta\gamma}u^{\alpha}u^{\beta}u^{\gamma}
+\frac{1}{80}\nabla_{\alpha}\nabla_{\beta}
R_{\gamma\delta}u^{\alpha}u^{\beta}u^{\gamma}u^{\delta}
\nonumber\\
&+&
\frac{1}{288}R_{\alpha\beta}R_{\gamma\delta}u^{\alpha}u^{\beta}u^{\gamma}u^{\delta}
+\frac{1}{360}R_{\mu\alpha\nu\beta}
R^{\mu}{}_{\gamma}{}^{\nu}{}_{\delta}u^{\alpha}u^{\beta}u^{\gamma}u^{\delta}
+O(u^{5})\;,
\end{eqnarray}
\begin{eqnarray}\label{37}
X^{\mu\nu}&=&
g^{\mu\nu}
+\frac{1}{3}R^{\mu}{}_{\alpha}{}^{\nu}{}_{\beta}u^{\alpha}u^{\beta}
-\frac{1}{6}\nabla_{\alpha}
R^{\mu}{}_{\beta}{}^{\nu}{}_{\gamma}u^{\alpha}u^{\beta}u^{\gamma}
+\frac{1}{20}\nabla_{\alpha}\nabla_{\beta}
R^{\mu}{}_{\gamma}{}^{\nu}{}_{\delta}
u^{\alpha}u^{\beta}u^{\gamma}u^{\delta}
\nonumber\\
&+&
{1\over 15}
R^{\mu}{}_{\alpha\lambda\beta}
R^\lambda{}_{\gamma}{}^{\nu}{}_{\delta}u^{\alpha}u^{\beta}u^{\gamma}u^{\delta}
+O(u^{5})\;,
\end{eqnarray}
\begin{eqnarray}
\mathscr{A}_{\mu}
&=&
-{1\over 2}\mathcal{R}_{\mu\alpha}u^\alpha
+\frac{1}{3}\nabla_{\alpha}\mathcal{R}_{\mu\beta}u^{\alpha}u^{\beta}
+
{1\over 24}R_{\mu\alpha\nu\beta}
\mathcal{R}^\nu{}_\gamma u^\alpha u^\beta u^\gamma
\nonumber\\
&-&\frac{1}{8}\nabla_{\alpha}\nabla_{\beta}
\mathcal{R}_{\mu\gamma}u^{\alpha}u^{\beta}u^{\gamma}
+O(u^6)\;,
\end{eqnarray}
where the above quantities have been defined, respectively, in (\ref{3c}), (\ref{36b}), (\ref{3e}), (\ref{newwilly1}) and (\ref{310}).
We would like to stress that all coefficients of such expansions are evaluated
at the point $x^{\prime}$.

\bigskip

Here, we will briefly describe the covariant Fourier transform which we will use in Chapter 6
to evaluate the heat kernel asymptotic coefficients for a non-Laplace type operator.
The Fourier integral can be defined by using the two-point functions that we introduced
earlier. Let $f$ be a function defined on $\mathcal{U}$. Its covariant
Fourier transform is defined as follows \cite{avramidi91,avramidi00}
\begin{equation}\label{newwilly6}
f(x)=\int_{\mathbb{R}^{n}}\frac{dk}{(2\pi)^{n}}g^{-1/2}(x^{\prime})e^{-ik_{\mu}u^{\mu}}\hat{f}(k;x^{\prime})\;,
\end{equation}
where $k_{\mu}$ represents the coordinate in the momentum space. The inverse Fourier transform is written as
\begin{equation}
\hat{f}(k;x^{\prime})=\int_{M}du\,g^{1/2}(x^{\prime})e^{ik_{\mu}u^{\mu}}f(x)\;.
\end{equation}
We can transform the last integral from normal coordinates to local coordinates.
It is not difficult to show, by making a change of variables, that
\be
du=g^{1/2}(x)g^{-1/2}(x^{\prime})\Delta(x,x^{\prime})dx\;.
\ee
By using this formula one gets \cite{avramidi91,avramidi00}
\be\label{newwilly5}
\hat{f}(k;x^{\prime})=\int_{M}dx\,\Delta(x,x^{\prime})g^{1/2}(x)e^{ik_{\mu}u^{\mu}}f(x)\;.
\ee
We would like to mention, here, that by multiplying this expression by the parallel transport operators, as many as needed, we can
find an expression for the covariant Fourier transform of arbitrary tensors.
At this point it is convenient to derive a representation
of the covariant Dirac delta function in terms of the Fourier integral. It is not
difficult to show, by substituting (\ref{newwilly5}) into (\ref{newwilly6}), that one obtains
\begin{equation}
\label{612}
\delta(x,x^{\prime})=
\Delta^{1/2}(x,x^{\prime})
\int_{\mathbb{R}^{n}}\frac{dk}{(2\pi)^{n}}\;\exp{\{\imath
k_{\mu}u^{\mu}\}}\;.
\end{equation}
The covariant Fourier transform represents another important tool for the
evaluation of the heat kernel asymptotic expansion which we will mainly use in Chapter 6.

\subsection{Perturbation Theory for the Heat Semigroup}

The operators for which
we want to evaluate the heat kernel asymptotic expansion are known in terms
of power series in a small formal parameter $\varepsilon$, namely
\be
\mathscr{L}=\mathscr{L}_{0}+\sum_{k=1}^{\infty}\mathscr{L}_{k}\;,
\ee
where $\mathscr{L}_{0}$ represents the unperturbed part and $\mathscr{L}_{k}$ is of order $\varepsilon^{k}$.

In order to evaluate the heat kernel for the operator $\mathscr{L}$ we need to compute,
first, the heat semigroup for $\mathscr{L}$. Since $\mathscr{L}$ is given in terms
of a perturbative series, we are faced with the problem of finding an expansion
for the exponent of two non-commuting operators.
The expansion for the exponent of two non-commuting operators, known in the literature,
is called Volterra series.

Let $A$ and $B$ be arbitrary, non-commuting operators, then the following can be proved \cite{avramidi07,taylor96}
\begin{equation}
\label{635}
e^{A+B}=e^{A}+\sum_{k=1}^{\infty}\int\limits_{0}^{1}d\tau_{k}\int\limits_{0}^{\tau_{k}}d\tau_{k-1}\cdots\int\limits_{0}^{\tau_{2}}d\tau_{1}e^{(1-\tau_{k})A}Be^{(\tau_{k}-\tau_{k-1})A}\cdots
 e^{(\tau_{2}-\tau_{1})A}Be^{\tau_{1}A}\;.
\end{equation}
By considering only few of the low order terms (the ones we will use in this work)
we write
\begin{eqnarray}
e^{A+B}&=&e^{A}+\int_{0}^{1}d\tau_{1}e^{(1-\tau_{1})A}Be^{\tau_{1}A}\nonumber\\
&+&\int_{0}^{1}d\tau_{2}\int_{0}^{\tau_{2}}d\tau_{1}e^{(1-\tau_{2})A}Be^{(\tau_{2}-\tau_{1})A}Be^{\tau_{1}A}+\cdots\;.
\end{eqnarray}

This series can be also written in another way. Let us suppose that
the operator $A$ is of zeroth order (unperturbed) and the operator $B$ represents higher orders
in the perturbation theory. The Volterra series can be written
in terms of a specific operator $T$ acting on the unperturbed semigroup as follows \cite{avramidi07,taylor96}
\be
\exp(A+B)=T\exp A\,,
\ee
where
\be
T=I
+\sum_{k=1}^{\infty}\int\limits_{0}^{1}d\tau_{k}
\int\limits_{0}^{\tau_{k}}d\tau_{k-1}\cdots
\int\limits_{0}^{\tau_{2}}d\tau_{1}\;
\tilde B(\tau_1)\tilde B(\tau_2)
\cdots
\tilde B(\tau_k)
\ee
and
\be
\tilde B(\tau)=e^{\tau A}Be^{-\tau A}\;.
\ee
This particular form of the Volterra series will be utilized in Chapter 3.


\section{Mathematical Framework in Matrix Gravity}

\subsection{Motivation and Discussion}

In the rest of the present Chapter, we will describe the motivations and the mathematical framework of the second
major topic in this Dissertation, namely Matrix Gravity.
The main idea in Matrix Gravity is to describe the
gravitational field as a matrix-valued symmetric two-tensor field.
It is well known that General Relativity is nothing
but the dynamical theory of the metric 2-tensor field which is,
basically, an isomorphism between tangent and cotangent bundles. The
dynamics of the metric is described by the Hilbert-Einstein action,
\begin{equation}\label{newwilly10}
S_{{\rm HE}}=\frac{1}{16\pi G}\int_{M}dx\,g^{1/2}(R-2\Lambda)\;,
\end{equation}
where $G$ is the Newtonian gravitational constant, $\Lambda$ is the
cosmological constant and $R$ is the scalar curvature.

In Matrix Gravity, the metric 2-tensor field $g^{\mu\nu}$, is replaced by a
endomo-rphism-valued 2-tensor field $a^{\mu\nu}$ which represents
an isomorphism of more general bundles over the manifold $M$. The
main idea here, similar to General Relativity, is to develop a
dynamical theory of this endomorphism-valued 2-tensor field
$a^{\mu\nu}$. This generality brings a much richer structure and
content to the model.

We would like to stress, at this point, that the dynamical equations
that we will derive in this Dissertation for Matrix Gravity,
are \emph{classical} and therefore they
should be studied from the classical point of view.

The motivation for such a deformation of General Relativity is explained
in detail in \cite{avramidi04}.
The very basic physical concepts are the notions of event and the spacetime.
An event is a collection of variables that specifies the location of a point
in space at a certain time. To assign a time to each point in space one needs
to place clocks at every point (say on a lattice in space) and to
synchronize these clocks. Once the position of the clocks is fixed the only way
to synchronize the clocks is by transmitting the information from a fixed point
(say, the origin of the coordinate system in space) to all other points. This can be
done by sending a signal through space from one point to another. Therefore,
the synchronization procedure depends on the propagation of the signal through
space, and, as a result, on the properties of the space it propagates through,
in particular, on the presence of any physical background fields in space.
The propagation of signals is described by a wave equation (a hyperbolic
partial differential equation of second order). Therefore, the propagation
of a signal depends on the matrix of the coefficients (a symmetric 2-tensor) $g^{\mu\nu}(x)$ of the second
derivatives in the wave equation which must be non-degenerate and
have the signature $(-+\dots+)$. This matrix
can be interpreted as a pseudo-Riemannian metric, which defines the geodesic flow,
the curvature and the Einstein equations of General Relativity
(for more details, see \cite{avramidi04}).

The picture described above applies to the propagation of light, which is
described by a single wave equation. However, now we know that at microscopic scales
there are other fields that could be used to transmit a signal.
In particular, the propagation of a multiplet of $N$ gauge fields is described
not by a single wave equation but by a {\it hyperbolic system}
of second order partial differential equations.
We would like to stress, here, that the number $N$ of gauge fields that one should
use in order to describe the gravitational field is not known at this time.
It is possible that a quantized version of this theory could shed light on the precise
number of gauge fields to use, however this problem needs further research.
For simplicity of calculations, in Chapter 7 we analyze a model of Matrix Gravity where
two gauge fields are taken into account. For now, we will leave $N$ arbitrary.

The coefficients at the second derivatives of such a system are not given by just a 2-tensor like $g^{\mu\nu}(x)$
but by a $N\times N$ {\it matrix-valued} symmetric $2$-tensor $a^{\mu\nu}(x)$ as in (\ref{newwilly11}).
If $a^{\mu\nu}$ does not factorize as $a^{\mu\nu}\ne \Xi g^{\mu\nu}$, where $\Xi$
is some non-degenerate matrix, then there is no geometric
interpretation of this hyperbolic system in terms of a single Riemannian metric.
Instead, one obtains a new kind of geometry that is called \emph{Matrix Geometry}, which is
equivalent to a collection of Finsler geometries. In this theory,
instead of a single Riemannian geodesic flow, there is a system of $N$ Finsler
geodesic flows. Moreover, a gravitating particle is described not by one mass
parameter but by $N$ mass parameters (which could be different).
The general idea is similar to the concept of colors in quantum chromodynamics.
In Matrix Gravity each particle is considered to be composed of $N$ different ``colors''
each of them described by a different mass parameter. Each of these colors follows its own
Finsler geodesic. In this sense, the introduction of a matrix-valued metric $a^{\mu\nu}$ leads to
the splitting of a single Riemannian geodesic to a system of $N$ close Finsler geodesics.
We argue that at microscopic distances, and high energies, a single Riemannian geodesic
is described by a system of $N$ Finsler geodesics.
As we will see in Chapter 7 the equations of the geodesics in Matrix Gravity are
non-linear with respect to the mass parameters $\mu_{i}$. This makes it not possible to
write the geodesics in Matrix Gravity as a weighted sum (with weight $\mu_{i}$)
of all the Finsler geodesics for each mass parameter. However, if one consider the following splitting
$\mu_{i}=1/N+\alpha_{i}$ where $\alpha_{i}$ are assumed to be small, one could linearize
the equations for the geodesics and obtain, in the first order, a description
of the geodesics in Matrix Gravity as weighted sum (with weights $\alpha_{i}$) of the
single Finsler geodesics.

Notice that because the tensor $a^{\mu\nu}$ is matrix-valued, various components of this
tensor do not commute, that is, $[a^{\mu\nu}, a^{\alpha\beta}]\ne 0$. In this sense,
such geometry may be also called {\it non-commutative Riemannian geometry}.
In the commutative limit, $a^{\mu\nu}\to g^{\mu\nu}$, and the
standard Riemannian geometry with all its ingredients is recovered. Only the
total mass of a gravitating particle is observed.
For more details and discussions see
\cite{avramidi04,avramidi04b}.

We would like to mention, at this point, that Matrix Gravity contains two important
physical consequences. The first is the presence, in the theory, of a \emph{new} non-geodesic
acceleration. In General Relativity, the motion of a test particle in the gravitational
field is described by the equation
\begin{equation}
\frac{d^{2}x^{\mu}}{dt^{2}}+\Gamma^{\mu}{}_{\alpha\beta}\dot{x}^{\alpha}\dot{x}^{\beta}=0\;,
\end{equation}
where $\dot{x}$ represents the tangent vector to the geodesic and $\Gamma^{\mu}{}_{\alpha\beta}$ are
the Christoffel symbols. This equation describes the motion of a test particle free from external forces.
As we will see in Chapter 7, in Matrix Gravity the equation for the geodesics becomes
the following
\begin{equation}
\frac{d^{2}x^{\mu}}{dt^{2}}+\Gamma^{\mu}{}_{\alpha\beta}\dot{x}^{\alpha}\dot{x}^{\beta}=A^{\mu}_{\rm anom}(x,\dot{x})\;.
\end{equation}
In this equation a non-geodesic acceleration is present which is written only in terms
of the non-commutative part of the metric. In other words, the test particles in Matrix Gravity,
do not follow any Riemannian geodesic.
The second important physical consequence of this model is the violation of the equivalence principle.
In General Relativity, test particles move along specific geodesics of a Riemannian metric
independently of their masses. In Matrix Gravity, instead, test particles exhibit a non-geodesic motion $A^{\mu}_{\rm anom}$
which depends on the different mass parameters. Therefore, test particles that are described by different mass parameters
will follow different trajectories in the spacetime. This is the origin of the violation of the
equivalence principle.

\subsubsection{Main Differences Between Matrix Gravity and Non-commutative Gravity}

It is important to stress that this approach for deforming
General Relativity is different from the ones proposed in the
framework of non-commutative geometry \cite{fulling89,konechny02,madore94,harikumar06}, where the
coordinates do not commute and the standard product between
functions is replaced by the Moyal product \cite{madore99}.
In flat space one usually introduces
non-commutative coordinates satisfying the commutation relations
\begin{equation}
[x^{\mu},x^{\nu}]=\theta^{\mu\nu}\;.
\end{equation}
Here, $\theta^{\mu\nu}$ is a real constant anti-symmetric matrix,
and one replaces the standard algebra of functions with the non-commutative
algebra with the Moyal star product
\begin{equation}
f(x)\;\star\;g(x)=\exp\left(\frac{i}{2}\theta^{\mu\nu}\frac{\partial}{\partial y^{\mu}}\frac{\partial}{\partial z^{\nu}}\right)f(x+y)g(x+z)\Big|_{y=z=0}\;.
\end{equation}
An extensive review of different realizations of gravity in the framework
of non-commutative geometry, especially in connection with string theory, can be found in \cite{szabo06}.
An extension of the star product and noncommutativity from flat to curved
spacetime can be found in \cite{aschieri05,harikumar06}.

We list below the most relevant differences
between the two approaches;
a more detailed and extensive discussion can be found in
\cite{avramidi04,avramidi04b,avramidi08d,fucci08}.
The biggest problem with the curved manifolds is the nature of the object
$\theta^{\mu\nu}$. All these models are defined, strictly speaking, only in
perturbation theory in the deformation parameter. That is, one takes
$\theta^{\mu\nu}$ as a {\it formal} parameter and considers {\it formal power series} in $\theta^{\mu\nu}$.
In the approach of \cite{avramidi04,avramidi04b} to Matrix Gravity the deformation parameters are
not formal and the theory is defined for all finite
values of the deformation
parameter.

In the standard non-commutative approach the coordinates themselves are non-commutative. This condition
raises the questions of whether the spacetime has the structure of a manifold, and how one can
define analysis on such spaces. Moreover, one needs a way to relate the non-commutative coordinates
with the usual (commutative) coordinates.
In Matrix Gravity one does not have non-com\-mu\-ta\-tive coordinates.
The spacetime, here, is a proper {\it smooth manifold} with the standard analysis defined on it.

In non-commutative geometry approach the deformation parameter $\theta^{\mu\nu}$ is {\it non-dynamical},
therefore there are no dynamical equations for it. This poses the question of what kind of physical,
or mathematical, conditions can be used in order to determine it. In addition, in many models of non-commutative gravity
(as in \cite{aschieri05}), $\theta^{\mu\nu}$ is a
\emph{non-tensorial} object
which makes it dependent on the choice of the system of coordinates.
This feature leads the theory to be not invariant under the usual group of diffeomorphisms.
In \cite{aschieri05} the authors construct all the relevant geometric
quantities (such as connection, curvature, etc.) in terms of the non-commutative
deformation parameter. In this framework, they obtain an
expansion of these quantities and of
the action up to second order. In their approach the object
$\theta^{\mu\nu}$ is a constant anti-symmetric matrix ({\it not a tensor}).
This violates
the usual diffeomorphism invariance, Lorentz invariance, etc.
for which there exist very strict experimental bounds.
The main result in \cite{aschieri05} is the derivation of the deformed Einstein equations.
The zero-order part (Einstein) is diffeomorphism-invariant, and the
corrections (quadratic in theta) are not. Therefore the theory contains some preferred
system of coordinates and its whole content depends on it.
Of course, the theory needs
to justify the choice of such system of coordinates.

In the approach of \cite{avramidi04,avramidi04b} there is no need to introduce any non-tensorial objects. As a result the theory is
{\it diffeomorphism-invariant}. So, there are no problems related to the violation
of Lorenz invariance, etc. and there are no preferred systems of coordinates. Moreover,
the non-commutative part of the metric in this approach is {\it dynamical}.
There are non-commutative Einstein equations for it. The goal of this chapter is,
in particular, to derive these dynamical equations in the perturbation
theory.

In \cite{harikumar06} the authors
assume that $\theta^{\mu\nu}$
is a covariantly constant tensor.
But then, there are strict
algebraic constraints on the Riemann curvature tensor of the commutative
metric (obtained by a commutator of second covariant derivatives).
In Matrix Gravity such algebraic constraints are absent---
the commutative metric is arbitrary.

In the usual approach of non-commutative geometry
(as in, for example, \cite{aschieri05}),
when one defines the affine connection,
the covariant derivative, the curvature and the torsion
the {\it ordering of factors is not unique}.
There is no natural reason why
one should prefer one ordering over the other. That is, the connection
coefficients can be placed on the left, or on the right, (or one
could symmetrize over
these two possibilities) from the object of
differentiation.
Another aspect of the ordering problem is the fact that
there is no unique way to raise and lower indices. One
can act with the metric from the left or from the right.
The approach in \cite{avramidi04b}, instead,
is pretty much {\it unique}. There is no need to define
the affine connection, the covariant derivative, the curvature and the
torsion. There is no ordering problem.

The definition of a ``measure", in standard non-commutative geometry, as a star determinant (as in \cite{aschieri05})
does not
guarantee its positivity. It only guarantees the
positivity in the zero order of the perturbation theory.
In the definition of \cite{avramidi04,avramidi04b}, the measure is positive even in strongly
non-com\-mu\-ta\-tive regime.

Moreover, the Moyal star product is non-local which makes the whole theory
{\it non-local} with possible unitarity problems.
In the approach \cite{avramidi04b} the action functional is
a usual local functional of sigma-model
type (like General Relativity, but with additional
non-commutative degrees of freedom).
There may be problems with the renormalizability
(which requires further study)
but not with unitarity.

One should also mention the relation of Matrix Gravity to so called
``analog models of gravity''. In particular, the analysis in \cite{barcelo02}
is surprisingly similar to the analysis of the papers \cite{avramidi03,avramidi04}.
The authors of \cite{barcelo02}
consider a hyperbolic system of second order partial differential equations,
the corresponding Hamilton-Jacobi equations and the Hamiltonian system
as in \cite{avramidi03,avramidi04}.
In fact, their ${\bf f}^{\mu\nu}$ is equivalent to the matrix-valued
tensor $a^{\mu\nu}$,
However, their goal was very different---they impose the commutativity conditions
on ${\bf f}^{\mu\nu}$ (eq. (44) in \cite{barcelo02}) to enforce a unique effective metric for the
compatibility with the Equivalence Principle.
They barely mention the general geometric interpretation in terms
of Finsler geometries as it ``does not seem to be immediately relevant for either particle physics or gravitation''.
The motivation of the authors of
\cite{barcelo02} is also very different from the approach of \cite{avramidi04,avramidi04b}. Their idea is that gravity
is not fundamental so that the effective metric simply
reflects the properties of an underlying physics (such as fluid mechanics and
condensed matter theory). They just need to have enough fields to be able to parameterize
an arbitrary effective metric. In the approach of Matrix Gravity, the matrix-valued field
$a^{\mu\nu}$ is fundamental; it is: i) non-commutative and ii) dynamical.

The action of Matrix Gravity
can be constructed in two different ways. One
approach, developed in \cite{avramidi03,avramidi04}, called \emph{Matrix General Relativity}, is to try to
extend all standard concepts of differential geometry to the
non-commutative setting and to construct a matrix-valued
connection and a matrix-valued curvature. We will be mainly interested
in this approach in Chapter 5.

The second approach,
developed in \cite{avramidi04b}, called \emph{Spectral Matrix Gravity},
is based on constructing the action form the coefficients of the spectral asymptotics
of a non-Laplace type self-adjoint elliptic partial differential operator $L$ of second
order with a positive definite leading
symbol. We will analyze this particular approach in Chapter 6.

\subsection{Matrix General Relativity}

We will describe, in this section, the construction of
the action for Matrix General Relativity following
\cite{avramidi03,avramidi04}.
The formalism that we are going to
describe is related to the algebra-valued formulation of Mann
\cite{mann84} and Wald \cite{wald87}. In these papers the authors
introduce algebra-valued tensor fields and generalize the
formalism of differential geometry to the algebra-valued case.
More precisely they were studying a consistent theory to describe
the interaction of a collection of massless spin-$2$ fields. The
authors found that in order to have a consistent theory, the
algebra to consider must be associative and
commutative. In this case the theory simply becomes a sum of usual
Hilbert-Einstein actions for the fields without cross-interaction
terms. In the approach of \cite{avramidi04,avramidi04b}, the algebra is associative but
\emph{non-commutative} where the gauge group is simply the product
of the group of diffeomorphism of a real manifold with the
internal group. Because of this form of the gauge transformations
one can allow the algebra to be non-commutative which leads to a
different dynamics from the one described in \cite{mann84} and
\cite{wald87}.

Let ${\cal V}$ be an $N$-dimensional Hermitian vector bundle over $M$,
let $\mathscr{T}=TM\otimes {\cal V}$ be the bundle constructed by taking
the tensor product of the tangent bundle to the manifold $M$ with
the vector bundle ${\cal V}$, and let
$\mathscr{T^{\ast}}=T^{\ast}M\otimes{\cal V}$, where $T^{\ast}M$ is the
cotangent bundle to $M$.
Let $a$ be a symmetric self-adjoint element of $TM\otimes
TM\otimes\textrm{End}({\cal V})$, that is,
\be
a^{\mu\nu}=a^{\nu\mu},\qquad (a^{\mu\nu})^*=a^{\mu\nu}\,.
\ee
Suppose that $a^{\mu\nu}$
is an isomorphism between $\mathscr{T}$ and $\mathscr{T}^{\ast}$, then the inverse
isomorphism $b_{\mu\nu}$ satisfies the equation
\begin{equation}\label{54}
a^{\rho\nu}b_{\nu\mu}=b_{\mu\nu}a^{\nu\rho}=\delta_{\mu}^{\rho}\cdot\mathbb{I}\;.
\end{equation}
There are some properties of the matrix $b_{\mu\nu}$ that need
attention. The first property is the following: the matrix
$b_{\mu\nu}$ satisfies the equation
\begin{equation}\label{55}
b_{\mu\nu}^{\ast}=b_{\nu\mu}\;,
\end{equation}
but it is not necessarily a self-adjoint matrix symmetric in its
tensor indices. Moreover, one can use $a^{\mu\nu}$ and
$b_{\mu\nu}$ to lower and raise indices, although particular care
is required in these operations because, in general, $a^{\mu\nu}$
and $b_{\mu\nu}$ do not commute and $b_{\mu\nu}$ is not symmetric
in its tensorial indices \cite{avramidi04}.

Let $\mathscr{A}^{\alpha}{}_{\lambda\mu}$ be the matrix-valued Christoffel symbol
defined as \cite{avramidi04}
\begin{equation}\label{515}
\mathscr{A}^{\alpha}{}_{\lambda\mu}=\frac{1}{2}b_{\lambda\sigma}(a^{\alpha\gamma}\partial_{\gamma}a^{\rho\sigma}-a^{\rho\gamma}\partial_{\gamma}a^{\sigma\alpha}-a^{\sigma\gamma}\partial_{\gamma}a^{\alpha\rho})b_{\rho\mu}\;,
\end{equation}
it is not difficult to prove that this quantity transforms as a connection coefficient.
It is important to notice, at this point, that in matrix geometry the connection
(\ref{515}) is not symmetric in the two lower indices.

In complete analogy with the ordinary Riemannian geometry, by using the matrix-valued Christoffel symbol,
one can define the matrix-valued Riemann tensor as follows \cite{avramidi04}
\begin{equation}\label{510}
\mathcal{R}^{\lambda}{}_{\alpha\mu\nu}
=\partial_{\mu}\mathscr{A}^{\lambda}{}_{\alpha\nu}
-\partial_{\nu}\mathscr{A}^{\lambda}{}_{\alpha\mu}
+\mathscr{A}^{\lambda}{}_{\beta\mu}\mathscr{A}^{\beta}{}_{\alpha\nu}
-\mathscr{A}^{\lambda}{}_{\beta\nu}\mathscr{A}^{\beta}{}_{\alpha\mu}\;.
\end{equation}

Once the matrix curvature (Riemann) tensor is defined one can
construct the matrix Ricci tensor, namely
\begin{equation}\label{512}
\mathcal{R}_{\mu\nu}=\mathcal{R}^{\alpha}{}_{\mu\alpha\nu}\;.
\end{equation}
In order to write the action for Matrix Gravity, one needs to
introduce the matrix scalar curvature $\mathcal{R}$. Since the
metric $a^{\mu\nu}$ and the Ricci tensor $\mathcal{R}_{\mu\nu}$
are matrices, they do not commute in general and the definition of
the scalar curvature, obtained by contracting the metric tensor
with the Ricci tensor from the left, would be different if the
contraction would be performed with the metric tensor on the
right. In order to avoid this choice, we use a symmetrized
definition of the matrix-valued scalar curvature as follows
\begin{equation}\label{513}
\mathcal{R}=\frac{1}{2}\left(a^{\mu\nu}\mathcal{R}_{\mu\nu}+\mathcal{R}_{\mu\nu}a^{\mu\nu}\right)\;.
\end{equation}


In order to write an action for the model under consideration
a generalization of the concept of measure is needed. As a guiding
principle, any generalization of the measure $\mu$ has to lead, in
the commutative limit, to the ordinary Riemannian measure
$\sqrt{|\det g_{\mu\nu}|}$.
Let $\rho$ be a matrix-valued scalar density, which can be
defined, for example, as follows
\begin{equation}\label{520}
\rho=\int_{\mathbb{R}^{n}}\frac{d\xi}{\pi^{\frac{n}{2}}}
\exp(-a^{\mu\nu}\xi_{\mu}\xi_{\nu})\;.
\end{equation}
Then $\rho$ only depends on the metric $a$ and transforms in the
correct way under diffeomorphisms of $M$. We would like to stress,
here, that the choice of the measure is not unique. However, the
definition (\ref{520}) seems to be the
most natural because it represents the quantity that appears as the
$A_{0}$ coefficient in the heat kernel asymptotic of a generalized
Laplace operator with matrix-valued symbol defined on the manifold
$M$ under consideration. Of course different choices of the
measure would lead to different non-commutative limits of the
theory. More precisely, the zeroth order of the expansion in the
deformation parameter of the action always gives the usual General Relativity. The
second order term, which gives the dynamical equations for the
non-commutative part of the metric, instead, changes if the
definition of measure is different. Further studies are required
in order to fully understand how the choice of the measure affects
the dynamics of the non-commutative part of the metric.

Now that all the relevant geometric quantities have been described,
one can construct the action functional for the field $a^{\mu\nu}$
following \cite{avramidi04}. This functional has to be invariant under
both diffeomorphisms of $M$ and gauge transformations.
By using the matrix-valued scalar curvature, defined in (\ref{513}), and the
matrix-valued scalar density (\ref{520}), one obtains, by analogy to (\ref{newwilly10}),
\cite{avramidi04}
\begin{equation}\label{523}
S_{\textrm{MGR}}(a)=\frac{1}{16\pi
G}\int_{M}dx\frac{1}{N}\textrm{Tr}_{\;V}[\rho(\mathcal{R}-2\Lambda)]\;.
\end{equation}
It is worth noticing that because of the cyclic property of the
trace, the relative position of $\rho$ and the scalar curvature is
irrelevant, moreover it is easily shown that the action functional
(\ref{523}) is invariant under the diffeomorphisms of $M$ and
under the gauge transformations. Of course, as $a^{\mu\nu}\to g^{\mu\nu}$
this action reproduces the Hilbert-Einstein action (\ref{newwilly10}) of
General Relativity.

The field equations for the tensor
$a^{\mu\nu}$, that we call non-commutative Einstein
equations are obtained by varying the action with respect to
$a^{\mu\nu}$. In the vacuum we have,
\be
\frac{\partial {\cal L}}{\partial a^{\alpha\beta}}
-\partial_\mu
\frac{\partial {\cal L}}{\partial a^{\alpha\beta}{}_{,\mu}}=0\,,
\ee
where $a^{\alpha\beta}{}_{,\mu}=\partial_\mu a^{\alpha\beta}$ and ${\cal L}$ is the Lagrangian density.

The action  has an additional new {\it global} gauge symmetry
\be
a^{\mu\nu}(x) \mapsto U a^{\mu\nu}(x) U^{-1}\,,
\ee
where $U$ is a
constant unitary
matrix (for more details, see the papers cited above).
By the Noether theorem
this symmetry leads to the conserved
currents (vector densities)
$$
{\cal J}^\mu=\left[a^{\alpha\beta},\;{\partial {\cal L}\over\partial
(a^{\alpha\beta}{}_{,\mu})}\right]\,,
\qquad
\partial_\mu {\cal J}^\mu=0\,.
$$
In other words, this suggests the existence of
{\it new physical charges}
$$
Q=\int\limits d\hat x\, {\cal J}^0\,,
$$
where $d\hat x$ denotes the integration over the space coordinates only.
These charges have purely non-commutative origin and vanish in the
commutative limit.


This model may be viewed as a ``non-commutative deformation'' of Einstein
gravity, which
describes, in the weak deformation limit, General
Relativity, and a multiplet
of self-interacting massive two-tensor fields of spin $2$ that interact
also with gravity.

\subsection{Spectral Matrix Gravity}

By using the equations (\ref{3120igaw}), (\ref{3120igaw1}) and (\ref{newwilly3}), it is easy to see that the
Hilbert-Einstein action (\ref{newwilly10}) is nothing but a linear
combination of $A_{0}$ and $A_{1}$ for a Laplace type operator.
In full analogy, the action of Spectral Matrix Gravity proposed in
\cite{avramidi04b} is a linear combination of the global heat kernel
coefficients $A_{0}$ and $A_{1}$ for a general second order non-Laplace type operator,
more precisely
\begin{equation}\label{62}
S=\frac{1}{16\pi GN}\left[6A_{1}-2\Lambda
A_{0}\right]\;.
\end{equation}
We would like to point out here that the above action can be also thought of
as a particular case of the Spectral Action Principle introduced in the framework
of non-commutative geometry in \cite{connes} and \cite{chamseddine93}.
For the Laplace operator, $L=-\Delta$,
the heat kernel coefficients are (\ref{3120igaw}) and (\ref{3120igaw1})
and, therefore, the action of Spectral Matrix Gravity reduces to the
standard Hilbert-Einstein action (\ref{newwilly10}).

We would like to stress, here, that we are interested in a much more complicated general case of an arbitrary
non-Laplace type operator (with a non-scalar leading symbol). In
this case there is no preferred Riemannian metric and the whole
language of Riemannian geometry is not very helpful in computing
the heat kernel asymptotics. That is why, until now, there are no
explicit general formulas for the coefficient $A_{1}$. A class of
so-called natural non-Laplace type operators was studied in
\cite{avramidi01,avramidi02a} where this coefficient was computed
explicitly.

We would like to mention, here, that similar calculations have been
performed in non-commutative geometry regarding heat kernel asymptotics
expansion. In \cite{sasakura04,sasakura05}, the author evaluates the relevant
geometric quantities from an approximate power expansion of the trace of the
heat kernel for a Laplace operator on a compact fuzzy space. In
\cite{vassilevich}, the author studies the quantization of non-commutative
gravity in two dimensions by considering a non-commutative deformation (using
the Moyal product) of the Jackiw-Teitelboim model for gravity. In this case the
path integral can be evaluated exactly and the operator for the quantum
fluctuations can be found. Once the operator is known one can study the first
two heat kernel asymptotic coefficients and obtain information about the
conformal anomaly and the Polyakov action.

\chapter[NON-PERTURBATIVE HEAT KERNEL ASYMPTOTICS
\newheadline ON HOMOGENEOUS ABELIAN BUNDLES]
{NON-PERTURBATIVE HEAT KERNEL ASYMPTOTICS
\newheadline ON HOMOGENEOUS ABELIAN BUNDLES\footnotemark[1]}

\footnotetext[1]{The material in this chapter has been published in \emph{Communications in Mathematical Physics}:
I. G. Avramidi and G. Fucci, Non-Perturbative Heat Kernel Asymptotics on Homogeneous Abelian Bundles, \emph{Comm. Math. Phys.} (2009) doi: 10.1007/s00220-009-0804-6}

\begin{chapabstract}

We study the heat kernel for a Laplace type partial differential operator
acting on smooth sections of a complex vector bundle with the structure
group $G\times U(1)$
over a Riemannian manifold $M$ without boundary. The total
connection on the vector bundle naturally splits into a $G$-connection and a
$U(1)$-connection, which is assumed to have a parallel curvature $F$. We find
a new local short time asymptotic expansion of the off-diagonal heat kernel
$U(t|x,x')$ close to the diagonal of $M\times M$ assuming the curvature $F$ to
be of order $t^{-1}$. The coefficients of this expansion are polynomial
functions in the Riemann curvature tensor (and the curvature of the
$G$-connection) and its derivatives with universal coefficients depending in a
non-polynomial but analytic way on the curvature $F$, more precisely, on $tF$.
These functions generate all terms quadratic and linear in the Riemann
curvature and of arbitrary order in $F$ in the usual heat kernel coefficients.
In that sense, we effectively sum up the usual short time heat kernel
asymptotic expansion to all orders of the curvature $F$. We compute the first
three coefficients (both diagonal and off-diagonal) of this new asymptotic
expansion.
\end{chapabstract}

\section{Introduction}

The heat kernel is one of the most powerful tools in quantum field theory and
quantum gravity as well as mathematical physics and differential geometry (see
for example
\cite{gilkey95,vassile03,avramidi00,avramidi02,avramidi99,kirsten01,vandeven98,
hurt83} and further references therein). It is of particular importance because
the heat kernel methods give a framework for manifestly covariant calculation
of a wide range of relevant quantities in quantum field theory like one-loop
effective action, Green's functions, effective potential etc.

Unfortunately the exact computation of the heat kernel can be carried out only
for exceptional highly symmetric cases when the spectrum of the operator is
known exactly, (see \cite{camporesi90,hurt83,kirsten01} and the references in
\cite{avramidi96,avramidi08a,avramidi08,avramidi08b}). Although these special
cases are very important, in quantum field theory we need the effective action,
and, therefore, the heat kernel for general background fields. For this reason
various approximation schemes have been developed. One of the oldest methods is
the Minackshisundaram-Pleijel short-time asymptotic expansion (\ref{newwilly7}), (\ref{newwilly2}) of the heat
kernel as $t\to 0$ (see the references in
\cite{gilkey95,avramidi91,vassile03}).

Despite its enormous importance, this method is essentially perturbative. It is
an expansion in powers of the curvatures $R$ and their derivatives and, hence,
is inadequate for large curvatures when $tR\sim 1$. To be able to describe the
situation when at least some of the curvatures are large one needs an
essentially {\it non-perturbative approach}, which effectively sums up in the
short time asymptotic expansion of the heat kernel an infinite series of terms
of certain structure that contain large curvatures (for a detailed analysis see
\cite{avramidi94,avramidi97} and reviews \cite{avramidi99,avramidi02}).
For example, the partial summation of higher derivatives enables one to obtain
a non-local expansion of the heat kernel in powers of curvatures (high-energy
approximation in physical terminology). This is still an essentially
perturbative approach since the curvatures (but not their derivatives) are
assumed to be small and one expands in powers of curvatures.

On another hand to study the situation when curvatures (but not their
derivatives) are large (low energy approximation) one needs an essentially {\it
non-per\-tur\-bative approach}. A promising approach to the calculation of the
low-energy heat kernel expansion was developed in non-Abelian gauge theories
and quantum gravity in
\cite{avramidi93,avramidi94,avramidi94a,avramidi95,avramidi95a,avramidi96,
avramidi08b,avramidi08,avramidi08a}. While the papers
\cite{avramidi93,avramidi94,avramidi95,avramidi95a} dealt with the parallel
$U(1)$-curvature (that is, constant electromagnetic field) in flat space, the
papers \cite{avramidi94a,avramidi96,avramidi08b} dealt with symmetric spaces
(pure gravitational field in absence of an electromagnetic field). The
difficulty of combining the gauge fields and gravity was finally overcome in
the papers \cite{avramidi08,avramidi08a}, where homogeneous bundles
with parallel curvature on symmetric spaces were studied.

In this chapter we compute the heat kernel for the covariant Laplacian with a
large parallel $U(1)$ curvature $F$ in a Riemannian manifold (that is, strong
covariantly constant electromagnetic field in an arbitrary gravitational
field). Our aim is to evaluate the first three coefficients of the heat kernel
asymptotic expansion in powers of Riemann curvature $R$ but \emph{in all
orders} of the $U(1)$ curvature $F$. This is equivalent to a partial summation
in the heat kernel asymptotic expansion as $t\to 0$ of all powers of $F$ in
terms which are linear and quadratic in Riemann curvature $R$.

\section{Setup of the Problem}

Let $M$ be a $n$-dimensional compact Riemannian manifold without boundary and
${\cal S}$ be a complex vector bundle over $M$ realizing a representation of
the group $G\otimes U(1)$. Let $\varphi$ be a section of the bundle ${\cal S}$
and $\nabla$ be the total connection on the bundle $\mathcal{S}$ (including the
$G$-connection as well as the ${\rm U}(1)$-connection). Then the commutator of
covariant derivatives defines the curvatures
\begin{equation}\label{30}
[\nabla_\mu,\nabla_\nu]\varphi=({\cal R}_{\mu\nu}+i F_{\mu\nu})\varphi\;,
\end{equation}
where ${\cal R}_{\mu\nu}$ is the curvature of the $G$-connection
and $F_{\mu\nu}$ is the curvature of the $U(1)$-connection (which will be also
called the electromagnetic field).

In the present chapter we consider the Laplacian
\begin{equation}\label{31}
\mathscr{L}=-\Delta\;. 
\end{equation}
The asymptotic expansion of the heat kernel $U(t|x,x^{\prime})$ for the Laplacian has the form
(\ref{newwilly2}) and its coefficients are (\ref{newwilly3}).
The diagonal heat kernel coefficients $a^{\rm diag}_k$ are polynomials in the
jets of the metric, the $G$- connection and the $U(1)$-connection; in other
words, in the curvature tensors and their derivatives. Let us symbolically
denote the jets of the metric and the $G$-connection by
\be
R_{(n)}=\left\{\nabla_{(\mu_1}\cdots\nabla_{\mu_{n}}
R^a{}_{\mu_{n+1}}{}^b{}_{\mu_{n+2})}\,,
\;
\nabla_{(\mu_1}\cdots\nabla_{\mu_{n}}
{\cal R}^a{}_{\mu_{n+1})}
\right\}
\;,
\ee
and the jets of the $U(1)$ connection by
\be
F_{(n)}=\nabla_{(\mu_1}\cdots\nabla_{\mu_{n}}F^a{}_{\mu_{n+1})}\,.
\ee
Here and everywhere below the parenthesis indicate complete symmetrization over
all indices included.

By counting the dimensions it is easy to describe the general structure of the
coefficients $a^{\rm diag}_k$. Let us introduce the multi-indices of nonnegative integers
\be
{\bf i}=(i_1,\dots,i_m), \qquad
{\bf j}=(j_1,\dots,j_l)\,.
\ee
Let us also denote
\be
|{\bf i}|=i_1+\cdots+i_m,\qquad
|{\bf j}|=j_1+\cdots+j_l\,.
\ee
Then symbolically
\be
a^{\rm diag}_k=
\sum_{N=1}^k
\sum_{l=0}^N\sum_{m=0}^{N-l}
\;
\sum\limits_{{{\bf i},{\bf j}\ge 0}\atop{|{\bf i}|+|{\bf j}|+2N=2k}}
C_{(k,l,m),{\bf i},{\bf j}} F_{(j_1)}\cdots F_{(j_l)}\;
R_{(i_1)}\cdots R_{(i_m)}\,,
\label{3216zza}
\ee
where $C_{(k,l,m),{\bf i},{\bf j}}$ are some universal constants.
The lower order diagonal heat kernel asymptotic coefficients are (\ref{3120igaw})-(\ref{3120iga}).

In the present chapter we study the case of a {\it parallel $U(1)$ curvature}
(covariantly constant electromagnetic field), i.e.
\begin{equation}
\label{32x}
\nabla_{\mu}F_{\alpha\beta}=0\;.
\end{equation}
That is, all jets $F_{(n)}$ are set to zero except the one of order zero, which
is $F$ itself. In this case eq. (\ref{3216zza}) takes the form
\be
a^{\rm diag}_k=
\sum_{N=1}^k
\sum_{l=0}^N\sum_{m=0}^{N-l}
\;
\sum\limits_{{{\bf i}\ge 0}\atop{|{\bf i}|+2N=2k}}
C_{(k,l,m),{\bf i}} F^l\;
R_{(i_1)}\cdots R_{(i_m)}\,,
\ee
where $C_{(k,l,m),{\bf i}}$ are now some (other) numerical coefficients.

Thus, by summing up all powers of $F$ in the asymptotic expansion
of the heat kernel diagonal we obtain a {\it new (non-perturbative)
asymptotic expansion}
\be
U^{\rm diag}(t)\sim(4\pi t)^{-n/2}
\sum_{k=0}^{\infty}t^{k}\tilde a^{\rm diag}_{k}(t)\;,
\label{3222zza}
\ee
where the coefficients $\tilde a^{\rm diag}_k(t)$ are polynomials in the jets
$R_{(n)}$
\be
\tilde a^{\rm diag}_k(t)=
\sum_{N=1}^k
\sum_{m=0}^{N}
\;
\sum\limits_{{{\bf i}\ge 0}\atop{|{\bf i}|+2N=2k}}
f^{(k)}_{(m,{\bf i})}(t)\;
R_{(i_1)}\cdots R_{(i_m)}\,,
\label{3121iga}
\ee
and  $f^{(k)}_{(m,{\bf i})}(t)$ are some universal dimensionless tensor-valued
analytic functions that depend on $F$ only in the dimensionless combination
$tF$.

For the heat trace we obtain then a new asymptotic expansion of the form
\begin{equation}
\textrm{Tr}\;\exp(-t\mathscr{L})
\sim(4\pi t)^{-n/2}\sum_{k=0}^{\infty}t^{k}
\tilde A_{k}(t)\;,
\end{equation}
where
\begin{equation}
\tilde A_{k}(t)=\int\limits_{M}d\vol\,
\textrm{tr}\, \tilde a^{\rm diag}_{k}(t)\;.
\end{equation}

This expansion can be described more rigorously as follows. We rescale the
$U(1)$-curvature $F$ by
\be
F \mapsto F(t)=t^{-1}\tilde F\,,
\ee
so that $tF(t)=\tilde F$ is independent of $t$. Then the operator
$\mathscr{L}(t)$ becomes dependent on $t$ (in a singular way!). However, the
heat trace still has a nice asymptotic expansion as $t\to 0$
\begin{equation}
\textrm{Tr}\;\exp[-t\mathscr{L}(t)]
\sim(4\pi t)^{-n/2}\sum_{k=0}^{\infty}t^{k}
\tilde A_{k}\;,
\end{equation}
where the coefficients $\tilde A_k$ are expressed in terms of $\tilde F=tF(t)$,
and, therefore, are independent of $t$. Thus, what we are doing is the {\it
asymptotic expansion of the heat trace for a particular case of a singular (as
$t\to 0$) time-dependent operator $\mathscr{L}(t)$}.

Let us stress once again that the eq. (\ref{3121iga}) should not be taken
literally; it only represents the general structure of the coefficients $\tilde
a^{\rm diag}_k(t)$. To avoid confusion we list below the general structure of
the low-order coefficients in more detail
\bea
\tilde a^{\rm diag}_0(t)&=&f^{(0)}(t)\,,
\\[10pt]
\tilde a^{\rm diag}_1(t)&=&
f^{(1)}_{(1,1)}{}^{\alpha\beta\mu\nu}(t)R_{\alpha\beta\mu\nu}
+f^{(1)}_{(1,2)}{}^{\mu\nu}(t){\cal R}_{\mu\nu}\,,
\\[10pt]
\tilde a^{\rm diag}_2(t)&=&
f^{(2)}_{(1,1)}{}^{\alpha\beta\mu\nu\sigma\rho}(t)
\nabla_{(\alpha}\nabla_{\beta)} R_{\mu\nu\sigma\rho}
+f^{(2)}_{(1,2)}{}^{\alpha\beta\mu\nu}(t)\nabla_{(\alpha}
\nabla_{\beta)}{\cal R}_{\mu\nu}
\nonumber\\[5pt]
&&
+f^{(2)}_{(2,1)}{}^{\alpha\beta\gamma\delta\mu\nu\sigma\rho}(t)
R_{\alpha\beta\gamma\delta}
R_{\mu\nu\sigma\rho}
+f^{(2)}_{(2,2)}{}^{\alpha\beta\mu\nu}(t)
{\cal R}_{\alpha\beta}
{\cal R}_{\mu\nu}
\nonumber\\[5pt]
&&
+f^{(2)}_{(2,3)}{}^{\alpha\beta\mu\nu\sigma\rho}(t){\cal R}_{\alpha\beta}
R_{\mu\nu\sigma\rho}
\,,
\eea
with obvious enumeration of the functions.
It is the {\it universal tensor functions} $f^{(i)}_{(l,m)}(t)$ that are of
prime interest in this chapter. Our main goal is to compute the functions
$f^{(i)}_{(l,m)}(t)$ for the coefficients $\tilde a^{\rm diag}_0(t)$, $\tilde
a^{\rm diag}_1(t)$ and
$\tilde a^{\rm diag}_2(t)$.

Of course, for $t=0$ (or $F=0$) the coefficients $\tilde a_k(t)$ are equal to
the usual diagonal heat kernel coefficients
\be
\tilde a^{\rm diag}_k(0)=a_k^{\rm diag}\,.
\ee
Therefore, by using the explicit form of the coefficients $a_k^{\rm diag}$
given by (\ref{3120iga}) we obtain the initial values for the functions
$f^{(i)}_{(j,k)}$. Moreover, by analyzing the corresponding terms in the
coefficients $a_3^{\rm diag}$ and $a_4^{\rm diag}$ (which are known,
\cite{gilkey95,avramidi91,vandeven98}), one can obtain partial information
about some lower order Taylor coefficients of the functions
$f^{(i)}_{(j,k)}(t)$:
\bea
f^{(0)}(t)&=&1-\frac{1}{12}t^2F_{\mu\nu}F^{\mu\nu}+O(t^3)\,,
\\[10pt]
f^{(1)}_{(1,1)}{}^{\alpha\beta}{}_{\mu\nu}(t)
&=&
\frac{1}{6}\delta^{\alpha}_{[\mu}\delta^\beta_{\nu]}
+O(t)\,,
\\[10pt]
f^{(1)}_{(1,2)}{}^{\mu\nu}(t)&=&
\frac{1}{6}tiF^{\mu\nu}+O(t^2)\,,
\eea
\bea
f^{(2)}_{(1,1)}{}^{\alpha\beta\mu\nu}{}_{\sigma\rho}(t)
&=&
\frac{1}{30}g^{\alpha\beta}\delta^{\mu}_{[\sigma}\delta^\nu_{\rho]}
+O(t)\,,
\\[10pt]
f^{(2)}_{(1,2)}{}^{\alpha\beta}{}_{\mu\nu}(t)&=&
-\frac{1}{15}tiF^{(\alpha}{}_{[\nu}\delta^{\beta)}_{\mu]}
+O(t^2)\,,
\\[10pt]
f^{(2)}_{(2,1)}{}_{\alpha\beta}{}^{\gamma\delta}{}_{\mu\nu}{}^{\sigma\rho}(t)
&=&
\frac{1}{180}g_{\mu[\alpha}g_{\beta]\nu}g^{\sigma[\gamma}g^{\delta]\rho}
-\frac{1}{180}\delta^{[\gamma}_{[\alpha}g_{\beta][\nu}g^{\delta][\rho}
\delta^{\sigma]}_{\mu]}
\nonumber\\
&&+\frac{1}{72}\delta^{\gamma}_{[\alpha}\delta^\delta_{\beta]}
\delta^{\sigma}_{[\mu}\delta^\rho_{\nu]}
+O(t)
\,,
\nonumber\\
&&\\
f^{(2)}_{(2,2)}{}^{\alpha\beta}{}_{\mu\nu}(t)
&=&
\frac{1}{12}\delta^{\alpha}_{[\mu}\delta^\beta_{\nu]}
+O(t)\,,
\\[10pt]
f^{(2)}_{(2,3)}{}^{\alpha\beta\mu\nu}{}_{\sigma\rho}(0)&=&
-\frac{1}{36}tiF^{\alpha\beta}\delta^{\mu}_{[\sigma}\delta^\nu_{\rho]}
-\frac{1}{30}tiF^{\mu\nu}\delta^{\alpha}_{[\sigma}\delta^\beta_{\rho]}
+\frac{1}{9}\delta^{[\mu}_{[\sigma}tiF^{\nu][\alpha}\delta^{\beta]}_{\rho]}
+O(t^2)
\nonumber\,.
\\[10pt]
\eea
This information can be used to check our final results.

Notice that the global coefficients $\tilde A_k(t)$ have exactly the same form
as the local ones; the only difference is that the terms with the derivatives
of the Riemann curvature do not contribute to the integrated coefficients since
they can be eliminated by integrating by parts and taking into account that $F$
is covariantly constant.

Moreover, we study even more general non-perturbative asymptotic expansion for
the {\it off-diagonal} heat kernel and compute the coefficients of zero, first
and second order in the Riemann curvature. We will show that there is a
{\it new non-perturbative asymptotic expansion} of the off-diagonal heat kernel
as $t\to 0$ (and $F=t^{-1}\tilde F$, so that $tF$ is fixed) of the form
\be
U(t|x,x')\sim {\cal P}(x,x')\Delta^{1/2}(x,x')U_0(t|x,x')
\sum_{k=0}^\infty t^{k/2} b_k(t|x,x')\;,
\ee
where $U_0$ is an analytic function of $F$ such that for $F=0$
\be
U_0(t|x,x')\Bigg|_{F=0}=(4\pi t)^{-n/2}
\exp\left[-\frac{\sigma(x,x')}{2t}\right]\,.
\ee
Here $b_k(t|x,x')$ are analytic functions of $t$ that depend on $F$ only in the
dimensionless combination $tF$. Of course, for $t=0$ they are equal to the
usual heat kernel coefficients, that is,
\be
b_{2k}(0|x,x')=a_k(x,x')\,,
\qquad
b_{2k+1}(0|x,x')=0\;.
\ee
Moreover, we will show below that the odd-order coefficients vanish
not only for $t=0$ and any $x\ne x'$ but also for any $t$ and $x=x'$,
that is, on the diagonal,
\be
b_{2k+1}^{\rm diag}(t)=0\,.
\ee
Thus, the heat kernel diagonal has the asymptotic expansion
(\ref{3222zza}) as $t\to 0$ with
\be
\tilde a^{\rm diag}_k(t)=(4\pi t)^{n/2}U_0^{\rm diag}(t)b_{2k}^{\rm diag}(t)\,.
\ee

In what follows we will consider the operators
\begin{equation}\label{310aa}
\mathcal{D}_{\mu}=\bar{\nabla}_{\mu}
-\frac{1}{2}i F_{\mu\alpha}u^{\alpha}\;.
\end{equation}
Obviously, they form the algebra
\begin{equation}\label{316a}
\left[\mathcal{D}_{\mu},\mathcal{D}_{\nu}\right]=iF_{\mu\nu}\;,\qquad \left[\mathcal{D}_{\mu},u^{\nu}\right]=\delta_{\mu}{}^{\nu}\;.
\end{equation}
For a covariantly constant electromagnetic field, considered in this work, the following relation holds \cite{avramidi00,avramidi91}
\begin{equation}
\label{32xx}
\nabla_{\mu}F_{\alpha\beta}=0\;.
\end{equation}
In this case we find it useful to decompose the quantity $\mathscr{A}_{\mu}$
as
\begin{equation}\label{39a}
\mathscr{A}_{\mu}=-{1\over 2}i F_{\mu\alpha}u^\alpha+\bar{\mathscr{A}}_{\mu}\;.
\end{equation}
It can be easily shown that $\bar{\mathscr{A}}_{\mu}$ has the following Taylor expansion
\begin{eqnarray}\label{39}
\bar{\mathscr{A}}_{\mu}
&=&
-{1\over 2}\mathcal{R}_{\mu\alpha}u^\alpha
+{1\over 24}R_{\mu\alpha\nu\beta}i F^\nu{}_{\gamma}u^\alpha u^\beta u^\gamma
+\frac{1}{3}\nabla_{\alpha}\mathcal{R}_{\mu\beta}u^{\alpha}u^{\beta}
\nonumber\\
&+&
{1\over 24}R_{\mu\alpha\nu\beta}
\mathcal{R}^\nu{}_\gamma u^\alpha u^\beta u^\gamma
-\frac{1}{8}\nabla_{\alpha}\nabla_{\beta}
\mathcal{R}_{\mu\gamma}u^{\alpha}u^{\beta}u^{\gamma}
\nonumber\\
&-&{1\over 720}R_{\mu\alpha\nu\beta}R^\nu{}_{\gamma\lambda\delta}
i F^\lambda{}_\epsilon u^\alpha u^\beta u^\gamma u^\delta u^\epsilon
+O(u^6)\;.
\end{eqnarray}

We would like to stress that, the expansion for
$\bar{\mathscr{A}_{\mu}}$ is valid in the case of a covariantly constant
electromagnetic field.

By utilizing the Taylor expansion for all the relevant quantities we
are able to find an expansion for the heat kernel. First of all, the heat kernel can be
presented in the form
\begin{equation}\label{35}
U(t|x,x^{\prime})=\exp\left({-t\mathscr{L}}\right)\mathcal{P}(x,x^{\prime})
\delta(x,x^{\prime})\;,
\end{equation}
which can also be written as
\begin{equation}\label{35a}
U(t|x,x^{\prime})
=\mathcal{P}(x,x^{\prime})\Delta^{\frac{1}{2}}(x,x^{\prime})
\exp(-t\tilde{\mathscr{L}})\delta(u)\;,
\end{equation}
where $\delta(u)$ is the usual delta-function in the normal coordinates
$u^\mu$ (recall that $u^\mu$ depends on $x$ and $x'$ and $u=0$ when $x=x'$)
and
$\tilde{\mathscr{L}}$ is an operator defined by
\begin{equation}
\tilde{\mathscr{L}}=
\mathcal{P}^{-1}(x,x^{\prime})\Delta^{-\frac{1}{2}}(x,x^{\prime})\mathscr{L}
\Delta^{\frac{1}{2}}(x,x^{\prime})\mathcal{P}(x,x^{\prime})\;.
\end{equation}
As is shown in \cite{avramidi91,avramidi00}
the operator $\tilde{\mathscr{L}}$ can be written in the form
\begin{eqnarray}\label{310a}
\tilde{\mathscr{L}}=-\left({\cal D}_{\mu}
+\bar{\mathscr{A}}_{\mu}-\zeta_{\mu}\right)X^{\mu\nu}
\left({\cal D}_{\nu}
+\bar{\mathscr{A}}_{\nu}+\zeta_{\nu}\right)
\;,
\end{eqnarray}
where
$
\zeta_{\mu}
=\bar{\nabla}_{\mu}\zeta
$.

Now, by using these equations and by recalling the formula in (\ref{39a}),
one can rewrite the operator in (\ref{310a}) in another way as follows
\begin{equation}
\tilde{\mathscr{L}}
=-\left(X^{\mu\nu}\mathcal{D}_{\mu}\mathcal{D}_{\nu}
+Y^{\mu}\mathcal{D}_{\mu}+Z\right)\;,
\end{equation}
where $X^{\mu\nu}$ is defined in (\ref{newwilly1}) and
\begin{eqnarray}
\label{36a}
Y^{\mu}&=&
(\bar{\nabla}_{\mu}X^{\mu\nu})
+2X^{\mu\nu}\bar{\mathscr{A}}_{\mu}\;,
\\[10pt]
Z&=&
\bar{\mathscr{A}}_{\mu}X^{\mu\nu}\bar{\mathscr{A}}_{\nu}
-\zeta_{\mu}X^{\mu\nu}\zeta_{\nu}
+(\bar{\nabla}_{\mu}X^{\mu\nu})\bar{\mathscr{A}}_{\nu}
+(\bar{\nabla}_{\mu}X^{\mu\nu})\zeta_{\nu}
\nonumber\\
&+&X^{\mu\nu}\bar{\nabla}_{\mu}\bar{\mathscr{A}}_{\nu}
+X^{\mu\nu}\bar{\nabla}_{\mu}\zeta_{\nu}\;.
\end{eqnarray}

By using the covariant Taylor expansion of the two-point quantities that
we described in Chapter 2, we obtain an expansion for the coefficients $X^{\mu\nu}$, $Y^{\mu}$ and $Z$ of
the operator $\mathscr{L}$ up to the fifth order 


\section{Perturbation Theory}

Our goal is now to develop the perturbation theory for the heat kernel. We need
to identify a small expansion parameter $\varepsilon$ in which the perturbation
theory will be organized as $\varepsilon\to 0$. First of all, we assume that
$t$ is small, more precisely, we require $t\sim\varepsilon^2$. Also, since we
will work close to the diagonal, that is, $x$ is close to $x'$, we require that
$u^\mu\sim \varepsilon$. This will also mean that $\bar\nabla\sim
\varepsilon^{-1}$ and $\partial_t\sim\varepsilon^{-2}$. Finally, we assume that
$F$ is large, that is, of order $F\sim \varepsilon^{-2}$. To summarize,
\be
t\sim \varepsilon^2,\qquad
u^\mu\sim \varepsilon,\qquad
F\sim \varepsilon^{-2}\,.
\ee


\subsection{Perturbation Theory for the Operator $\mathscr{L}$}

Now, we expand the operator $\tilde{\mathscr{L}}$ in a formal power series in
$\varepsilon$ (recall that ${\cal D}\sim \varepsilon^{-1}$ and $u\sim
\varepsilon$) to obtain
\begin{equation}\label{314}
\mathscr{L}
\sim-\sum_{k=0}^\infty \mathscr{L}_{k}
\;,
\end{equation}
where $\mathscr{L}_k$ are operators of order $\varepsilon^{k-2}$.
In particular,
\bea
\label{315}
\mathscr{L}_0
&=&
{\cal D}^2\;,
\\
\mathscr{L}_1&=&0\,,
\\
\label{316}
\mathscr{L}_{k}
&=&
X_{k}^{\mu\nu}\mathcal{D}_{\mu}\mathcal{D}_{\nu}
+Y_{k}^{\mu}\mathcal{D}_{\mu}
+Z_{k}\;,
\qquad
k\ge 2\,.
\eea
where
\be
{\cal D}^2=g^{\mu\nu}\mathcal{D}_{\mu}\mathcal{D}_{\nu}\;,
\ee
and $X^{\mu\nu}_k$, $Y^\mu_k$ and $Z_k$ are some tensor-valued polynomials in
normal coordinates $u^\mu$.

Note that $X^{\mu\nu}_k$ are homogeneous
polynomials in normal coordinates $u^\mu$ and $F$ of order $\varepsilon^{k}$.
Similarly, $Y^\mu_k\sim \varepsilon^{k-1}$ and $Z_k\sim \varepsilon^{k-2}$. Of
course, here the terms $Fuu$ are counted as of order zero. That is,
they have the form
\bea
X_k^{\mu\nu}
&=&
P_{(1),\;k}^{\mu\nu}(u)\,,
\\
Y_k^\mu
&=&
P_{(2),\;k-1}^\mu
+F_{\alpha\beta}P_{(3),\;k+1}^{\mu\alpha\beta}(u)\,,
\\
Z_k
&=&
P_{(4),\;k-2}
+F_{\alpha\beta}P_{(5),\;k}^{\alpha\beta}(u)
+F_{\alpha\beta}F_{\rho\sigma}P^{\alpha\beta\rho\sigma}_{(6),\;k+2}(u)\,,
\eea
where $P_{(j),\;k}(u)$ are homogeneous tensor valued
polynomials of degree $k$.

By using the covariant Taylor expansions in (\ref{37}), (\ref{39}) and (\ref{38})
we find the explicit expression of the coefficients
\begin{eqnarray}
\label{317}
X^{\mu\nu}_2
&=&
C^{\mu\nu}_{2}{}_{\alpha\beta} u^\alpha u^\beta\;,
\\
Y^\mu_2
&=&
E^{\mu}_{2}{}_{\alpha}u^\alpha
+G^{\mu}_{2}{}_{\alpha\beta\gamma}u^\alpha u^\beta u^\gamma\;,
\\
Z_2
&=&
H_{2\;\alpha\beta}u^\alpha u^\beta+L_{2}\;,
\label{317a}\\[8pt]
X^{\mu\nu}_3&=&
C^{\mu\nu}_{3}{}_{\alpha\beta\gamma} u^\alpha u^\beta u^{\gamma}\;,
\\
Y^\mu_3&=&
E^{\mu}_{3}{}_{\alpha\beta}u^\alpha u^{\beta}\;,
\\
Z_3&=&
H_{3\;\alpha}u^\alpha\;,
\label{317aa}
\\[8pt]
X^{\mu\nu}_4&=&
C^{\mu\nu}_{4}{}_{\alpha\beta\gamma\delta}
u^\alpha
u^\beta u^\gamma u^\delta\;,
\label{318d}
\\
Y^\mu_4&=&
E^{\mu}_{4}{}_{\;\alpha\beta\gamma}u^\alpha u^\beta u^\gamma
+G^{\mu}_{4}{}_{\;\alpha\beta\gamma\delta\epsilon}
u^\alpha u^\beta u^\gamma u^\delta u^\epsilon\;,
\\
Z_4&=&
H_{4\;\alpha\beta}u^\alpha u^\beta
+L_{4\;\alpha\beta\gamma\delta} u^\alpha u^\beta u^\gamma u^\delta
+O_{4\;\alpha\beta\gamma\delta\epsilon\kappa}
u^\alpha u^\beta u^\gamma u^\delta u^\epsilon u^\kappa\;,
\label{318a}
\end{eqnarray}
where
\begin{eqnarray}
\label{318b}
C^{\mu\nu}_{2}{}_{\alpha\beta}
&=&
{1\over 3}R^{\mu}{}_{(\alpha}{}^{\nu}{}_{\beta)}\;,
\nonumber\\
E^{\mu}_{2}{}_{\alpha}
&=&
-{1\over 3}R^{\mu}{}_{\alpha}
-{\cal R}^\mu{}_\alpha\;,
\nonumber\\
G^{\mu}_{2}{}_{\alpha\beta\gamma}
&=&
-{1\over 12}R^\mu{}_{(\alpha}{}^\nu{}_{\beta} i F_{\gamma)\nu}\;,
\nonumber\\
H_{2\;\alpha\beta}
&=&
-{1\over 24}R_{\mu(\alpha} i F^{\mu}{}_{\beta)}\;,
\nonumber\\
L_{2}
&=&
\frac{1}{6}R\;,
\end{eqnarray}
\begin{eqnarray}\label{318c}
C^{\mu\nu}_{3}{}_{\alpha\beta\gamma}
&=&
-{1\over 6}\nabla_{(\alpha}R^{\mu}{}_{\beta}{}^{\nu}{}_{\gamma)}\;,
\nonumber\\
E^{\mu}_{3}{}_{\alpha\beta}
&=&
\frac{1}{3}\nabla_{(\alpha}R^{\mu}{}_{\beta)}
-\frac{1}{6}\nabla^{\mu}R_{\alpha\beta}
+\frac{2}{3}\nabla_{(\alpha}\mathcal{R}^{\mu}{}_{\beta)}\;,
\nonumber\\
H_{3\;\alpha}
&=&
\frac{1}{3}\nabla_{\mu}\mathcal{R}^{\mu}{}_{\alpha}
-\frac{1}{6}\nabla_{\alpha}R\;,
\end{eqnarray}
\begin{eqnarray}
C^{\mu\nu}_{4}{}_{\alpha\beta\gamma\delta}
&=&
{1\over 15}R^{\mu}{}_{(\alpha|\lambda|\beta}
R^\lambda{}_{\gamma}{}^{\nu}{}_{\delta)}
+\frac{1}{20}\nabla_{(\alpha}\nabla_{\beta}
R^{\mu}{}_{\gamma}{}^{\nu}{}_{\delta)}\;,
\nonumber\\
E^{\mu}_{4}{}_{\;\alpha\beta\gamma}
&=&
-{1\over 15}R^\mu{}_{\nu(\alpha|\lambda|}R^\nu{}_\beta{}^{\lambda}{}_{\gamma)}
-{1\over 60} R^\mu{}_{(\alpha}{}^\nu{}_\beta R_{\gamma)\nu}
-{1\over 4}R^{\mu}{}_{(\alpha}{}^\nu{}_\beta {\cal R}_{|\nu|\gamma)}
\nonumber\\
&+&
\frac{1}{10}\nabla_{(\alpha}\nabla^{\mu}R_{\beta\gamma)}
-\frac{3}{20}\nabla_{(\alpha}\nabla_{\beta}R^{\mu}{}_{\gamma)}
-\frac{1}{4}\nabla_{(\alpha}\nabla_{\beta}\mathcal{R}^{\mu}{}_{\gamma)}\;,
\nonumber\\
G^{\mu}_{4}{}_{\;\alpha\beta\gamma\delta\epsilon}
&=&
{1\over 40}R^\mu{}_{(\alpha|\nu|\beta}R^{\nu}{}_{\gamma}{}^{\lambda}{}_{\delta} i F_{|\lambda|\epsilon)}\;,
\nonumber\\
H_{4\;\alpha\beta}
&=&
{1\over 4}{\cal R}_{\mu(\alpha}{\cal R}^{\mu}{}_{\beta)}
-\frac{1}{30}R_{\mu\alpha}R^{\mu}{}_{\beta}
-\frac{1}{4}\nabla_{(\alpha}\nabla_{|\mu|}\mathcal{R}^{\mu}{}_{\beta)}
+\frac{1}{60}R_{\mu\nu}R^{\mu}{}_{\alpha}{}^{\nu}{}_{\beta}
\nonumber\\
&+&
\frac{1}{60}R_{\mu\lambda\gamma\alpha}R^{\mu\lambda\gamma}{}_{\beta}
+\frac{1}{40}\Delta R_{\alpha\beta}
+\frac{3}{40}\nabla_{\alpha}\nabla_{\beta}R\;,
\nonumber\\
L_{4\;\alpha\beta\gamma\delta}
&=&
-{1\over 80}R_{\mu(\alpha}{}^{\nu}{}_{\beta}
R^{\mu}{}_{\gamma}i F_{|\nu|\delta)}
-{1\over 80}R_{\mu(\alpha|\lambda|\beta}
R^{\lambda}{}_{\gamma}{}^{\mu\nu} i F_{|\nu|\delta)}
-{1\over 24}{\cal R}_{\mu(\alpha}R^\mu{}_\beta{}^\nu{}_\gamma i F_{|\nu|\delta)}\;,
\nonumber\\
O_{4\;\alpha\beta\gamma\delta\epsilon\kappa}
&=&
{1\over 576}R_{\mu(\alpha}{}^{\nu}{}_{\beta} R^{\mu}{}_{\gamma}{}^{\lambda}{}_{\delta}
i F_{|\nu|\epsilon}i F_{|\lambda|\kappa)}\;.
\end{eqnarray}
Here and everywhere below the parenthesis denote the complete symmetrization
over all indices enclosed; the vertical lines indicate the indices excluded
from the symmetrization.

\subsection{Perturbation Theory for the Heat Semigroup}

Now, by using the perturbative expansion (\ref{314}) of the operator
$\tilde{\mathscr{L}}$ and recalling that $\mathcal{D}^{2}\sim \varepsilon^{-2}$
and $t\sim \varepsilon^2$, we see that the operator $t\mathcal{D}^{2}$ is of
zero order and the operator $t\mathscr{L}_k$, $k\ge 2$, is of (higher) order
$\varepsilon^{k}$. Therefore, we can consider the terms $t{\mathscr{L}}_k$ with
$k\ge 2$ as a perturbation.

By using the Volterra series for the operator in (\ref{314})
we obtain
\begin{equation}
\label{327a}
\exp(-t\tilde{\mathscr{L}})
=T(t)\;\exp(t\mathcal{D}^{2})\;,
\end{equation}
where $T(t)$ is an operator defined by a formal perturbative
expansion
\begin{eqnarray}
\label{327b}
T(t)\sim
\sum_{k=0}^\infty T_k(t)
\;,
\end{eqnarray}
with $T_k(t)$ being of order $\varepsilon^k$.
Explicitly, up to terms of fifth order we obtain
\bea
T_0(t)&=&I\,,
\\[10pt]
T_1(t)&=&0\,,
\\
\label{327c}
T_2(t)&=&
t\int\limits_{0}^{1}d\tau_{1}\;
V_2(t\tau_1)\,,
\\
T_3(t)&=&
t\int\limits_{0}^{1}d\tau_{1}\;
V_3(t\tau_1)
\\
T_4(t)&=&
t\int\limits_{0}^{1}d\tau_{1}\;
V_4(t\tau_1)
+t^2\int\limits_{0}^{1}d\tau_{2}
\int\limits_{0}^{\tau_{2}}d\tau_{1}\;
V_2(t\tau_1)V_2(t\tau_2)
\;,
\label{327d}
\eea
and
\be
V_k(s)=e^{s\mathcal{D}^{2}}
\mathscr{L}_{k}
e^{-s\mathcal{D}^{2}}\,.
\label{331iga}
\ee

\subsection{Perturbation Theory for the Heat Kernel}

As we already mentioned above the heat kernel can be computed from the heat
semigroup by using the equation (\ref{35a}). By using the heat semigroup
expansion from the previous section we now obtain the heat kernel in the form
\bea
\label{325a}
U(t|x,x^{\prime})
&\sim &
\mathcal{P}(x,x^{\prime})\Delta^{1/2}(x,x')U_{0}(t|x,x^{\prime})
\sum_{k=0}^\infty t^{k/2}b_k(t|x,x')\;,
\label{354}
\eea
where
\begin{equation}
\label{323}
U_{0}(t|x,x^{\prime})=\exp(t\mathcal{D}^{2})
\delta(u)\;,
\end{equation}
and
\begin{equation}
\label{354az}
b_{k}(t|x,x^{\prime})=t^{-k/2}
U^{-1}_0(t|x,x')T_{k}(t)U_{0}(t|x,x^{\prime})\;.
\end{equation}
Thus, the calculation of the heat kernel coefficients reduces to the
evaluation of the zero-order heat kernel $U_0(t|x,x')$ and to the action of the
differential operators $T_k(t)$ on it.

The zero order heat kernel $U_{0}(t|x,x^{\prime})$ can be evaluated by using
the algebraic method developed in \cite{avramidi93,avramidi94}. First, the heat
semigroup $\exp(t{\cal D}^2)$ can be represented as an average over the
(nilpotent) Lie group (\ref{316a}) with a Gaussian measure
\begin{eqnarray}
\label{322}
\exp(t\mathcal{D}^{2})
=(4\pi t)^{-n/2}J(t)
\int\limits_{\mathbb{R}^{n}} dk\;
\exp\left\{-\frac{1}{4}k^{\mu}M_{\mu\nu}(t)k^{\nu}
+k^{\mu}\mathcal{D}_{\mu}\right\}\;,
\end{eqnarray}
where
\be
J(t)=\det\left(\frac{tiF}{\sinh(tiF)}\right)^{1/2}\;,
\label{325zz}
\ee
and $M(t)$ is a symmetric matrix defined by
\be
M(t)=iF\coth(tiF)\,.
\label{3338xzx}
\ee
We would like to stress, at this point, that here and everywhere below all the
functions of the $2$-form $F$ are analytic and should be understood in terms of
a power series in $F$.

Then by using the relation
\begin{equation}
\exp(k^{\mu}\mathcal{D}_{\mu})\delta(u)
=\delta(u+k)\;,
\end{equation}
one obtains
\begin{equation}\label{325}
U_{0}(t|x,x^{\prime})=(4\pi t)^{-n/2}J(t)
\exp\left\{-\frac{1}{4}u^{\mu}M_{\mu\nu}(t)u^{\nu}\right\}\;,
\end{equation}
which is nothing but the Schwinger kernel for an electromagnetic field on
$\RR^n$ \cite{schwinger51}.

To obtain the asymptotic expansion of the heat kernel diagonal we just
need to set $x=x'$ (or $u=0$).
At this point, we notice the following interesting fact.
The operators $t{\cal L}_k$, $tV_k(t\tau)$ and
$T_k(t)$ are differential operators with
homogeneous polynomial coefficients (in $u^\mu$)
of order $\varepsilon^k$.
Recall that $u\sim \varepsilon$, $t\sim \varepsilon^2$ and $F\sim
\varepsilon^{-2}$, so that $tF$ and $Fuu$ are counted as of order zero.
Since the zero order
heat kernel $U_0$ is Gaussian,
then the off-diagonal coefficients $b_k(t|x,x')$ are
polynomials in $u$. The point we want to make now is the following.

\begin{lemma}
The off-diagonal odd-order coefficients $b_{2k+1}$ are odd order polynomials
in $u^\mu$, that is, they satisfy
\be
b_{2k+1}(t|x,x')\Big|_{u\mapsto -u}=-b_{2k+1}(t|x,x')\,,
\ee
and, therefore, vanish on the diagonal,
\be
b^{\rm diag}_{2k+1}(t)=0\,.
\ee
\end{lemma}
\begin{proof}
We discuss the transformation properties of various quantities
under the reflection of the coordinates, $u\mapsto -u$.
First, we note that the operator
${\cal D}$ changes sign, and, therefore, the
operator ${\cal L}_0=-{\cal D}^2$ is invariant.
Next, from the general form of the operator ${\cal L}_k$ discussed above
we see that ${\cal L}_k \mapsto (-1)^k{\cal L}_k$. Therefore,
the same is true for the operator $V_k(t\tau)$, that is,
$V_k \mapsto (-1)^k V_k$.

Now, the operator $T_k(t)$ has the following general form
\be
T_k=t^k\sum_{m=1}^{[k/2]}
\int\limits_0^1 d\tau_1 \cdots \int\limits_0^{\tau_{m-1}} d\tau_m
\sum_{|{\bf j}|=k} C_{m,\, {\bf j}}
V_{j_1}(t\tau_1)\cdots V_{j_m}(t\tau_m)\,,
\ee
where the summation goes over multiindex ${\bf j}=(j_1,\dots,j_m)$
of integers $j_1,\dots,j_m\ge 2$
such that $|{\bf j}|=j_1+\cdots+j_m=k$,
and $C_{m,\, {\bf j}}$ are some numerical
coefficients.
Therefore, the operator $T_k$ transforms as $T_k \mapsto (-1)^k T_k$.

Since the zero-order heat kernel $U_0$ is invariant under the reflection
of coordinates $u\mapsto -u$, we finally find that the coefficients
$b_k$ transform according to $b_k \mapsto (-1)^k b_k$.
Thus, $b_{2k}$ are even polynomials and $b_{2k+1}$ are odd-order
polynomials.

\end{proof}

By using this lemma and by
setting $x=x'$ we obtain the asymptotic expansion of the
heat kernel diagonal
\begin{equation}\label{367}
U^{\rm diag}(t)\sim
(4\pi t)^{-n/2}J(t)\sum_{k=0}^\infty t^k b_{2k}^{\rm diag}(t)
\;,
\end{equation}
where the function $J(t)$ is defined in (\ref{325zz}).
Thus, we obtain
\be
\tilde a_k^{\rm diag}(t)=J(t)b_{2k}^{\rm diag}(t)\,.
\ee


\subsection{Algebraic Framework}

As we have shown above the evaluation of the heat semigroup is reduced
to the calculation of the operators $V_k(s)$ defined by (\ref{331iga}),
which reduces, in turn, to the computation of general expressions
\begin{equation}
\label{329}
e^{s\mathcal{D}^{2}}u^{\nu_{1}}\cdots u^{\nu_{n}}
\mathcal{D}_{\mu_{1}}\cdots\mathcal{D}_{\mu_{m}}
e^{-s\mathcal{D}^{2}}
=Z^{\nu_{1}}(s)\cdots Z^{\nu_{n}}(s)
A_{\mu_1}(s)\cdots A_{\mu_m}(s)
\;,
\end{equation}
where
\begin{equation}
\label{334}
Z^{\nu}(s)=e^{s\mathcal{D}^{2}}u^{\nu}e^{-s\mathcal{D}^{2}}\;,
\end{equation}
\begin{equation}
\label{331}
A_{\mu}(s)=e^{s\mathcal{D}^{2}}\mathcal{D}_{\mu}e^{-s\mathcal{D}^{2}}\;.
\end{equation}
Obviously, the operators $A_\mu$ and $Z_\nu$ form the algebra
\be
[A_\mu(s),Z^\nu(s)]=\delta^\nu_\mu\;,\qquad
[A_\mu(s),A_\nu(s)]=iF_{\mu\nu}\,\;,\qquad
[Z^\mu(s),Z^\nu(s)]=0\;.
\ee

The operators $A_\mu(s)$ and $Z^\nu(s)$ can be computed as follows.
First, we notice that $A_\mu(s)$ satisfies
the differential equation
\begin{equation}
\label{332}
\partial_{s}A_{\mu}(s)=\textrm{Ad}_{\mathcal{D}^{2}}A_{\mu}(s)\;,
\end{equation}
with the initial condition
\begin{displaymath}
A_{\mu}(0)=\mathcal{D}_{\mu}\;.
\end{displaymath}

Hereafter $\textrm{Ad}_{\mathcal{D}^{2}}$ is an operator acting as a commutator,
that is,
\begin{equation}
\textrm{Ad}_{\mathcal{D}^{2}}A_{\mu}(s)
\equiv[\mathcal{D}^{2},A_{\mu}(s)]\;.
\end{equation}

The solution of eq. (\ref{332}) is
\begin{equation}
A_{\mu}(s)
=\exp(s\textrm{Ad}_{\mathcal{D}^{2}})\mathcal{D}_{\mu}\;,
\end{equation}
which can be written in terms of series as
\begin{equation}\label{333}
A_{\mu}(s)=\sum_{k=0}^{\infty}\frac{s^{k}}{k!}
\left(\textrm{Ad}_{\mathcal{D}^{2}}\right)^{k}\mathcal{D}_{\mu}\;.
\end{equation}
Now, by using the algebra (\ref{316a}) we first obtain the commutator
\begin{equation}
[\mathcal{D}^{2},\mathcal{D}_{\mu}]=-2iF_{\mu\alpha}\mathcal{D}^{\alpha}\;,
\end{equation}
and then, by induction,
\begin{equation}
\left(\textrm{Ad}_{\mathcal{D}^{2}}\right)^{k}\mathcal{D}_{\mu}
=(-2i)^{k}F_{\mu\alpha_{1}}F^{\alpha_{1}}{}_{\alpha_{2}}\cdots
F^{\alpha_{k-1}\alpha_{k}}\mathcal{D}_{\alpha_{k}}
=[(-2i F)^{k}]_{\mu\alpha}\mathcal{D}^{\alpha}\;.
\end{equation}
By substituting this result in the series (\ref{333}) we finally find that
\begin{equation}\label{336}
A_{\mu}(s)=\Psi_{\mu}{}^{\alpha}(s)\mathcal{D}_{\alpha}\;,
\end{equation}
where
\begin{equation}\label{335aa}
\Psi(s)=\exp(-2s iF)\;.
\end{equation}

Similarly, for the operators $Z^\nu(s)$ we find
\begin{equation}
Z^{\mu}(s)
=\exp(s\textrm{Ad}_{\mathcal{D}^{2}})u^{\mu}
=\sum_{k=0}^{\infty}\frac{s^{k}}{k!}
\left(\textrm{Ad}_{\mathcal{D}^{2}}\right)^{k}u^{\mu}\;.
\end{equation}
Now, by using the commutators in (\ref{316a}), we find
\begin{equation}
\textrm{Ad}_{\mathcal{D}^{2}}u^{\mu}
=\left[\mathcal{D}^{2},u^{\mu}\right]=2\mathcal{D}^{\mu}\;,
\end{equation}
and then, by induction, we obtain, for $k\geq 2$,
\begin{equation}
\left(\textrm{Ad}_{\mathcal{D}^{2}}\right)^{k}u^{\mu}=2[(-2i F)^{k-1}]^{\mu\alpha}\mathcal{D}_{\alpha}\;.
\end{equation}
Thus the operator $Z^\mu(s)$ in (\ref{331}) takes the form
\begin{equation}
Z^{\mu}(s)
=u^{\mu}-2s\mathcal{D}^{\mu}
+2\sum_{k=2}^{\infty}\frac{s^{k}}{k!}
[(-2i F)^{k-1}]^{\mu\alpha}\mathcal{D}_{\alpha}\;.
\end{equation}
This series can be easily summed up to give
\begin{equation}\label{335a}
Z^{\mu}(s)=u^{\mu}+\Omega^{\mu\alpha}(s)\mathcal{D}_{\alpha}\;,
\end{equation}
where
\begin{equation}\label{335aaa}
\Omega(s)=\frac{1-\exp(-2siF)}{iF}
=2\exp(-siF)\frac{\sinh(siF)}{iF}
\;.
\end{equation}

Now, by using (\ref{335aa}) and (\ref{335aaa}) we obtain
\begin{equation}\label{3360}
\Omega^{-1}(s)=\frac{1}{2}iF\left[\coth(siF)+1\right]
=\frac{1}{2}\left[M(s)+iF\right]\;.
\end{equation}
We will need the symmetric and the antisymmetric parts of $\Omega^{-1}(s)$.
By recalling that the matrix $F$ is anti-symmetric it is easy to show
\begin{equation}
\Omega^{-1}_{(\mu\nu)}(s)=\frac{1}{2}M_{\mu\nu}(s)\;.
\end{equation}
\begin{equation}
\Omega^{-1}_{[\mu\nu]}(s)=\frac{1}{2}iF_{\mu\nu}\;,
\label{3363iga}
\end{equation}
Here and everywhere below the square brackets denote the complete
antisymmetrization over all indices included.

For the future reference we also notice that
\begin{equation}\label{3w}
\Omega^{-1}(s)\Omega^{T}(s)=\Psi^{-1}(s)=\exp(2siF)\;,
\end{equation}
Finally, we define another function
\begin{equation}
\Phi(s)=\Psi(s)\Omega^{-1}(s)
=\left(\Omega^{-1}(s)\right)^T=\frac{1}{2}\left[M(s)-iF\right]\;.
\label{3365iga}
\end{equation}
It is useful to remember that the functions $\Psi$,
$F\Omega$ and $\Phi\Omega$ are dimensionless.


\subsection{Flat Connection}

Next, we transform the operators $Z^\mu$ to define new
(time-dependent) derivative operators by
\bea
D_{\mu}(s)&=&\Omega^{-1}_{\mu\nu}(s)Z^\nu(s)
\;.
\label{336aaz}
\eea
By using the explicit form of the operators $Z^\mu$ and ${\cal D}_\mu$
we have
\bea
D_{\mu}(s)&=&
\mathcal{D}_{\mu}+\Omega^{-1}_{\mu\rho}(s)u^{\rho}
\nonumber\\
&=&\bar{\nabla}_{\mu}+\frac{1}{2}M_{\mu\rho}(s)u^\rho
\;.
\label{336aa}
\eea

Since the operators $Z^\mu$ commute, the operators $D_\mu(s)$ obviously commute
as well. In other words the connection $D_{\mu}$ is flat. Therefore, it can
also be written as
\bea
\label{33600}
D_{\mu}(s)&=&
e^{-\Theta(s)}\bar{\nabla}_{\mu}e^{\Theta(s)}\,,
\eea
where,
\begin{equation}\label{33601}
\Theta(s)=\frac{1}{4}u^{\mu}M_{\mu\nu}(s)u^{\nu}\;.
\end{equation}

Now, we can rewrite the operators
$A_\mu(s)$ and $Z^\mu(s)$
in (\ref{336}) and (\ref{335a}) in terms of the operators
$D_\mu(s)$
\begin{eqnarray}\label{336a}
A_{\mu}(s)&=&\Psi_{\mu}{}^{\alpha}(s)\left(D_{\alpha}(s)
-\Omega^{-1}_{\alpha\rho}(s)u^{\rho}\right)\;,
\nonumber\\
Z^{\mu}(s)&=&\Omega^{\mu\alpha}(s)D_{\alpha}(s)\;.
\end{eqnarray}
It is useful, for future calculations, to prove the following
\begin{lemma}
Let $D_{\mu}$ and $u^\nu$ be operators satisfying the algebra
\be
[D_\mu, u^\nu]=\delta^\nu_\mu\,,\qquad
[D_\mu,D_\nu]=[u^\mu,u^\nu]=0\,.
\label{3800}
\ee
Then
\begin{eqnarray}
\left[D_{\mu_{1}}\cdots D_{\mu_{n}},u^{\rho}\right]
&=&n\;\delta^{\rho}{}_{(\mu_{1}}D_{\mu_{2}}
\cdots D_{\mu_{n})}\;,
\label{336b}
\\
\left[D_{\mu_{1}}\cdots D_{\mu_{n}},u^{\rho}u^{\sigma}\right]&=&n(n-1)
\delta^{\rho}{}_{(\mu_{1}}
\delta^{\sigma}{}_{\mu_{2}}D_{\mu_{3}}\cdots D_{\mu_{n})}
\label{336c}
\nonumber\\
&+&2n\;u^{(\rho}\delta^{\sigma)}{}_{(\mu_{1}}D_{\mu_{2}}
\cdots D_{\mu_{n})}\;.
\end{eqnarray}
\end{lemma}

\begin{proof}
Let $\mathcal{X}(\xi)=\xi^{\mu}D_{\mu}$ and
\begin{equation}
\varphi^{\rho}(t)
=\left[e^{t\mathcal{X}(\xi)},u^{\rho}\right]
=\left(e^{t\mathcal{X}(\xi)}u^{\rho}
e^{-t\mathcal{X}(\xi)}
-u^{\rho}\right)e^{t\mathcal{X}(\xi)}\;,
\end{equation}
Then
\begin{equation}
e^{t\mathcal{X}(\xi)}u^{\rho}e^{-t\mathcal{X}(\xi)}
=\sum_{k=0}^{\infty}\frac{t^{k}}{k!}
\left(\textrm{Ad}_{\mathcal{X}(\xi)}\right)^{k}u^{\rho}\;.
\end{equation}
By using the commutation relation in (\ref{3800}) we have
\be
[\mathcal{X}(\xi),u^\rho]=\xi^\rho\;,
\ee
and, therefore,
\begin{equation}
e^{t\mathcal{X}(\xi)}u^{\rho}e^{-t\mathcal{X}(\xi)}=u^{\rho}+t\xi^{\rho}\;.
\end{equation}
Thus
\begin{equation}
\varphi^{\rho}(t)=t\xi^{\rho}e^{t\mathcal{X}(\xi)}\;.
\end{equation}
By expanding in Taylor series both sides of the last equation we obtain
\be
\sum_{k=0}^{\infty}
\frac{t^{k+1}}{(k+1)!}\xi^{\mu_{1}}\cdots\xi^{\mu_{k+1}}
\left[D_{(\mu_{1}}\cdots D_{\mu_{k+1})},u^{\rho}\right]
=\sum_{k=0}^{\infty}\frac{t^{k+1}}{k!}\xi^{\mu_{1}}
\cdots\xi^{\mu_{k+1}}\delta^{\rho}{}_{(\mu_{1}}D_{\mu_{2}}
\cdots D_{\mu_{k+1})}\;.
\ee
Now by equating the same powers of $t$ in both series we obtain the
claim (\ref{336b}).

The second relation can be proved in a similar manner. We introduce, in this
case, the following generating function
\begin{equation}
\varphi^{\rho\sigma}(t)=\left[e^{t\mathcal{X}(\xi)},u^{\rho}u^{\sigma}\right]
\;.
\end{equation}
By the same argument used in the proof of the first relation we obtain that
\begin{equation}
\varphi^{\rho\sigma}(t)
=\left[e^{t\mathcal{X}(\xi)},u^{\rho}u^{\sigma}\right]
=2t\xi^{(\rho}u^{\sigma)}e^{t\mathcal{X}(\xi)}
+t^{2}\xi^{\rho}\xi^{\sigma}\;.
\end{equation}
Now, as before, by expanding the last equation in Taylor series and equating
the same powers of $t$
we obtain the claim (\ref{336c}).
\end{proof}

\section{Evaluation of the Operator $T$}

The perturbative expansion of the operator $T$ is given by the eq. (\ref{327b}),
with the operators $T_k$ being integrals of the operators $V_k(s)$ and their
product. Thus, according to (\ref{327c})-(\ref{327d}), to compute the operator
$T$ up to the fourth order we need to compute the operators $V_2(s)$, $V_3(s)$,
$V_4(s)$ and $V_2(s_1)V_2(s_2)$.

\subsection{Second Order}

Now, by using the explicit expression for $\mathscr{L}_{2}$ given by eqs.
(\ref{316}), (\ref{317a}) and (\ref{318b}), utilizing the results of the Section
3, exploiting eqs. (\ref{336a}), (\ref{336b}) and (\ref{336c}), using eqs.
(\ref{335aa}), (\ref{335aaa}), (\ref{3w}) and (\ref{3365iga}) after some
straightforward but cumbersome calculations we obtain
\begin{eqnarray}
\label{343}
V_{2}(s)
&=&
\frac{1}{6}R
+N_{(2)}^{\sigma}D_{\sigma}
+P_{(2)}^{\gamma\delta}D_{\gamma}D_{\delta}
+W_{(2)}^{\sigma\gamma\delta}D_{\sigma}D_{\gamma}D_{\delta}
\nonumber\\
&+&
Q_{(2)}^{\rho\sigma\gamma\delta}
D_{\rho}D_{\sigma}D_\gamma D_\delta\;,
\end{eqnarray}
where
\begin{eqnarray}
N_{(2)}^{\sigma}
&=&\left(\mathcal{R}^{\mu}{}_{\alpha}
-\frac{1}{3}R^{\mu}{}_{\alpha}\right)
\Omega^{\alpha\sigma}\Phi_{\mu\eta}u^{\eta}\;,\label{344a}
\\
P_{(2)}^{\gamma\delta}&=&
\frac{1}{3}R^{\mu}{}_{\alpha}{}^\nu{}_{\beta}
\Omega^{\alpha(\gamma}\Omega^{|\beta|\delta)}\left[\Phi_{\mu\kappa}
\Phi_{\nu\sigma}u^{\kappa}u^{\sigma}
-\frac{1}{2}M_{\mu\nu}\right]
\nonumber\\
&+&
\frac{1}{24}R^{\nu}{}_{\rho}\Omega^{\rho(\gamma}
\left[\delta_{\nu}{}^{\delta)}
+7\Psi_{\nu}{}^{\delta)}\right]
-\mathcal{R}^{\nu}{}_{\beta}
\Omega^{\beta(\gamma}\Psi_{\nu}{}^{\delta)}\;,
\\
W_{(2)}^{\sigma\delta\gamma}
&=&-\frac{1}{12}
R^{\mu}{}_{\alpha}{}^{\nu}{}_{\beta}\Omega^{\alpha(\sigma}
\Omega^{|\beta|\delta}
\left[\delta_{\nu}{}^{\gamma)}
+7\Psi_{\nu}{}^{\gamma)}\right]\Phi_{\mu\kappa}u^{\kappa}\;,
\\
Q_{(2)}^{\rho\sigma\delta\gamma}&=&
\frac{1}{12}
R^{\mu}{}_{\alpha}{}^{\nu}{}_{\beta}
\Omega^{\alpha(\rho}\Omega^{|\beta|\sigma}
\Psi_{\mu}{}^{\delta}
\left[\delta_{\nu}{}^{\gamma)}
+3\Psi_{\nu}{}^{\gamma)}\right]\;.\label{344b}
\end{eqnarray}
Note that all these coefficients as well as the operators $D_\mu$ depend on the
time variable $s$. We will indicate explicitly the dependence of various
quantities on the time parameter only in the cases when it causes confusion, in
particular, when there are two time parameters.


\subsection{Third Order}

Similarly, by using the explicit expression for $\mathscr{L}_{3}$ given by
(\ref{316}), (\ref{317aa}) and (\ref{318c}), utilizing the results of the Section
3, exploiting eqs. (\ref{336a}), (\ref{336b}) and (\ref{336c}), using eqs.
(\ref{335aa}), (\ref{335aaa}), (\ref{3w}) and (\ref{3365iga}) after some
straightforward but cumbersome calculations we obtain
\begin{eqnarray}\label{348a}
V_3(s)
&=&
N_{(3)}^{\sigma}D_{\sigma}+P_{(3)}^{\sigma\rho}D_{\sigma}D_{\rho}
+W_{(3)}^{\sigma\rho\iota}D_{\sigma}D_{\rho}D_{\iota}
\nonumber\\
&&\qquad
+Q_{(3)}^{\sigma\rho\iota\epsilon}D_{\sigma}D_{\rho}D_{\iota}D_{\epsilon}
+Y_{(3)}^{\sigma\rho\iota\epsilon\kappa}D_{\sigma}
D_{\rho}D_{\iota}D_{\epsilon}D_{\kappa}\;,
\end{eqnarray}
where
\bea
\label{348b}
N_{(3)}^{\sigma}
&=&
-\frac{1}{6}\left(\nabla_{\alpha}R
+2\nabla_{\mu}\mathcal{R}^{\mu}{}_{\alpha}\right)\Omega^{\alpha\sigma}\;,
\\[5pt]
P_{(3)}^{\gamma\delta}
&=&
-\frac{1}{6}\left(\nabla^{\mu}R_{\alpha\beta}
-2\nabla_{\alpha}R^{\mu}{}_{\beta}
+4\nabla_{\alpha}\mathcal{R}^{\mu}{}_{\beta}\right)
\Omega^{\alpha(\gamma}\Omega^{|\beta|\delta)}\Phi_{\mu\kappa}u^{\kappa}\;,
\\[5pt]
W_{(3)}^{\sigma\gamma\delta}
&=&
-\frac{1}{6}\nabla_{\alpha}R^{\mu}{}_{\beta}{}^{\nu}{}_{\rho}
\Omega^{\alpha(\sigma}\Omega^{|\beta|\gamma}\Omega^{|\rho|\delta)}
\left[\Phi_{\mu\kappa}\Phi_{\nu\epsilon}u^{\kappa}u^{\epsilon}
-\frac{1}{2}M_{\mu\nu}\right]
\nonumber\\[5pt]
&+&
\frac{1}{6}\left(\nabla^{\mu}R_{\alpha\beta}
-2\nabla_{(\alpha}R^{\mu}{}_{\beta)}
+4\nabla_{(\alpha}\mathcal{R}^{\mu}{}_{\beta)}\right)
\Omega^{\alpha(\sigma}\Omega^{|\beta|\gamma}\Psi_{\mu}{}^{\delta)}\;,
\\[5pt]
Q_{(3)}^{\rho\sigma\gamma\delta}
&=&\frac{1}{3}\nabla_{\alpha}
R^{\mu}{}_{\beta}{}^{\nu}{}_{\epsilon}\Omega^{\alpha(\rho}
\Omega^{|\beta|\sigma}\Omega^{|\epsilon|\gamma}
\Psi_{\nu}^{\delta)}\Phi_{\mu\kappa}u^{\kappa}\;,
\\[5pt]
Y_{(3)}^{\rho\sigma\gamma\delta\epsilon}
&=&
-\frac{1}{6}\nabla_{(\alpha}R^{\mu}{}_{\beta}{}^{\nu}{}_{\eta)}
\Omega^{\alpha(\rho}\Omega^{|\beta|\sigma}\Omega^{|\eta|\gamma}
\Psi_{\mu}^{\delta}\Psi_{\nu}^{\epsilon)}
\;.
\label{348c}
\eea
Here again, for simplicity, we omitted the dependence of the
coefficient functions and the derivatives on the time variable
$s$.


\subsection{Fourth Order}

\subsubsection{Operator $V_4(s)$}

By taking into account the definition of $\mathscr{L}_{4}$ in (\ref{316}) by
using eqs. (\ref{318d})-(\ref{318a}),  (\ref{336a}), (\ref{336b}) and (\ref{336c}),
and the explicit form of the functions $\Psi$ and $\Omega$, we obtain
\begin{eqnarray}\label{348}
V_4(s)
&=&
P_{(4)}^{\sigma\rho}D_{\sigma}D_{\rho}
+W_{(4)}^{\sigma\rho\iota}D_{\sigma}D_{\rho}D_{\iota}
+Q_{(4)}^{\sigma\rho\iota\epsilon}D_{\sigma}D_{\rho}D_{\iota}D_{\epsilon}
\nonumber\\
&+&Y_{(4)}^{\sigma\rho\iota\epsilon\kappa}
D_{\sigma}D_{\rho}D_{\iota}D_{\epsilon}D_{\kappa}
+S_{(4)}^{\sigma\rho\iota\epsilon\kappa\lambda}
D_{\sigma}D_{\rho}D_{\iota}D_{\epsilon}D_{\kappa}D_{\lambda}
\;,
\end{eqnarray}
where
\begin{eqnarray}
P_{(4)}^{\sigma\rho}
&=&
\frac{1}{60}\left[
R_{\mu\nu}R^{\mu}{}_{\alpha}{}^{\nu}{}_{\beta}
+R_{\mu\nu\lambda\alpha}R^{\mu\nu\lambda}{}_{\beta}
-2R^{\mu}{}_{\alpha}R_{\mu\beta}\right]
\Omega^{\alpha(\rho}\Omega^{|\beta|\sigma)}
\nonumber\\
&+&
\frac{1}{40}\left[\Delta R_{\alpha\beta}
+3\nabla_{\alpha}\nabla_{\beta}R\right]\Omega^{\alpha(\rho}
\Omega^{|\beta|\sigma)}\nonumber\\
&+&
\frac{1}{4}\left[\mathcal{R}_{\mu\alpha}\mathcal{R}^{\mu}{}_{\beta}
+\nabla_{\alpha}\nabla_{\mu}\mathcal{R}^{\mu}{}_{\beta}\right]
\Omega^{\alpha(\rho}\Omega^{|\beta|\sigma)}\;,
\end{eqnarray}
\begin{eqnarray}
W_{(4)}^{\sigma\rho\iota}
&=&\frac{1}{60}
\Big[6\nabla_{\alpha}\nabla^{\mu}R_{\beta\gamma}
+15\nabla_{\alpha}\nabla_{\beta}\mathcal{R}^{\mu}{}_{\gamma}
+15R^{\mu}{}_{\alpha}{}^{\nu}{}_{\beta}\mathcal{R}_{\gamma\nu}
-9\nabla_{\alpha}\nabla_{\beta}R^{\mu}{}_{\gamma}
\nonumber\\
&-&R^{\mu}{}_{\alpha}{}^{\nu}{}_{\beta}R_{\gamma\nu}
-4R^{\mu}{}_{\nu\alpha\lambda}R^{\nu}{}_{\beta}{}^{\lambda}{}_{\gamma}\Big]
\Omega^{\alpha(\sigma}\Omega^{|\beta|\rho}\Omega^{|\gamma|\iota)}
\Phi_{\mu\xi}u^{\xi}\;,
\nonumber\\
\end{eqnarray}
\begin{eqnarray}
Q_{(4)}^{\sigma\rho\iota\epsilon}
&=&\frac{1}{300}
\left[20R^{\mu}{}_{\alpha\lambda\beta}
R^\lambda{}_{\gamma}{}^{\nu}{}_{\delta}
+15\nabla_{\alpha}\nabla_{\beta}
R^{\mu}{}_{\gamma}{}^{\nu}{}_{\delta}\right]
\Omega^{\alpha(\sigma}\Omega^{|\beta|\rho}\Omega^{|\gamma|\iota}
\Omega^{|\delta|\epsilon)}
\nonumber\\
&\times&\left[\Phi_{\mu\xi}\Phi_{\nu\varsigma}u^{\xi}u^{\varsigma}
-\frac{1}{2}M_{\mu\nu}\right]
\nonumber\\
&+&
\frac{1}{240}R^{\alpha}{}_{\nu}R^{\mu\beta\nu\gamma}
\Omega_{\alpha}{}^{(\sigma}\Omega_{\beta}{}^{\rho}\Omega_{\gamma}{}^{\iota}
\left[3\delta_{\mu}{}^{\epsilon)}
+\Psi_{\mu}{}^{\epsilon)}\right]
\nonumber\\
&+&
\frac{1}{240}R_{\lambda}{}^{\mu}{}_{\nu}{}^{\alpha}
R^{\lambda\beta\nu\gamma}
\Omega_{\alpha}{}^{(\sigma}\Omega_{\beta}{}^{\rho}\Omega_{\gamma}{}^{\iota}
\left[3\delta_{\mu}{}^{\epsilon)}
+13\Psi_{\mu}{}^{\epsilon)}\right]
\nonumber\\
&-&
\frac{1}{24}\mathcal{R}_{\nu}{}^{\alpha}
R^{\mu\beta\nu\gamma}
\Omega_{\alpha}{}^{(\sigma}\Omega_{\beta}{}^{\rho}\Omega_{\gamma}{}^{\iota}
\left[\delta_{\mu}{}^{\epsilon)}
+5\Psi_{\mu}{}^{\epsilon)}\right]
\nonumber\\
&+&
\frac{1}{20}\left[3\nabla^{\alpha}\nabla^{\beta}
R^{\mu\gamma}-2\nabla^{\alpha}\nabla^{\mu}R^{\beta\gamma}
-5\nabla^{\alpha}\nabla^{\beta}\mathcal{R}^{\mu\gamma}\right]
\Omega_{\alpha}{}^{(\sigma}\Omega_{\beta}{}^{\rho}\Omega_{\gamma}{}^{\iota}
\Psi_{\mu}{}^{\epsilon)}\;,
\nonumber\\
\end{eqnarray}
\begin{eqnarray}
Y_{(4)}^{\sigma\rho\iota\epsilon\kappa}
&=&
-\frac{1}{10}\nabla^{\alpha}\nabla^{\beta}
R^{\mu\gamma\nu\delta}
\Omega_{\alpha}{}^{(\sigma}\Omega_{\beta}{}^{\rho}
\Omega_{\gamma}{}^{\iota}
\Omega_{\delta}{}^{\epsilon}
\Psi_{\nu}{}^{\kappa)}\Phi_{\mu\xi}u^{\xi}
\nonumber\\
&-&
\frac{1}{120}R_{\lambda}{}^{\alpha\mu\beta}
R^{\lambda\gamma\nu\delta}
\Omega_{\alpha}{}^{(\sigma}\Omega_{\beta}{}^{\rho}
\Omega_{\gamma}{}^{\iota}
\Omega_{\delta}{}^{\epsilon}
\left[3\delta_{\nu}{}^{\kappa)}
+13\Psi_{\nu}{}^{\kappa)}\right]\Phi_{\mu\xi}u^{\xi}\;,
\nonumber\\
\end{eqnarray}
\begin{eqnarray}
S_{(4)}^{\sigma\rho\iota\epsilon\kappa\lambda}
&=&
\frac{1}{20}\nabla^{\alpha}\nabla^{\beta}
R^{\mu\gamma\nu\delta}\Omega_{\alpha}{}^{(\sigma}
\Omega_{\beta}{}^{\rho}\Omega_{\gamma}{}^{\iota}\Omega_{\delta}{}^{\epsilon}
\Psi_{\mu}{}^{\kappa}\Psi_{\nu}{}^{\lambda)}
\nonumber\\
&+&
\frac{1}{2880}R^{\eta\alpha\mu\beta}
R_{\eta}{}^{\gamma\nu\delta}
\Omega_{\alpha}{}^{(\sigma}\Omega_{\beta}{}^{\rho}
\Omega_{\gamma}{}^{\iota}\Omega_{\delta}{}^{\epsilon}
\Big[62\Psi_{(\mu}{}^{\kappa}\delta_{\nu)}{}^{\lambda)}
\nonumber\\
&+&
125\Psi_{\mu}{}^{\kappa}\Psi_{\nu}{}^{\lambda)}
+5\delta_{\mu}{}^{\kappa}\delta_{\nu}{}^{\lambda)}\Big]\;.
\end{eqnarray}

\subsubsection{Operator $V_2(s_1)V_2(s_2)$}

Next, we need to compute the product of two operators $V_2(s)$ depending on
different times $s_1$ and $s_2$ by using the eq. (\ref{343}). To simplify the
notation we denote the derivatives $D_\mu(s_k)$ depending on different times
$s_k$ simply by $D^{(k)}_\mu$. To present the product $V_2(s_1)V_2(s_2)$ in the
``normal'' form we need to move all derivative operators $D^{(1)}_\mu$ to the
right and all coordinates $u^\nu$ to the left. In order to perform this task we
need the commutator of the derivative operator $D^{(1)}_{\mu}$ with the
coefficients of the operator $V_{2}(s_2)$. First, by using the commutators
found earlier we obtain the relevant commutators
\begin{eqnarray}
\label{350}
\left[D^{(1)}_{\mu_{1}}\cdots D^{(1)}_{\mu_{n}},
N^{\iota}_{(2)}(s_2)\right]
&=&
n f^{\iota}{}_{(\mu_{1}}(s_2)
D^{(1)}_{\mu_{2}}\cdots D^{(1)}_{\mu_{n})}\;,
\\
\label{351}
\left[D^{(1)}_{\mu_{1}}\cdots D^{(1)}_{\mu_{n}},
P_{(2)}^{\iota\eta}(s_2)\right]
&=&
n(n-1)
g^{\iota\eta}{}_{(\mu_{1}\mu_{2}}(s_2)
D^{(1)}_{\mu_{3}}\cdots D^{(1)}_{\mu_{n})}
\nonumber\\
&&
+nh^{\iota\eta}{}_{(\mu_{1}}(s_2)
D^{(1)}_{\mu_{2}}\cdots D^{(1)}_{\mu_{n})}\;,
\label{352}
\\
\left[D^{(1)}_{\mu_{1}}\cdots D^{(1)}_{\mu_{n}},
W_{(2)}^{\iota\eta\kappa}(s_2)\right]
&=&
np^{\iota\eta\kappa}{}_{(\mu_{1}}(s_2)
D^{(1)}_{\mu_{2}}\cdots D^{(1)}_{\mu_{n})}\;,
\end{eqnarray}
where
\begin{eqnarray}
f^{\iota}{}_{\lambda}
&=&
\left(\mathcal{R}^{\mu}{}_{\beta}
-\frac{1}{3}R^{\mu}{}_{\beta}\right)\Omega^{\beta\iota}\Phi_{\mu\lambda}\;,
\\
g^{\iota\eta}{}_{\lambda\kappa}
&=&
\frac{1}{3}
R^{\mu}{}_{(\alpha}{}^{\nu}{}_{\beta)}
\Omega^{\alpha\iota}\Omega^{\beta\eta}\Phi_{\mu\lambda}\Phi_{\nu\kappa}\;,
\nonumber\\
h^{\iota\eta}{}_{\lambda}
&=&
\frac{2}{3}R^{\mu}{}_{(\alpha}{}^{\nu}{}_{\beta)}
\Omega^{\alpha\iota}\Omega^{\beta\eta}\Phi_{\mu\kappa}
\Phi_{\nu\lambda}u^{\kappa}\;,
\\
p^{\iota\eta\kappa}{}_{\lambda}
&=&
-\frac{1}{12}R^{\mu}{}_{\alpha}{}^{\nu}{}_{\beta}
\Omega^{\alpha(\iota}\Omega^{|\beta|\eta}
\left[
\delta_{\nu}{}^{\kappa)}
+7\Psi_{\nu}{}^{\kappa)}\right]\Phi_{\mu\lambda}\;.
\end{eqnarray}

Next, by using the expression for the operator $V_2(s)$ in
(\ref{343}) and the non-vanishing
commutators in (\ref{350})-(\ref{352}) we obtain
\begin{eqnarray}\label{353}
V_{2}(s_{1})V_{2}(s_{2})
&=&\frac{1}{36}R^{2}
+\frac{1}{6}R\;\Big[V_{2}(s_{1})
+V_{2}(s_{2})\Big]
+L(s_{1},s_{2})\;,
\end{eqnarray}
where
\begin{equation}\label{353a}
L(s_{1},s_{2})
=\sum_{k=1}^{4}\sum_{n=0}^{4}
C_{(n,k)}^{\mu_{1}\cdots\mu_{n}\nu_{1}\cdots\nu_{k}}(s_{1},s_{2})
D_{\mu_{1}}^{(1)}\cdots D_{\mu_{n}}^{(1)}D_{\nu_{1}}^{(2)}\cdots D_{\nu_{k}}^{(2)}\;,
\end{equation}
and
\begin{eqnarray}
\label{354a}
C_{(0,1)}^{\rho}
&=&
N^{\alpha}_{(2)}(s_{1})f^{\rho}{}_{\alpha}(s_{2})\;,
\nonumber\\
C_{(1,1)}^{\alpha\rho}
&=&
2N^{\alpha}_{(2)}(s_1)N^{\rho}_{(2)}(s_2)
+2P_{(2)}^{\iota\alpha}(s_1)f^{\rho}{}_{\iota}(s_2)\;,
\nonumber\\
C_{(2,1)}^{\alpha\beta\rho}
&=&
2P^{\alpha\beta}_{(2)}(s_1)N^{\rho}_{(2)}(s_2)
+3W_{(2)}^{\kappa\alpha\beta}(s_1)f^{\rho}{}_{\kappa}(s_2)\;,
\nonumber\\
C_{(3,1)}^{\alpha\beta\gamma\rho}
&=&
2W_{(2)}^{\alpha\beta\gamma}(s_1)N^{\rho}_{(2)}(s_2)
+4Q_{(2)}^{\lambda\alpha\beta\gamma}(s_1)f^{\rho}{}_{\lambda}(s_2)\;,
\nonumber\\
C_{(4,1)}^{\alpha\beta\gamma\delta\rho}
&=&
2Q_{(2)}^{\alpha\beta\gamma\delta}(s_1)N^{\rho}_{(2)}(s_2)\;,
\end{eqnarray}
\begin{eqnarray}
C_{(0,2)}^{\rho\sigma}
&=&
N^{\alpha}_{(2)}(s_1)h^{\rho\sigma}{}_{\alpha}(s_2)
+2P_{(2)}^{\alpha\beta}(s_1)g^{\rho\sigma}{}_{\alpha\beta}(s_2)\;,
\nonumber\\
C_{(1,2)}^{\alpha\rho\sigma}
&=&
2N^{\alpha}_{(2)}(s_1)P_{(2)}^{\rho\sigma}(s_2)
+2P_{(2)}^{\alpha\beta}(s_1)h^{\rho\sigma}{}_{\beta}(s_2)
+6W_{(2)}^{\alpha\beta\gamma}(s_1)g^{\rho\sigma}{}_{\beta\gamma}(s_2)\;,
\nonumber\\
C_{(2,2)}^{\alpha\beta\rho\sigma}
&=&
2P^{\alpha\beta}_{(2)}(s_1)P_{(2)}^{\rho\sigma}(s_2)
+3W_{(2)}^{\alpha\beta\gamma}(s_1)h^{\rho\sigma}{}_{\gamma}(s_2)
+12Q_{(2)}^{\alpha\beta\gamma\delta}(s_1)g^{\rho\sigma}{}_{\gamma\delta}(s_2)\;,
\nonumber\\
C_{(3,2)}^{\alpha\beta\gamma\rho\sigma}
&=&
2W^{\alpha\beta\gamma}_{(2)}(s_1)P_{(2)}^{\rho\sigma}(s_2)
+4Q_{(2)}^{\alpha\beta\gamma\delta}(s_1)h^{\rho\sigma}{}_{\delta}(s_2)\;,
\nonumber\\
C_{(4,2)}^{\alpha\beta\gamma\delta\rho\sigma}
&=&
2Q_{(2)}^{\alpha\beta\gamma\delta}(s_1)P_{(2)}^{\rho\sigma}(s_2)\;,
\end{eqnarray}
\begin{eqnarray}
C_{(0,3)}^{\rho\sigma\upsilon}
&=&
N^{\alpha}_{(2)}(s_1)p^{\rho\sigma\upsilon}{}_{\alpha}(s_2)\;,
\nonumber\\
C_{(1,3)}^{\alpha\rho\sigma\upsilon}
&=&
2N^{\alpha}_{(2)}(s_1)W^{\rho\sigma\upsilon}_{(2)}(s_2)
+2P_{(2)}^{\mu\alpha}(s_1)p^{\rho\sigma\upsilon}{}_{\mu}(s_2)\;,
\nonumber\\
C_{(2,3)}^{\alpha\beta\rho\sigma\upsilon}
&=&
2P^{\alpha\beta}_{(2)}(s_1)W^{\rho\sigma\upsilon}_{(2)}(s_2)
+3W_{(2)}^{\mu\alpha\beta}(s_1)p^{\rho\sigma\upsilon}{}_{\mu}(s_2)\;,
\nonumber\\
C_{(3,3)}^{\alpha\beta\gamma\rho\sigma\upsilon}
&=&
2W^{\alpha\beta\gamma}_{(2)}(s_1)W^{\rho\sigma\upsilon}_{(2)}(s_2)
+4Q_{(2)}^{\mu\alpha\beta\gamma}(s_1)p^{\rho\sigma\upsilon}{}_{\mu}(s_2)\;,
\nonumber\\
C_{(4,3)}^{\alpha\beta\gamma\delta\rho\sigma\upsilon}
&=&
2Q^{\alpha\beta\gamma\delta}_{(2)}(s_1)W^{\rho\sigma\upsilon}_{(2)}(s_2)\;,
\end{eqnarray}
\begin{eqnarray}\label{354b}
C_{(1,4)}^{\alpha\rho\sigma\upsilon\chi}
&=&
N^{\alpha}_{(2)}(s_1)Q^{\rho\sigma\upsilon\chi}_{(2)}(s_2)\;,
\nonumber\\
C_{(2,4)}^{\alpha\beta\rho\sigma\upsilon\chi}
&=&
P^{\alpha\beta}_{(2)}(s_1)Q^{\rho\sigma\upsilon\chi}_{(2)}(s_2)\;,
\nonumber\\
C_{(3,4)}^{\alpha\beta\gamma\rho\sigma\upsilon\chi}
&=&
W^{\alpha\beta\gamma}_{(2)}(s_1)Q^{\rho\sigma\upsilon\chi}_{(2)}(s_2)\;,
\nonumber\\
C_{(4,4)}^{\alpha\beta\gamma\delta\rho\sigma\upsilon\chi}
&=&
Q^{\alpha\beta\gamma\delta}_{(2)}(s_1)Q^{\rho\sigma\upsilon\chi}_{(2)}(s_2)\;.
\end{eqnarray}

\section{Generalized Hermite Polynomials}

Thus, we reduced the calculation of the asymptotic expansion of the heat kernel
to the calculation of the derivatives $D_{\mu}(s)$ of the zero order heat
kernel $U_{0}(t|x,x^{\prime})$ given by (\ref{325}). The needed derivatives of
the zero order heat kernel can be expressed in terms of the following symmetric
tensors
\begin{equation}
\label{358}
\mathcal{H}_{\mu_{1}\cdots\mu_{n}}(s)
=U_{0}^{-1}(t|x,x^{\prime})D_{\mu_{1}}(s)\cdots
D_{\mu_{n}}(s)U_{0}(t|x,x^{\prime})\;,
\end{equation}
and
\begin{equation}
\label{363b}
\Xi_{\nu_{1}\cdots\nu_{m}\mu_{1}\cdots\mu_{n}}(s_1,s_2)
=U_{0}^{-1}(t|x,x^{\prime})D_{\nu_{1}}^{(1)}
\cdots D_{\nu_{m}}^{(1)}D_{\mu_{1}}^{(2)}\cdots
D_{\mu_{n}}^{(2)}U_{0}(t|x,x^{\prime})\;,
\end{equation}
where we denoted as before $D^{(k)}_\mu=D_\mu(s_k)$.

We recall that the derivatives $D^{(1)}_\mu$ and $D^{(2)}_\nu$ do not commute!
Also, $U_0$ is a scalar function that depends on $x$ and $x'$ only through the
normal coordinates $u^\mu$. The derivative operator $D_\mu(s)$ is defined by
(\ref{336aa}), and, when acting on a scalar function is equal to
\bea
D_\mu(s)&=&\frac{\partial}{\partial u^\mu}
+\frac{1}{2}M_{\mu\nu}(s)u^\nu
\nonumber\\
&=&
e^{-\Theta(s)}\frac{\partial}{\partial u^\mu}e^{\Theta(s)}\,,
\eea
where the tensor $M_{\mu\nu}(s)$ is defined by (\ref{3338xzx})
and the function $\Theta(s)$ is a quadratic form defined by (\ref{33601}).

Therefore, by using the explicit form of the zero order heat kernel
(\ref{325}) we see that
the tensors $\mathcal{H}_{\mu_{1}\cdots\mu_{n}}(s)$
can be written in the form
\begin{equation}
\label{358xzx}
\mathcal{H}_{\mu_{1}\cdots\mu_{n}}(s)
=
\exp\{\Theta(t)-\Theta(s)\}
\frac{\partial}{\partial u^{\mu_1}}\cdots
\frac{\partial}{\partial u^{\mu_n}}
\exp\{\Theta(s)-\Theta(t)\}
\;,
\end{equation}

The tensors $\mathcal{H}_{\mu_{1}\cdots\mu_{n}}(s)$ are polynomials in $u^\mu$.
They differ from the usual Hermite polynomials of several variables (see, for
example, \cite{bateman53}) by some normalization. That is why, we call them
just Hermite polynomials. The generating function for Hermite polynomials
\be
{\cal H}(\xi,s)=\sum_{n=0}^\infty \frac{1}{n!}\xi^{\mu_1}
\cdots\xi^{\mu_n}{\cal H}_{\mu_1\dots\mu_n}(s)\,,
\ee
can be computed as follows
\bea
{\cal H}(\xi,s)
&=&
\exp{\left\{\Theta(t)-\Theta(s)\right\}}
\exp\left(\xi^\mu\frac{\partial}{\partial u^\mu}\right)
\exp{\left\{\Theta(s)-\Theta(t)\right\}}
\nonumber\\
&=&
\exp\left\{\frac{1}{2}\xi^{\alpha}\Lambda_{\alpha\beta}(s)
\left[\xi^{\beta}+2u^{\sigma}\right]\right\}\;,
\eea
where
\begin{eqnarray}
\label{3590}
\Lambda(s)&=&\frac{1}{2}\Big[M(s)-M(t)\Big]
\nonumber\\[5pt]
&=&
\frac{1}{2}\frac{iF}{\sinh(tiF)}\frac{\sinh[(t-s)iF]}{\sinh(siF)}
\;.
\end{eqnarray}

By expanding the exponent in $\xi$ we obtain the Hermite
polynomials explicitly. They can be read off from the expression
\begin{equation}
\label{361z}
\xi^{\mu_{1}}\cdots\xi^{\mu_{n}}\mathcal{H}_{\mu_{1}\cdots\mu_{n}}(s)
=\sum_{k=0}^{\left[\frac{n}{2}\right]}
\frac{(2k)!}{2^{k}k!}{n\choose 2k}
\left(\xi^\alpha\Lambda_{\alpha\beta}(s)\xi^\beta\right)^{k}
\left(\xi^\rho\Lambda_{\rho\sigma}(s)u^\sigma\right)^{n-2k}\;.
\end{equation}
For convenience some low-order Hermite polynomials are given explicitly
in tensorial form in the next section.

Similarly, the tensors
$\Xi_{\nu_{1}\cdots\nu_{m}\mu_{1}\cdots\mu_{n}}(s_1,s_2)$ can be written in the
form
\bea
&&\Xi_{\nu_{1}\cdots\nu_{m}\mu_{1}\cdots\mu_{n}}(s_1,s_2)
=
\exp\left[\Theta(t)-\Theta(s_1)\right]
\\[5pt]
&&
\qquad\times
\frac{\partial}{\partial u^{\nu_1}}\cdots
\frac{\partial}{\partial u^{\nu_m}}
\exp\left[\Theta(s_1)-\Theta(s_2)\right]
\frac{\partial}{\partial u^{\mu_1}}\cdots
\frac{\partial}{\partial u^{\mu_n}}
\exp\left[\Theta(s_2)-\Theta(t)\right]\;.
\nonumber
\eea
They are obviously polynomial in $u^\mu$ as well. We call them
Hermite polynomials of second kind.
The generating function for these polynomials is defined by
\begin{equation}\label{363c}
\Xi(\xi,\eta,s_1,s_2)=\sum_{m,n=0}^{\infty}\frac{1}{m!n!}
\xi^{\nu_{1}}\cdots\xi^{\nu_{m}}
\eta^{\mu_{1}}\cdots\eta^{\mu_{n}}\Xi_{\nu_{1}
\cdots\nu_{m}\mu_{1}\cdots\mu_{n}}(s_1,s_2)\;,
\end{equation}
and can be computed as follows
\bea
&&\Xi(\xi,\eta,s_1,s_2)
=\exp\left\{\Theta(t)-\Theta(s_{1})\right\}
\exp\left(\xi^\mu\frac{\partial}{\partial u^\mu}\right)
\exp\left\{\Theta(s_{1})-\Theta(s_{2})\right\}
\nonumber\\
&&\qquad\times
\exp\left(\eta^\nu\frac{\partial}{\partial u^\nu}\right)
\exp\left\{\Theta(s_{2})-\Theta(t)\right\}\;,
\\
&&\qquad
=
\label{363d}
\exp\left\{
\frac{1}{2}\xi^\alpha\Lambda_{\alpha\beta}(s_{1})(\xi^\beta+2u^\beta)
+\frac{1}{2}\eta^\mu\Lambda_{\mu\nu}(s_{2})(\eta^\nu+2u^\nu)
+\xi^\rho\Lambda_{\rho\sigma}(s_{2})\eta^\sigma\right\}\;.
\nonumber
\eea

Notice that
\be
\Xi(\xi,\eta,s_1,s_2)={\cal H}(\xi,s_1){\cal H}(\eta,s_2)
\exp\left\{
\xi^\rho\Lambda_{\rho\sigma}(s_{2})\eta^\sigma\right\}\;.
\ee
This enables one to express all Hermite polynomials of second kind
$\Xi_{(n)}(s_1,s_2)$
in terms of the Hermite polynomials
${\cal H}_{(m)}(s_1)$, ${\cal H}_{(l)}(s_2)$, and the matrix
$\Lambda(s_2)$.
Namely, they can be read off from the expression
\bea
&&
\xi^{\nu_{1}}\cdots\xi^{\nu_{m}}
\eta^{\mu_{1}}\cdots\eta^{\mu_{n}}\Xi_{\nu_{1}
\cdots\nu_{m}\mu_{1}\cdots\mu_{n}}(s_1,s_2)
=
\sum_{k=0}^{\min(m,n)}
k!{m\choose k}{n\choose k}
\label{3613iga}
\\
&&
\qquad
\times
\xi^{\nu_{1}}\cdots\xi^{\nu_{m-k}}\mathcal{H}_{\nu_{1}\cdots\nu_{m-k}}(s_1)
\eta^{\mu_{1}}\cdots\eta^{\mu_{n-k}}\mathcal{H}_{\mu_{1}\cdots\mu_{n-k}}(s_2)
\left(\xi^\rho\Lambda_{\rho\sigma}(s_2)\eta^\sigma\right)^{k}\;.
\nonumber
\eea

\subsection{Calculation of Hermite Polynomials}

The Hermite polynomials are defined by
\bea
\mathcal{H}_{\mu_{1}\cdots\mu_{n}}
&=&\exp\left\{-\frac{1}{2}u^\alpha \Lambda_{\alpha\beta}u^\beta\right\}
\frac{\partial}{\partial u^{\mu_1}}\cdots
\frac{\partial}{\partial u^{\mu_n}}
\exp\left\{\frac{1}{2}u^\alpha \Lambda_{\alpha\beta}u^\beta\right\}
\nonumber\\
&=&
\left(\frac{\partial}{\partial u^{\mu_1}}+\Lambda_{\mu_1\nu_1}
u^{\nu_1}\right)\cdots
\left(\frac{\partial}{\partial u^{\mu_n}}+\Lambda_{\mu_n\nu_n}
u^{\nu_n}\right)
\cdot 1\,.
\eea
They can be computed explicitly as follows.
First, let
\begin{equation}\label{359aa}
\mathcal{H}_{(n)}(\xi)
=\xi^{\mu_{1}}\cdots\xi^{\mu_{n}}\mathcal{H}_{\mu_{1}\cdots\mu_{n}}\;,
\end{equation}
and
\begin{equation}\label{359c}
B=\xi^{\mu}\frac{\partial}{\partial u^\mu}\;,\qquad
A=\xi^\mu\Lambda_{\mu\nu}u^\nu\;.
\end{equation}
Then
\begin{equation}\label{360aa}
\mathcal{H}_{(n)}(\xi)
=(A+B)^n\cdot 1\;.
\end{equation}

Finally, let
\be
C=[B,A]=\xi^\mu\Lambda_{\mu\nu}\xi^\nu\,.
\ee
Obviously, the operators $A$, $B$, $C$ form the Heisenberg algebra
\begin{displaymath}
[B,A]=C\;,\quad [A,C]=[B,C]=0\;.
\end{displaymath}

\begin{lemma}\label{lemma1}
There holds,
\begin{equation}\label{360}
(A+B)^{n}=\sum_{k=0}^{\left[\frac{n}{2}\right]}
\sum_{m=0}^{n-2k}\frac{(2k)!}{2^{k}k!}
{n\choose 2k}
\;{n-2k \choose m}
C^{k}A^{n-2k-m}B^{m}\;.
\end{equation}
\end{lemma}
\begin{proof}
Notice that $e^{t(A+B)}$ is the generating functional for $(A+B)^{n}$.
Now, by using the Baker-Hausdorff-Campbell formula
\begin{displaymath}
e^{t(A+B)}=e^{\frac{t^{2}}{2}C}e^{tA}e^{tB}\;,
\end{displaymath}
expanding both sides in $t$ and
computing the Taylor coefficients of the right hand side
we obtain the eq. (\ref{360}).
\end{proof}

By using this result we
obtain an explicit expression for (\ref{360aa})
\begin{equation}
\label{361}
\mathcal{H}_{(n)}(\xi)=
\xi^{\mu_{1}}\cdots\xi^{\mu_{n}}\mathcal{H}_{\mu_{1}\cdots\mu_{n}}
=\sum_{k=0}^{\left[\frac{n}{2}\right]}
\frac{n!}{2^{k}k!(n-2k)!}C^{k}A^{n-2k}\;.
\end{equation}
By setting $A=0$ we immediately obtain the (diagonal)
values of Hermite polynomials
at $u=0$
\bea
\left[\mathcal{H}_{\mu_{1}\cdots\mu_{2n+1}}\right]^{\rm diag}&=&0\,,
\\
\left[\mathcal{H}_{\mu_{1}\cdots\mu_{2n}}\right]^{\rm diag}
&=&\frac{(2n)!}{2^n n!}
\Lambda_{(\mu_{1}\mu_{2}}\cdots\Lambda_{\mu_{2n-1}\mu_{2n})}
\,.
\eea

We list below a few low order Hermite polynomials needed for our calculation
\begin{eqnarray}
\mathcal{H}_{(0)}&=&1\;,
\\[10pt]
\mathcal{H}_{\mu_{1}}
&=&\Lambda_{\mu_{1}\alpha}u^{\alpha}\;,
\\[10pt]
\mathcal{H}_{\mu_{1}\mu_{2}}
&=&\Lambda_{(\mu_{1}\mu_{2})}
+\Lambda_{\mu_{1}\alpha}
\Lambda_{\mu_{1}\beta}u^{\alpha}u^{\beta}\;,
\\[10pt]
\mathcal{H}_{\mu_{1}\mu_{2}\mu_{3}}
&=&3\Lambda_{(\mu_{1}\mu_{2}}
\Lambda_{\mu_{3})\alpha}u^{\alpha}
+\Lambda_{\mu_{1}\alpha}
\Lambda_{\mu_{2}\beta}\Lambda_{\mu_{3}\gamma}
u^{\alpha}u^{\beta}u^{\gamma}\;,
\\[10pt]
\mathcal{H}_{\mu_{1}\mu_{2}\mu_{3}\mu_{4}}
&=&3\Lambda_{(\mu_{1}\mu_{2}}\Lambda_{\mu_{3}\mu_{4})}
+3\Lambda_{(\mu_{1}\mu_{2}}
\Lambda_{\mu_{3}|\alpha|}\Lambda_{\mu_{4})\beta}u^{\alpha}u^{\beta}
\nonumber\\
&+&\Lambda_{\mu_{1}\alpha}
\Lambda_{\mu_{2}\beta}\Lambda_{\mu_{3}\gamma}
\Lambda_{\mu_{4}\delta}u^{\alpha}u^{\beta}u^{\gamma}u^{\delta}\;,
\end{eqnarray}
\begin{eqnarray}
\mathcal{H}_{\mu_{1}\mu_{2}\mu_{3}\mu_{4}\mu_{5}}
&=&15\Lambda_{(\mu_{1}\mu_{2}}
\Lambda_{\mu_{3}\mu_{4}}\Lambda_{\mu_{5})\alpha}u^{\alpha}
+5\Lambda_{(\mu_{1}\mu_{2}}
\Lambda_{\mu_{3}|\alpha|}\Lambda_{\mu_{4}|\beta|}
\Lambda_{\mu_{5})\gamma}u^{\alpha}u^{\beta}u^{\gamma}
\nonumber\\
&+&\Lambda_{\mu_{1}\alpha}
\Lambda_{\mu_{2}\beta}\Lambda_{\mu_{3}\gamma}
\Lambda_{\mu_{4}\delta}\Lambda_{\mu_{5}\eta}
u^{\alpha}u^{\beta}u^{\gamma}u^{\delta}u^{\eta}\;,
\\[10pt]
\mathcal{H}_{\mu_{1}\mu_{2}\mu_{3}\mu_{4}\mu_{5}\mu_{6}}
&=&15\Lambda_{(\mu_{1}\mu_{2}}
\Lambda_{\mu_{3}\mu_{4}}\Lambda_{\mu_{5}\mu_{6})}
+45\Lambda_{(\mu_{1}\mu_{2}}
\Lambda_{\mu_{3}\mu_{4}}\Lambda_{\mu_{5}|\alpha|}
\Lambda_{\mu_{6})\beta}u^{\alpha}u^{\beta}
\nonumber\\
&+&
15\Lambda_{(\mu_{1}\mu_{2}}
\Lambda_{\mu_{3}|\alpha|}\Lambda_{\mu_{4}|\beta|}
\Lambda_{\mu_{5}|\gamma|}\Lambda_{\mu_{6})\delta}
u^{\alpha}u^{\beta}u^{\gamma}u^{\delta}
\nonumber\\
&+&
\Lambda_{(\mu_{1}|\alpha|}
\Lambda_{\mu_{2}|\beta|}\Lambda_{\mu_{3}|\gamma|}
\Lambda_{\mu_{4}|\delta|}\Lambda_{\mu_{5}|\eta|}
\Lambda_{\mu_{6})\iota}u^{\alpha}u^{\beta}u^{\gamma}
u^{\delta}u^{\eta}u^{\iota}\;.
\end{eqnarray}

We list below some of the generalized Hermite polynomials of second kind.
Now we have two sets of Hermite polynomials that depend on the
quadratic forms $\Lambda$ at two different times, $s_1$ and
$s_2$.
Let us define
\begin{equation}
\mathcal{H}_{(n)}(s_{1})=
\xi^{\mu_{1}}\cdots\xi^{\mu_{n}}\mathcal{H}_{\mu_{1}\cdots\mu_{n}}(s_{1})\;,
\end{equation}
\begin{equation}
\mathcal{H}_{(n)}(s_{2})=
\eta^{\mu_{1}}\cdots\eta^{\mu_{n}}\mathcal{H}_{\mu_{1}\cdots\mu_{n}}(s_{2})\;,
\end{equation}
and
\begin{equation}
\Lambda(s_{2})=\xi^{\alpha}\Lambda_{\alpha\beta}(s_{2})\eta^{\beta}\;.
\end{equation}
Then from eq. (\ref{3613iga}) we obtain
the quantities $\Xi_{(m,n)}$ that we need in our calculations
\begin{eqnarray}
\Xi_{(0,1)}(s_{1},s_{2})
&=&\mathcal{H}_{(1)}(s_{2})\;,
\\
\Xi_{(1,1)}(s_{1},s_{2})
&=&\Lambda(s_{2})
+\mathcal{H}_{(1)}(s_{1})\mathcal{H}_{(1)}(s_{2})\;,
\\
\Xi_{(2,1)}(s_{1},s_{2})
&=&2\Lambda(s_{2})\mathcal{H}_{(1)}(s_{1})
+\mathcal{H}_{(1)}(s_{2})\mathcal{H}_{(2)}(s_{1})\;,
\\
\Xi_{(3,1)}(s_{1},s_{2})
&=&3\Lambda(s_{2})\mathcal{H}_{(2)}(s_{1})
+\mathcal{H}_{(1)}(s_{2})\mathcal{H}_{(3)}(s_{1})\;,
\\
\Xi_{(4,1)}(s_{1},s_{2})
&=&4\Lambda(s_{2})\mathcal{H}_{(3)}(s_{1})
+\mathcal{H}_{(1)}(s_{2})\mathcal{H}_{(4)}(s_{1})\;,
\end{eqnarray}
\begin{eqnarray}
\Xi_{(0,2)}(s_{1},s_{2})
&=&\mathcal{H}_{(2)}(s_{2})
\\
\Xi_{(1,2)}(s_{1},s_{2})
&=&2\Lambda(s_{2})\mathcal{H}_{(1)}(s_{2})
+\mathcal{H}_{(2)}(s_{2})\mathcal{H}_{(1)}(s_{1})\;,
\\
\Xi_{(2,2)}(s_{1},s_{2})
&=&2\Lambda^{2}(s_{2})
+4\Lambda(s_{2})\mathcal{H}_{(1)}(s_{2})\mathcal{H}_{(1)}(s_{1})
+\mathcal{H}_{(2)}(s_{2})\mathcal{H}_{(2)}(s_{1})\;,
\\
\Xi_{(3,2)}(s_{1},s_{2})
&=&6\Lambda^{2}(s_{2})\mathcal{H}_{(1)}(s_{1})
+6\Lambda(s_{2})\mathcal{H}_{(1)}(s_{2})\mathcal{H}_{(2)}(s_{1})
\nonumber\\
&+&\mathcal{H}_{(2)}(s_{2})\mathcal{H}_{(3)}(s_{1})\;,
\\
\Xi_{(4,2)}(s_{1},s_{2})
&=&12\Lambda^{2}(s_{2})\mathcal{H}_{(2)}(s_{1})
+8\Lambda(s_{2})\mathcal{H}_{(1)}(s_{2})\mathcal{H}_{(3)}(s_{1})
\nonumber\\
&+&\mathcal{H}_{(2)}(s_{2})\mathcal{H}_{(4)}(s_{1})\;,
\end{eqnarray}
\begin{eqnarray}
\Xi_{(0,3)}(s_{1},s_{2})
&=&\mathcal{H}_{(3)}(s_{2})
\\
\Xi_{(1,3)}(s_{1},s_{2})
&=&3\Lambda(s_{2})\mathcal{H}_{(2)}(s_{2})
+\mathcal{H}_{(3)}(s_{2})\mathcal{H}_{(1)}(s_{1})\;,
\\
\Xi_{(2,3)}(s_{1},s_{2})
&=&6\Lambda^{2}(s_{2})\mathcal{H}_{(1)}(s_{2})
+6\Lambda(s_{2})\mathcal{H}_{(2)}(s_{2})\mathcal{H}_{(1)}(s_{1})
\nonumber\\
&+&\mathcal{H}_{(3)}(s_{2})\mathcal{H}_{(2)}(s_{1})\;,
\\
\Xi_{(3,3)}(s_{1},s_{2})
&=&6\Lambda^{3}(s_{2})
+18\Lambda^{2}(s_{2})\mathcal{H}_{(1)}(s_{2})\mathcal{H}_{(1)}(s_{1})
\nonumber\\
&+&
9\Lambda(s_{2})\mathcal{H}_{(2)}(s_{2})\mathcal{H}_{(2)}(s_{1})
+\mathcal{H}_{(3)}(s_{2})\mathcal{H}_{(3)}(s_{1})\;,
\\
\Xi_{(4,3)}(s_{1},s_{2})
&=&24\Lambda^{3}(s_{2})\mathcal{H}_{(1)}(s_{1})
+36\Lambda^{2}(s_{2})\mathcal{H}_{(1)}(s_{2})\mathcal{H}_{(2)}(s_{1})
\nonumber\\
&+&
12\Lambda(s_{2})\mathcal{H}_{(2)}(s_{2})\mathcal{H}_{(3)}(s_{1})
+\mathcal{H}_{(3)}(s_{2})\mathcal{H}_{(4)}(s_{1})\;,
\end{eqnarray}
\begin{eqnarray}
\Xi_{(0,4)}(s_{1},s_{2})
&=&\mathcal{H}_{(4)}(s_{2})\;,
\\
\Xi_{(1,4)}(s_{1},s_{2})
&=&4\Lambda(s_{2})\mathcal{H}_{(3)}(s_{2})
+\mathcal{H}_{(4)}(s_{2})\mathcal{H}_{(1)}(s_{1})\;,
\\
\Xi_{(2,4)}(s_{1},s_{2})
&=&12\Lambda^{2}(s_{2})\mathcal{H}_{(2)}(s_{2})
+8\Lambda(s_{2})\mathcal{H}_{(3)}(s_{2})\mathcal{H}_{(1)}(s_{1})
\nonumber\\
&+&\mathcal{H}_{(4)}(s_{2})\mathcal{H}_{(2)}(s_{1})\;,
\\
\Xi_{(3,4)}(s_{1},s_{2})
&=&24\Lambda^{3}(s_{2})\mathcal{H}_{(1)}(s_{2})
+36\Lambda^{2}(s_{2})\mathcal{H}_{(2)}(s_{2})\mathcal{H}_{(1)}(s_{1})
\nonumber\\
&+&
12\Lambda(s_{2})\mathcal{H}_{(3)}(s_{2})\mathcal{H}_{(2)}(s_{1})
+\mathcal{H}_{(4)}(s_{2})\mathcal{H}_{(3)}(s_{1})\;,
\\
\Xi_{(4,4)}(s_{1},s_{2})
&=&24\Lambda^{4}(s_{2})
+96\Lambda^{3}(s_{2})\mathcal{H}_{(1)}(s_{2})\mathcal{H}_{(1)}(s_{1})
+72\Lambda^{2}(s_{2})\mathcal{H}_{(2)}(s_{2})\mathcal{H}_{(2)}(s_{1})
\nonumber\\
&+&16\Lambda(s_{2})\mathcal{H}_{(3)}(s_{2})\mathcal{H}_{(3)}(s_{1})
+\mathcal{H}_{(4)}(s_{2})\mathcal{H}_{(4)}(s_{1})\;.
\end{eqnarray}

The coincidence limit of the quantities $\Xi_{(m,n)}$ ,with $m+n$ odd, vanishes identically
\be
\left[\Xi_{(m,n)}(s_{1},s_{2})\right]^{\rm diag}=0\,,
\qquad \mbox{if $(m+n)$ is odd}\,.
\ee
By recalling the coincidence limits of the Hermite polynomials we obtain the following
\begin{eqnarray}
\left[\Xi_{(1,1)}(s_{1},s_{2})\right]^{\rm diag}
&=&\Lambda(s_{2})\;,
\\
\left[\Xi_{(3,1)}(s_{1},s_{2})\right]^{\rm diag}
&=&3\Lambda(s_{1})\Lambda(s_{2})\;,
\\
\left[\Xi_{(0,2)}(s_{1},s_{2})\right]^{\rm diag}
&=&\Lambda(s_{2})\;,
\\
\left[\Xi_{(2,2)}(s_{1},s_{2})\right]^{\rm diag}
&=&\Lambda(s_{1})\Lambda(s_{2})
+2\Lambda^{2}(s_{2})\;,
\\
\left[\Xi_{(4,2)}(s_{1},s_{2})\right]^{\rm diag}
&=&3\Lambda^{2}(s_{1})\Lambda(s_{2})
+12\Lambda(s_{1})\Lambda^{2}(s_{2})\;,
\\
\left[\Xi_{(1,3)}(s_{1},s_{2})\right]^{\rm diag}
&=&3\Lambda^{2}(s_{2})\;,
\\
\left[\Xi_{(3,3)}(s_{1},s_{2})\right]^{\rm diag}
&=&9\Lambda(s_{1})\Lambda^{2}(s_{2})
+6\Lambda^{3}(s_{2})\;,
\\
\left[\Xi_{(2,4)}(s_{1},s_{2})\right]^{\rm diag}
&=&3\Lambda(s_{1})\Lambda^{2}(s_{2})
+12\Lambda^{3}(s_{2})\;,
\\
\left[\Xi_{(4,4)}(s_{1},s_{2})\right]^{\rm diag}
&=&9\Lambda^{2}(s_{1})\Lambda^{2}(s_{2})
+72\Lambda(s_{1})\Lambda^{3}(s_{2})
+24\Lambda^{4}(s_{2})\;.
\end{eqnarray}

\section{Off-diagonal Coefficients $b_k$}

By using the machinery developed above, we can now write the coefficients of
the asymptotic expansion of the heat kernel in terms of generalized Hermite
polynomials. We define the following quantity
\bea
b_{2, (1)}(t|x,x')
&=&
\int\limits_{0}^{1}d\tau\Big[N_{(2)}^{\sigma}(t\tau)\mathcal{H}_{\sigma}(t\tau)
+P_{(2)}^{\gamma\delta}(t\tau)\mathcal{H}_{\gamma\delta}(t\tau)
+W_{(2)}^{\sigma\gamma\delta}(t\tau)\mathcal{H}_{\sigma\gamma\delta}(t\tau)
\nonumber\\
&&
+Q_{(2)}^{\rho\sigma\gamma\delta}(t\tau)
\mathcal{H}_{\rho\sigma\gamma\delta}(t\tau)\Big]\;.
\eea
Then, by referring to the formulas (\ref{343}), (\ref{348a}), (\ref{348}) and
(\ref{353}) and by using the following formula for multiple integrals
\begin{equation}
\int\limits_{a}^{b}d\tau_{n}
\int\limits_{a}^{\tau_{n}}d\tau_{n-1}
\cdots\int\limits_{a}^{\tau_{2}}d\tau_1 f(\tau_1)
=\frac{1}{(n-1)!}\int\limits_{a}^{b}d\tau\;(b-\tau)^{n-1}f(\tau)\,,
\end{equation}
we obtain
\bea
\label{363}
b_{2}(t|x,x^{\prime})
&=&\frac{1}{6}R+b_{2, (1)}(t|x,x')\,,
\eea
\begin{eqnarray}\label{363a}
b_{3}(t|x,x^{\prime})
&=&t^{-1/2}\int\limits_{0}^{1}d\tau\Big[
N_{(3)}^{\sigma}(t\tau)\mathcal{H}_{\sigma}(t\tau)
+P_{(3)}^{\gamma\delta}(t\tau)\mathcal{H}_{\gamma\delta}(t\tau)
+W_{(3)}^{\sigma\gamma\delta}(t\tau)\mathcal{H}_{\sigma\gamma\delta}(t\tau)
\nonumber\\
&&+
Q_{(3)}^{\rho\sigma\gamma\delta}(t\tau)
\mathcal{H}_{\rho\sigma\gamma\delta}(t\tau)
+Y_{(3)}^{\iota\rho\sigma\gamma\delta}(t\tau)
\mathcal{H}_{\iota\rho\sigma\gamma\delta}(t\tau)\Big]\;,
\end{eqnarray}
\begin{eqnarray}
\label{364}
b_{4}(t|x,x^{\prime})
&=&
\frac{1}{72}R^{2}
+\frac{1}{6}R b_{2, (1)}(t|x,x')
\nonumber\\
&&
+t^{-1}\int\limits_{0}^{1}d\tau\Big[
P^{\iota\epsilon}_{(4)}(t\tau)\mathcal{H}_{\iota\epsilon}(t\tau)
+W^{\iota\epsilon\kappa}_{(4)}(t\tau)
\mathcal{H}_{\iota\epsilon\kappa}(t\tau)
\nonumber\\
&&
+Q^{\iota\epsilon\kappa\lambda}_{(4)}(t\tau)
\mathcal{H}_{\iota\epsilon\kappa\lambda}(t\tau)
+Y^{\iota\epsilon\kappa\lambda\eta}_{(4)}(t\tau)
\mathcal{H}_{\iota\epsilon\kappa\lambda\eta}(t\tau)
+S^{\iota\epsilon\kappa\lambda\eta\gamma}_{(4)}(t\tau)
\mathcal{H}_{\iota\epsilon\kappa\lambda\eta\gamma}(t\tau)\Big]
\nonumber\\
&&+
\sum_{k=1}^{4}\sum_{n=0}^{4}\int\limits_{0}^{1}d\tau_{2}
\int\limits_{0}^{\tau_{2}}d\tau_{1}\;
C_{(n,k)}^{\mu_{1}\cdots\mu_{n}\nu_{1}\cdots\nu_{k}}(t\tau_{1},t\tau_{2})
\Xi_{\mu_{1}\cdots\mu_{n}\nu_{1}\cdots\nu_{k}}(t\tau_{1},t\tau_{2})\;.
\nonumber\\
&&
\end{eqnarray}

\section{Diagonal Coefficients $b_k$}

In order to obtain the diagonal values $b_k^{\rm diag}(t)$ of the coefficients
$b_k(t|x,x')$ we just need to set $u=0$ in eqs. (\ref{363}), (\ref{363a}) and
(\ref{364}). For the rest of this section we will employ the usual convention of
denoting the coincidence limit by square brackets, that is,
\begin{equation}
\left[f(u)\right]^{\rm diag}=f(0).
\end{equation}

By inspection of the equation defining the generalized Hermite polynomials,
one can easily notice that, in the coincidence limit, all the ones
with an odd number of indices vanish identically, namely
\begin{equation}
\label{369}
\left[\mathcal{H}_{\mu_{1}\cdots\mu_{2n+1}}\right]^{\rm diag}=0\;.
\end{equation}
By using the last remark we have the following expression for the
coincidence limit of (\ref{363}), i.e.
\begin{equation}
\label{370}
b^{\rm diag}_{2}(t)
=\frac{1}{6}R
+b^{\rm diag}_{2, (1)}(t)\,,
\ee
where
\be
b^{\rm diag}_{2, (1)}(t)
=\int\limits_{0}^{1}d\tau\left[P_{(2)}^{\gamma\delta}(t\tau)
\mathcal{H}_{\gamma\delta}(t\tau)
+Q_{(2)}^{\rho\sigma\gamma\delta}(t\tau)
\mathcal{H}_{\rho\sigma\gamma\delta}(t\tau)\right]^{\rm diag}\;.
\end{equation}
By using the explicit form of the coefficients $P_{(2)}$,
$Q_{(2)}$ and the generalized Hermite polynomials,
we obtain
\begin{equation}
b^{\rm diag}_{2, (1)}(t)
=J_{(1)}{}^{\alpha\beta}{}_{\mu\nu}(t)
R^{\mu}{}_{\alpha}{}^{\nu}{}_{\beta}
+J_{(2)}{}^{\mu\nu}(t)R_{\mu\nu}
+J_{(3)}{}^{\mu\nu}(t)\mathcal{R}_{\mu\nu}\;,
\end{equation}
where
\begin{eqnarray}
J_{(1)}{}^{\alpha\beta}{}_{\mu\nu}(t)
&=&
\int\limits_{0}^{1}d\tau\;
\Bigg\{-\frac{1}{6}
\Omega^{\alpha\gamma}\Omega^{\beta\delta}
M_{\mu\nu}\Lambda_{\gamma\delta}
\nonumber\\
&&
+\frac{1}{4}\left(
\delta_{\nu}{}^{\gamma}
+3\Psi_{\nu}{}^{\gamma}\right)
\Omega^{\alpha\rho}\Omega^{\beta\sigma}
\Psi_{\mu}{}^{\delta}
\Lambda_{(\rho\sigma}\Lambda_{\delta\gamma)}\Bigg\}\;,
\\
J_{(2)}{}^{\mu\nu}(t)
&=&
\frac{1}{24}\int\limits_{0}^{1}d\tau
\left(\delta^{(\nu}{}_{\delta}
+7\Psi^{(\nu}{}_{\delta}\right)
\Omega^{\mu)\gamma}
\Lambda_{\gamma}{}^{\delta}\;,
\\
J_{(3)}{}^{\mu\nu}(t)
&=&
\int\limits_{0}^{1}d\tau\;
\Omega^{[\mu}{}_{\gamma}\Psi^{\nu]\delta}\Lambda^{\gamma}{}_{\delta}\;.
\label{370b}
\end{eqnarray}
Here all functions in the integrals depend on $t\tau$.

Next, we introduce the following matrices
\begin{equation}
\label{371}
\mathcal{A}(s)
=\Omega(s)\Lambda(s)
=\frac{1}{2}\frac{\exp[(t-2s)iF]-\exp(-tiF)}{\sinh(tiF)}\;,
\end{equation}
\begin{equation}\label{371b}
\mathcal{B}(s)
=\Omega(s)\Lambda(s)\Omega(s)^{T}
=\frac{\coth(tiF)}{iF}
-\frac{\cosh[(t-2s)iF]}{iF\sinh(tiF)}\;,
\end{equation}
\bea
\Gamma(s)
&=&\Omega^{-1}(s)
-\frac{1}{4}\Psi(s)\Lambda(s)
-\frac{3}{4}\Lambda(s)
\nonumber\\
&&
=\frac{1}{8}\left(3iF\coth(tiF)
+\frac{iF}{\sinh(tiF)}\cosh[(t-2s)iF]\right)\;.
\eea
Then, by using the relation
\begin{equation}\label{371a}
\Omega(s)\Lambda(s)\Psi(s)^{T}
=\Omega^{T}(s)\Lambda(s)=\mathcal{A}^{T}(s)\;,
\end{equation}
we obtain
\bea
\label{372c}
J_{(1)}{}^{\alpha\beta}{}_{\mu\nu}(t)
&=&\int\limits_{0}^{1}d\tau
\Bigg\{
-\frac{1}{3}\mathcal{B}^{\alpha\beta}(t\tau)\Gamma_{(\mu\nu)}(t\tau)
\nonumber\\
&&
+\frac{1}{6}\left(\mathcal{A}_{\mu}{}^{(\alpha}(t\tau)
\mathcal{A}^{\beta)}{}_{\nu}(t\tau)
+3\mathcal{A}_{(\mu}{}^{\alpha}(t\tau)
\mathcal{A}_{\nu)}{}^{\beta}(t\tau)\right)\Bigg\}\;,
\\
\label{372a}
J_{(2)}{}^{\mu\nu}(t)
&=&\frac{1}{3}\int\limits_{0}^{1}
d\tau\mathcal{A}^{(\mu\nu)}(t\tau)
=\frac{1}{6}\delta^{\mu\nu}\;,
\\
\label{372b}
J_{(3)}{}^{\mu\nu}(t)
&=&-\int\limits_{0}^{1}d\tau\mathcal{A}^{[\mu\nu]}(t\tau)
=
-\frac{1}{2}\left(\frac{1}{tiF}-\coth(tiF)\right)^{[\mu\nu]}\;.
\eea
Unfortunately the integral $J_{(1)}{}^{\alpha\beta}{}_{\mu\nu}$
cannot be computed explicitly, in general.

As we already mentioned above all odd order coefficients $b_{2k+1}$ have zero
diagonal values. We see this directly for the coefficient $b_3$, which is given
by (\ref{363a}). That is, by recalling the formulas in (\ref{348b}) through
(\ref{348c}) and the remark (\ref{369}) we have
\begin{equation}\label{372f}
b^{\rm diag}_{3}(t)=0\;.
\end{equation}

Finally, we evaluate the diagonal values of fourth order coefficient $b_4$
given by (\ref{364}). It can be written as follows
\begin{equation}\label{372}
b^{\rm diag}_{4}(t)
=\frac{1}{72}R^{2}
+\frac{1}{6}Rb^{\rm diag}_{2, (1)}(t)
+b^{\rm diag}_{4, (2)}(t)
+b^{\rm diag}_{4, (3)}(t)\;.
\end{equation}
By noticing that for odd $n+k$, the diagonal values of the coefficients
$C_{(n,k)}$ vanish,
\begin{equation}
\left[C_{(n,k)}^{\mu_{1}\cdots\mu_{n}\nu_{1}\cdots\nu_{k}}
\right]^{\rm diag}=0\;,
\end{equation}
and by using the explicit form of Hermite polynomials and the generating
function (\ref{363d}) we obtain
\begin{eqnarray}\label{374}
b^{\rm diag}_{4, (2)}(t)
&=&t^{-1}\int\limits_{0}^{1}d\tau
\bigg\{P^{\iota\epsilon}_{(4)}(t\tau)\Lambda_{\iota\epsilon}(t\tau)
+3\left[Q^{\iota\epsilon\kappa\lambda}_{(4)}(t\tau)\right]^{\rm diag}
\Lambda_{(\iota\epsilon}(t\tau)\Lambda_{\kappa\lambda)}(t\tau)
\nonumber\\
&&
+15S^{\iota\epsilon\kappa\lambda\eta\gamma}_{(4)}(t\tau)
\Lambda_{(\iota\epsilon}\Lambda_{\kappa\lambda}(t\tau)
\Lambda_{\eta\gamma)}(t\tau)\bigg\}\;,
\end{eqnarray}
\begin{eqnarray}\label{375}
b^{\rm diag}_{4, (3)}(t)
&=&
\int\limits_{0}^{1}d\tau_{2}\int\limits_{0}^{\tau_{2}}d\tau_{1}
\bigg\{2\Big[P_{(2)}^{\iota\alpha}(\tau_{1})\Big]^{\rm diag}
f^{\rho}{}_{\iota}(\tau_{2})\Lambda^{(2)}_{\alpha\rho}
+2\Big[P_{(2)}^{\alpha\beta}(\tau_{1})\Big]^{\rm diag}
g^{(\rho\sigma)}{}_{\alpha\beta}(\tau_{2})\Lambda^{(2)}_{\rho\sigma}
\nonumber\\
&+&
12Q_{(2)}^{\lambda\alpha\beta\gamma}(\tau_{1})
f^{\rho}{}_{\lambda}(\tau_{2})
\Lambda^{(1)}_{\alpha\beta}\Lambda^{(2)}_{\gamma\rho}
+\Big(2\Big[P^{\alpha\beta}_{(2)}(\tau_{1})\Big]^{\rm diag}
\Big[P_{(2)}^{\rho\sigma}(\tau_{2})\Big]^{\rm diag}
\nonumber\\[4pt]
&+&
12Q_{(2)}^{\alpha\beta\gamma\delta}(\tau_{1})
g^{(\rho\sigma)}{}_{\gamma\delta}(\tau_{2})\Big)
\left(\Lambda^{(1)}_{\alpha\beta}\Lambda^{(2)}_{\rho\sigma}
+2\Lambda^{(2)}_{\alpha\rho}\Lambda^{(2)}_{\beta\sigma}\right)
\nonumber\\[4pt]
&+&
6\Big[P_{(2)}^{\mu\alpha}(\tau_{1})\Big]^{\rm diag}
p^{(\rho\sigma\nu)}{}_{\mu}(\tau_{2})
\Lambda^{(2)}_{\alpha\rho}\Lambda^{(2)}_{\sigma\nu}
\nonumber\\[4pt]
&+&
2Q_{(2)}^{\alpha\beta\gamma\delta}(\tau_{1})
\Big[P_{(2)}^{\rho\sigma}(\tau_{2})\Big]^{\rm diag}
\left(3\Lambda^{(1)}_{\alpha\beta}
\Lambda^{(1)}_{\gamma\delta}\Lambda^{(2)}_{\rho\sigma}
+12\Lambda^{(1)}_{\alpha\beta}
\Lambda^{(2)}_{\gamma\rho}\Lambda^{(2)}_{\delta\sigma}\right)
\nonumber\\[4pt]
&+&
4Q_{(2)}^{\mu\alpha\beta\gamma}(\tau_{1})
p^{(\rho\sigma\nu)}{}_{\mu}(\tau_{2})
\left(9\Lambda^{(1)}_{\alpha\beta}
\Lambda^{(2)}_{\gamma\rho}
\Lambda^{(2)}_{\sigma\nu}
+6\Lambda^{(2)}_{\alpha\rho}
\Lambda^{(2)}_{\beta\sigma}\Lambda^{(2)}_{\gamma\nu}\right)
\nonumber\\[4pt]
&+&
\Big[P^{\alpha\beta}_{(2)}(\tau_{1})\Big]^{\rm diag}
Q^{\rho\sigma\nu\chi}_{(2)}(\tau_{2})
\left(3\Lambda^{(1)}_{\alpha\beta}
\Lambda^{(2)}_{\rho\sigma}\Lambda^{(2)}_{\nu\chi}
+12\Lambda^{(2)}_{\alpha\rho}
\Lambda^{(2)}_{\beta\sigma}\Lambda^{(2)}_{\nu\chi}\right)
\nonumber\\[4pt]
&+&
Q^{\alpha\beta\gamma\delta}_{(2)}(\tau_{1})
Q^{\rho\sigma\nu\chi}_{(2)}(\tau_{2})
\Big(9\Lambda^{(1)}_{\alpha\beta}
\Lambda^{(1)}_{\gamma\delta}\Lambda^{(2)}_{\rho\sigma}
\Lambda^{(2)}_{\nu\chi}
+72\Lambda^{(1)}_{\alpha\beta}\Lambda^{(2)}_{\rho\gamma}
\Lambda^{(2)}_{\sigma\delta}
\Lambda^{(2)}_{\nu\chi}
\nonumber\\[4pt]
&+&
24\Lambda^{(2)}_{\alpha\rho}
\Lambda^{(2)}_{\beta\sigma}
\Lambda^{(2)}_{\gamma\nu}\Lambda^{(2)}_{\delta\chi}\Big)\bigg\}\;,
\end{eqnarray}
where the superscript on the matrix $\Lambda$ denotes its dependence on
either $t\tau_{1}$ or $t\tau_{2}$.

We see that the scalar curvature appears only in the term $b^{\rm diag}_{2,
(1)}(t)$. Now, the term $b^{\rm diag}_{4, (2)}(t)$ only contains derivatives of
the curvature and quantities which are quadratic in the curvature with some of
their indices contracted. It has the following form
\begin{eqnarray}
b^{\rm diag}_{4, (2)}(t)
&=&\frac{1}{60}
B_{\alpha\beta}(t)R_{\mu\nu\lambda}{}^{\alpha}
R^{\mu\nu\lambda\beta}
+\mathrm{A}^{(1)}_{\lambda\alpha\gamma\beta}(t)
R_{\mu}{}^{\lambda}{}_{\nu}{}^{\alpha}
R^{\mu\gamma\nu\beta}
+\mathrm{A}^{(2)}_{\alpha\mu\beta\gamma\nu\delta}(t)
R_{\eta}{}^{\alpha\mu\beta}
R^{\eta\gamma\nu\delta}
\nonumber\\
&+&\frac{1}{60}
B_{\alpha\beta}(t)R_{\mu\nu}
R^{\mu\alpha\nu\beta}
+\mathrm{A}^{(3)}_{\alpha\mu\beta\gamma}(t)R^{\alpha}{}_{\nu}
R^{\mu\beta\nu\gamma}
-\frac{1}{30}B_{\alpha\beta}(t)R_{\mu}{}^{\alpha}R^{\mu\beta}
\nonumber\\
&+&
\mathrm{A}^{(4)}_{\alpha\mu\beta\gamma}(t)
\mathcal{R}_{\nu}{}^{\alpha}R^{\mu\beta\nu\gamma}
+\frac{1}{4}B_{\alpha\beta}(t)\mathcal{R}_{\nu}{}^{\alpha}
\mathcal{R}^{\nu\beta}
\nonumber\\
&+&
\mathrm{A}^{(5)}_{\alpha\beta\mu\gamma\nu\delta}(t)
\nabla^{\alpha}\nabla^{\beta}R^{\mu\gamma\nu\delta}
+
\mathrm{A}^{(6)}_{\alpha\beta\mu\nu}(t)
\nabla^{\alpha}\nabla^{\beta}R^{\mu\nu}
+\frac{1}{40}B_{\alpha\beta}(t)\Delta R^{\alpha\beta}
\nonumber\\
&+&
\frac{3}{40}B_{\alpha\beta}(t)\nabla^{\alpha}\nabla^{\beta}R\;
+\mathrm{A}^{(7)}_{\alpha\beta\mu\nu}(t)
\nabla^{\alpha}\nabla^{\beta}\mathcal{R}^{\mu\nu}
+\frac{1}{4}B_{\alpha\beta}(t)\nabla^{\alpha}\nabla_{\mu}
\mathcal{R}^{\mu\beta}\;.
\end{eqnarray}
Here the tensors $a^{(i)}(s)$ are functions that only depend
on $F$ (but not on the Riemann curvature) defined by
\begin{eqnarray}
a^{(1)}{}_{\lambda\alpha\gamma\beta}(s)
&=&
\frac{3}{80}\mathcal{B}_{(\alpha\gamma}\mathcal{A}_{\beta)\lambda}
+\frac{13}{80}\mathcal{B}_{(\alpha\gamma}\mathcal{A}_{|\lambda|\beta)}\;,
\\
a^{(2)}{}_{\alpha}{}^{\mu}{}_{\beta\gamma}{}^{\nu}{}_{\delta}(s)
&=&
\frac{1}{480}\mathcal{B}_{\alpha(\beta}\mathcal{B}_{\gamma\delta)}
\left(31(\Psi\Lambda)^{(\mu\nu)}
+65\Lambda^{\mu\nu}\right)
-\frac{1}{10}M^{\mu\nu}\mathcal{B}_{\alpha(\beta}
\mathcal{B}_{\gamma\delta)}
\nonumber\\
&+&\frac{187}{480}\mathcal{B}_{\alpha(\beta}
\mathcal{A}_{\gamma}{}^{(\mu}\mathcal{A}^{\nu)}{}_{\delta)}
+\frac{31}{240}\mathcal{B}_{(\beta\gamma}
\mathcal{A}^{(\nu}{}_{\delta)}\mathcal{A}_{\alpha}{}^{\mu)}
\nonumber\\
&+&\frac{25}{96}
\mathcal{B}_{(\beta\gamma}
\mathcal{A}^{(\nu}{}_{\delta)}\mathcal{A}^{\mu)}{}_{\alpha}
+\frac{1}{96}\left(\mathcal{B}_{\alpha(\beta}
\mathcal{A}_{\gamma}{}^{(\mu}\mathcal{A}_{\delta)}{}^{\nu)}
+\mathcal{B}_{(\beta\gamma}
\mathcal{A}_{\delta)}{}^{(\nu}\mathcal{A}_{\alpha}{}^{\mu)}\right)\;,\;\;\;\;\;\;\;\;\;\;
\\
a^{(3)}{}_{\alpha\mu\beta\gamma}(s)
&=&\frac{3}{80}\mathcal{B}_{(\alpha\beta}\mathcal{A}_{\gamma)\mu}
+\frac{1}{80}\mathcal{B}_{(\alpha\beta}\mathcal{A}_{|\mu|\gamma)}\;,
\\
a^{(4)}{}_{\alpha\mu\beta\gamma}(s)
&=&
-\frac{1}{8}\mathcal{B}_{(\alpha\beta}
\mathcal{A}_{\gamma)\mu}
-\frac{5}{8}\mathcal{B}_{(\alpha\beta}\mathcal{A}_{|\mu|\gamma)}\;,
\\
a^{(5)}{}_{\alpha\beta}{}^{\mu}{}_{\gamma}{}^{\nu}{}_{\delta}(s)
&=&
-\frac{3}{40}\mathcal{B}_{\alpha(\beta}
\mathcal{B}_{\gamma\delta)}M^{\mu\nu}(t)
+\frac{3}{10}\mathcal{B}_{\alpha(\beta}
\mathcal{A}^{(\mu}{}_{\gamma}\mathcal{A}^{\nu)}{}_{\delta)}
\nonumber\\
&+&\frac{3}{10}\mathcal{B}_{(\beta\gamma}
\mathcal{A}^{(\mu}{}_{\delta)}\mathcal{A}^{\nu)}{}_{\alpha}\;,
\end{eqnarray}
\begin{eqnarray}
a^{(6)}{}_{\alpha\beta\mu\nu}(s)
&=&\frac{9}{20}\mathcal{B}_{(\alpha\beta}
\mathcal{A}_{|\mu|\nu)}
-\frac{3}{10}\mathcal{B}_{\mu(\alpha}
\mathcal{A}_{\beta\nu)}\;,\\
a^{(7)}{}_{\alpha\beta\mu\nu}(s)
&=&
-\frac{3}{4}\mathcal{B}_{(\alpha\beta}\mathcal{A}_{|\mu|\nu)}\;.
\end{eqnarray}
All functions here are evaluated at the time $s$
(unless  specified otherwise).

The term $b^{\rm diag}_{4, (3)}(t)$ only contains quantities which are
quadratic in the curvature with none of their indices contracted.
It has the form
\begin{eqnarray}
b^{\rm diag}_{4, (3)}(t)&=&
\mathrm{D}^{(1)}_{\alpha\beta\mu\nu\gamma\delta\rho\sigma}(t)
R^{\alpha\beta\mu\nu}R^{\gamma\delta\rho\sigma}
+
\mathrm{D}^{(2)}_{\mu\nu\alpha\beta\rho\sigma}(t)
R^{\mu\nu}R^{\alpha\beta\rho\sigma}
+\mathrm{D}^{(3)}_{\mu\nu\alpha\beta}(t)
R^{\mu\nu}R^{\alpha\beta}
\nonumber\\
&+&
\mathrm{D}^{(4)}_{\mu\nu\alpha\beta\rho\sigma}(t)
\mathcal{R}^{\mu\nu}R^{\alpha\beta\rho\sigma}
+
\mathrm{D}^{(5)}_{\mu\nu\alpha\beta}(t)
\mathcal{R}^{\mu\nu}R^{\alpha\beta}
+
\mathrm{D}^{(6)}_{\mu\nu\alpha\beta}(t)
\mathcal{R}^{\mu\nu}\mathcal{R}^{\alpha\beta}
\;,
\end{eqnarray}
where $\mathrm{D}^{(i)}_{\mu_{1}\cdots\mu_{n}}(t)$ are some tensor-valued
functions that depend on $tF$.
They have the form
\begin{equation}
\mathrm{D}^{(i)}_{\mu_{1}\cdots\mu_{n}}(t)
=\int\limits_{0}^{1}d\tau_{2}\int\limits_{0}^{\tau_{2}}
d\tau_{1}\;d^{(i)}_{\mu_{1}\cdots\mu_{n}}(t\tau_{1},t\tau_{2})\;.
\end{equation}


To describe our results for the tensors $d^{(k)}$
we define new tensors
\begin{eqnarray}
\mathcal{E}_{(p)\mu}{}^{\nu}
&=&
\delta_{\mu}{}^{\nu}+p\;\Psi_{\mu}{}^{\nu}\;,
\\
\label{376}
\mathcal{S}_{\alpha\beta\rho\sigma\iota\kappa}
&=&
\mathcal{B}_{\beta\sigma}\Phi_{\alpha\iota}
\Phi_{\rho\kappa}
-\mathcal{A}_{\beta(\iota}\mathcal{A}_{|\sigma|\kappa)}
M_{\alpha\rho}
\nonumber\\
&-&\frac{3}{4}\Omega_{\beta}{}^{(\eta}\Omega
_{\sigma}{}^{\chi}{}\mathcal{E}_{(1)\rho}{}^{\epsilon)}
\Phi_{\alpha\iota}\Lambda_{\kappa\eta}\Lambda_{\chi\epsilon}
+\frac{3}{2}\Omega_{\beta}{}^{(\epsilon}
\Omega_{\sigma}{}^{\lambda}\Psi_{\alpha}{}^{\eta}\mathcal{E}_{(3)\rho}{}^{\chi)}
\Lambda_{\iota\epsilon}\Lambda_{\kappa\lambda}
\Lambda_{\eta\chi}\;,\;\;\;\;
\end{eqnarray}
\begin{eqnarray}
\mathcal{V}_{\gamma\delta\rho\sigma\iota\kappa\eta\chi}
(t\tau_{1},t\tau_{2})
&=&
\Lambda_{\eta\chi}(t\tau_{1})
\Big(\mathcal{B}_{\delta\sigma}\Phi_{\gamma\iota}
\Phi_{\rho\kappa}\Big)(t\tau_{2})
+2\Big(\mathcal{A}_{\delta\iota}\mathcal{A}_{\sigma\kappa}
\Phi_{\gamma\eta}\Phi_{\rho\chi}\Big)(t\tau_{2})
\nonumber\\
&-&\frac{1}{4}\Big(\Lambda_{\iota\kappa}\Lambda_{\eta\chi}\Big)
(t\tau_{1})\Big(\mathcal{B}_{\delta\sigma}M_{\gamma\rho}\Big)
(t\tau_{2})
\nonumber\\
&-&\Lambda_{\iota\kappa}(t\tau_{1})
\Big(\mathcal{A}_{\delta(\chi}\mathcal{A}_{|\sigma|\eta)}
M_{\gamma\rho}\Big)(t\tau_{2})
\nonumber\\
&-&\frac{3}{4}\left\{\Lambda_{\kappa\eta}(t\tau_{1})
\Big(\Lambda_{\chi\epsilon}\Lambda_{\omega\tau}\big)
(t\tau_{2})+\frac{2}{3}\Lambda_{\kappa\epsilon}(t\tau_{1})
\Big(\Lambda_{\eta\omega}
\Lambda_{\chi\tau}\Big)(t\tau_{2})\right\}
\nonumber\\
&\times&\Big(\Omega_{\delta}{}^{(\epsilon}\Omega_{\sigma}
{}^{\omega}{}\mathcal{E}_{(1)\rho}{}^{\tau)}\Phi_{\gamma\iota}\Big)(t\tau_{2})
\nonumber\\
&+&\frac{3}{16}\Bigg\{\Big(\Lambda_{\iota\kappa}
\Lambda_{\eta\chi}\Big)(t\tau_{1})
\Big(\Lambda_{\epsilon\tau}\Lambda_{\omega\lambda}\Big)(t\tau_{2})
+8\Lambda_{\iota\kappa}(t\tau_{1})\Big(\Lambda_{\epsilon\eta}
\Lambda_{\tau\chi}\Lambda_{\omega\lambda}\Big)(t\tau_{2})
\nonumber\\
&+&\frac{8}{3}\Big(\Lambda_{\iota\epsilon}
\Lambda_{\kappa\tau}\Lambda_{\eta\omega}
\Lambda_{\chi\lambda}\Big)(t\tau_{2})\Bigg\}
\left(\Omega_{\delta}{}^{(\epsilon}\Omega_{\sigma}{}^{\tau}
\Psi_{\gamma}{}^{\omega}\mathcal{E}_{(3)\rho}
{}^{\lambda)}\right)(t\tau_{2})\;.
\end{eqnarray}

Then the tensors $d^{(k)}$ have the form
\begin{eqnarray}
d^{(1)}_{\alpha\beta\mu\nu\gamma\delta\rho\sigma}(t\tau_{1},t\tau_{2})
&=&
-\frac{1}{9}\Big(\Omega_{\beta}{}^{(\iota}\Omega_{\nu}{}^{\kappa)}
M_{\alpha\mu}\Big)(t\tau_{1})\Big(\mathcal{B}_{\delta\sigma}
\Phi_{\gamma\iota}\Phi_{\rho\kappa}\Big)(t\tau_{2})
\nonumber\\
&+&\frac{1}{9}\Big(\mathcal{B}_{\beta\nu}
M_{\alpha\mu}\Big)(t\tau_{1})
\Big(\mathcal{B}_{\delta\sigma}\Omega_{(\gamma\rho)}^{-1}\Big)(t\tau_{2})
\nonumber\\
&+&\frac{1}{9}\Big(\Omega_{\beta}{}^{(\iota}\Omega_{\nu}{}^{\kappa)}
M_{\alpha\mu}\Big)(t\tau_{1})
\Big(\mathcal{A}_{(\delta|\iota|}\mathcal{A}_{\sigma)\kappa}
M_{\gamma\rho}^{-1}\Big)(t\tau_{2})
\nonumber\\
&+&\frac{1}{12}\Big(\Omega_{(\beta}{}^{\iota}
\mathcal{A}_{\nu)\eta}M_{\alpha\mu}\Big)
(t\tau_{1})\Big(
\Omega_{\sigma}{}^{(\eta}\Omega_{\delta}{}^{\epsilon}{}\mathcal{E}_{(1)\rho}{}^{\chi)}
\Phi_{\gamma\iota}\Lambda_{\epsilon\chi}\Big)(t\tau_{2})
\nonumber\\
&-&\frac{1}{24}\Big\{\Big(\mathcal{B}_{\delta\sigma}
M_{\gamma\rho}\Big)(t\tau_{1})
\Big(\Lambda_{\iota\kappa}\Lambda_{\eta\chi}\Big)(t\tau_{2})
\nonumber\\
&+&4\Big(\Omega_{\delta}{}^{(\omega}\Omega_{\sigma}{}^{\lambda)}
M_{\gamma\rho}\Big)(t\tau_{1})
\Big(\Lambda_{\omega\iota}\Lambda_{\lambda\kappa}
\Lambda_{\eta\chi}\Big)(t\tau_{2})\Big\}
\Big(\Omega_{\beta}{}^{(\iota}\Omega_{\nu}{}^{\kappa}
\Psi_{\alpha}{}^{\eta}\mathcal{E}_{(3)\mu}{}^{\chi)}\Big)(t\tau_{2})
\nonumber\\
&+&\frac{1}{3}
\Big(\Omega_{\beta}{}^{(\iota}\Omega_{\nu}{}^{\kappa}\Psi_{\alpha}
{}^{\eta}\mathcal{E}_{(3)\mu}{}^{\chi)}\Big)
(t\tau_{1})\mathcal{V}_{\gamma\delta\rho\sigma\iota\kappa\eta\chi}
(t\tau_{1},t\tau_{2})\;,
\end{eqnarray}
\begin{eqnarray}
d^{(2)}_{\mu\nu\alpha\beta\rho\sigma}(t\tau_{1},t\tau_{2})
&=&
\frac{1}{9}\Big(\Omega_{\beta}{}^{(\iota}\Omega_{\sigma}{}^{\kappa)}
M_{\alpha\rho}\Big)(t\tau_{1})
\Big(\Phi_{\mu\iota}\mathcal{A}_{\nu\kappa}\Big)(t\tau_{2})
\nonumber\\
&-&
\frac{1}{9}\Big(\mathcal{B}_{\beta\sigma}
M_{\alpha\rho}\Big)(t\tau_{1})\mathcal{A}_{(\mu\nu)}(t\tau_{2})
\nonumber\\
&-&\frac{1}{9}\mathcal{A}_{(\mu\nu)}(t\tau_{1})
\Big(\mathcal{B}_{\beta\sigma}M_{\alpha\rho}\Big)(t\tau_{2})
\nonumber\\
&+&
\frac{1}{12}\Big(\Omega_{\beta}{}^{(\iota}
\Omega_{\sigma}{}^{\kappa}\Psi_{\alpha}{}^{\eta}\mathcal{E}_{(3)\rho}{}^{\chi)}
\mathcal{A}_{(\mu\nu)}\Lambda_{\iota\kappa}\Lambda_{\eta\chi}\Big)(t\tau_{2})
\nonumber\\
&-&\frac{1}{36}\Big(\Omega_{\beta}{}^{(\iota}
\Omega_{\sigma}{}^{\kappa)}M_{\alpha\rho}\Big)
(t\tau_{1})\Big(\mathcal{A}_{\nu\iota}\Lambda_{\kappa\eta}
\mathcal{E}_{(7)\mu}{}^{\eta}\Big)(t\tau_{2})
\nonumber\\
&+&\frac{1}{3}\Big(\Omega_{\beta}{}^{(\iota}\Omega_{\sigma}{}^{\kappa}
\Psi_{\alpha}{}^{\eta}\mathcal{E}_{(3)\rho}{}^{\chi)}\Big)(t\tau_{1})
\Bigg\{-\Lambda_{\kappa\eta}(t\tau_{1})\Big(\mathcal{A}_{\nu\chi}
\Phi_{\mu\iota}\Big)(t\tau_{2})
\nonumber\\
&+&\frac{1}{2}\Big(\mathcal{A}_{(\mu\nu)}\Lambda_{\iota\kappa}
\Lambda_{\eta\chi}\Big)(t\tau_{1})
+\frac{1}{4}\Lambda_{\iota\kappa}(t\tau_{1})\Big(
\mathcal{A}_{\nu(\eta}\Lambda_{\chi)\epsilon}\mathcal{E}_{(7)\mu}{}^{\epsilon}\Big)(t\tau_{2})\Bigg\}
\nonumber\\
&+&\frac{1}{36}\Omega_{\nu}{}^{(\iota}\mathcal{E}_{(7)\mu}{}^{\kappa)}
(t\tau_{1})\mathcal{S}_{\alpha\beta\rho\sigma\iota\kappa}(t\tau_{2})\;,
\end{eqnarray}
\begin{eqnarray}
d^{(3)}_{\mu\nu\alpha\beta}(t\tau_{1},t\tau_{2})
&=&
-\frac{1}{36}\Big(\Omega_{\nu}{}^{(\iota}\mathcal{E}_{(7)\mu}
{}^{\kappa)}\Big)(t\tau_{1})\Big(\Phi_{\alpha\iota}
\mathcal{A}_{\beta\kappa}\Big)(t\tau_{2})
+\frac{2}{9}\mathcal{A}_{(\mu\nu)}(t\tau_{1})
\mathcal{A}_{(\alpha\beta)}(t\tau_{2})
\nonumber\\
&+&\frac{1}{144}\Omega_{\nu}{}^{(\iota}\mathcal{E}_{(7)\mu}{}^{\kappa)}
(t\tau_{1})\Big(\mathcal{A}_{\beta\iota}\Lambda_{\kappa\sigma}\mathcal{E}_{(7)\alpha}{}^{\sigma}\Big)(t\tau_{2})\;,
\end{eqnarray}
\begin{eqnarray}
d^{(4)}_{\mu\nu\alpha\beta\rho\sigma}(t\tau_{1},t\tau_{2})
&=&
-\frac{1}{3}\Big(\Omega_{\beta}{}^{(\iota}
\Omega_{\sigma}{}^{\kappa)}M_{\alpha\rho}\Big)
(t\tau_{1})\Big(\Phi_{\mu\iota}\mathcal{A}_{\nu\kappa}\Big)(t\tau_{2})
\nonumber\\
&-&\frac{2}{3}\Big(\Omega_{\nu}{}^{(\iota}
\Psi_{\mu}{}^{\kappa)}\Big)(t\tau_{1})
\Big(\mathcal{B}_{\beta\sigma}\Phi_{\alpha\iota}
\Phi_{\rho\kappa}\Big)(t\tau_{2})
+\frac{1}{3}\Big(\mathcal{B}_{\beta\sigma}
M_{\alpha\rho}\Big)(t\tau_{1})\mathcal{A}_{\mu\nu}(t\tau_{2})
\nonumber\\
&+&\frac{1}{3}\mathcal{A}_{\mu\nu}(t\tau_{1})
\Big(\mathcal{B}_{\beta\sigma}M_{\alpha\rho}\Big)(t\tau_{2})
\nonumber\\
&+&\frac{2}{3}\Big(\Omega_{\beta}{}^{(\iota}
\Omega_{\sigma}{}^{\kappa)}
M_{\alpha\rho}\Big)(t\tau_{1})
\Big(\Psi_{\mu}{}^{\eta}\mathcal{A}_{\nu\iota}\Lambda_{\kappa\eta}
\Big)(t\tau_{2})
\nonumber\\
&+&\frac{2}{3}\Big(\Omega_{\nu}{}^{(\iota}
\Psi_{\mu}{}^{\kappa)}\Big)(t\tau_{1})
\Big(\mathcal{A}_{\beta\iota}\mathcal{A}_{\sigma\kappa}
M_{\alpha\rho}\Big)(t\tau_{2})
\nonumber\\
&+&\frac{1}{2}\Big(\Omega_{\nu}{}^{(\epsilon}
\Psi_{\mu}{}^{\lambda)}\Big)(t\tau_{1})
\Big(\Omega_{\beta}{}^{(\iota}\Omega_{\sigma}
{}^{\kappa}\mathcal{E}_{(1)\rho}{}^{\eta)}\Phi_{\alpha\epsilon}
\Lambda_{\lambda\iota}\Lambda_{\kappa\eta}\Big)(t\tau_{2})
\nonumber\\
&+&\frac{1}{2}\Big(\Omega_{\beta}{}^{(\iota}
\Omega_{\sigma}{}^{\kappa}\Psi_{\alpha}{}^{\eta}\mathcal{E}_{(3)\rho}
{}^{\epsilon)}\Big)(t\tau_{1})\Big\{2\Lambda_{\kappa\eta}(t\tau_{1})
\Big(\mathcal{A}_{\nu\epsilon}\Phi_{\mu\iota}\Big)(t\tau_{2})
\nonumber\\
&-&\Big(\mathcal{A}_{\mu\nu}\Lambda_{\iota\kappa}
\Lambda_{\eta\epsilon}\Big)(t\tau_{1})
-4\Lambda_{\iota\kappa}(t\tau_{1})
\Big(\Psi_{\mu}{}^{\lambda}\mathcal{A}_{\nu\epsilon}\Lambda_{\eta
\lambda}\Big)(t\tau_{2})\Big\}
\nonumber\\
&-&\frac{1}{4}\Big\{\Big(\mathcal{A}_{\mu\nu}
\Lambda_{\iota\kappa}\Lambda_{\eta\epsilon}\Big)(t\tau_{2})
+4\Big(\Omega_{\nu}{}^{(\omega}\Psi_{\mu}{}^{\lambda)}\Big)(t\tau_{1})
\Big(\Lambda_{\omega\iota}\Lambda_{\lambda\kappa}
\Lambda_{\eta\epsilon}\Big)(t\tau_{2})\Big\}
\nonumber\\
&\times&\Big(\Omega_{\beta}{}^{(\iota}\Omega_{\sigma}
{}^{\kappa}\Psi_{\alpha}{}^{\eta}\mathcal{E}_{(3)\rho}{}^{\epsilon)}\Big)(t\tau_{2})\;,
\end{eqnarray}
\begin{eqnarray}
d^{(5)}_{\mu\nu\alpha\beta}(t\tau_{1},t\tau_{2})
&=&
\frac{1}{12}\Big(\Omega_{\beta}{}^{(\gamma}\mathcal{E}_{(7)\alpha}
{}^{\delta)}\Big)(t\tau_{1})\Big(\Phi_{\mu\gamma}
\mathcal{A}_{\nu\delta}\Big)(t\tau_{2})
\nonumber\\
&+&
\frac{2}{3}\Big(\Omega_{\nu}{}^{(\gamma}
\Psi_{\mu}{}^{\delta)}\Big)(t\tau_{1})
\Big(\Phi_{\alpha\gamma}\mathcal{A}_{\beta\delta}\Big)(t\tau_{2})
\nonumber\\
&-&\frac{2}{3}\mathcal{A}_{(\alpha\beta)}(t\tau_{1})
\mathcal{A}_{\mu\nu}(t\tau_{2})
-\frac{2}{3}\mathcal{A}_{(\alpha\beta)}(t\tau_{2})
\mathcal{A}_{\mu\nu}(t\tau_{1})
\nonumber\\
&-&\frac{1}{6}\Big(\Omega_{\beta}{}^{(\iota}
\mathcal{E}_{(7)\alpha}{}^{\kappa)}\Big)(t\tau_{1})\Big(\Psi_{\mu}{}^{\epsilon}
\mathcal{A}_{\nu\kappa}
\Lambda_{\iota\epsilon}\Big)(t\tau_{2})
\nonumber\\
&-&\frac{1}{6}\Big(\Omega_{\nu}{}^{(\iota}
\Psi_{\mu}{}^{\kappa)}\Big)(t\tau_{1})
\Big(\mathcal{C}_{\beta}{}^{(\eta}{}_{\alpha}
{}^{\epsilon)}\Lambda_{\iota\eta}\Lambda_{\kappa\epsilon}\Big)(t\tau_{2})\;,
\end{eqnarray}
\begin{eqnarray}
d^{(6)}_{\mu\nu\alpha\beta}(t\tau_{1},t\tau_{2})&=&
-2\Big(\Omega_{\nu}{}^{(\gamma}\Psi_{\mu}{}^{\delta)}\Big)
(t\tau_{1})\Big(\Phi_{\alpha\gamma}\mathcal{A}_{\beta\delta}\Big)(t\tau_{2})
+2\mathcal{A}_{\mu\nu}(t\tau_{1})\mathcal{A}_{\alpha\beta}(t\tau_{2})
\nonumber\\
&+&4\Big(\Omega_{\nu}{}^{(\gamma}\Psi_{\mu}{}^{\delta)}
\Big)(t\tau_{1})\Big(\Psi_{\alpha}{}^{\sigma}
\mathcal{A}_{\beta\delta}\Lambda_{\gamma\sigma}\Big)(t\tau_{2})\;.
\end{eqnarray}

\section{Conclusions}

In this chapter we studied the heat kernel expansion for a Laplace operator
acting on sections of a complex vector bundle over a smooth compact
Riemannian manifold without boundary. We assumed that the curvature $F$ of
the $U(1)$ part of the total connection (the electromagnetic field) is
covariantly constant and large, so that $tF\sim 1$, that is, $F$ is of order
$t^{-1}$. In this situation the standard asymptotic expansion of the heat
kernel as $t\to 0$ does not apply since the electromagnetic field cannot be
treated as a perturbation.

In order to calculate the heat kernel asymptotic expansion we use an algebraic
approach in which the nilpotent algebra of the operators ${\cal D}_\mu$ plays
a major role. In this approach the calculation of the asymptotic expansion of
the heat kernel is reduced to the calculation of the asymptotic expansion of
the heat semigroup and, then, to the action of differential operators on the
zero-order heat kernel. Since the zero-order heat kernel has the Gaussian form
the heat kernel asymptotics are expressed in terms of generalized Hermite
polynomials.

The main result of this work is establishing the existence of a new
non-per\-tur\-bative asymptotic expansion of the heat kernel and the explicit
calculation of the first three coefficients of this expansion (both
off-diagonal and the diagonal ones). As far as we know, such an asymptotic
expansion and the explicit form of these modified heat kernel coefficients are
new.

We presented our result as explicitly as possible. Unfortunately, some of
the integrals of the tensor-valued functions cannot be evaluated explicitly in
full generality. They can be evaluated, in principle, by using the spectral
decomposition of the two-form $F$,
\be
F=\sum_{k=1}^{[n/2]} B_k E_k\,,
\qquad
F^2=-\sum_{k=1}^{[n/2]} B_k^2 \Pi_k\,,
\ee
where $B_k$ are the eigenvalues, $E_k$ are the (2-dimensional) eigen-two-forms,
and $\Pi_k=-E_k^2$ are the corresponding eigen-projections onto 2-dimensional
eigenspaces. Then for any analytic function of $tiF$ we have
\be
f(tiF)=\sum_{k=1}^{[n/2]} f(tB_k) \frac{1}{2}(\Pi_k+iE_k)
+\sum_{k=1}^{[n/2]} f(-tB_k) \frac{1}{2}(\Pi_k-iE_k)\,.
\ee
However, this seems impractical in general case in $n$ dimensions.
It would simplify substantially in the following cases: i) there is only one eigenvalue
(one magnetic field) in a corresponding two-dimensional subspace, that is,
$F=B_1 E_1$ (which is essentially 2-dimensional), and ii) all eigenvalues are
equal so that $F^2=-I$ (which is only possible in even dimensions).




\endchapterapp{}


\chapter[LOW-ENERGY EFFECTIVE ACTION IN NON-PERTURBATIVE
\newheadline ELECTRODYNAMICS  IN CURVED SPACETIME]
{LOW-ENERGY EFFECTIVE ACTION IN NON-PERTURBATIVE
\newheadline ELECTRODYNAMICS  IN CURVED SPACETIME\footnotemark[2]}

\footnotetext[2]{The material in this chapter has been submitted for peer review to \emph{Journal of Mathematical Physics}: I.G. Avramidi
and G. Fucci, Low-Energy Effective Action in Non-Perturbative Electrodynamics in Curved Spacetime, arXiv: 0902.1541 [hep-th]}

\begin{chapabstract}

We study the heat kernel for the Laplace type partial differential operator
acting on smooth sections of a complex spin-tensor bundle over a generic
$n$-dimensional Riemannian manifold. Assuming that the curvature of the $U(1)$
connection (that we call the electromagnetic field) is constant we compute the
first two coefficients of the non-perturbative asymptotic expansion of the heat
kernel which are of zero and the first order in Riemannian curvature and of
arbitrary order in the electromagnetic field. We apply these results to the
study of the effective action in non-perturbative electrodynamics in four
dimensions and derive a generalization of the Schwinger's result for the
creation of scalar and spinor particles in electromagnetic field induced by the
gravitational field. We discover a new infrared divergence in the imaginary
part of the effective action due to the gravitational corrections, which seems
to be a new physical effect.
\end{chapabstract}

\section{Introduction}

Schwinger used, in \cite{schwinger51}, the heat kernel asymptotic expansion technique to evaluate the one-loop effective action
in quantum electrodynamics.
In particular he solved, exactly, the case of a constant electromagnetic
field and derived an heat kernel integral representation for the effective
action. He showed that the heat kernel becomes a meromorphic function and a
careful evaluation of the integral leads to an imaginary part of the effective
action. Schwinger computed the imaginary part of the effective action and
showed that it describes the effect of creation of electron-positron pairs by
the electric field. This effect is now called the Schwinger mechanism. This is
an essentially non-perturbative effect (non-analytic in electric field) that
vanishes exponentially for weak electric fields.

Therefore its evaluation requires non-perturbative techniques for the
calculation of the heat kernel in the situation when curvatures (but not their
derivatives) are large (low energy approximation).

In \cite{avramidi08e} we computed the heat kernel for the covariant Laplacian
with a strong covariantly constant electromagnetic field in an arbitrary
gravitational field. We evaluated the first three coefficients of the heat
kernel asymptotic expansion in powers of Riemann curvature $R$ but \emph{in all
orders} of the electromagnetic field $F$. 
In the
present chapter we use those results to compute {\it explicitly} the terms linear
in the Riemann curvature in the non-perturbative heat kernel expansion for the
scalar and the spinor fields and compute their contribution to the imaginary
part of the effective action. In other words, we generalize the Schwinger
mechanism to the case of a strong electromagnetic field in a gravitational
field and compute the {\it gravitational corrections to the original Schwinger
result}.

\section{Setup of the Problem}

Let $M$ be a $n$-dimensional compact Riemannian manifold
(with positive-definite metric $g_{\mu\nu}$) without boundary and
${\cal S}$ be a complex spin-tensor vector bundle over $M$ realizing a
representation of
the group ${\rm Spin}(n)\otimes U(1)$.
Let $\varphi$ be a section of the bundle ${\cal S}$
and $\nabla$ be the total connection on the bundle $\mathcal{S}$ (including the
spin connection as well as the ${\rm U}(1)$-connection). Then the commutator of
covariant derivatives defines the curvatures
\begin{equation}\label{40}
[\nabla_\mu,\nabla_\nu]\varphi=({\cal R}_{\mu\nu}+i F_{\mu\nu})\varphi\;,
\end{equation}
where $F_{\mu\nu}$ is the curvature of the $U(1)$-connection (which will be
also
called the electromagnetic field) and
${\cal R}_{\mu\nu}$ is the curvature of the spin connection defined by
\be
{\cal R}_{\mu\nu}=\frac{1}{2}R^{ab}{}_{\mu\nu}\Sigma_{ab}\,,
\ee
with $\Sigma_{ab}$ being the generators of the spin group ${\rm Spin}(n)$
satisfying
the commutation relations
\be
[\Sigma_{ab},\Sigma^{cd}]
=4\delta^{[c}{}_{[a}\Sigma^{d]}{}_{b]}\,.
\label{4280}
\ee
Note that for the scalar fields ${\cal R}_{\mu\nu}=0$ and for the spinor fields
\be
\Sigma_{ab}=\frac{1}{2}\gamma_{ab}\,,
\ee
where $\gamma_{ab}=\gamma_{[a}\gamma_{b]}$
(more generally, we define
$\gamma_{a_1\dots a_m}=\gamma_{[a_1}\cdots\gamma_{a_m]}$)
and $\gamma_a$ are
the Dirac matrices
generating the Clifford algebra
\be
\gamma_a\gamma_b+\gamma_b\gamma_a=2g_{ab}\II\,.
\ee

\subsection{Differential Operators}

In the present chapter we consider a second-order Laplace type partial
differential operator,
\be
\label{41}
L=-\Delta+\xi R + Q,
\ee
where $\Delta=g^{\mu\nu}\nabla_\mu\nabla_\nu$ is the Laplacian,
$\xi$ is a constant parameter,
and $Q$ is a smooth endomorphism of the bundle ${\cal S}$.
This operator is elliptic and self-adjoint and has a
positive-definite leading symbol.
Usually, for scalar fields we set
\be
Q^{\rm scalar}=0\,.
\ee
Moreover, for canonical scalar fields the coupling
\be
\xi^{\rm scalar}=
\left\{
\begin{array}{ll}
0 & \mbox{for canonical scalar fields}\,,\\[10pt]
\displaystyle
\frac{(n-2)}{4(n-1)} & \mbox{for conformal scalar fields}\,.
\end{array}
\right.
\ee
Another important case is the square of the Dirac operator acting on
spinor fields
\be
L=D^2\,,
\ee
where
\be
D=i\gamma^\mu\nabla_\mu\,.
\ee
It is easy to see that in this case
we have
\be
\xi^{\rm spinor}=\frac{1}{4}\;,
\ee
and
\be
Q^{\rm spinor}=-\frac{1}{2}iF_{\mu\nu}\gamma^{\mu\nu}\,.
\ee


The object of primary interest in quantum field theory is the
(Euclidean)
one-loop effective action determined by the functional
determinant
\be
\Gamma_{(1)}=\varrho\log\Det(L+m^2)\,,
\ee
where $\varrho$ is the fermion number of the field equal to $(+1)$ for
boson fields and $(-1)$ for fermion fields,
$m$ is a
mass parameter, which is assumed to be sufficiently
large so that the operator $(L+m^2)$ is positive.
Notice that the usual factor $\frac{1}{2}$ is missing because the field
is complex, which is equivalent to the contribution of two real fields.
Of course, this formal
expression is divergent.
To rigorously define the determinant of a differential operator
one needs to introduce some regularization and then to renormalize it.
One of the best ways to do it is via the heat kernel method.

\subsection{Spectral Functions}



The determinant of the operator $L+m^{2}$, considered above, can be defined within the so-called
zeta-function regularization as follows.
First, one defines the zeta function by
\be
\zeta(s)=\mu^{2s}\Tr \left(L+m^2\right)^{-s}
=\int\limits_M dx\;g^{1/2}Z(s)\,,
\ee
where
\be
Z(s)=\frac{\mu^{2s}}{\Gamma(s)}\int\limits_0^\infty
dt\; t^{s-1} e^{-tm^2}\textrm{tr}\;U^{\rm diag}(t)\,,
\ee
$\textrm{tr}$ denotes the fiber trace over the bundle ${\cal S}$ and $\mu$ is a renormalization parameter
introduced to preserve dimensions. 
Therefore, the zeta-regularized one-loop effective action is simply
\be
\Gamma_{(1)}=-\varrho\zeta'(0)\,,
\ee
and the one-loop effective Lagrangian is given by
\be
{\cal L}=-\varrho Z'(0)\,.
\ee
The effective Lagrangian can be also defined simply in the cut-off
regularization by
\be
{\cal L}=-\varrho\int\limits_{\varepsilon\mu^2}^\infty
\frac{dt}{t}e^{-tm^2}\textrm{tr}\;U^{\rm diag}(t)\,,
\ee
where $\varepsilon$ is a regularization parameter, which should be
set to zero after subtracting the divergent terms.
Another regularization is the dimensional regularization, in which one
simply defines the effective action by the formal integral
\be
{\cal L}=-\varrho\mu^{2\varepsilon}\int\limits_{0}^\infty
\frac{dt}{t}e^{-tm^2}\textrm{tr}\;U^{\rm diag}(t)\,,
\ee
where the heat trace is formally computed in complex dimension
$(n-2\varepsilon)$
with sufficiently large real part of $\varepsilon$ so that the integral is
finite. The renormalized effective action is obtained then by
subtracting the simple pole in $\varepsilon$.

For elliptic operators (in the Euclidean setup) the heat trace is a smooth
function of $t$; in many cases it is even an analytic function of $t$ in the
neighborhood of the positive real axis. However, in the physical case for
hyperbolic operators (in the Lorentzian setup) the heat trace can have
singularities even on the positive real axis of $t$. As we will show later
in the approximation under consideration (for constant electromagnetic field)
it becomes a meromorphic
function of $t$ with an essential singularity at $t=0$ and some poles
$t_k$, $k=1,2,\dots,$,
on the positive real axis.
It turns out that the imaginary part of the effective action does not depend
on the regularization method and is uniquely defined by the contribution
of these poles. These poles should be avoided from above, which gives
\be
{\rm Im}\;{\cal L}
=-\varrho \pi\sum_{k=1}^\infty
\Res\left\{t^{-1} e^{-t m^2} \textrm{tr}\;U^{\rm diag}(t); t_k\right\}\,.
\label{4232xxx}
\ee
This method was first elaborated and used by Schwinger
\cite{schwinger51} in quantum electrodynamics to calculate the
electron-positron pair production by a constant electric field.
One of the goal of our work is to {\it generalize the Schwinger results}
for the case of constant electromagnetic field in a gravitational field.
We will compute the {\it extra contribution} to the particle production by
a constant electromagnetic field {\it induced by
the gravitational field}.

\subsection{Heat Kernel Asymptotic Expansion}


In the previous Chapter
we studied the case of a {\it parallel $U(1)$ curvature}
(covariantly constant electromagnetic field), i.e. \cite{avramidi08b}
\begin{equation}
\label{42x}
\nabla_{\mu}F_{\alpha\beta}=0\;.
\end{equation}
In the present chapter we will also assume that the potential term $Q$
is covariantly constant
\be
\nabla_\mu Q=0\,.
\ee
By summing up all powers of $F$ in the asymptotic expansion
of the heat kernel diagonal we obtained a {\it new (non-perturbative)
asymptotic expansion}
\be
U^{\rm diag}(t)\sim(4\pi t)^{-n/2}\exp\left(-tQ\right)J(t)
\sum_{k=0}^{\infty}t^{k} b_{2k}(t)\;,
\label{4222zza}
\ee
where
\be
J(t)=\det\left(\frac{tiF}{\sinh(tiF)}\right)^{1/2}\;,
\label{425zz}
\ee
and $b_2k(t)$ are the modified heat kernel coefficients
which are analytic functions of $t$ at $t=0$ which depend on $F$ only in the
dimensionless combination $tF$.
Here and everywhere below all
functions of the $2$-form $F$ are analytic at $0$ and should be understood in
terms of a power series in the matrix $F=(F^\mu{}_\nu)$. Notice the
{\it position of indices} here, it is important! There is a difference here
between Euclidean case and the
Lorentzian one since the raising of indices by a Minkowski metric does change
the properties of the matrix $F$.
Also, here $\det$ denotes the determinant with respect to the tangent space
indices.

The fiber trace of the heat kernel diagonal
has then the asymptotic expansion
\begin{equation}
\textrm{tr}\;U^{\rm diag}(t)
\sim(4\pi t)^{-n/2}\Phi(t)\sum_{k=0}^{\infty}t^{k}
B_{2k}(t)\;,
\label{4225xx}
\end{equation}
where
\be
\Phi(t)=J(t)\textrm{tr}\, \exp\left(-tQ\right)\,,
\ee
\be
B_{2k}(t)=\frac{\textrm{tr}\, \exp\left(-tQ\right)b_{2k}(t)}
{\textrm{tr}\, \exp\left(-tQ\right)}\;,
\ee
are {\it new (non-perturbative) heat kernel coefficients}
of the operator $L$. The integrals $\int_M dx g^{1/2}B_{2k}(t)$ are then
the spectral invariants of the operator $L$.


\section{Calculation of the Coefficient $B_2(t)$ }

In Chapter 3 we obtained, in particular, the first three coefficients of the heat kernel asymptotic expansion, namely,
\cite{avramidi08b}
\bea
b_0^{\rm }(t)&=&1\,,
\\
b_{1}(t)&=&0\;,
\\
\label{470}
b^{\rm }_{2}(t)&=&
\left\{\Sigma_{\mu\alpha}W_{\nu\beta}(t)
+V_{\mu\alpha\nu\beta}(t)\right\}
R^{\mu\alpha\nu\beta}\;,
\eea
where
\bea
W(t)
&=&\frac{1}{2}\left(\coth(tiF)-\frac{1}{tiF}\right)\;,
\\
\label{472c}
V^{\mu\alpha}{}_{\nu\beta}(t)
&=&
\left(\frac{1}{3}-\xi\right)\delta^{[\mu}_\nu\delta^{\alpha]}_\beta
+\int\limits_{0}^{1}d\tau
\Bigg\{
-\frac{1}{24}\mathcal{B}^{[\mu}{}_{[\nu}(\tau)
{\cal Z}^{\alpha]}{}_{\beta]}(\tau)
+\frac{1}{6}\mathcal{A}^{[\mu\alpha]}(\tau)\mathcal{A}_{[\nu\beta]}(\tau)
\nonumber\\
&&
-\frac{1}{12}\mathcal{A}^{[\mu}{}_{[\nu}(\tau)
\mathcal{A}^{\alpha]}{}_{\beta]}(\tau)
-\frac{1}{4}\mathcal{A}^{[\mu}{}_{[\nu}(\tau)
\mathcal{A}_{\beta]}{}^{\alpha]}(\tau)
\Bigg\} \;,
\eea
and
\bea
\label{471}
\mathcal{A}(\tau)
&=&
\frac{1}{2}\frac{\exp[(1-2\tau)tiF]-\exp(-tiF)}{\sinh(tiF)}\;,
\\[5pt]
\label{471b}
\mathcal{B}(\tau)
&=&
\frac{\coth(tiF)}{tiF}
-\frac{1}{tiF\sinh(tiF)}\cosh[(1-2\tau)tiF]\;,
\\[5pt]
{\cal Z}(\tau)
&=&
3tiF\coth(tiF)
+\frac{tiF}{\sinh(tiF)}\cosh[(1-2\tau)tiF]\;.
\eea
The trace coefficients are then given by
\bea
B_0(t)&=&1\,,
\\
\label{470b}
B_{2}(t)
&=&
\bigg\{\Psi_{\mu\alpha}(t)W_{\nu\beta}(t)
+V_{\mu\alpha\nu\beta}(t)\bigg\}
R^{\mu\alpha\nu\beta}
\;,
\eea
where
\bea
\Psi(t)_{\mu\alpha}
&=&\;\frac{\textrm{tr}\, \exp\left(-tQ\right)\Sigma_{\mu\alpha}}
{\textrm{tr}\, \exp\left(-tQ\right)}\,.
\eea

\subsection{Spectral Decomposition}

To evaluate it we use the spectral
decomposition of the matrix $F=(F^\mu{}_\nu)$,
\be
F=\sum_{k=1}^{N} B_k E_k\,,
\ee
where $B_k$ are some real
invariants and $E_k=(E_k{}^\mu{}_\nu)$ are some
matrices
satisfying the equations
\be
E_k{}_{\mu\nu}=-E_k{}_{\nu\mu}\,,
\ee
\be
E^k_{\mu[\nu}E^k_{\alpha\beta]}=0
\,,
\ee
and for $k\ne m$
\be
E_k E_{m}=0\,.
\ee
Here, of course, $N\le [n/2]$.
The invariants $B_k$ (that we call ``magnetic fields'') should not be confused
with the heat trace coefficients
$B_0$ and $B_1$.

Next, we define the
matrices $\Pi_k=(\Pi_k{}^\mu{}_\nu)$ by
\be
\Pi_k=-E_k^2\,.
\ee
They satisfy the equations
\be
\Pi_k{}_{\mu\nu}=\Pi_k{}_{\nu\mu}\,,
\ee
\be
E_k\Pi_k=\Pi_kE_k=E_k\,,
\ee
and for $k\ne m$
\be
E_k \Pi_m=\Pi_m E_k=0\,, \qquad
\Pi_k\Pi_m=0\,.
\ee

To compute functions of the matrix $F$ we need to know its eigenvalues.
We distinguish two different cases.

\paragraph{Euclidean Case.} In this case the metric has Euclidean
signature $(++\cdots+)$ and the non-zero
eigenvalues of the matrix $F$ are
$\pm iB_1$, \dots, $\pm iB_N$, (which are all imaginary).
Of course, it may also have a number of zero eigenvalues.
In this case the matrices $\Pi_k$ are nothing but the projections
on $2$-dimensional eigenspaces satisfying
\be
\Pi_k^2=\Pi_k\,, \qquad
\Pi_k{}^\mu{}_\mu=2\,.
\ee
In this case we also have
\be
B_k=\frac{1}{2}E_k^{\mu\nu}F_{\mu\nu}\,.
\ee
Then we have
\be
(iF)^{2m}=\sum_{k=1}^{N}B_k^{2m}\Pi_k\,,\qquad (m\ge 1)
\ee
\be
(iF)^{2m+1}=\sum_{k=1}^{N}B_k^{2m+1}iE_k\,,\qquad
(m\ge 0)\,,
\ee
and, therefore, for
any analytic function of $tiF$ at $t=0$ we have
\be
f(tiF)=f(0)\II
+\sum_{k=1}^{N}\Bigg\{
\frac{1}{2}\bigg[f(tB_k)+f(-tB_k)-2f(0)\bigg]\Pi_k
+\frac{1}{2}\bigg[f(tB_k)-f(-tB_k)\bigg]iE_k
\Bigg\}\,.
\ee

\paragraph{Pseudo-Euclidean Case.}
This is the physically relevant case of pseudo-Eucli\-dean
(Lorentzian) metric with the signature $(-+\cdots+)$.
Then the non-zero
eigenvalues of the matrix $F$ are $\pm B_1$ (which are real)
and $\pm i B_2$, \dots, $\pm iB_N$, (which are imaginary).
We will call the invariant $B_1$,
determining the real eigenvalue, the ``electric field'' and denote it by
$B_1=E$,
and the invariants $B_k$, $k=2,\dots, N$,
 determining the imaginary eigenvalues, the
``magnetic fields''. So, in general, there is one electric field and
$(N-1)$ magnetic fields.
Again, there may be some zero eigenvalues as well.

In this case the matrices $\Pi_2$,\dots,$\Pi_N$ are the
orthogonal eigen-projections
as before, but the matrix $\Pi_1$ is equal to the negative of the
corresponding projection, in particular,
\be
\Pi_1^2=-\Pi_1\,,\qquad
\Pi_1 E_1=-E_1\,,\qquad
\Pi_1{}^\mu{}_\mu=-2\,.
\ee
Now, we have
\be
(iF)^{2m}
=-(iE)^{2m}\Pi_1
+\sum_{k=2}^{N}B_k^{2m}\Pi_k\,,\qquad (m\ge 1)
\ee
\be
(iF)^{2m+1}
=(iE)^{2m+1}E_1
+\sum_{k=2}^{N}B_k^{2m+1}iE_k\,,\qquad
(m\ge 0)\,.
\ee

Thus, to obtain the results for the pseudo-Euclidean case from the
result for the Euclidean case we should
just substitute formally
\be
B_1\mapsto iE, \qquad
iE_1\mapsto E_1,\qquad
\Pi_1\mapsto -\Pi_1\,.
\label{4327xxx}
\ee
In this way, we obtain for an analytic function of $itF$,
\bea
f(tiF)&=&f(0)\II
-\frac{1}{2}\bigg[f(itE)+f(-itE)-2f(0)\bigg]\Pi_1
+\frac{1}{2}\bigg[f(itE)-f(-itE)\bigg]E_1
\nonumber\\
&&
+\sum_{k=2}^{N}\Bigg\{
\frac{1}{2}\bigg[f(tB_k)+f(-tB_k)-2f(0)\bigg]\Pi_k
+\frac{1}{2}\bigg[f(tB_k)-f(-tB_k)\bigg]iE_k
\Bigg\}\,.
\nonumber\\
\eea

\subsection{Scalar and Spinor Fields}

First of all, we note that for scalar fields
\be
\Phi^{\rm scalar}(t)=J(t)\,,\qquad
\Psi^{\rm scalar}_{\mu\nu}(t)=0\,.
\ee

For the spinor fields we have
\be
\Phi^{\rm spinor}(t)
=J(t)\;\tr\exp\left(\frac{1}{2}tiF_{\mu\nu}\gamma^{\mu\nu}\right)\;,
\ee
\be
\Psi^{\rm spinor}_{\alpha\beta}(t)
=\frac{1}{2}\frac{\tr\;\gamma_{\alpha\beta}
\exp\left(\frac{1}{2}tiF_{\mu\nu}\gamma^{\mu\nu}\right)}
{\tr\exp\left(\frac{1}{2}tiF_{\rho\sigma}\gamma^{\rho\sigma}\right)}\,.
\ee
Here $\tr$ denotes the trace with respect to the spinor indices.

We will compute these functions as follows.
We define the matrices
\be
T_k=\frac{1}{2}iE_{k}^{\mu\nu}\gamma_{\mu\nu}\,.
\ee
Then by using the properties of the matrices $E_k$
and the product of the matrices $\gamma_{\mu\nu}$
\be
\gamma^{\mu\nu}\gamma_{\alpha\beta}=\gamma^{\mu\nu}{}_{\alpha\beta}
-4\delta^{[\mu}_{[\alpha}\gamma^{\nu]}{}_{\beta]}
-2\delta^{[\mu}_{\alpha}\delta^{\nu]}_{\beta}\II\,,
\ee
(and some other properties of Dirac matrices in $n$ dimensions)
one can show that these matrices are mutually commuting involutions,
that is,
\be
T_k^2=\II\,,
\ee
and
\be
[T_k,T_m]=0\,.
\ee
Also, the product of two different matrices is
(for $k\ne m$)
\be
T_kT_m=-\frac{1}{4}E_k^{\mu\nu}E_m^{\alpha\beta}\gamma_{\mu\nu\alpha\beta}\,.
\ee
More generally, the product of $m>1$ different matrices is
\be
T_{k_1}\cdots T_{k_m}
=\left(\frac{i}{2}\right)^m
E_{k_1}^{\mu_1\mu_2}\cdots E_{k_m}^{\mu_{2m-1}\mu_{2m}}
\gamma_{\mu_1\dots\mu_{2m}}\,.
\ee

It is well known that the matrices
$\gamma_{\mu_1\dots\mu_k}$ are traceless for any $k$ and the
trace of the product of two matrices $\gamma_{\mu_1\dots\mu_k}$
and $\gamma_{\nu_1\dots\nu_m}$ is non-zero only for $k=m$.
By using these properties we obtain the traces
\be
\tr T_k=0\,,
\ee
\be
\tr \gamma^{\alpha\beta}T_k=-2^{[n/2]}iE_k^{\alpha\beta}\,,
\ee
and for $m>1$:
\be
\tr T_{k_1}\cdots T_{k_m}=0\,,
\ee
\be
\tr \gamma^{\alpha\beta} T_{k_1}\cdots T_{k_m}=0\,,
\ee
when all indices $k_1$, \dots $k_m$ are different.

Now, by using the spectral decomposition of the matrix $F$ we easily
obtain first
\be
J(t)=\prod\limits_{k=1}^{N}\frac{tB_k}{\sinh(tB_k)}\,,
\ee
and
\be
\tr\exp\left(\frac{1}{2}tiF_{\mu\nu}\gamma^{\mu\nu}\right)
=\;\tr\prod_{k=1}^{N}
\exp\left(tT_kB_k\right)\,,
\ee
\be
\tr\gamma^{\alpha\beta}\exp\left(\frac{1}{2}tiF_{\mu\nu}\gamma^{\mu\nu}\right)
=\;\tr\gamma^{\alpha\beta}\prod_{k=1}^{N}
\exp\left(tT_kB_k\right)\,.
\ee
By using the properties of the matrices $T_k$ we get
\be
\exp\left(tT_kB_k\right)
=\cosh(tB_k)+T_k\sinh(tB_k)\,.
\ee
Therefore
\be
\tr\exp\left(\frac{1}{2}tiF_{\mu\nu}\gamma^{\mu\nu}\right)
=2^{[n/2]}
\prod\limits_{k=1}^{N}\cosh(tB_k)\,,
\ee
and
\bea
\tr \gamma^{\alpha\beta}
\prod\limits_{k=1}^{N}\exp(tT_kB_k)
&=&
\prod\limits_{j=1}^{N}\cosh(tB_j)
\sum\limits_{k=1}^{N}\tanh(tB_k)
\tr \gamma^{\alpha\beta}T_k
\nonumber\\
&=&
-2^{[n/2]}\prod\limits_{j=1}^{N}\cosh(tB_j)
\sum\limits_{k=1}^{N}\tanh(tB_k)iE_k^{\alpha\beta}\,.
\eea
Thus for the spinor fields
\be
\Phi^{\rm spinor}(t)=2^{[n/2]}\prod\limits_{k=1}^{N}
tB_k\coth(tB_k)\,,
\ee
and
\be
\Psi^{\rm spinor}_{\alpha\beta}(t)=
-\frac{1}{2}\sum\limits_{k=1}^{N}\tanh(tB_k)iE_k{}_{\alpha\beta}\,.
\ee
By the way, this simply means that
\be
\Psi^{\rm spinor}(t)=-\frac{1}{2}\tanh(tiF)\,.
\ee

\subsection{Calculation of the Tensor $V_{\mu\alpha\nu\beta}(t)$}

Next, we compute the tensor $V^{\mu\alpha}{}_{\nu\beta}(t)$.
First, we rewrite in the form
\bea
V^{\mu\alpha}{}_{\nu\beta}(t)
&=&
\left(\frac{1}{3}-\xi\right)\delta^{[\mu}_{\nu}\delta^{\alpha]}_{\beta}
+\int\limits_{0}^{1}d\tau
\Bigg\{
-\frac{1}{24}\mathcal{B}^{[\mu}{}_{[\nu}(\tau)
{\cal Z}^{\alpha]}{}_{\beta]}(\tau)
\nonumber\\
&&
+\frac{1}{16}{\cal X}^{\mu\alpha}(\tau)
\mathcal{X}_{\nu\beta}(\tau)
-\frac{1}{12}\mathcal{Y}^{[\mu}{}_{[\nu}(\tau)
\mathcal{Y}^{\alpha]}{}_{\beta]}(\tau)\Bigg\}\;,
\eea
where
\bea
{\cal X}(\tau)&=&
-\coth(tiF)+\frac{\cosh[(1-2\tau)tiF]}{\sinh(tiF)}\,,
\\[10pt]
{\cal Y}(\tau)&=&
\II+\frac{\sinh[(1-2\tau)tB_k]}{\sinh(tB_k)}\,.
\eea

Next, we parameterize these matrices as follows
\bea
\mathcal{B}(\tau)
&=&
2\tau(1-\tau)\II+
\sum_{k=1}^{N}f_{1,k}(\tau)\Pi_k\,,
\\
{\cal Z}(\tau)
&=&
4\II
+\sum_{k=1}^{N}f_{2,k}(\tau)\Pi_k\,,
\\
{\cal Y}(\tau)
&=&
2(1-\tau)\II
+\sum_{k=1}^{N}f_{3,k}(\tau)\Pi_k\,,
\\
{\cal X}(\tau)
&=&
\sum_{k=1}^{N}f_{4,k}(\tau)iE_k\,,
\\
W(t)&=&
\sum_{k=1}^{N}f_{5,k}(t)iE_k\,,
\eea
where
\bea
f_{1,k}(\tau)
&=&
\frac{\coth(tB_k)}{tB_k}
-\frac{1}{tB_k\sinh(tB_k)}\cosh[(1-2\tau)tB_k]
-2\tau(1-\tau)\;,
\\[5pt]
f_{2,k}(\tau)
&=&
3tB_k\coth(tB_k)
+\frac{tB_k}{\sinh(tB_k)}\cosh[(1-2\tau)tB_k]-4\,,
\\[5pt]
f_{3,k}(\tau)
&=&
\frac{\sinh[(1-2\tau)tB_k]}{\sinh(tB_k)}-(1-2\tau)\,,
\\[5pt]
f_{4,k}(\tau)
&=&
-\coth(tB_k)
+\frac{\cosh[(1-2\tau)tB_k]}{\sinh(tB_k)}\,,
\\[5pt]
f_{5,k}(t)
&=&
\frac{1}{2}\left(\coth(tB_k)-\frac{1}{tB_k}\right)\,.
\eea
This parametrization is convenient because all functions
$f_{m,k}(\tau)$ are analytic functions of $t$ at $t=0$ and
$f_{m,k}(\tau)\Big|_{t=0}=0$.

Then we obtain
\bea
V^{\mu\alpha}{}_{\nu\beta}(t)
&=&
\left(\frac{1}{6}-\xi\right)
\delta^{[\mu}{}{}_{[\nu}\delta^{\alpha]}{}_{\beta]}
+\sum_{k=1}^{N}
\varphi_k(t)\Pi_k{}^{[\mu}{}_{[\nu}\delta^{\alpha]}{}_{\beta]}
\nonumber\\
&&
+\sum_{k=1}^{N}\sum_{m=1}^{N}
\left[\rho_{km}(t)\Pi_k{}^{[\mu}{}_{[\nu}\Pi_{|m|}{}^{\alpha]}{}_{\beta]}
-\sigma_{km}(t)E_k{}^{\mu\alpha}E_{m\;}{}_{\nu\beta}
\right]\,,
\eea
where
\bea
\varphi_k(t)
&=&
-\frac{1}{12}\int\limits_0^1d\tau
\left[2f_{1,k}(\tau)+\tau(1-\tau)f_{2,k}(\tau)+4(1-\tau)f_{3,k}(\tau)
\right]\,,
\\
\rho_{km}(t)
&=&
-\frac{1}{48}\int\limits_0^1d\tau
\left[f_{1,k}(\tau)f_{2,m}(\tau)+f_{2,k}(\tau)f_{1,m}(\tau)
+4f_{3,k}(\tau)f_{3,m}(\tau)\right]\,,
\\
\sigma_{km}(t)
&=&
\frac{1}{16}\int\limits_0^1d\tau
f_{4,k}(\tau)f_{4,m}(\tau)\,.
\eea

\subsection{Calculation of the Coefficient Functions}

The remaining coefficient functions $\varphi_k(t)$, $\rho_{km}(t)$ and
$\sigma_{km}(t)$ are analytic functions of $t$ at $t=0$.
Here we give the solution the integrals above which have
the following general form
\begin{equation}
A(\alpha,x)=\int\limits_{0}^{1}d\tau\;\tau^{\alpha}\cosh[(1-2\tau)x]\;,
\end{equation}
and
\begin{equation}
B(\alpha,x)=\int\limits_{0}^{1}d\tau\;\tau^{\alpha}\sinh[(1-2\tau)x]\;.
\end{equation}
After a change of variables, it is not difficult to prove that for $\textrm{Re}(\alpha)>-1$ we get
\begin{equation}
A(\alpha,x)=\frac{\sinh(x)}{2x}+\frac{\alpha}{(2x)^{\alpha+1}}\left[e^{x}
\;\gamma(\alpha,2y)+(-1)^{\alpha+1}e^{-x}\;\gamma(\alpha,-2y)\right]\;,
\end{equation}
and
\begin{equation}
B(\alpha,x)=-\frac{\cosh(x)}{2x}+\frac{\alpha}{(2x)^{\alpha+1}}\left[e^{x}
\;\gamma(\alpha,2y)+(-1)^{\alpha}e^{-x}\;\gamma(\alpha,-2y)\right]\;,
\end{equation}
where $\gamma(\alpha,x)$ is the lower incomplete gamma function.

Moreover, if the coefficient $\alpha$ is an integer, as in our case, we can
write the formulas above as
\begin{equation}
A(\alpha,x)=\frac{\sinh(x)}{2x}
-\frac{\alpha!}{(2x)^{\alpha+1}}
\sum\limits_{k=1}^{\alpha-1}\frac{(2x)^{k}}{k!}
\left[\frac{e^{-x}+(-1)^{\alpha+k+1}e^{x}}{2}\right]\;,
\end{equation}
and
\begin{equation}
B(\alpha,x)=-\frac{\cosh(x)}{2x}
-\frac{\alpha!}{(2x)^{\alpha+1}}
\sum\limits_{k=1}^{\alpha-1}\frac{(2x)^{k}}{k!}
\left[\frac{e^{-x}+(-1)^{\alpha+k}e^{x}}{2}\right]\;.
\end{equation}

For even $\alpha=2m$ we have
\be
A(2m,x)=\frac{\sinh(x)}{2x}
-\frac{(2m)!}{(2x)^{2m+1}}
\Bigg\{\sinh x \sum\limits_{k=1}^{m}\frac {(2x)^{2k}}{(2k)!}
+\cosh x\sum\limits_{k=0}^{m-1}\frac{(2x)^{2k+1}}{(2k+1)!}\Bigg\}\;,
\ee
\begin{equation}
B(2m,x)=-\frac{\cosh(x)}{2x}
-\frac{(2m)!}{(2x)^{2m+1}}\left\{
\cosh x\sum\limits_{k=1}^{m}\frac{(2x)^{2k}}{(2k)!}
+\sinh x\sum\limits_{k=0}^{m-1}\frac{(2x)^{2k+1}}{(2k+1)!}\right\}\;.
\end{equation}
For odd $\alpha=2m+1$ we have
\be
A(2m+1,x)=\frac{\sinh(x)}{2x}
-\frac{(2m+1)!}{(2x)^{2m+2}}
\Bigg\{\cosh x \sum\limits_{k=1}^{m}\frac {(2x)^{2k}}{(2k)!}
+\sinh x\sum\limits_{k=0}^{m-1}\frac{(2x)^{2k+1}}{(2k+1)!}\Bigg\}\;,
\ee
\begin{equation}
B(2m+1,x)=-\frac{\cosh(x)}{2x}
-\frac{(2m+1)!}{(2x)^{2m+2}}\left\{
\sinh x\sum\limits_{k=1}^{m}\frac{(2x)^{2k}}{(2k)!}
+\cosh x\sum\limits_{k=0}^{m-1}\frac{(2x)^{2k+1}}{(2k+1)!}\right\}\;.
\end{equation}

From these last formulas we can compute $\varphi_k(t)$, $\rho_{km}(t)$ and
$\sigma_{km}(t)$ by using following particular case of the above integrals
\bea
\int\limits_0^1 d\tau \cosh[(1-2\tau)x]
&=&
\frac{\sinh x}{x}\,,
\\
\int\limits_0^1 d\tau \sinh[(1-2\tau)x]
&=&0\,.
\eea
By differentiating these integrals with respect to $x$ we obtain all other
integrals we need
\bea
\int\limits_{0}^{1}d\tau\;\tau
\cosh[(1-2\tau)x]
&=&
\frac{1}{2}\frac{\sinh x}{x}\,,
\\
\int\limits_{0}^{1}d\tau\;\tau^2
\cosh[(1-2\tau)x]
&=&
\frac{1}{2}\left(\frac{1}{x}+\frac{1}{x^3}\right)\sinh x
-\frac{1}{2}\frac{1}{x^2}\cosh x\,,
\\
\int\limits_0^1 d\tau\;\tau \sinh[(1-2\tau)x]
&=&
-\frac{1}{2}\frac{\cosh x}{x}+\frac{1}{2}\frac{\sinh x}{x^2}\,.
\eea
We also have the integrals
\bea
\int\limits_{0}^{1}d\tau
\cosh[(1-2\tau)x]\cosh[(1-2\tau)y]
&=&
\frac{1}{2}\Bigg\{
\frac{\sinh(x+y)}{x+y}
+\frac{\sinh(x-y)}{x-y}
\Bigg\}\,,
\nonumber
\\
\\
\int\limits_{0}^{1}d\tau
\cosh[(1-2\tau)x]\sinh[(1-2\tau)y]
&=&
0\,,
\\
\int\limits_{0}^{1}d\tau
\sinh[(1-2\tau)x]\sinh[(1-2\tau)y]
&=&
\frac{1}{2}\Bigg\{
\frac{\sinh(x+y)}{x+y}
-\frac{\sinh(x-y)}{x-y}
\Bigg\}\,.
\nonumber\\
\eea

By using these integrals we obtain
\bea
\varphi_k(t)&=&
\frac{1}{6}
+\frac{3}{8}\frac{1}{(tB_k)^2}
-\frac{1}{24}\coth(tB_k)\left(tB_k+9\frac{1}{(tB_k)}\right)\,,
\\[10pt]
%
\sigma_{km}(t)&=&
\frac{1}{16}\coth(tB_k)\coth(tB_m)
-\frac{1}{16}\frac{\coth(tB_k)}{tB_m}
-\frac{1}{16}\frac{\coth(tB_m)}{tB_k}
\nonumber\\[5pt]
&&
+\frac{1}{32}\frac{\coth(tB_m)+\coth(tB_k)}{t(B_k+B_m)}
+\frac{1}{32}\frac{\coth(tB_m)-\coth(tB_k)}{t(B_k-B_m)}\,,
\eea
\bea
\rho_{km}(t)
&=&
-\frac{1}{48}
\Bigg\{
4
+9\frac{1}{(tB_k)^2}
+9\frac{1}{(tB_m)^2}
-8\frac{1}{tB_k}\coth(tB_k)
-8\frac{1}{tB_m}\coth(tB_m)
\nonumber\\[5pt]
&&
-(tB_k)\coth(tB_k)
-(tB_m)\coth(tB_m)
-3\frac{B_k}{tB_m^2}\coth(tB_k)
\nonumber\\[5pt]
&&
-3\frac{B_m}{tB_k^2}\coth(tB_m)
+3\left(\frac{B_k}{B_m}+\frac{B_m}{B_k}\right)
\coth(tB_m)\coth(tB_k)
\nonumber\\[5pt]
&&
-\frac{1}{2}\Bigg[
\frac{B_m}{B_k}
+\frac{B_k}{B_m}-4
\Bigg]
\frac{\coth(tB_m)+\coth(tB_k)}{t(B_k+B_m)}
\nonumber\\[5pt]
&&
-\frac{1}{2}\Bigg[
\frac{B_k}{B_m}
+\frac{B_m}{B_k}+4
\Bigg]\frac{\coth(tB_m)-\coth(tB_k)}{t(B_k-B_m)}
\Bigg\}\,.
\eea

\subsection{Trace of the Heat Kernel Diagonal}

The trace of the heat kernel diagonal in the general case
within the considered approximation is given by
\be
\textrm{tr}\;U^{\rm diag}(t)\sim
(4\pi t)^{-n/2}\Phi(t)\left\{1+tB_2(t)
+\cdots\right\}\,,
\label{4379xxx}
\ee
where the function $\Phi(t)$ was computed above and the coefficient
$B_2$ is given by
\bea
B_2(t)&=&
\left(\frac{1}{6}-\xi\right)R
+\sum_{k=1}^{N}
\bigg\{
\Psi^{\mu\alpha}(t)f_{5,k}(t)iE_{k}^{\nu\beta}R_{\mu\alpha\nu\beta}
+\varphi_k(t)\Pi_{k}^{\mu\nu}R_{\mu\nu}
\bigg\}
\nonumber\\
&&
+\sum_{k=1}^{N}\sum_{m=1}^{N}\bigg\{
\rho_{km}(t)\Pi_k^{\mu\nu}\Pi_m^{\alpha\beta}R_{\mu\alpha\nu\beta}
-\sigma_{km}(t)E_k^{\mu\alpha}E_m^{\nu\beta}R_{\mu\alpha\nu\beta}
\bigg\}\,.
\eea

Let us
specify it for the two cases of interest.

\subsubsection{Scalar Fields}

For scalar fields we have
\bea
\Phi^{\rm scalar}(t)&=&\prod\limits_{k=1}^{N}
\frac{tB_k}{\sinh(tB_k)}\,,
\\[10pt]
B^{\rm scalar}_2(t)&=&
\left(\frac{1}{6}-\xi\right)R
+\sum_{k=1}^{N}
\varphi_k(t)\Pi_{k}^{\mu\nu}R_{\mu\nu}
\\
&&
+\sum_{k=1}^{N}\sum_{m=1}^{N}\bigg\{
\rho_{km}(t)\Pi_k^{\mu\nu}\Pi_m^{\alpha\beta}R_{\mu\alpha\nu\beta}
-\sigma_{km}(t)E_k^{\mu\alpha}E_m^{\nu\beta}R_{\mu\alpha\nu\beta}
\bigg\}\,.
\nonumber
\eea

\subsubsection{Spinor Fields}

For the spinor fields we obtain
\bea
\Phi^{\rm spinor}(t)&=&2^{[n/2]}\prod\limits_{k=1}^{N}
tB_k\coth(tB_k)\,,
\\[10pt]
B^{\rm spinor}_2(t)&=&
-\frac{1}{12}R
+\sum_{k=1}^{N}
\varphi_k(t)\Pi_{k}^{\mu\nu}R_{\mu\nu}
\\
&&
+\sum_{k=1}^{N}\sum_{m=1}^{N}\bigg\{
\rho_{km}(t)\Pi_k^{\mu\nu}\Pi_m^{\alpha\beta}R_{\mu\alpha\nu\beta}
-\lambda_{km}(t)E_k^{\mu\alpha}E_m^{\nu\beta}R_{\mu\alpha\nu\beta}
\bigg\}\,,
\nonumber
\eea
where
\bea
\lambda_{km}(t)&=&\sigma_{km}(t)
+\frac{1}{8}\frac{\tanh(tB_m)}{tB_k}
+\frac{1}{8}\frac{\tanh(tB_k)}{tB_m}
\nonumber\\[5pt]
&&
-\frac{1}{8}\tanh(tB_m)\coth(tB_k)
-\frac{1}{8}\tanh(tB_k)\coth(tB_m)
\,.
\eea

\subsection{Equal Magnetic Fields}

We will specify the obtained result for the case when
all magnetic invariants are equal to each other, that is,
\be
B_1=\cdots=B_N=B\,.
\ee

\subsubsection{Scalar Fields}

For scalar fields it takes the form
\bea
\Phi^{\rm scalar}(t)&=&
\left(\frac{tB}{\sinh(tB)}\right)^N\,,
\\[10pt]
B^{\rm scalar}_2(t)&=&
\left(\frac{1}{6}-\xi\right)R
+\varphi(t)H^{\mu\nu}_1 R_{\mu\nu}
\nonumber
\\[5pt]
&&
+\rho(t)H^{\mu\nu}_1H^{\alpha\beta}_1R_{\mu\alpha\nu\beta}
-\sigma(t)X^{\mu\alpha}_1X^{\nu\beta}_1 R_{\mu\alpha\nu\beta}\,,
\eea
where
\be
H^{\mu\nu}_1=\sum_{k=1}^{N}\Pi_{k}^{\mu\nu}\,,
\qquad
X^{\mu\nu}_1=\sum_{k=1}^{N}E_{k}^{\mu\nu}\,,
\ee
\bea
\varphi(t)&=&
\frac{1}{6}
+\frac{3}{8}\frac{1}{(tB)^2}
-\frac{1}{24}tB\coth(tB)
-\frac{3}{8}\frac{\coth(tB)}{tB}\,,
\\[5pt]
\sigma(t)&=&
\frac{1}{16}
-\frac{3}{32}\frac{\coth(tB)}{tB}
+\frac{3}{32}\frac{1}{\sinh^2(tB)}\;,
\\[5pt]
\rho(t)
&=&
-\frac{5}{24}
-\frac{3}{8}\frac{1}{(tB)^2}
+\frac{1}{24}tB\coth(tB)
+\frac{7}{16}\frac{\coth(tB)}{tB}
-\frac{1}{16}\frac{1}{\sinh^2(tB)}\,.
\eea

\subsubsection{Spinor Fields}

For the spinor fields we obtain
\bea
\Phi^{\rm spinor}(t)
&=&
2^{[n/2]}\left[tB\coth(tB)\right]^N\,,
\\[10pt]
B^{\rm spinor}_2(t)&=&
-\frac{1}{12}R
+\varphi(t)H^{\mu\nu}_1R_{\mu\nu}
\nonumber
\\[5pt]
&&
+\rho(t)H^{\mu\nu}_1 H^{\alpha\beta}_1R_{\mu\alpha\nu\beta}
-\lambda(t)X^{\mu\alpha}_1 X^{\nu\beta}_1
R_{\mu\alpha\nu\beta}
\,,
\eea
where
\bea
\lambda(t)&=&
-\frac{3}{16}
+\frac{3}{32}\frac{1}{\sinh^2(tB)}
+\frac{1}{4}\frac{\tanh(tB)}{tB}
-\frac{3}{32}\frac{\coth(tB)}{tB}
\,.
\eea

\subsection{Electric and Magnetic Fields}

Now we specify the above results for the
pseudo-Euclidean case when there
is one electric field and $(N-1)$ equal magnetic fields.
By using the recipe (\ref{4327xxx}) we obtain the following results.

\subsubsection{Scalar Fields}

For scalar fields we have
\bea
\Phi^{\rm scalar}(t)
&=&
\frac{tE}{\sin(tE)}
\left(\frac{tB}{\sinh(tB)}\right)^{N-1}\,,
\label{43107xxx}
\\[10pt]
B^{\rm scalar}_2(t)&=&
\left(\frac{1}{6}-\xi\right)R
-\tilde\varphi(t)\Pi_{1}^{\mu\nu}R_{\mu\nu}
+\varphi(t)H_2^{\mu\nu}R_{\mu\nu}
+\tilde\rho(t)\Pi_1^{\mu\nu}\Pi_1^{\alpha\beta}R_{\mu\alpha\nu\beta}
\nonumber
\\[5pt]
&&
+\tilde\sigma(t)E_1^{\mu\alpha}E_1^{\nu\beta}R_{\mu\alpha\nu\beta}
-2\rho_1(t)H^{\mu\nu}_2\Pi_1^{\alpha\beta}
R_{\mu\alpha\nu\beta}
+2\sigma_1(t)X^{\mu\alpha}_2 E_1^{\nu\beta}R_{\mu\alpha\nu\beta}
\nonumber\\[5pt]
&&
+\rho(t)H^{\mu\nu}_2 H^{\alpha\beta}_2 R_{\mu\alpha\nu\beta}
-\sigma(t)X^{\mu\alpha}_2 X^{\nu\beta}_2 R_{\mu\alpha\nu\beta}
\,,
\eea
where
\be
H_2^{\mu\nu}=\sum_{k=2}^{N}\Pi_{k}^{\mu\nu}\,,
\qquad
X_2^{\mu\nu}=\sum_{k=2}^{N}E_{k}^{\mu\nu}\,,
\ee
\bea
\tilde\varphi(t)&=&
\frac{1}{6}
-\frac{3}{8}\frac{1}{(tE)^2}
-\frac{1}{24}tE\cot(tE)
+\frac{3}{8}\frac{\cot(tE)}{tE}\,,
\\[5pt]
\tilde\rho(t)
&=&
-\frac{5}{24}
+\frac{3}{8}\frac{1}{(tE)^2}
+\frac{1}{24}tE\cot(tE)
-\frac{7}{16}\frac{\cot(tE)}{tE}
+\frac{1}{16}\frac{1}{\sin^2(tE)}\,,
\nonumber\\
\\[5pt]
\tilde\sigma(t)&=&
\frac{1}{16}
+\frac{3}{32}\frac{\cot(tE)}{tE}
-\frac{3}{32}\frac{1}{\sin^2(tE)}\,,
\\[5pt]
\sigma_1(t)
&=&
\frac{1}{16}\cot(tE)\coth(tB)
-\frac{1}{16}\frac{\cot(tE)}{tB}
-\frac{1}{16}\frac{\coth(tB)}{tE}
\nonumber\\[10pt]
&&
+\frac{1}{16}\frac{B\cot(tE)+E\coth(tB)}{t(B^2+E^2)}
\,,
\eea
\bea
\rho_1(t)
&=&
-\frac{1}{48}
\Bigg\{
4
-9\frac{1}{(tE)^2}
+9\frac{1}{(tB)^2}
+8\frac{1}{tE}\cot(tE)
-8\frac{1}{tB}\coth(tB)
\nonumber\\[10pt]
&&
-(tE)\cot(tE)
-(tB)\coth(tB)
\nonumber\\[10pt]
&&
-3\frac{E}{tB^2}\cot(tE)
+3\frac{B}{tE^2}\coth(tB)
+3\left(\frac{E}{B}-\frac{B}{E}\right)
\coth(tB)\cot(tE)
\nonumber\\[10pt]
&&
+\frac{5B^2-E^2}{tB(B^2+E^2)}\coth(tB)
-\frac{5E^2-B^2}{tE(B^2+E^2)}\cot(tE)
\Bigg\}\,.
\eea

\subsubsection{Spinor Fields}

For the spinor fields we obtain
\bea
\Phi^{\rm spinor}(t)
&=&2^{[n/2]}tE\cot(tE)
\left[tB\coth(tB)\right]^{N-1}\,,
\label{43112xxx}
\\[10pt]
B^{\rm spinor}_2(t)&=&
-\frac{1}{12}R
-\tilde\varphi(t)\Pi_{1}^{\mu\nu}R_{\mu\nu}
+\varphi(t)H^{\mu\nu}_2R_{\mu\nu}
+\tilde\rho(t)\Pi_1^{\mu\nu}\Pi_1^{\alpha\beta}R_{\mu\alpha\nu\beta}
\nonumber
\\[5pt]
&&
+\tilde\lambda(t)E_1^{\mu\alpha}E_1^{\nu\beta}R_{\mu\alpha\nu\beta}
-2\rho_1(t)H^{\mu\nu}_2\Pi_1^{\alpha\beta}R_{\mu\alpha\nu\beta}
+2\lambda_1(t) X^{\mu\alpha}_2 E_1^{\nu\beta}R_{\mu\alpha\nu\beta}
\nonumber\\[5pt]
&&
+\rho(t)H^{\mu\nu}_2 H^{\alpha\beta}_2 R_{\mu\alpha\nu\beta}
-\lambda(t)X^{\mu\alpha}_2 X^{\nu\beta}_2 R_{\mu\alpha\nu\beta}
\,,
\eea
where
\bea
\tilde\lambda(t)&=&
-\frac{3}{16}
-\frac{3}{32}\frac{1}{\sin^2(tE)}
+\frac{1}{4}\frac{\tan(tE)}{tE}
+\frac{3}{32}\frac{\cot(tE)}{tE}\,,
\\[10pt]
\lambda_1(t)
&=&
\frac{1}{16}\cot(tE)\coth(tB)
-\frac{1}{16}\frac{\cot(tE)}{tB}
-\frac{1}{16}\frac{\coth(tB)}{tE}
\nonumber\\[10pt]
&&
+\frac{1}{16}\frac{B\cot(tE)+E\coth(tB)}{t(B^2+E^2)}
+\frac{1}{8}\frac{\tanh(tB)}{tE}
-\frac{1}{8}\frac{\tan(tE)}{tB}
\nonumber\\
&&
-\frac{1}{8}\tanh(tB)\cot(tE)
+\frac{1}{8}\tan(tE)\coth(tB)
\,.
\eea

\section{Imaginary Part of the Effective Lagrangian}

Now, we can compute the imaginary part of the effective Lagrangian in the same
approximation
taking into account linear terms in the curvature.
The effective action is given by the integral over $t$ of the
trace of the heat kernel diagonal. Of course, it should be properly
regularized as discussed above. The most important point we want
to make is that in the presence of the electric field the heat kernel
is no longer a nice analytic function of $t$ but it becomes
a meromorphic function of $t$ in the complex plane of $t$ with
poles on the real axis determined by the trigonometric functions
in the coefficient functions computed above.
As was pointed out first by Schwinger these poles should be carefully
avoided by deforming the contour of integration which leads to an
imaginary part of the effective action determined by the contribution
of the residues of the poles. This imaginary part is always finite and does
not depend on the regularization.
We compute below the imaginary part of the effective Lagrangian for the scalar
and the spinor fields.

The trace of the heat kernel, $\textrm{tr}\;U^{\rm diag}(t)$, was computed above and is given
by (\ref{4379xxx}).
Now, by using (\ref{4232xxx})
the calculation of the
imaginary part of the effective Lagrangian is reduced to the calculation
of the residues of the functions $t^{-n/2-1}e^{-tm^2}\Phi(t)$
and $t^{-n/2}e^{-tm^2}\Phi(t)B_2(t)$ at the poles on the real line.
By using the result (\ref{43107xxx}) and (\ref{43112xxx}) for the function $\Phi$
it is not difficult to see that the function
$t^{-n/2-1}e^{-tm^2}\Phi(t)$
is a meromorphic function with
isolated simple poles at $t_{k}={k\pi}/E$
with $k=1,2\dots$. The function
$t^{-n/2}e^{-tm^2}\Phi(t)B_2(t)$ is also a meromorphic function with the same
poles but the poles could be double or even triple.
The imaginary part is, then, simply evaluated by summing the residues of the
integrand at the
poles.
It has the following form
\bea
\textrm{Im}\;\mathcal{L}
&=&\pi(4\pi)^{-n/2}E^{n/2}G_{0}(x,y)
+\pi(4\pi)^{-n/2}E^{n/2-1}
\bigg[G_{1}(x,y)R\nonumber\\
&+&G_{2}(x,y)\Pi_{1}^{\mu\nu}R_{\mu\nu}
+
G_{3}(x,y)H_2{}^{\mu\nu}R_{\mu\nu}
+G_{4}(x,y)\Pi_1^{\mu\nu}\Pi_1^{\alpha\beta}R_{\mu\alpha\nu\beta}
\nonumber\\[5pt]
&+&
G_{5}(x,y)E_1^{\mu\alpha}E_1^{\nu\beta}R_{\mu\alpha\nu\beta}
+G_{6}(x,y)H_2{}^{\mu\nu}\Pi_1^{\alpha\beta}R_{\mu\alpha\nu\beta}
\nonumber\\[5pt]
&+&
G_{7}(x,y)X_2{}^{\mu\alpha} E_1^{\nu\beta}R_{\mu\alpha\nu\beta}
+G_{8}(x,y)H_2{}^{\mu\nu}H_2{}^{\alpha\beta}R_{\mu\alpha\nu\beta}
\nonumber\\
&+&
G_{9}(x,y)X_2{}^{\mu\alpha}X_2{}^{\nu\beta}
R_{\mu\alpha\nu\beta}\bigg]\;,
\eea
where
\be
x=\frac{B}{E}\;,\qquad y=\frac{m^{2}}{E}\;,
\ee
and $G_i(x,y)$ are some functions computed below.

\subsection{Scalar Fields}

At this point it is useful to introduce some auxiliary functions so that the
final result for the quantities $G^{\rm scalar}_{i}(x,y)$ can be written in a
somewhat compact form, namely
\bea
f_{k}(x,y)
&=&\left[\frac{k\pi x}{\sinh\left(k\pi
x\right)}\right]^{N-1}\exp\left({-k\pi y}\right)\;,
\\[10pt]
g_{k}(x,y)&=&(N-1)(k\pi x)\coth(k\pi x)+k\pi y\;,
\\
h_{k}(x,y)&=&\frac{1}{2}N(N-1)(k\pi x)^{2}\coth^{2}(k\pi
x)+\left(\frac{n}{2}-N\right)k\pi y
\nonumber\\
&+&\frac{1}{2}(N-1)[(n-2N)+2k\pi y](k\pi x)\coth(k\pi x)
\nonumber\\
&+&\frac{1}{2}(k\pi)^{2}[1-(N-1)x^{2}+y^{2}]\;,
\\
l_{k}(x,y)
&=&
-k\pi x+\left[\left(\frac{n}{2}-N\right)
+k\pi y\right]\coth(k\pi x)
+N(k\pi x)\coth^{2}(k\pi x)\;,\;\;\;\;
\eea
\bea
\Omega_{1,k}(x,y)
&=&\frac{1}{8}
+\frac{n-2N}{48}
-\frac{3}{8(k\pi)^{2}}\left(\frac{n}{2}-N+2\right)
\nonumber\\[5pt]
&&
+\frac{1}{24}\left(1-\frac{9}{(k\pi)^{2}}\right)g_{k}(x,y)\;,
\\[10pt]
\Omega_{2,k}(x,y)
&=&-\frac{1}{6}
-\frac{n-2N}{48}
+\frac{1}{32(k\pi)^{2}}\left(\frac{n}{2}-N+2\right)
\left(\frac{n}{2}-N+13\right)
\nonumber\\[5pt]
&-&\frac{1}{24}\left(1-\frac{21}{2(k\pi)^{2}}\right)g_{k}(x,y)
+\frac{1}{16}\frac{h_{k}(x,y)}{(k\pi)^{2}}\;,
\\[10pt]
\Omega_{3,k}(x,y)
&=&\frac{1}{16}
-\frac{3}{64(k\pi)^{2}}\left(\frac{n}{2}-N+1\right)
\left(\frac{n}{2}-N+2\right)
\nonumber\\[5pt]
&-&\frac{3}{32(k\pi)^{2}}\Big[h_{k}(x,y)+g_{k}(x,y)\Big]\;,
\eea
\bea
\Omega_{4,k}(x,y)
&=&-\frac{1}{8}
-\frac{n-2N}{48}
+\frac{3}{8(k\pi)^{2}}\left(1-\frac{1}{x^{2}}\right)
+\frac{1}{8}\left(\frac{1}{x}-x\right)\frac{l_{k}(x,y)}{k\pi}
\nonumber\\[5pt]
&+&
\frac{1}{8(k\pi)^{2}}\left[\left(\frac{n}{2}-N+1\right)
+g_{k}(x,y)\right]\left[
\frac{3x^{4}-1}{x^{2}(x^{2}+1)}\right]
\nonumber\\[5pt]
&-&\frac{1}{8}\frac{\coth(k\pi x)}{k\pi x}
\left[\frac{x^{4}-3}{x^{2}+1}-\frac{(k\pi x)^{2}}{3}\right]
-\frac{1}{24}g_{k}(x,y)\;,
\\[10pt]
\Omega_{5,k}(x,y)
&=&\frac{1}{8(k\pi)^{2}}\left[\left(\frac{n}{2}-N+1\right)
+g_{k}(x,y)\right]\frac{1}{x(x^{2}+1)}
-\frac{1}{8}\frac{l_{k}(x,y)}{(k\pi)}
\nonumber\\[5pt]
&-&
\frac{1}{8}\frac{\coth(k\pi x)}{k\pi x}
\left(\frac{x^{3}}{x^{2}+1}\right)\;.
\eea

By using these quantities we obtain the
functions $G^{\rm scalar}_{i}(x,y)$ in the form of the
following series
\bea
G^{\rm scalar}_{0}(x,y)&=&\sum_{k=1}^{\infty}
\frac{(-1)^{k+1}}{(k\pi)^{n/2}}f_{k}(x,y)\;,
\\[10pt]
G^{\rm scalar}_{1}(x,y)&=&\left(\frac{1}{6}-\xi\right)
\sum_{k=1}^{\infty}\frac{(-1)^{k+1}}{(k\pi)^{n/2-1}}f_{k}(x,y)\;,
\\[10pt]
G^{\rm scalar}_{2}(x,y)&=&-\sum_{k=1}^{\infty}
\frac{(-1)^{k+1}}{(k\pi)^{n/2-1}}f_{k}(x,y)\Omega_{1,k}(x,y)\;,
\\[10pt]
G^{\rm scalar}_{3}(x,y)&=&\sum_{k=1}^{\infty}
\frac{(-1)^{k+1}}{(k\pi)^{n/2-1}}f_{k}(x,y)\varphi(k\pi x)\;,
\\[10pt]
G^{\rm scalar}_{4}(x,y)&=&\sum_{k=1}^{\infty}
\frac{(-1)^{k+1}}{(k\pi)^{n/2-1}}f_{k}(x,y)\Omega_{2,k}(x,y)\;,
\\[10pt]
G^{\rm scalar}_{5}(x,y)&=&\sum_{k=1}^{\infty}
\frac{(-1)^{k+1}}{(k\pi)^{n/2-1}}f_{k}(x,y)\Omega_{3,k}(x,y)\;,
\\[10pt]
G^{\rm scalar}_{6}(x,y)&=&-\sum_{k=1}^{\infty}
\frac{(-1)^{k+1}}{(k\pi)^{n/2-1}}f_{k}(x,y)\Omega_{4,k}(x,y)\;,
\end{eqnarray}
\begin{eqnarray}
G^{\rm scalar}_{7}(x,y)&=&\sum_{k=1}^{\infty}
\frac{(-1)^{k+1}}{(k\pi)^{n/2-1}}f_{k}(x,y)\Omega_{5,k}(x,y)\;,
\\[10pt]
G^{\rm scalar}_{8}(x,y)&=&\sum_{k=1}^{\infty}
\frac{(-1)^{k+1}}{(k\pi)^{n/2-1}}f_{k}(x,y)\rho(k\pi x)\;,
\\[10pt]
G^{\rm scalar}_{9}(x,y)&=&\sum_{k=1}^{\infty}
\frac{(-1)^{k+1}}{(k\pi)^{n/2-1}}f_{k}(x,y)\sigma(k\pi x)\;.
\eea

\subsection{Spinor Fields}

Exactly as we did in the previous section, we introduce, now, some auxiliary
functions that will be useful in the presentation of the final result,
namely
\bea
f_{S,k}(x,y)
&=&
\Big[(k\pi x)\coth(k\pi x)\Big]^{N-1}\exp\left({-k\pi y}\right)\;,
\\[10pt]
g_{S,k}(x,y)&=&
(N-1)(k\pi x)\coth(k\pi x)-(N-1)(k\pi x)\tanh(k\pi x)+k\pi y\,,
\\[10pt]
h_{S,k}(x,y)&=&\frac{1}{2}(k\pi y)^{2}
-(N-1)^{2}(k\pi x)^{2}
+\left(\frac{n}{2}-N\right)k\pi y
\\[5pt]
&+&
\frac{1}{2}(N-1)\left(n-2N
+2k\pi y\right)(k\pi x)\Big[\coth(k\pi x)
-\tanh(k\pi x)\Big]
\nonumber\\[5pt]
&+&
\frac{1}{2}N(N-1)(k\pi x)^{2}\coth^{2}(k\pi x)
+\frac{1}{2}(N-1)(N-2)(k\pi x)^{2}\tanh^{2}(k\pi x)\,,
\nonumber
\\[10pt]
l_{S,k}(x,y)
&=&
-Nk\pi x+\left(\frac{n}{2}-N+k\pi y\right)
\coth(k\pi x)+N(k\pi x)\coth^{2}(k\pi x)\;,
\\[10pt]
p_{S,k}(x,y)&=&
(N-2)k\pi x+\left(\frac{n}{2}-N+k\pi y\right)
\tanh(k\pi x)-(N-2)(k\pi x)\tanh^{2}(k\pi x)\;,
\nonumber\\
&&
\\
\Lambda_{1,k}(x,y)&=&\frac{1}{8}
+\frac{n-2N}{48}
-\frac{3}{8(k\pi)^{2}}\left(\frac{n}{2}-N+2\right)
+\frac{1}{24}\left(1
-\frac{9}{(k\pi)^{2}}\right)g_{S,k}(x,y)\;,
\nonumber\\
\eea
\begin{eqnarray}
\Lambda_{2,k}(x,y)&=&-\frac{1}{6}
-\frac{n-2N}{48}
+\frac{1}{32(k\pi)^{2}}\left(\frac{n}{2}-N+2\right)
\left(\frac{n}{2}-N+13\right)
\nonumber\\[5pt]
&-&\frac{1}{24}\left(1-\frac{21}{2(k\pi)^{2}}\right)g_{S,k}(x,y)
+\frac{1}{16}\frac{h_{S,k}(x,y)}{(k\pi)^{2}}\;,
\\[10pt]
\Lambda_{3,k}(x,y)&=&-\frac{3}{16}
-\frac{3}{64(k\pi)^{2}}\left(\frac{n}{2}-N+1\right)
\left(\frac{n}{2}-N+2\right)
\nonumber\\[5pt]
&-&\frac{3}{32(k\pi)^{2}}\Big[g_{S,k}(x,y)+h_{S,k}(x,y)\Big]\;,
\eea
\bea
\Lambda_{4,k}(x,y)&=&-\frac{1}{8}
-\frac{n-2N}{48}
+\frac{3}{8(k\pi)^{2}}\left(1-\frac{1}{x^{2}}\right)
+\frac{1}{8}\left(\frac{1}{x}-x\right)\frac{l_{S,k}(x,y)}{k\pi}
\nonumber\\[5pt]
&+&\frac{1}{8(k\pi)^{2}}\left(\frac{n}{2}-N+1
+g_{S,k}(x,y)\right)
\frac{3x^{4}-1}{x^{2}(x^{2}+1)}
\nonumber\\[5pt]
&-&\frac{1}{8}\frac{\coth(k\pi x)}{k\pi x}
\left[\frac{x^{4}-3}{x^{2}+1}-\frac{(k\pi x)^{2}}{3}\right]
-\frac{1}{24}g_{S,k}(x,y)\;,
\\[10pt]
\Lambda_{5,k}(x,y)
&=&\frac{1}{8(k\pi)^{2}}\left(\frac{n}{2}-N+1
+g_{S,k}(x,y)\right)
\frac{1}{x(x^{2}+1)}
-\frac{1}{8}\frac{l_{S,k}(x,y)}{k\pi}
\nonumber\\
&-&\frac{1}{8}\frac{\coth(k\pi x)}{k\pi x}
\left(\frac{x^{3}}{x^{2}+1}\right)
+\frac{1}{4(k\pi)}\Big[\tanh(k\pi x)
+p_{S,k}(x,y)\Big]\;.
\eea

By using the above functions we can write the explicit expression for the
quantities
$G^{\rm spinor}_{i}(x,y)$
\bea
G^{\rm spinor}_{0}(x,y)&=&2^{\left[\frac{n}{2}\right]}
\sum_{k=1}^{\infty}
\frac{1}{(k\pi)^{n/2}}f_{S,k}(x,y)\;,
\\[10pt]
G^{\rm spinor}_{1}(x,y)&=&-\frac{2^{\left[\frac{n}{2}\right]}}{12}
\sum_{k=1}^{\infty}
\frac{1}{(k\pi)^{n/2-1}}f_{S,k}(x,y)\;,
\\[10pt]
G^{\rm spinor}_{2}(x,y)&=&-2^{\left[\frac{n}{2}\right]}
\sum_{k=1}^{\infty}
\frac{1}{(k\pi)^{n/2-1}}f_{S,k}(x,y)\Lambda_{1,k}(x,y)\;,
\end{eqnarray}
\begin{eqnarray}
G^{\rm spinor}_{3}(x,y)&=&2^{\left[\frac{n}{2}\right]}
\sum_{k=1}^{\infty}
\frac{1}{(k\pi)^{n/2-1}}f_{S,k}(x,y)\varphi(k\pi x)\;,
\\[10pt]
G^{\rm spinor}_{4}(x,y)&=&2^{\left[\frac{n}{2}\right]}
\sum_{k=1}^{\infty}
\frac{1}{(k\pi)^{n/2-1}}f_{S,k}(x,y)\Lambda_{2,k}(x,y)\;,
\\[10pt]
G^{\rm spinor}_{5}(x,y)&=&2^{\left[\frac{n}{2}\right]}
\sum_{k=1}^{\infty}
\frac{1}{(k\pi)^{n/2-1}}f_{S,k}(x,y)\Lambda_{3,k}(x,y)\;,
\\[10pt]
G^{\rm spinor}_{6}(x,y)&=&-2^{\left[\frac{n}{2}\right]}
\sum_{k=1}^{\infty}
\frac{1}{(k\pi)^{n/2-1}}f_{S,k}(x,y)\Lambda_{4,k}(x,y)\;,
\eea
\bea
G^{\rm spinor}_{7}(x,y)&=&2^{\left[\frac{n}{2}\right]}
\sum_{k=1}^{\infty}
\frac{1}{(k\pi)^{n/2-1}}f_{S,k}(x,y)\Lambda_{5,k}(x,y)\;,
\\[10pt]
G^{\rm spinor}_{8}(x,y)&=&2^{\left[\frac{n}{2}\right]}
\sum_{k=1}^{\infty}
\frac{1}{(k\pi)^{n/2-1}}f_{S,k}(x,y)\rho(k\pi x)\;,
\\[10pt]
G^{\rm spinor}_{9}(x,y)&=&-2^{\left[\frac{n}{2}\right]}
\sum_{k=1}^{\infty}
\frac{1}{(k\pi)^{n/2-1}}f_{S,k}(x,y)\lambda(k\pi x)\;.
\eea

Notice that because of the infrared cutoff factor $e^{-k\pi y}$ the functions
$G_i(x,y)$ are exponentially small for massive fields in weak electric fields
when the parameter is large, $y>>1$ (that is, $m^2>>E$), independently on $x$.
In this case, all
these functions are
approximated by just the first term of the series corresponding to
$k=1$.


\section{Strong Electric Field in Four Dimensions}

The formulas obtained in the previous section are very general and are valid in
any dimensions.
In this section we will present some particular cases of major interest.

\subsection{Four Dimensions}

In this section we will consider the physical case when $n=4$. Obviously in
four dimensions
we only have two invariants, and, therefore, $N=2$.
The imaginary part of the effective Lagrangian reads now
\bea\label{gf1}
\textrm{Im}\;\mathcal{L}
&=&\pi(4\pi)^{-2}E^{2}G_{0}(x,y)
+\pi(4\pi)^{-2}E
\bigg[G_{1}(x,y)R\nonumber\\
&+&G_{2}(x,y)\Pi_{1}^{\mu\nu}R_{\mu\nu}
+
G_{3}(x,y)\Pi_2{}^{\mu\nu}R_{\mu\nu}
+G_{4}(x,y)\Pi_1^{\mu\nu}\Pi_1^{\alpha\beta}R_{\mu\alpha\nu\beta}
\nonumber\\[5pt]
&+&
G_{5}(x,y)E_1^{\mu\alpha}E_1^{\nu\beta}R_{\mu\alpha\nu\beta}
+G_{6}(x,y)\Pi_2{}^{\mu\nu}\Pi_1^{\alpha\beta}R_{\mu\alpha\nu\beta}
\nonumber\\[5pt]
&+&
G_{7}(x,y)E_2{}^{\mu\alpha} E_1^{\nu\beta}R_{\mu\alpha\nu\beta}
+G_{8}(x,y)\Pi_2{}^{\mu\nu}\Pi_2{}^{\alpha\beta}R_{\mu\alpha\nu\beta}
\nonumber\\
&+&
G_{9}(x,y)E_2{}^{\mu\alpha}E_2{}^{\nu\beta}
R_{\mu\alpha\nu\beta}\bigg]\;.
\eea

For scalar fields in four dimensions the functions $G^{\rm scalar}_{i}(x,y)$
take the form
\bea
G^{\rm scalar}_{0}(x,y)
&=&\frac{x}{\pi}
\sum_{k=1}^{\infty}
\frac{e^{-k\pi y}}{k\,\sinh(k\pi x)}\;,
\\[10pt]
G^{\rm scalar}_{1}(x,y)
&=&\left(\frac{1}{6}-\xi\right)x\sum_{k=1}^{\infty}
\frac{e^{-k\pi y}}{\sinh(k\pi x)}\;,
\\[10pt]
G^{\rm scalar}_{2}(x,y)
&=&-x\sum_{k=1}^{\infty}
\frac{e^{-k\pi y}}{\sinh(k\pi x)}
\Bigg\{\frac{1}{8}
-\frac{3}{4(k\pi)^{2}}
+\frac{1}{24}\left(k\pi-\frac{9}{k\pi}\right)
\left[y+x \coth(k\pi x)\right]\Bigg\}\;,
\nonumber\\
\\
G^{\rm scalar}_{3}(x,y)
&=&
x\sum_{k=1}^{\infty}
\frac{e^{-k\pi y}}{\sinh(k\pi x)}
\Bigg\{\frac{1}{6}
+\frac{3}{8}\frac{1}{(k\pi x)^2}
-\frac{1}{24}k\pi x\coth(k\pi x)
-\frac{3}{8}\frac{\coth(k\pi x)}{k\pi x}
\Bigg\}
\;,
\nonumber\\
\eea
\bea
G^{\rm scalar}_{4}(x,y)
&=&x\sum_{k=1}^{\infty}
\frac{e^{-k\pi y}}{\sinh(k\pi x)}
\Bigg\{-\frac{13}{96}
+\frac{13}{16(k\pi)^{2}}
-\frac{x^{2}}{32}
+\frac{y^{2}}{32}
+\left(\frac{7}{16(k\pi)}-\frac{k\pi}{24}\right)y
\nonumber\\[5pt]
&&
+\left(\frac{y}{16}+\frac{7}{16 k\pi}-\frac{k\pi}{24}\right)
x\coth(k\pi x)
+\frac{x^2}{16}\coth^2(k\pi x)\Bigg\}\,,
\\[10pt]
G^{\rm scalar}_{5}(x,y)&=&
x\sum_{k=1}^{\infty}
\frac{e^{-k\pi y}}{\sinh(k\pi x)}
\Bigg\{\frac{1}{64}
-\frac{3}{32(k\pi)^{2}}
+\frac{3x^{2}}{64}
-\frac{3y^{2}}{64}
-\frac{3}{32k\pi}y
\nonumber\\[5pt]
&-&\frac{3}{32}
\left(y+\frac{1}{k\pi}\right)\coth(k\pi x)
-\frac{3}{32}x^2\coth^2(k\pi x)\Bigg\}\,,
\eea
\bea
G^{\rm scalar}_{6}(x,y)&=&
-\frac{x}{x^2+1}\sum_{k=1}^{\infty}
\frac{e^{-k\pi y}}{\sinh(k\pi x)}\Bigg\{-\frac{1}{4}
-\frac{1}{2(k\pi x)^{2}}
+\frac{3x^{2}}{4(k\pi)^{2}}
+\frac{x^{2}}{8}(x^{2}-1)
\nonumber\\[5pt]
&-&\frac{k\pi}{24}y(x^2+1)
-\frac{1}{8k\pi}y\left(\frac{1}{x^{2}}-3x^{2}\right)
-\frac{1}{4}(x^{4}-1)\coth^{2}(k\pi x)
\nonumber\\[5pt]
&+&\left[\frac{1}{4k\pi}\left(x^{3}+\frac{1}{x}\right)
+\frac{y}{8}\left(\frac{1}{x}-x^{3}\right)\right]\coth(k\pi x)
\Bigg\}\;,
\eea
\bea
G^{\rm scalar}_{7}(x,y)&=&
\frac{1}{x^2+1}
\sum_{k=1}^{\infty}
\frac{e^{-k\pi y}}{\sinh(k\pi x)}
\Bigg\{
\frac{1}{8(k\pi)^{2}}
+\frac{x^{2}}{8}(x^{2}+1)
+\frac{y}{8k\pi}
\nonumber\\[5pt]
&-&\frac{x}{8}\left[y(1+x^2)
-\frac{1}{k\pi}(1-x^2)\right]\coth(k\pi x)
\nonumber\\[5pt]
&-&\frac{1}{4}x^{2}(x^{2}+1)\coth^{2}(k\pi x)\Bigg\}\;,
\\[10pt]
G^{\rm scalar}_{8}(x,y)
&=&
x\sum_{k=1}^{\infty}
\frac{e^{-k\pi y}}{\sinh(k\pi x)}
\Bigg\{
-\frac{7}{48}
-\frac{3}{8}\frac{1}{(k\pi x)^2}
-\frac{1}{16}\coth^2(k\pi x)
\nonumber\\[5pt]
&&
+\left(\frac{k\pi}{24}x
+\frac{7}{16k\pi x}\right)\coth(k\pi x)
\Bigg\}
\;,
\\[10pt]
G^{\rm scalar}_{9}(x,y)
&=&x\sum_{k=1}^{\infty}
\frac{e^{-k\pi y}}{\sinh(k\pi x)}
\Bigg\{
-\frac{1}{32}
-\frac{3}{32 k\pi x}\coth(k\pi x)
+\frac{3}{32}\coth^2(k\pi x)
\Bigg\}
\;.
\nonumber\\
\eea

For spinor fields in four dimensions the functions
$G^{\rm spinor}_{i}(x,y)$ take the form
\bea
G^{\rm spinor}_{0}(x,y)
&=&\frac{4x}{\pi}
\sum_{k=1}^{\infty}
\frac{1}{k}\coth(k\pi x)e^{-k\pi y}\;,
\\[10pt]
G^{\rm spinor}_{1}(x,y)
&=&-\frac{x}{3}
\sum_{k=1}^{\infty}
\coth(k\pi x)e^{-k\pi y}\;,
\eea
\bea
G^{\rm spinor}_{2}(x,y)
&=&-4x
\sum_{k=1}^{\infty}
\coth(k\pi x)e^{-k\pi y}\Bigg\{\frac{1}{8}
-\frac{3}{4(k\pi)^{2}}
\nonumber\\[5pt]
&+&\frac{1}{24}\left(k\pi
-\frac{9}{k\pi}\right)\Big[y+x\coth(k\pi x)-x
\tanh(k\pi x)\Big]\Bigg\}\;,
\\[10pt]
G^{\rm spinor}_{3}(x,y)
&=&4
x\sum_{k=1}^{\infty}
\coth(k\pi x)e^{-k\pi y}
\Bigg\{\frac{1}{6}
+\frac{3}{8}\frac{1}{(k\pi x)^2}
-\frac{1}{24}k\pi x\coth(k\pi x)
\nonumber\\[5pt]
&&
-\frac{3}{8}\frac{\coth(k\pi x)}{k\pi x}
\Bigg\}
\;,
\eea
\bea
G^{\rm spinor}_{4}(x,y)&=&
4x\sum_{k=1}^{\infty}
\coth(k\pi x)e^{-k\pi y}\Bigg\{-\frac{1}{6}
+\frac{13}{16(k\pi)^{2}}
-\frac{x^{2}}{16}
+\frac{y^{2}}{32}
\nonumber\\[5pt]
&-&\frac{y}{24}\left(k\pi-\frac{21}{2k\pi}\right)
+\frac{x^{2}}{16}\coth^{2}(k\pi x)
\nonumber\\[5pt]
&-&\frac{x}{24}\left(k\pi
-\frac{21}{2k\pi}-\frac{3y}{2}\right)\Big[\coth(k\pi x)
-\tanh(k\pi x)\Big]\Bigg\}\;,
\\[10pt]
%
G^{\rm spinor}_{5}(x,y)
&=&4x
\sum_{k=1}^{\infty}
\coth(k\pi x)e^{-k\pi y}
\Bigg\{-\frac{3}{16}
-\frac{3}{32(k\pi)^{2}}
+\frac{3x^{2}}{32}
\nonumber\\[5pt]
&-&\frac{3y^{2}}{64}
-\frac{3y}{32(k\pi)}
-\frac{3x^{2}}{32}\coth^{2}(k\pi x)
\nonumber\\[5pt]
&-&\frac{3x}{32}\left(\frac{1}{k\pi}+y\right)
\Big[\coth(k\pi x)-\tanh(k\pi x)\Big]\Bigg\}\;,
\eea
\bea
G^{\rm spinor}_{6}(x,y)
&=&-\frac{4x}{x^2+1}
\sum_{k=1}^{\infty}
\coth(k\pi x)e^{-k\pi y}\Bigg\{-\frac{3}{8}
+\frac{3}{4(k\pi)^{2}}\left(x^{2}-\frac{2}{3x^{2}}\right)
\nonumber\\[5pt]
&+&\frac{x^{2}}{8}(2x^{2}-1)
-\frac{x^{2}y}{24}\left(k\pi-\frac{9}{k\pi}\right)
-\frac{y}{24}\left(k\pi+\frac{3}{k\pi x^{2}}\right)
\nonumber\\[5pt]
&+&\left[\frac{1}{4k\pi x}+\frac{y}{8x}-\frac{x^{3}}{8}
\left(y-\frac{2}{k\pi}
\right)\right]\coth(k\pi x)
\nonumber\\[5pt]
&+&\left[\frac{1}{8k\pi x}+\frac{k\pi x}{24}
+\frac{x^{3}}{24}\left(k\pi-\frac{9}{k\pi}\right)
\right]\tanh(k\pi x)
\nonumber\\[5pt]
&-&\frac{1}{4}(x^{4}-1)\coth^{2}(k\pi x)
\Bigg\}\;,
\\[10pt]
G^{\rm spinor}_{7}(x,y)
&=&\frac{4}{x^2+1}
\sum_{k=1}^{\infty}
\coth(k\pi x)
e^{-k\pi y}\Bigg\{\frac{1}{8(k\pi)^{2}}
+\frac{x^{2}}{4}(x^{2}+1)
\nonumber\\[5pt]
&+&\frac{y}{8k\pi}
-\frac{x^{2}}{4}(x^{2}+1)\coth^{2}(k\pi x)
\nonumber\\
&+&\frac{x}{8}\left[\frac{1}{k\pi}(1-x^{2})-y(1+x^{2})
\right]\coth(k\pi x)
\nonumber\\[5pt]
&+&\frac{x}{8}\left[\frac{1}{k\pi}(1+2x^{2})+2y(1+x^{2})
\right]\tanh(k\pi x)\Bigg\}\;,
\eea
\bea
G^{\rm spinor}_{8}(x,y)
&=&4x
\sum_{k=1}^{\infty}
\coth(k\pi x)e^{-k\pi y}
\Bigg\{
-\frac{7}{48}
-\frac{3}{8}\frac{1}{(k\pi x)^2}
+\frac{1}{24}k\pi x\coth(k\pi x)
\nonumber\\[5pt]
&&
+\frac{7}{16}\frac{\coth(k\pi x)}{k\pi x}
-\frac{1}{16}\coth^2(k\pi x)
\Bigg\}
\;,
\\[10pt]
G^{\rm spinor}_{9}(x,y)
&=&-4x
\sum_{k=1}^{\infty}
\coth(k\pi x)e^{-k\pi y}
\Bigg\{
-\frac{9}{32}
+\frac{3}{32}\coth^2(k\pi x)
+\frac{1}{4}\frac{\tanh(k\pi x)}{k\pi x}
\nonumber\\[5pt]
&&
-\frac{3}{32}\frac{\coth(k\pi x)}{k\pi x}
\Bigg\}
\;.
\eea
%

\subsection{Supercritical Electric Field}

As we already mentioned above, the functions $G_i(x,y)$ are exponentially
small for massive fields in weak electric fields for large $y=m^2/E$, as $y\to
\infty$. Now we are considering the opposite case of light (or massless)
fields in strong (supercritical) electric fields, when $y\to 0$ with a fixed
$x$. This corresponds to the regime
\be
m^2<<B, E\,.
\ee

\subsubsection{Scalar Fields}

The infrared (massless) limit for scalar fields is regular---there
are no infrared divergences. This is due
to the presence of the hyperbolic sine $\sinh(k\pi x)$ in the denominator, which gives a cut-off for large $k$ in the series, and
therefore, assures its convergence. The result for the massless limit in the
scalar case can be simply obtained by setting $y=0$ in the above formulas
for the functions $G_i(x,y)$.

\subsubsection{Spinor Fields}

The spinor case is quite different. The presence of the hyperbolic cotangent
$\coth(k\pi x)$ does not provide a cut-off for the
convergence of the series as $k\to\infty$. This leads, in the spinor case in
four dimensions, to the presence of infrared divergences as $y=m^2/E\to 0$.
By carefully studying the behavior of the series as $k\to \infty$ for a finite
$y$ and then letting $y\to 0$ we compute the asymptotic expansion of the
functions $G_{i}^{\rm spinor}(x,y)$ as $y\to 0$.

We obtain
\bea
G_{0}^{\rm spinor}(x,y)
&=&\frac{2x}{3}+O(y)\;,
\label{4564xxx}
\\[5pt]
G_{1}^{\rm spinor}(x,y)
&=&-\frac{1}{3\pi}\frac{x}{y}
+\frac{x}{8}+O(y)\;,
\\[5pt]
G_{2}^{\rm spinor}(x,y)
&=&-\frac{2}{3\pi}\frac{x}{y}
+\frac{3x}{4}+O(y)\;,
\\[5pt]
G_{3}^{\rm spinor}(x,y)
&=&-\frac{1}{6\pi}\frac{x^{2}}{y^{2}}
+\frac{2}{3\pi}\frac{x}{y}
+\frac{3}{2\pi}\log(\pi y)
+\frac{x^{2}(\pi x-24)+18}{72x}+O(y)\;,
\nonumber\\
\\
G_{4}^{\rm spinor}(x,y)
&=&-\frac{5}{6\pi}\frac{x}{y}
+\frac{7x}{8}+O(y)\;,
\\[5pt]
G_{5}^{\rm spinor}(x,y)
&=&-\frac{3}{4\pi}\frac{x}{y}
+\frac{5x}{16}+O(y)\;,
\\[5pt]
G_{6}^{\rm spinor}(x,y)
&=&\frac{x^{2}-1}{6\pi(x^{2}+1)}\frac{x^{2}}{y^{2}}
+\frac{2}{3\pi}\frac{x}{y}
-\frac{x^{4}-3}{2\pi(x^{2}+1)}\log(\pi y)
\\[5pt]
&-&\frac{6\pi(9x^{4}+3x^{2}-4)
-36x(x^{4}-1)+\pi^{2}x^{3}(x^{2}-1)}{72\pi
x(x^{2}+1)}+O(y)\;,
\nonumber
\\[5pt]
G_{7}^{\rm spinor}(x,y)
&=&-\frac{x(x^{2}+2)}{2\pi(x^{2}+1)}\log(\pi y)
+\frac{6x(x^{2}+1)+\pi}{12\pi(x^{2}+1)}+O(y)\;,
\\[5pt]
G_{8}^{\rm spinor}(x,y)
&=&\frac{1}{6\pi}\frac{x^{2}}{y^{2}}
-\frac{5}{6\pi}\frac{x}{y}
-\frac{7}{4\pi}\log(\pi y)
-\frac{x^{2}(\pi x-30)+18}{72x}+O(y)\;,
\nonumber\\
\\[5pt]
G_{9}^{\rm spinor}(x,y)
&=&\frac{3}{4\pi}\frac{x}{y}
+\frac{5}{8\pi}\log(\pi y)
-\frac{3x}{8}+O(y)\;.
\label{4573xxx}
\eea

Thus, we clearly see the infrared divergences of order $x^2/y^2=B^2/m^{4}$,
$x/y=B/m^{2}$ and $\log y=\log(m^2/E)$.

\subsection{Pure Electric Field}

We analyze now the case of pure electric field without a
magnetic field, that is, $B=0$, which corresponds to the limit $x\to 0$
with fixed $y$. This corresponds to the physical regime when
\be
B<<m^2, E\,.
\ee
In this discussion we present the results
in arbitrary
dimension first and then we specialize them to the physical dimension $n=4$.

\subsubsection{Scalar Fields}

We now evaluate the functions $G_i(x,y)$ for $x=0$ and a finite $y$.
In this limit we are presented with
series of the following general form
\be
\chi_{n}^{\rm scalar}(y)=\sum_{k=1}^{\infty}
\frac{(-1)^{k+1}e^{-k\pi y}}{k^{n/2}}\;.
\ee
This series can be expressed in terms of the
polylogarithmic
function defined by
\be
\textrm{Li}_{j}(z)=\sum_{k=1}^{\infty}\frac{z^{k}}{k^{j}}\;,
\ee
so that, we have
\be\label{47}
\chi_{n}^{\rm scalar}(y)
=-\textrm{Li}_{\frac{n}{2}}(-e^{-\pi y})\;.
\ee

It is not difficult to notice that the limit as $x\rightarrow 0$ of the
functions
$G_{3}^{\rm scalar}$, $G_{6}^{\rm scalar}$, $G_{7}^{\rm scalar}$,
$G_{8}^{\rm scalar}$ and $G_{9}^{\rm scalar}$ vanish
identically, that is,
\be
G_{3}^{\rm scalar}(0,y)=G_{6}^{\rm scalar}(0,y)=
G_{7}^{\rm scalar}(0,y)=G_{8}^{\rm scalar}(0,y)=
G_{9}^{\rm scalar}(0,y)=0\,.
\ee
The explicit expression for the remaining non-vanishing $G_{i}^{\rm scalar}$
for pure
electric field in $n$ dimensions is
\bea
G_{0}^{\rm scalar}(0,y)&=&
-\pi^{-n/2}
\textrm{Li}_{\frac{n}{2}}(-e^{-\pi y})\;,
\\[10pt]
G_{1}^{\rm scalar}(0,y)&=&
-\left(\frac{1}{6}-\xi\right)\frac{1}{\pi^{n/2-1}}
\textrm{Li}_{\frac{n}{2}-1}(-e^{-\pi y})\;,
\\
G_{2}^{\rm scalar}(0,y)&=&
-\frac{1}{48\pi^{n/2+1}}
\bigg\{2\pi^{3}y\textrm{Li}_{\frac{n}{2}-2}(-e^{-\pi y})
+(n+4)\pi^{2}\textrm{Li}_{\frac{n}{2}-1}(-e^{-\pi y})
\nonumber\\[10pt]
&-&18\pi y\textrm{Li}_{\frac{n}{2}}(-e^{-\pi y})
-9(n+2)\textrm{Li}_{\frac{n}{2}+1}(-e^{-\pi y})\bigg\}\;,
\eea
\bea
G_{4}^{\rm scalar}(0,y)&=&
\frac{1}{384\pi^{n/2+1}}
\bigg\{-16\pi^{3}y\textrm{Li}_{\frac{n}{2}-2}(-e^{-\pi y})
-4\pi^{2}(2n+9-3y^{2})\textrm{Li}_{\frac{n}{2}-1}(-e^{-\pi y})
\nonumber\\[5pt]
&+&12(n+12)\pi y\textrm{Li}_{\frac{n}{2}}(-e^{-\pi y})
+3(n+2)(n+24)\textrm{Li}_{\frac{n}{2}+1}(-e^{-\pi y})\bigg\}\;,
\\[10pt]
G_{5}^{\rm scalar}(0,y)&=&\frac{1}{256\pi^{n/2+1}}
\bigg\{4\pi^{2}(1-3y^{2})\textrm{Li}_{\frac{n}{2}-1}(-e^{-\pi y})
-12n\pi y\textrm{Li}_{\frac{n}{2}}(-e^{-\pi y})
\nonumber\\[5pt]
&-&3n(n+2)\textrm{Li}_{\frac{n}{2}+1}(-e^{-\pi y})\bigg\}\;.
\eea

In the physical case of $n=4$ some of the polylogarithmic functions
can be expressed in terms of elementary functions. In this case we have
\bea
G_{0}^{\rm scalar}(0,y)
&=&-\frac{1}{\pi^{2}}\textrm{Li}_{2}(-e^{-\pi
y})\;,
\\[10pt]
G_{1}^{\rm scalar}(0,y)&=&
-\left(\frac{1}{6}-\xi\right)
\frac{1}{\pi}\ln(1+e^{-\pi y})\;,
\\[10pt]
G_{2}^{\rm scalar}(0,y)&=&
\frac{1}{48\pi^{3}}
\bigg\{\frac{2\pi^{3}y e^{-\pi y}}{1+e^{-\pi y}}
+8\pi^{2}\ln(1+e^{-\pi y})
\nonumber\\[5pt]
&+&18\pi y\textrm{Li}_{2}(-e^{-\pi y})
+54\textrm{Li}_{3}(-e^{-\pi y})\bigg\}\;,
\eea
\bea
G_{4}^{\rm scalar}(0,y)&=&
\frac{1}{384\pi^{3}}
\bigg\{\frac{16\pi^{3}y e^{-\pi y}}{1+e^{-\pi y}}
+4\pi^{2}(17-3y^{2})\ln(1+e^{-\pi y})
\nonumber\\[5pt]
&+&192\pi y\textrm{Li}_{2}(-e^{-\pi y})
+504\textrm{Li}_{3}(-e^{-\pi y})\bigg\}\;,
\\[10pt]
G_{5}^{\rm scalar}(0,y)&=&
-\frac{1}{256\pi^{3}}
\bigg\{4\pi^{2}(1-3y^{2})\ln(1+e^{-\pi y})
+48\pi y\textrm{Li}_{2}(-e^{-\pi y})
\nonumber\\[5pt]
&+&72\textrm{Li}_{3}(-e^{-\pi y})\bigg\}\;.
\eea


We study now the behavior of these functions as $y\to 0$, which corresponds to
the limit
\be
B=0\,,\qquad m^2<<E\,.
\ee
By taking the
limit as $y\rightarrow 0$ of the expression (\ref{47}) and by noticing that
\be\label{438}
\textrm{Li}_{n}(-1)=-(1-2^{1-n})\zeta\left(n\right)\;,
\ee
where $\zeta(x)$ denotes the Riemann zeta function, we obtain
\be\label{476}
G_{0}^{\rm scalar}(0,0)
=\frac{(1-2^{1-n/2})}{\pi^{n/2}}
\zeta\left(\frac{n}{2}\right)\;.
\ee
Next, by taking the limit
as $y\rightarrow 0$
and by using the formula
(\ref{438}),
it is not difficult to obtain
\bea
G_{1}^{\rm scalar}(0,0)&=&
-\left(\frac{1}{6}-\xi\right)\pi^{1-n/2}(1-2^{2-n/2})
\zeta\left(\frac{n}{2}-1\right)\;,
\\[10pt]
G_{2}^{\rm scalar}(0,0)
&=&-\frac{1}{48\pi^{n/2+1}}
\bigg\{-(n+4)\pi^{2}(1-2^{2-n/2})\zeta\left(\frac{n}{2}-1\right)
\nonumber\\[5pt]
&+&9(n+2)(1-2^{-n/2})\zeta\left(\frac{n}{2}+1\right)\bigg\}\;,
\eea
\bea
G_{4}^{\rm scalar}(0,0)
&=&\frac{1}{384\pi^{n/2+1}}
\bigg\{4\pi^{2}(2n+9)(1-2^{2-n/2})\zeta\left(\frac{n}{2}-1\right)
\nonumber\\[5pt]
&-&3(n+2)(n+24)(1-2^{-n/2})\zeta\left(\frac{n}{2}+1\right)\bigg\}\;,
\\[10pt]
G_{5}^{\rm scalar}(0,0)
&=&\frac{1}{256\pi^{n/2+1}}
\bigg\{-4\pi^{2}(1-2^{2-n/2})\zeta\left(\frac{n}{2}-1\right)
\nonumber\\[5pt]
&+&3n(n+2)(1-2^{-n/2})\zeta\left(\frac{n}{2}+1\right)\bigg\}\;.
\eea

We consider, at this point, the physical case of four dimensions.
By setting $n=4$ in (\ref{476}) we obtain
\be
G_{0}^{\rm scalar}(0,0)=\frac{1}{12}\;.
\ee
Now, we notice the following
relation
\be
(1-2^{2-n/2})\zeta\left(\frac{n}{2}-1\right)=\eta\left(\frac{n}{2}-1\right)\;,
\ee
where $\eta(x)$ is the Dirichlet eta function. In the particular case of four
dimensions we have that
\be
\lim_{n\rightarrow 4}(1-2^{2-n/2})
\zeta\left(\frac{n}{2}-1\right)=\eta(1)=\ln
2\;.
\ee
By using the last remark we obtain the values of the functions
$G_{i}(n,y)$ in four dimensions
\bea
G_{1}^{\rm scalar}(0,0)
&=&-\left(\frac{1}{6}-\xi\right)\frac{1}{\pi}\ln 2\;,
\\[10pt]
G_{2}^{\rm scalar}(0,0)
&=&\frac{1}{6\pi}\ln 2 -\frac{27}{32 \pi^{3}}\zeta(3)\;,
\\[10pt]
G_{4}^{\rm scalar}(0,0)
&=&\frac{17}{96\pi}\ln 2-\frac{63}{64\pi^{3}}\zeta(3)\;,
\\[10pt]
G_{5}^{\rm scalar}(0,0)
&=&-\frac{1}{64\pi}\ln 2+\frac{27}{128\pi^{3}}\zeta(3)\;.
\eea

\subsubsection{Spinor Fields}

For spinor fields the expressions
for the non-vanishing $G_{i}^{\rm spinor}$ in the limit $x\to 0$ are
\bea\label{472}
G_{0}^{\rm spinor}(0,y)&=&
2^{[n/2]}\pi^{-n/2}\textrm{Li}_{n/2}(e^{-\pi y})\;,
\\[10pt]
G_{1}^{\rm spinor}(0,y)&=&
-\frac{2^{[n/2]}}{12}\frac{1}{\pi^{n/2-1}}
\textrm{Li}_{\frac{n}{2}-1}(e^{-\pi y})\;,
\\[10pt]
G_{2}^{\rm spinor}(0,y)
&=&
-\frac{2^{[n/2]}}{48\pi^{n/2+1}}\bigg\{2\pi^{3}y
\textrm{Li}_{\frac{n}{2}-2}(e^{-\pi
y})
+(n+4)\pi^{2}\textrm{Li}_{\frac{n}{2}-1}(e^{-\pi y})
\nonumber\\[5pt]
&-&18\pi y\textrm{Li}_{\frac{n}{2}}(e^{-\pi y})
-9(n+2)\textrm{Li}_{\frac{n}{2}+1}(e^{-\pi y})\bigg\}\;,
\\[10pt]
G_{4}^{\rm spinor}(0,y)
&=&\frac{2^{[n/2]}}{384\pi^{n/2+1}}
\bigg\{-16\pi^{3}y\textrm{Li}_{\frac{n}{2}-2}(e^{
-\pi y})
-4\pi^{2}(2n+12-3y^{2})\textrm{Li}_{\frac{n}{2}-1}(e^{-\pi y})
\nonumber\\[5pt]
&+&12(n+12)\pi y\textrm{Li}_{\frac{n}{2}}(e^{-\pi y})
+3(n+2)(n+24)\textrm{Li}_{\frac{n}{2}+1}(e^{-\pi y})\bigg\}\;,
\eea
\bea
G_{5}^{\rm spinor}(0,y)&=&
-2^{[n/2]}\frac{3}{256\pi^{n/2+1}}
\bigg\{4\pi^{2}(4+y^{2})\textrm{Li}_{\frac{n}{2}-1}(e^{-\pi y})
+4n\pi y\textrm{Li}_{\frac{n}{2}}(e^{-\pi y})
\nonumber\\[5pt]
&+&n(n+2)\textrm{Li}_{\frac{n}{2}+1}(e^{-\pi y})\bigg\}\;.
\eea

In the particular case of $n=4$ the above results read
\bea
G_{0}^{\rm spinor}(0,y)&=&
\frac{4}{\pi^{2}}\textrm{Li}_{2}(e^{-\pi
y})\;,
\\[10pt]
G_{1}^{\rm spinor}(0,y)&=&
\frac{1}{3\pi}\ln(1-e^{-\pi y})\;,
\label{4565xxx}
\\[10pt]
G_{2}^{\rm spinor}(0,y)
&=&-\frac{1}{12\pi^{3}}\bigg\{\frac{2\pi^{3}y
e^{-\pi y}}{1-e^{-\pi y}}
-8\pi^{2}\ln(1-e^{-\pi y})
\nonumber\\[5pt]
&-&18\pi y\textrm{Li}_{2}(e^{-\pi y})
-54\textrm{Li}_{3}(e^{-\pi y})\bigg\}\;,
\eea
\bea
G_{4}^{\rm spinor}(0,y)&=&
-\frac{1}{96\pi^{3}}\bigg\{\frac{16\pi^{3}y e^{-\pi y}}{1-e^{-\pi y}}
-4\pi^{2}(20-3y^{2})\ln(1-e^{-\pi y})
\nonumber\\[5pt]
&-&192\pi y\textrm{Li}_{2}(e^{-\pi y})
-504\textrm{Li}_{3}(e^{-\pi y})\bigg\}\;,
\\[10pt]
G_{5}^{\rm spinor}(0,y)&=&\frac{3}{16\pi^{3}}
\bigg\{\pi^{2}(4+y^{2})\ln(1-e^{-\pi y})
-4\pi y\textrm{Li}_{2}(e^{-\pi y})
-6\textrm{Li}_{3}(e^{-\pi y})\bigg\}\;.
\nonumber\\
&&
\label{4568xxx}
\eea


In the case of spinor fields, for $n>4$, there is a well defined limit as
$y\rightarrow 0$.
In fact, by taking the massless limit,
$y\rightarrow 0$, of the expression (\ref{472}) and noticing
that
\be\label{48}
\textrm{Li}_{n}(1)=\zeta(n)\;,
\ee
we obtain
\be
\label{4803}
G_{0}^{\rm spinor}(0,0)
=\frac{2^{[n/2]}}{\pi^{n/2}}
\zeta\left(\frac{n}{2}\right)\;.
\ee
Analogously, in the limit as $y\rightarrow 0$ the result for
the remaining $G_{i}^{\rm spinor}$ can be written as follows
\bea
G_{1}^{\rm spinor}(0,0)
&=&-\frac{2^{[n/2]}}{12}\pi^{1-n/2}\zeta\left(\frac{n}{2}-1\right)\;,
\\[10pt]
G_{2}^{\rm spinor}(0,0)
&=&-\frac{2^{[n/2]}}{48\pi^{n/2+1}}
\bigg\{(n+4)\pi^{2}\zeta\left(\frac{n}{2}-1\right)
-9(n+2)\zeta\left(\frac{n}{2}+1\right)\bigg\}\;,
\nonumber\\
\eea
\bea
G_{4}^{\rm spinor}(0,0)
&=&\frac{2^{[n/2]}}{384\pi^{n/2+1}}
\bigg\{-4\pi^{2}(2n+12)\zeta\left(\frac{n}{2}-1\right)
\nonumber\\[5pt]
&&
+3(n+2)(n+24)\zeta\left(\frac{n}{2}+1\right)\bigg\}\;,
\\[10pt]
G_{5}^{\rm spinor}(0,0)
&=&-2^{[n/2]}\frac{3}{256\pi^{n/2+1}}
\bigg\{16\pi^{2}\zeta\left(\frac{n}{2}-1\right)
+n(n+2)\zeta\left(\frac{n}{2}+1\right)\bigg\}\;.
\nonumber\\
\eea

We turn our attention, now, to the physical case of $n=4$. From the expression
in
(\ref{4803}) we obtain the following result
\be
G_{0}^{\rm spinor}(0,0)=\frac{2}{3}\;.
\ee
It is evident, from the expressions in
(\ref{4565xxx})-(\ref{4568xxx}), that the functions
\\
$G_{i}^{\rm spinor}(0,y)$ in
four dimensions represent a special case since there is an infrared
divergence as $m\rightarrow 0$ (or $y\rightarrow 0$). This means that there
is no well-defined value for the massless limit $y\rightarrow 0$. Instead,
we find a logarithmic divergence, $\log (\pi y)$. In order to
analyze this case we set $n=4$ from the beginning in the expressions
for finite $y$, and then we examine the asymptotics as $y\rightarrow 0$.
By using the equations (\ref{4565xxx})-(\ref{4568xxx}) we obtain
\bea
G_{1}^{\rm spinor}(0,y)
&=&\frac{1}{3\pi}\log(\pi y)+O(y)\;,
\label{45120xxx}
\\[10pt]
G_{2}^{\rm spinor}(0,y)
&=&\frac{2}{3\pi}\log(\pi y)
-\frac{1}{6\pi}
+\frac{9}{2\pi^{3}}\zeta(3)+O(y)\;,
\\[10pt]
G_{4}^{\rm spinor}(0,y)
&=&\frac{5}{6\pi}\log(\pi y)
-\frac{1}{6\pi}
+\frac{21}{4\pi^{3}}\zeta(3)+O(y)\;,
\\[10pt]
G_{5}^{\rm spinor}(0,y)
&=&\frac{3}{4\pi}\log(\pi y)
+\frac{9}{8\pi^{3}}\zeta(3)+O(y)\;.
\label{45123xxx}
\eea

Notice that, in four dimensions the functions
$G_i^{\rm spinor}(x,y)$ are singular at
the point $x=y=0$. In particular, the limits $x\to 0$
and
$y\to 0$ are not commutative, that is, the limits
as $x\to 0$
of the eqs.
(\ref{4564xxx})-(\ref{4573xxx})
(obtained as $y\to 0$ for a finite $x$)
 are different from the eqs.
(\ref{45120xxx})-(\ref{45123xxx})
(obtained as $y\to 0$ for $x=0$).

\section{Concluding Remarks}

In this chapter we have continued the study of the heat kernel and the effective
action for complex (scalar and spinor) quantum fields in a strong constant
electromagnetic field and a gravitational field initiated in
\cite{avramidi08e}. We study here an
{\it essentially non-perturbative regime} when
the electromagnetic field is so strong that one has to take into account all its
orders. In this situation the standard asymptotic expansion of the heat kernel
does not apply since the electromagnetic field cannot be treated as a
perturbation. In \cite{avramidi08e} we established the existence of a new
non-perturbative asymptotic expansion of the heat kernel and computed
explicitly the first three coefficients of this expansion.

We computed the first two coefficients (of zero and the first order in the
Riemann curvature) explicitly in $n$-dimensions by using the spectral
decomposition of the electromagnetic field tensor. We applied this result for
the calculation of the effective action in the physical pseudo-Euclidean
(Lorentzian) case and computed explicitly the imaginary part of the effective
action both in the general case and in the cases of physical interest. We
also computed the asymptotics of the obtained results for supercritical
electric fields.

We have discovered a {\it new infrared divergence} in the
imaginary part of the effective action for massless spinor fields
in four dimensions (or supercritical electric field), which is
induced purely by the gravitational corrections. This means physically that
the creation of massless spinor particles
(or massive particles in supercritical electric field)
is magnified substantially by the
presence of the gravitational field. Further analysis shows that a
similar effect occurs for any massless fields (also scalar fields) in the
second order in the Riemann curvature. This effect could have important
consequences for theories with spontaneous symmetry breakdown when the mass of
charged particles is generated by a Higgs field. Such theories would exhibit a
significant amount of created particles (in the massless limit an infinite
amount) at the phase transition point when the symmetry is restored and the
massive charged particles become massless.
That is why this seems to be an interesting {\it new physical
effect} that deserves further investigation.





\chapter[NONCOMMUTATIVE EINSTEIN EQUATIONS IN MATRIX \newline GENERAL RELATIVITY]
{NONCOMMUTATIVE EINSTEIN EQUATIONS IN MATRIX \newline
GENERAL RELATIVITY\footnotemark[3]}

\footnotetext[3]{The material in this chapter has been published in \emph{Classical and Quantum Gravity}: G. Fucci and I. G. Avramidi, Noncommutative Einstein Equations, \emph{Class. Quant. Grav.} {\bf 25} (2008) 025005  }

\begin{chapabstract}

We study a non-commutative deformation of General Relativity where
the gravitational field is described by a matrix-valued symmetric
two-tensor field. The equations of motion are derived in the
framework of this new theory by varying a diffeomorphisms and
gauge invariant action constructed by using a matrix-valued scalar
curvature. Interestingly the genuine non-commutative part of the
dynamical equations is described only in terms of a particular
tensor density that vanishes identically in the commutative limit.
A non-commutative generalization of the energy-momentum tensor for
the matter field is studied as well.
\end{chapabstract}

\section{Introduction}


The purpose of this chapter is to derive the equations of motion for
the field $a^{\mu\nu}$ that generalizes the role played by
$g^{\mu\nu}$ in the general theory of relativity. Since this model
is a non-commutative extension of Einstein's General Relativity we
will call the corresponding equations of motions non-commutative
Einstein's equations.



\section{Variation of the Action}

The action functional for Matrix Gravity has been introduced in (\ref{523})
By varying the action functional, (\ref{523}), we can derive the equations
of motion for the field $a^{\mu\nu}$, which is the {\it main goal
and the main result of the present chapter}. These equations will be
matrix-valued and they will constitute a generalization of the
ordinary Einstein's equations that we will call {\it
non-commutative Einstein equations}. In order to find the dynamics
of the model we vary the action (\ref{523}) with respect to the
field $a^{\mu\nu}$ considered as independent variable, namely
\begin{displaymath}
a^{\mu\nu}\longrightarrow a^{\mu\nu}+\delta a^{\mu\nu}\;.
\end{displaymath}
By doing so we obtain, for the variation of the action, the
following
\begin{equation}\label{524}
\delta S=S(a^{\mu\nu}+\delta
a^{\mu\nu})-S(a^{\mu\nu})=\frac{1}{16\pi
G}\int_{M}dx\frac{1}{N}\textrm{Tr}_{\;V}(\mathcal{G}_{\mu\nu}\delta
a^{\mu\nu})\;,
\end{equation}
where $\mathcal{G}_{\mu\nu}$ is some matrix valued symmetric
tensor density. Then, of course, the desired equations of motion
are
\begin{equation}\label{525}
\mathcal{G}_{\mu\nu}=0\;.
\end{equation}
It is important to notice that the matrix-valued tensor density
(\ref{525}) has to coincide with the Einstein tensor in the
commutative limit, more precisely we need that, in the commutative
limit, the following relation holds
\begin{equation}\label{526}
\frac{1}{N}\textrm{Tr}\;\mathcal{G}_{\mu\nu}=\sqrt{g}\left(R_{\mu\nu}-\frac{1}{2}g_{\mu\nu}R\right)\;.
\end{equation}

Our main task, then, is to find the explicit form of the equations
of motion that result from the variation of the action (\ref{523}).
In all the calculations that will follow the {\it order of the
terms is important}, unless explicitly stated, due to the matrix
nature of them.

First of all, we rewrite the action in a more explicit form which
is more suitable for the subsequent variation, namely
\begin{equation}\label{527}
S_{\textrm{MGR}}(a)=\frac{1}{16\pi
G}\int_{M}dx\frac{1}{N}\textrm{Tr}_{\;V}
\left[\rho\frac{1}{2}(a^{\mu\nu}\mathcal{R}_{\mu\nu}
+\mathcal{R}_{\mu\nu}a^{\mu\nu})\right]\;.
\end{equation}
By varying the terms in (\ref{527}) with respect to the independent
field $a^{\mu\nu}$, and by using the cyclic property of the trace
we get
\begin{eqnarray}\label{528}
\delta S_{\textrm{MGR}}(a)=\frac{1}{16\pi
G}\int_{M}dx\frac{1}{N}\textrm{Tr}_{\;V}\left[\delta\rho\mathcal{R}+\frac{1}{2}\{\mathcal{R}_{\mu\nu},\rho\}\delta
a^{\mu\nu}+\frac{1}{2}\{\rho,
a^{\mu\nu}\}\delta\mathcal{R}_{\mu\nu}\right]\;,
\end{eqnarray}
where the curly brackets $\{\;,\;\}$ denote anti-commutation,
namely $\{A,B\}=AB+BA$.

From the expressions (\ref{510}) and (\ref{512}), we can evaluate
the variation of the matrix-valued Ricci tensor, more precisely we
have
\begin{eqnarray}\label{529}
\delta\mathcal{R}_{\mu\nu}&=&\partial_{\alpha}(\delta\mathscr{A}^{\alpha}{}_{\mu\nu})-\partial_{\nu}(\delta\mathscr{A}^{\alpha}{}_{\mu\alpha})+\delta\mathscr{A}^{\alpha}{}_{\lambda\alpha}\mathscr{A}^{\lambda}{}_{\mu\nu}+\mathscr{A}^{\alpha}{}_{\lambda\alpha}\delta\mathscr{A}^{\lambda}{}_{\mu\nu}+\nonumber\\
&-&\delta\mathscr{A}^{\alpha}{}_{\lambda\nu}\mathscr{A}^{\lambda}{}_{\mu\alpha}-\mathscr{A}^{\alpha}{}_{\lambda\nu}\delta\mathscr{A}^{\lambda}{}_{\mu\alpha}\;.
\end{eqnarray}
From now on, for simplicity of notation, we set \be
B^{\mu\nu}\equiv\{\rho,a^{\mu\nu}\}\,. \ee By substituting
(\ref{529}) in (\ref{528}), and by using the cyclic property of the
trace we obtain
\begin{eqnarray}\label{530}
\delta S_{\textrm{MGR}}(a)&=&\frac{1}{16\pi
G}\int_{M}dx\frac{1}{N}\textrm{Tr}_{\;V}\bigg[\delta\rho\mathcal{R}+\frac{1}{2}\{\mathcal{R}_{\mu\nu},\rho\}\delta
a^{\mu\nu}+\frac{1}{2}B^{\mu\nu}\partial_{\alpha}(\delta\mathscr{A}^{\alpha}{}_{\mu\nu})+\nonumber\\
&-&\frac{1}{2}B^{\mu\nu}\partial_{\nu}(\delta\mathscr{A}^{\alpha}{}_{\mu\alpha})+\frac{1}{2}\mathscr{A}^{\lambda}{}_{\mu\nu}B^{\mu\nu}\delta\mathscr{A}^{\alpha}{}_{\lambda\alpha}+\frac{1}{2}B^{\mu\nu}\mathscr{A}^{\alpha}{}_{\lambda\alpha}\delta\mathscr{A}^{\lambda}{}_{\mu\nu}+\nonumber\\
&-&\frac{1}{2}\mathscr{A}^{\lambda}{}_{\mu\alpha}B^{\mu\nu}\delta\mathscr{A}^{\alpha}{}_{\lambda\nu}-\frac{1}{2}B^{\mu\nu}\mathscr{A}^{\alpha}{}_{\lambda\nu}\delta\mathscr{A}^{\lambda}{}_{\mu\alpha}\bigg]\;.
\end{eqnarray}
By integrating by parts and by collecting similar terms we get
\begin{eqnarray}\label{531}
\delta S_{\textrm{MGR}}(a)&=&\frac{1}{16\pi
G}\int_{M}dx\frac{1}{N}\textrm{Tr}_{\;V}\bigg[\delta\rho\mathcal{R}+\frac{1}{2}\{\mathcal{R}_{\mu\nu},\rho\}\delta
a^{\mu\nu}-\frac{1}{2}\bigg(B^{\mu\nu}{}_{,\alpha}-B^{\mu\nu}\mathscr{A}^{\lambda}{}_{\alpha\lambda}+\nonumber\\
&+&\mathscr{A}^{\mu}{}_{\lambda\alpha}B^{\lambda\nu}+B^{\mu\lambda}\mathscr{A}^{\nu}{}_{\alpha\lambda}\bigg)\delta\mathscr{A}^{\alpha}{}_{\mu\nu}+\frac{1}{2}\bigg(B^{\mu\nu}{}_{,\nu}+\mathscr{A}^{\mu}{}_{\lambda\nu}B^{\lambda\nu}\bigg)\delta\mathscr{A}^{\alpha}{}_{\mu\alpha}\bigg]\;.
\end{eqnarray}
We can rewrite the last expression in a more compact form, namely
\begin{equation}\label{532}
\delta S_{\textrm{MGR}}(a)=\frac{1}{16\pi
G}\int_{M}dx\frac{1}{N}\textrm{Tr}_{\;V}\bigg[\delta\rho\mathcal{R}+\frac{1}{2}\{\mathcal{R}_{\mu\nu},\rho\}\delta
a^{\mu\nu}-\frac{1}{2}C^{\mu\nu}{}_{\alpha}\delta\mathscr{A}^{\alpha}{}_{\mu\nu}+\frac{1}{2}D^{\mu}\delta\mathscr{A}^{\alpha}{}_{\mu\alpha}\bigg]\;,
\end{equation}
where the matrix-valued tensor densities $C^{\mu\nu}{}_{\alpha}$
and $D^{\mu}$ have the explicit expression
\begin{eqnarray}\label{533}
C^{\mu\nu}{}_{\alpha}&=&\{a^{\mu\nu},\rho_{,\alpha}-\rho\mathscr{A}^{\lambda}{}_{\alpha\lambda}\}-\rho[a^{\mu\nu},\mathscr{A}^{\lambda}{}_{\alpha\lambda}]-[\rho,\mathscr{A}^{\mu}{}_{\rho\alpha}]a^{\rho\nu}+\nonumber\\
&-&\{\rho,[\mathscr{A}^{\nu}{}_{\lambda\alpha},a^{\mu\lambda}]\}-a^{\mu\lambda}[\mathscr{A}^{\nu}{}_{\lambda\alpha},\rho]+2\{\rho,a^{\mu\lambda}\}\mathscr{A}^{\nu}{}_{[\alpha\lambda]}\;,
\end{eqnarray}
and
\begin{equation}\label{534}
D^{\mu}=\{a^{\mu\nu},\rho_{,\nu}-\mathscr{A}^{\rho}{}_{\nu\rho}\rho\}-[\rho,\mathscr{A}^{\mu}{}_{\rho\nu}]a^{\rho\nu}-[\rho,\mathscr{A}^{\rho}{}_{\nu\rho}]a^{\mu\nu}\;.
\end{equation}
It is worth noticing that in the commutative limit, or, in other
words, when all the matrices commute, the tensor densities
$C^{\mu\nu}{}_{\alpha}$ and $D^{\mu}$ are identically zero, and
the variation of the action $\delta S_{\textrm{MGR}}$ simply
reduces to the standard result of the general theory of
relativity.

We can write, now, the variation of the connection coefficients.
By using the expression (\ref{515}), and by noticing that
\begin{displaymath}
\delta b_{\mu\nu}=-b_{\mu\rho}(\delta
a^{\rho\sigma})b_{\sigma\nu}\;,
\end{displaymath}
we obtain the following
\begin{eqnarray}\label{535}
\delta\mathscr{A}^{\alpha}{}_{\lambda\mu}&=&-b_{\lambda\nu}\delta
a^{\nu\beta}\mathscr{A}^{\alpha}{}_{\beta\mu}-\mathscr{A}^{\alpha}{}_{\lambda\nu}\delta
a^{\nu\beta}b_{\beta\mu}+\frac{1}{2}b_{\lambda\sigma}\delta
a^{\alpha\gamma}(\partial_{\gamma}a^{\rho\sigma})b_{\rho\mu}+\nonumber\\
&-&\frac{1}{2}b_{\lambda\sigma}\delta
a^{\rho\gamma}(\partial_{\gamma}a^{\sigma\alpha})b_{\rho\mu}
-\frac{1}{2}b_{\lambda\sigma}\delta
a^{\sigma\gamma}(\partial_{\gamma}a^{\rho\alpha})b_{\rho\mu}+\frac{1}{2}b_{\lambda\sigma}a^{\alpha\gamma}(\partial_{\gamma}\delta
a^{\rho\sigma})b_{\rho\mu}+\nonumber\\
&-&\frac{1}{2}b_{\lambda\sigma}a^{\rho\gamma}(\partial_{\gamma}\delta
a^{\sigma\alpha})b_{\rho\mu}-\frac{1}{2}b_{\lambda\sigma}a^{\sigma\gamma}(\partial_{\gamma}\delta
a^{\rho\alpha})b_{\rho\mu}\;.
\end{eqnarray}
Once we have the explicit expression for the variation of the
connection coefficients, we can evaluate the last two terms that
appear in the variation of the action (\ref{532}). We start with
the first of the two
\begin{eqnarray}\label{536}
&-&\frac{1}{2}\int_{M}dx\;\textrm{Tr}_{\;V}(C^{\mu\nu}{}_{\alpha}\delta\mathscr{A}^{\alpha}{}_{\mu\nu})=\nonumber\\
&=&\frac{1}{2}\int_{M}dx\;\textrm{Tr}_{\;V}\bigg[\mathscr{A}^{\alpha}{}_{\beta\nu}C^{\mu\nu}{}_{\alpha}b_{\mu\lambda}\delta
a^{\lambda\beta}
+b_{\beta\nu}C^{\mu\nu}{}_{\alpha}\mathscr{A}^{\alpha}{}_{\mu\lambda}\delta
a^{\lambda\beta}+\nonumber\\
&-&\frac{1}{2}(\partial_{\gamma}a^{\rho\sigma})b_{\rho\nu}C^{\mu\nu}{}_{\alpha}b_{\mu\sigma}\delta
a^{\alpha\gamma}
+\frac{1}{2}(\partial_{\gamma}a^{\alpha\sigma})b_{\rho\nu}C^{\mu\nu}{}_{\alpha}b_{\mu\sigma}\delta
a^{\rho\gamma}+\nonumber\\
&+&\frac{1}{2}(\partial_{\gamma}a^{\rho\alpha})b_{\rho\nu}C^{\mu\nu}{}_{\alpha}b_{\mu\sigma}\delta
a^{\sigma\gamma}-\frac{1}{2}b_{\rho\nu}C^{\mu\nu}{}_{\alpha}b_{\mu\sigma}a^{\alpha\gamma}(\partial_{\gamma}\delta
a^{\rho\sigma})+\nonumber\\
&+&\frac{1}{2}b_{\rho\nu}C^{\mu\nu}{}_{\alpha}b_{\mu\sigma}a^{\rho\gamma}(\partial_{\gamma}\delta
a^{\alpha\sigma})+\frac{1}{2}b_{\rho\nu}C^{\mu\nu}{}_{\alpha}b_{\mu\sigma}a^{\sigma\gamma}(\partial_{\gamma}\delta
a^{\rho\alpha})\bigg]\;,
\end{eqnarray}
where in this last expression we used the cyclic property of the
trace.

We introduce the following definition, which will be useful in
order to simplify the notation,
\begin{equation}\label{537}
F_{\beta\alpha\rho}=b_{\beta\nu}C^{\mu\nu}{}_{\alpha}b_{\mu\rho}\;.
\end{equation}
By using the above definition, the expression in (\ref{536}) can be
rewritten as follows
\begin{eqnarray}\label{538}
&-&\frac{1}{2}\int_{M}dx\;\textrm{Tr}_{\;V}(C^{\mu\nu}{}_{\alpha}\delta\mathscr{A}^{\alpha}{}_{\mu\nu})=\nonumber\\
&=&\frac{1}{2}\int_{M}dx\;\textrm{Tr}_{\;V}\bigg[\mathscr{A}^{\alpha}{}_{\beta\gamma}a^{\gamma\rho}F_{\rho\alpha\lambda}\delta
a^{\lambda\beta}+F_{\beta\alpha\rho}a^{\rho\gamma}\mathscr{A}^{\alpha}{}_{\gamma\lambda}\delta
a^{\lambda\beta}+\nonumber\\
&-&\frac{1}{2}(\partial_{\gamma}a^{\sigma\rho})F_{\rho\alpha\sigma}\delta
a^{\alpha\gamma}+\frac{1}{2}(\partial_{\gamma}a^{\alpha\sigma})F_{\rho\alpha\sigma}\delta
a^{\rho\gamma}+\frac{1}{2}(\partial_{\gamma}a^{\alpha\rho})F_{\rho\alpha\sigma}\delta
a^{\sigma\gamma}+\nonumber\\
&-&\frac{1}{2}F_{\rho\alpha\sigma}a^{\alpha\gamma}(\partial_{\gamma}\delta
a^{\rho\sigma})+\frac{1}{2}F_{\rho\alpha\sigma}a^{\rho\gamma}(\partial_{\gamma}\delta
a^{\alpha\sigma})+\frac{1}{2}F_{\rho\alpha\sigma}a^{\sigma\gamma}(\partial_{\gamma}\delta
a^{\rho\alpha})\bigg]\;,
\end{eqnarray}
where the first two terms in the last expression has been derived
by using the relation
\begin{equation}\label{539}
\mathscr{A}^{\alpha}{}_{\beta\nu}C^{\mu\nu}{}_{\alpha}b_{\mu\lambda}\delta
a^{\lambda\beta}=\mathscr{A}^{\alpha}{}_{\beta\gamma}a^{\gamma\rho}b_{\rho\nu}C^{\mu\nu}{}_{\alpha}b_{\mu\lambda}\delta
a^{\lambda\beta}=\mathscr{A}^{\alpha}{}_{\beta\gamma}a^{\gamma\rho}F_{\rho\alpha\lambda}\delta
a^{\lambda\beta}\;.
\end{equation}
By integrating by parts and by relabeling dummy indices we find
the final expression for (\ref{538}), namely
\begin{eqnarray}\label{540}
-\frac{1}{2}\int_{M}dx\;\textrm{Tr}_{\;V}(C^{\mu\nu}{}_{\alpha}\delta\mathscr{A}^{\alpha}{}_{\mu\nu})&=&\frac{1}{2}\int_{M}dx\;\textrm{Tr}_{\;V}\bigg[\mathscr{A}^{\alpha}{}_{\beta\gamma}a^{\gamma\rho}F_{\rho\alpha\gamma}+F_{\beta\alpha\rho}a^{\rho\gamma}\mathscr{A}^{\alpha}{}_{\gamma\lambda}+\nonumber\\
&-&\frac{1}{2}(\partial_{\beta}a^{\rho\sigma})\bigg(F_{\rho\lambda\sigma}-F_{\lambda\sigma\rho}-F_{\sigma\rho\lambda}\bigg)\bigg]\delta
a^{\lambda\beta}+\nonumber\\
&+&\frac{1}{4}\bigg\{\partial_{\gamma}\bigg[\bigg(F_{\rho\lambda\sigma}-F_{\lambda\sigma\rho}-F_{\sigma\rho\lambda}\bigg)a^{\lambda\gamma}\bigg]\bigg\}\delta
a^{\rho\sigma}\;.
\end{eqnarray}

For the last term in the variation of the action (\ref{532}), we
use similar arguments which lead us to the expression (\ref{540}).
In this case we introduce the following definition:
\begin{equation}\label{541}
G_{\beta\alpha\rho}=b_{\beta\alpha}D^{\mu }b_{\mu\rho}\;.
\end{equation}
By using the definition above and the cyclic property of the trace
we obtain
\begin{eqnarray}\label{541a}
&\phantom{-}&\frac{1}{2}\int_{M}dx\;\textrm{Tr}_{\;V}(D^{\mu}\delta\mathscr{A}^{\alpha}{}_{\mu\alpha})=\nonumber\\
&-&\frac{1}{2}\int_{M}dx\;\textrm{Tr}_{\;V}\bigg[\mathscr{A}^{\alpha}{}_{\beta\rho}a^{\rho\gamma}G_{\gamma\alpha\lambda}\delta
a^{\lambda\beta}+G_{\beta\alpha\gamma}a^{\gamma\rho}\mathscr{A}^{\alpha}{}_{\rho\lambda}\delta
a^{\lambda\beta}+\nonumber\\
&-&\frac{1}{2}(\partial_{\gamma}a^{\rho\sigma})G_{\rho\alpha\sigma}\delta
a^{\alpha\gamma}+\frac{1}{2}(\partial_{\gamma}a^{\alpha\sigma})G_{\rho\alpha\sigma}\delta
a^{\rho\gamma}+\frac{1}{2}(\partial_{\gamma}a^{\rho\alpha})G_{\rho\alpha\sigma}\delta
a^{\sigma\gamma}+\nonumber\\
&-&\frac{1}{2}G_{\rho\alpha\sigma}a^{\alpha\gamma}(\partial_{\gamma}\delta
a^{\rho\sigma})+\frac{1}{2}G_{\rho\alpha\sigma}a^{\rho\gamma}(\partial_{\gamma}\delta
a^{\alpha\sigma})+\frac{1}{2}G_{\rho\alpha\sigma}a^{\sigma\gamma}(\partial_{\gamma}\delta
a^{\rho\alpha})\bigg]\;.
\end{eqnarray}
By integrating by parts and relabeling dummy indices we get
\begin{eqnarray}\label{542}
\frac{1}{2}\int_{M}dx\;\textrm{Tr}_{\;V}(D^{\mu}\delta\mathscr{A}^{\alpha}{}_{\mu\alpha})&=&-\frac{1}{2}\int_{M}dx\;\textrm{Tr}_{\;V}\bigg[\mathscr{A}^{\alpha}{}_{\beta\rho}a^{\rho\gamma}G_{\gamma\alpha\lambda}+G_{\beta\alpha\gamma}a^{\rho\gamma}\mathscr{A}^{\alpha}{}_{\rho\lambda}+\nonumber\\
&-&\frac{1}{2}(\partial_{\beta}a^{\rho\sigma})\bigg(G_{\rho\lambda\sigma}-G_{\lambda\sigma\rho}-G_{\sigma\rho\lambda}\bigg)\bigg]\delta
a^{\lambda\beta}+\nonumber\\
&-&\frac{1}{4}\bigg\{\partial_{\gamma}\bigg[\bigg(G_{\rho\lambda\sigma}-G_{\lambda\sigma\rho}-G_{\sigma\rho\lambda}\bigg)a^{\lambda\gamma}\bigg]\bigg\}\delta
a^{\rho\sigma}\;.
\end{eqnarray}

It is worth noticing that in the above expressions, (\ref{540}) and
(\ref{542}), the tensor densities $F$ and $G$ always appear in the
same combination. This observation justifies the following
definitions
\begin{equation}\label{543}
X_{\rho\lambda\sigma}=F_{\rho\lambda\sigma}-F_{\lambda\sigma\rho}-F_{\sigma\rho\lambda}\;,
\end{equation}
and
\begin{equation}\label{544}
Y_{\rho\lambda\sigma}=G_{\rho\lambda\sigma}-G_{\lambda\sigma\rho}-G_{\sigma\rho\lambda}\;.
\end{equation}
By using the two definitions above we can rewrite the arguments of
the traces in (\ref{540}) and in (\ref{542}) respectively as
\begin{eqnarray}\label{545}
-\frac{1}{2}C^{\mu\nu}{}_{\alpha}\delta\mathscr{A}^{\alpha}{}_{\mu\nu}&=&\frac{1}{2}\bigg[\mathscr{A}^{\alpha}{}_{\beta\gamma}a^{\gamma\rho}F_{\rho\alpha\lambda}+F_{\beta\alpha\rho}a^{\rho\gamma}\mathscr{A}^{\alpha}{}_{\gamma\lambda}-\frac{1}{2}(\partial_{\beta}a^{\rho\sigma})X_{\rho\lambda\sigma}+\nonumber\\
&+&\frac{1}{2}\partial_{\gamma}(X_{\lambda\rho\beta}a^{\rho\gamma})\bigg]\delta
a^{\lambda\beta}\;,
\end{eqnarray}
and
\begin{eqnarray}\label{546}
\frac{1}{2}D^{\mu}\delta\mathscr{A}^{\alpha}{}_{\mu\alpha}&=&-\frac{1}{2}\bigg[\mathscr{A}^{\alpha}{}_{\beta\rho}a^{\rho\gamma}G_{\gamma\alpha\lambda}+G_{\beta\alpha\gamma}a^{\rho\gamma}\mathscr{A}^{\alpha}{}_{\rho\lambda}-\frac{1}{2}(\partial_{\beta}a^{\rho\sigma})Y_{\rho\lambda\sigma}+\nonumber\\
&+&\frac{1}{2}\partial_{\gamma}(Y_{\lambda\rho\beta}a^{\rho\gamma})\bigg]\delta
a^{\lambda\beta}\;.
\end{eqnarray}
By combining the results (\ref{545}) and (\ref{546}) we obtain the
expression for the last two terms in the variation of the action,
namely
\begin{eqnarray}\label{547}
-\frac{1}{2}C^{\mu\nu}{}_{\alpha}\delta\mathscr{A}^{\alpha}{}_{\mu\nu}+\frac{1}{2}D^{\mu}\delta\mathscr{A}^{\alpha}{}_{\mu\alpha}=\frac{1}{2}\bigg\{\mathscr{A}^{\alpha}{}_{\beta\rho}a^{\rho\gamma}(F_{\gamma\alpha\lambda}-G_{\gamma\alpha\lambda})+\nonumber\\
+(F_{\beta\alpha\gamma}-G_{\beta\alpha\gamma})a^{\gamma\rho}\mathscr{A}^{\alpha}{}_{\rho\gamma}+\frac{1}{2}(\partial_{\beta}a^{\rho\sigma})(Y_{\rho\lambda\sigma}-X_{\rho\lambda\sigma})+\nonumber\\
-\frac{1}{2}\partial_{\gamma}\bigg[(Y_{\lambda\rho\beta}-X_{\lambda\rho\beta})a^{\rho\gamma}\bigg]\bigg\}\delta
a^{\lambda\beta}\;.
\end{eqnarray}

\section{Noncommutative Einstein Equations}

With the expression (\ref{547}) for the last two terms in
(\ref{532}), the variation of the action has the form (\ref{524})
which is suitable for the derivation of the dynamical equations of
the model. Before writing the complete dynamical equations, we
will simplify further the expression (\ref{547}).

The definition (\ref{543}) gives a linear relation between the
matrix-valued tensor density $X$ and a particular combination of
matrix-valued tensor density $F$, a similar linear relation
between $Y$ and $G$ is given in (\ref{544}). By using simple tensor
algebra, it can be easily shown that those relations can be
inverted, namely we can write
\begin{equation}\label{548}
F_{\rho\lambda\sigma}=-\frac{1}{2}(X_{\lambda\sigma\rho}+X_{\sigma\rho\lambda})\;,
\end{equation}
and
\begin{equation}\label{549}
G_{\rho\lambda\sigma}=-\frac{1}{2}(Y_{\lambda\sigma\rho}-Y_{\sigma\rho\lambda})\;.
\end{equation}
By substituting the equations (\ref{548}) and (\ref{549}) in the
expression (\ref{547}) we obtain the following
\begin{eqnarray}\label{550}
-\frac{1}{2}C^{\mu\nu}{}_{\alpha}\delta\mathscr{A}^{\alpha}{}_{\mu\nu}&+&\frac{1}{2}D^{\mu}\delta\mathscr{A}^{\alpha}{}_{\mu\alpha}=\frac{1}{4}\bigg\{\mathscr{A}^{\alpha}{}_{\beta\rho}a^{\rho\gamma}[(Y_{\alpha\lambda\gamma}-X_{\alpha\lambda\gamma})+(Y_{\lambda\gamma\alpha}-X_{\lambda\gamma\alpha})]+\nonumber\\
&+&[(Y_{\alpha\gamma\lambda}-X_{\alpha\gamma\lambda})+(Y_{\gamma\beta\alpha}-X_{\gamma\beta\alpha})]a^{\rho\gamma}\mathscr{A}^{\alpha}{}_{\rho\lambda}+\nonumber\\
&+&(\partial_{\beta}a^{\rho\sigma})(Y_{\rho\lambda\sigma}-X_{\rho\lambda\sigma})-\partial_{\gamma}[(Y_{\lambda\rho\beta}-X_{\lambda\rho\beta})a^{\rho\gamma}]\bigg\}\delta
a^{\lambda\beta}\;.
\end{eqnarray}

We can see, in the last formula, that the tensor densities $X$ and
$Y$ enter always in the same combination. It is useful, therefore,
to define the following tensor density
\begin{equation}\label{551}
H_{\mu\nu\rho}=Y_{\mu\nu\rho}-X_{\mu\nu\rho}\;.
\end{equation}
With this last definition we can rewrite (\ref{550}) as
\begin{eqnarray}\label{552}
-\frac{1}{2}C^{\mu\nu}{}_{\alpha}\delta\mathscr{A}^{\alpha}{}_{\mu\nu}+\frac{1}{2}D^{\mu}\delta\mathscr{A}^{\alpha}{}_{\mu\alpha}=\frac{1}{4}\bigg[\mathscr{A}^{\alpha}{}_{\beta\rho}a^{\rho\gamma}H_{\alpha\lambda\gamma}+\mathscr{A}^{\alpha}{}_{\beta\rho}a^{\rho\gamma}H_{\lambda\gamma\alpha}+\nonumber\\
+H_{\alpha\gamma\beta}a^{\rho\gamma}\mathscr{A}^{\alpha}{}_{\rho\lambda}+H_{\gamma\beta\alpha}a^{\rho\gamma}\mathscr{A}^{\alpha}{}_{\rho\lambda}+(\partial_{\beta}a^{\rho\sigma})H_{\rho\lambda\sigma}-\partial_{\gamma}(H_{\lambda\rho\beta}a^{\rho\gamma})\bigg]\delta
a^{\lambda\beta}\;.
\end{eqnarray}
By using the compatibility condition of the metric tensor $a^{\mu\nu}$ with the connection coefficients $\mathscr{A}^{\alpha}{}_{\mu\nu}$,
we can write that
\begin{equation}\label{553}
\partial_{\beta}a^{\rho\sigma}=-\mathscr{A}^{\rho}{}_{\gamma\beta}a^{\gamma\sigma}-\mathscr{A}^{\sigma}{}_{\gamma\beta}a^{\rho\gamma}\;,
\end{equation}
moreover we obtain that
\begin{equation}\label{554}
-\partial_{\gamma}(H_{\lambda\rho\beta}a^{\rho\gamma})=-(\partial_{\gamma}H_{\lambda\rho\beta})a^{\rho\gamma}+H_{\lambda\rho\beta}\mathscr{A}^{\rho}{}_{\sigma\gamma}a^{\sigma\gamma}+H_{\lambda\rho\beta}\mathscr{A}^{\gamma}{}_{\sigma\gamma}a^{\sigma\rho}\;.
\end{equation}
Since $H_{\mu\nu\rho}$ is a tensor density, we can write
\begin{equation}\label{555}
\mathcal{D}_{\gamma}H_{\lambda\rho\beta}=\partial_{\gamma}H_{\lambda\rho\beta}-\mathscr{A}^{\alpha}{}_{\lambda\gamma}H_{\alpha\rho\beta}-\mathscr{A}^{\alpha}{}_{\rho\gamma}H_{\lambda\alpha\beta}-\mathscr{A}^{\alpha}{}_{\beta\gamma}H_{\lambda\rho\alpha}-\mathscr{A}^{\alpha}{}_{\gamma\alpha}H_{\lambda\rho\beta}\;.
\end{equation}
By using the results obtained in (\ref{553}), (\ref{554}) and
(\ref{555}) we can express (\ref{552}) as follows
\begin{eqnarray}\label{556}
-\frac{1}{2}C^{\mu\nu}{}_{\alpha}\delta\mathscr{A}^{\alpha}{}_{\mu\nu}+\frac{1}{2}D^{\mu}\delta\mathscr{A}^{\alpha}{}_{\mu\alpha}=\frac{1}{4}\bigg\{2\mathscr{A}^{\alpha}{}_{[\beta\rho]}a^{\rho\gamma}H_{\alpha\lambda\gamma}+2H_{\alpha\lambda\beta}a^{\rho\gamma}\mathscr{A}^{\alpha}{}_{[\rho\lambda]}+\nonumber\\
-(\mathcal{D}_{\gamma}H_{\lambda\rho\beta})a^{\rho\gamma}-[\mathscr{A}^{\alpha}{}_{\lambda\rho},H_{\alpha\gamma\beta}]a^{\rho\gamma}-H_{\alpha\gamma\beta}[\mathscr{A}^{\alpha}{}_{\lambda\rho},a^{\rho\gamma}]-[\mathscr{A}^{\alpha}{}_{\rho\gamma},H_{\lambda\alpha\beta}]a^{\rho\gamma}+\nonumber\\
-[\mathscr{A}^{\alpha}{}_{\gamma\alpha},H_{\lambda\rho\beta}]a^{\rho\gamma}-\mathscr{A}^{\alpha}{}_{\beta\gamma}[H_{\lambda\rho\alpha},a^{\rho\gamma}]\bigg\}\delta
a^{\lambda\beta}\;.
\end{eqnarray}
At this point we introduce the operator $P$ defined as
\begin{equation}\label{557}
P_{\gamma}H_{\lambda\rho\beta}=\mathcal{D}H_{\lambda\rho\beta}+[\mathscr{A}^{\alpha}{}_{\lambda\gamma},H_{\alpha\rho\beta}]+[\mathscr{A}^{\alpha}{}_{\rho\gamma},H_{\lambda\alpha\beta}]+[\mathscr{A}^{\alpha}{}_{\beta\gamma},H_{\lambda\rho\alpha}]+[\mathscr{A}^{\alpha}{}_{\gamma\alpha},H_{\lambda\rho\beta}]\;.
\end{equation}
By using the last definition in (\ref{556}) one obtains
\begin{eqnarray}\label{558}
-\frac{1}{2}C^{\mu\nu}{}_{\alpha}\delta\mathscr{A}^{\alpha}{}_{\mu\nu}+\frac{1}{2}D^{\mu}\delta\mathscr{A}^{\alpha}{}_{\mu\alpha}=\frac{1}{4}\bigg\{2\mathscr{A}^{\alpha}{}_{[\beta\rho]}a^{\rho\gamma}H_{\alpha\lambda\gamma}+2H_{\alpha\lambda\beta}a^{\rho\gamma}\mathscr{A}^{\alpha}{}_{[\rho\lambda]}+\nonumber\\
-(P_{\gamma}H_{\lambda\rho\beta})a^{\rho\gamma}+[\mathscr{A}^{\alpha}{}_{\beta\gamma},H_{\lambda\rho\alpha}]a^{\rho\gamma}-H_{\alpha\gamma\beta}[\mathscr{A}^{\alpha}{}_{\lambda\rho},a^{\rho\gamma}]-\mathscr{A}^{\alpha}{}_{\beta\gamma}[H_{\lambda\rho\alpha},a^{\rho\gamma}]\bigg\}\delta
a^{\lambda\beta}\;.\nonumber\\
\;
\end{eqnarray}

We finally have all the ingredients that we need in order to write
the dynamical equations of the theory. Now we only have to find an
expression for the variation $\delta\rho$. The definition of
$\rho$ is given in (\ref{520}), and its variation can be
straightforwardly evaluated as follows
\begin{equation}\label{559}
\delta\rho=-\int\limits_{\RR^n}
\frac{d\xi}{\pi^{\frac{n}{2}}}\int_{0}^{1}ds\;
e^{-(1-s)A(\xi)}\delta a^{\mu\nu}\xi_{\mu}\xi_{\nu}
e^{-sA(\xi)}\;,
\end{equation}
where \be A(\xi)=a^{\mu\nu}\xi_\mu\xi_\nu\,. \ee Once we have the
expression (\ref{559}) for the variation, we can use the cyclic
property of the trace to write that
\begin{equation}\label{560}
\textrm{Tr}_{\;V}(\delta\rho\;\mathcal{R}) = \textrm{Tr}_{\;V}
\left[-\int\limits_{\RR^n}
\frac{d\xi}{\pi^{\frac{n}{2}}}\int_{0}^{1}ds\;
e^{-sA(\xi)}\mathcal{R}e^{-(1-s)A(\xi)}\xi_{\mu}\xi_{\nu}\right]\delta
a^{\mu\nu}\,.
\end{equation}

By combining (\ref{560}), (\ref{558}) and (\ref{532}) we obtain the
{\it non-commutative Einstein equations} in absence of matter,
namely
\begin{equation}\label{560a}
\mathcal{G}_{\mu\nu}=0\;,
\end{equation}
where
\begin{eqnarray}\label{561}
\mathcal{G}_{\mu\nu}&=& \frac{1}{2}\{\rho,\mathcal{R}_{\mu\nu}\}
+\mathcal{F}_{\mu\nu}+\frac{1}{2}\mathscr{A}^{\alpha}{}_{[\mu\rho]}
a^{\rho\gamma}H_{\alpha\nu\gamma}
+\frac{1}{2}H_{\alpha\lambda\nu}a^{\rho\gamma}
\mathscr{A}^{\alpha}{}_{[\rho\mu]}
-\frac{1}{4}(P_{\gamma}H_{\mu\rho\nu})a^{\rho\gamma}
+\nonumber\\
&+&\frac{1}{4}[\mathscr{A}^{\alpha}{}_{\nu\gamma},
H_{\mu\rho\alpha}]a^{\rho\gamma} -\frac{1}{4}H_{\alpha\gamma\nu}
[\mathscr{A}^{\alpha}{}_{\mu\rho},a^{\rho\gamma}]
-\frac{1}{4}\mathscr{A}^{\alpha}{}_{\nu\gamma}
[H_{\mu\rho\alpha},a^{\rho\gamma}]\;,
\end{eqnarray}
is the {\it non-commutative Einstein tensor},
$\mathcal{F}_{\mu\nu}$ is defined by \be \mathcal{F}_{\mu\nu} =
-\int\limits_{\RR^n}
\frac{d\xi}{\pi^{\frac{n}{2}}}\int_{0}^{1}ds\;
e^{-sA(\xi)}\mathcal{R}e^{-(1-s)A(\xi)}\xi_{\mu}\xi_{\nu}, \ee and
the tensor density $H$ has the explicit form
\begin{equation}\label{562}
H_{\alpha\lambda\gamma}=b_{\alpha\nu}(\delta^{\nu}{}_{\lambda}D^{\mu}-C^{\mu\nu}{}_{\lambda})b_{\mu\gamma}-b_{\lambda\nu}(\delta^{\nu}{}_{\gamma}D^{\mu}-C^{\mu\nu}{}_{\gamma})b_{\mu\alpha}-b_{\gamma\nu}(\delta^{\nu}{}_{\alpha}D^{\mu}-C^{\mu\nu}{}_{\alpha})b_{\mu\lambda}\;.
\end{equation}

These equations are the {\it main result of the present chapter}.
One can show that the first two terms in the equations (\ref{561})
represent a straightforward generalization of Einstein's equation
to endomorphism-valued objects and the rest of the terms can be
considered as a genuine non-commutative part which is not present
in Einstein's equation. It is interesting to note that the pure
non-commutative part is completely described by the tensor density
$H_{\mu\nu\rho}$ defined in (\ref{562}).

Moreover the equation (\ref{560a}) satisfies the requirement
(\ref{526}), which, in words, expresses the necessity that our
model reduces, in the commutative limit, to the standard theory of
General Relativity. In fact, the trace of the pure non-commutative
terms vanishes, because of the presence of the commutators, and
the first two terms just give
\begin{equation}\label{562a}
\frac{1}{N}\textrm{Tr}_{V}\left(\frac{1}{2}\{\rho,\mathcal{R}_{\mu\nu}\}+\mathcal{F}_{\mu\nu}\right)=\sqrt{g}\left(R_{\mu\nu}-\frac{1}{2}g_{\mu\nu}R\right)\;.
\end{equation}

For an arbitrary matrix algebra the equation (\ref{560a}) becomes
more complicated than the ordinary Einstein's equation due to
presence of the new tensor density $H_{\mu\nu\rho}$. We mention,
now, a particular case in which (\ref{560a}) simplifies. The
formalism used so far deals with geometric quantities which are
endomorphism-valued, namely they take values in $\textrm{End}(V)$.
By choosing a basis in the vector space $V$ we can represent
$\textrm{End}(V)$ by means of matrices. Let us suppose that the
algebra under consideration is Abelian, in this case all the
elements commute with each other and the tensor density
$H_{\mu\nu\rho}$ vanishes identically and the equation (\ref{560a})
becomes
\begin{equation}\label{562b}
\mathcal{R}_{\mu\nu}-\frac{1}{2}b_{\mu\nu}\mathcal{R}=0\;.
\end{equation}
Therefore, in case of a commutative matrix algebra, the equation
of motion of our model have the same form as Einstein's equation,
with the only difference that (\ref{562b}) is matrix-valued.

\section{The Action for the Matter Field}

In order to have a complete theory for the gravitational field we
need to describe the dynamics of the matter field in the framework
of matrix general relativity. The main idea is to extend the
general results of classical field theory. We will consider, in
the following, the dynamics of a multiplet of free scalar fields
propagating on a manifold $M$. We can construct an invariant
action by using the matrix valued metric $a^{\mu\nu}$ and the
measure $\rho$. A typical action is \be
S_{\textrm{matter}}(a,\varphi) =\frac{1}{4}\int_M
dx\;\left\{-\left<\partial_\mu\varphi, \{\rho, a^{\mu\nu}\}
\partial_\nu\varphi\right>
-\left<\varphi,\{\rho, Q\}\varphi\right> \right\}\,, \label{568}
\ee where $\left<\;,\;\right>$ denotes the fiber inner product on
the vector bundle $V$, and $Q$ is a constant mass matrix
determining the masses of the scalar fields. The equations of
motion of the scalar fields are then obviously \be
\left[-\partial_\mu\{\rho, a^{\mu\nu}\}\partial_\nu+\{\rho,Q\}
\right]\varphi=0\,. \ee

The complete action of the gravity and matter is described then by
\begin{equation}\label{570a}
{S}(a,\varphi)=S_{\textrm{MGR}}(a)
+S_{\textrm{matter}}(a,\varphi)\;.
\end{equation}
By varying the above action with respect to $a^{\mu\nu}$ one
obtains the non-commutative Einstein equation in presence of matter
\be \mathcal{G}_{\mu\nu}=8\pi G N {\cal T}_{\mu\nu}\,, \ee where
${\cal T}_{\mu\nu}$ is the matrix energy-momentum tensor defined
by
\begin{equation}\label{566}
{\cal T}_{\mu\nu}=-\frac{1}{2}\frac{\delta
S_{\textrm{matter}}}{\delta a^{\mu\nu}}\;.
\end{equation}
By using the explicit lagrangian (\ref{568}) for the matter field,
we obtain the expression for the energy-momentum tensor
\begin{equation}\label{570}
{\cal
T}_{\mu\nu}=\frac{1}{8}\left[\{\rho,\partial_{\mu}\varphi\otimes\partial_{\nu}\varphi\}+\mathcal{M}_{\mu\nu}+\mathcal{N}_{\mu\nu}\right]+(\mu\leftrightarrow\nu)\;,
\end{equation}
where the explicit form of $\mathcal{M}_{\mu\nu}$ and
$\mathcal{N}_{\mu\nu}$ is obtained by using the variation of the
scalar density $\rho$ in (\ref{559}), namely
\begin{equation}\label{571}
\mathcal{M}_{\mu\nu}= -\int\limits_{\RR^n}
\frac{d\xi}{\pi^{\frac{n}{2}}}\int_{0}^{1}ds\;
e^{-sA(\xi)}\{a^{\alpha\beta},\partial_{\alpha}\varphi\otimes\partial_{\beta}\varphi\}e^{-(1-s)A(\xi)}\xi_{\mu}\xi_{\nu}\;,
\end{equation}
and
\begin{equation}\label{572}
\mathcal{N}_{\mu\nu}=-\int\limits_{\RR^n}
\frac{d\xi}{\pi^{\frac{n}{2}}}\int_{0}^{1}ds\;
e^{-sA(\xi)}\{Q,\varphi\otimes\varphi\}e^{-(1-s)A(\xi)}\xi_{\mu}\xi_{\nu}\;.
\end{equation}

It is worth remarking, here, that the above formula (\ref{570}) for
the energy-momentum tensor ${\cal T}_{\mu\nu}$ reduces, in the
commutative limit, to the standard result, e.g. \cite{fulling89}.

\section{Conclusions}

The main idea of this new model is to describe the gravitational
field by a multiplet of gauge fields with some internal structure.
For this purpose the metric field $g^{\mu\nu}$, which describes
gravity in General Relativity, is replaced by a matrix-valued
2-tensor field $a^{\mu\nu}$. This allows the model to have a much
richer content in describing gravitational phenomena. A more
general geometric picture is developed by allowing the metric to
be matrix-valued. Most of the geometric quantities, used in
describing gravity, can be generalized to be endomorphism-valued.
In this framework it is possible to introduce an action for the
gravitational field which is diffeomorphisms and gauge invariant,
that leads, after performing the variation with respect to
$a^{\mu\nu}$, to the modified (non-commutative) Einstein equation.
It is interesting that the non-commutative part of the modified
equations only depends on a specific tensor density
$H_{\mu\nu\rho}$ and on a linear combination of its commutators.


An important question is related to the quantization of the
present model. The analysis developed in this chapter is purely
classical and the theory is represented by nothing but a
generalized sigma model. The problems of quantization of the
present theory, then, are the same that we encounter in performing
the quantization of a sigma model.

We would like to make a final remark. In our model all the
geometric quantities that we need to develop the formalism are
endomorphism-valued. Once a basis for the vector bundle $V$ has
been fixed, we can represent elements of $\textrm{End}(V)$ by
matrices. Of course the description of physical phenomena has to
be independent from the particular realization of the
representation. This is, ultimately, related to the gauge
invariance of the theory. We believe that by an opportune choice
of gauge, namely an opportune representation of $\textrm{End}(V)$
by matrices, the dynamical equation (\ref{560a}) could be
simplified further. The search for such particular gauge, if it
exists, requires further studies in matrix differential geometry
and matrix General Relativity.


\chapter[NONCOMMUTATIVE CORRECTIONS IN SPECTRAL \newheadline MATRIX GRAVITY]
{NONCOMMUTATIVE CORRECTIONS IN
\newheadline SPECTRAL MATRIX GRAVITY\footnotemark[4]}

\footnotetext[4]{The material in this chapter has been published in \emph{Classical and Quantum Gravity}: G. Fucci and I. G. Avramidi, Non-Commutative Corrections in Spectral Matrix Gravity, \emph{Class. Quant. Grav.} (2009) {\bf 26} 045019 (24pp)  }

\begin{chapabstract}

We study a non-commutative deformation of General Relativity based
on spectral invariants of a partial differential operator acting
on sections of a vector bundle over a smooth manifold. We compute
the first non-commutative corrections to Einstein equations in the
weak deformation limit and analyze the spectrum of the theory.
Related topics are discussed as well.
\end{chapabstract}

\section{Introduction}

The main goal of this chapter is to study the action of Spectral
Matrix Gravity in the weak deformation limit and to describe the
corresponding corrections to Einstein equations. 

We will describe a method for the calculation of the heat kernel
developed in \cite{avramidi94,avramidi01}, which is based on the
covariant Fourier transform proposed in
\cite{avramidi91,avramidi00}. In what follows we specialize the discussion to
second order partial differential operators with non-scalar leading symbol which
naturally arise in the framework of Matrix Gravity. These operators can be written
in a manifestly self-adjoint form as follows \cite{avramidi04b}
\begin{equation}
\label{618}
L=-\rho^{-1}\nabla_{\mu}\rho
a^{\mu\nu}\rho\nabla_{\nu}\rho^{-1}+\bar{Q}\;,
\end{equation}
where $a^{\mu\nu}$ is a matrix-valued symmetric tensor of type
$(2,0)$, $\rho$ is a matrix-valued density of weight $1/2$ and
$\bar{Q}$ is a matrix-valued function.

The heat kernel for a general non-Laplace type second order partial differential operator $L$
is the kernel of the heat semigroup, that is,
\begin{equation}\label{611}
U(t|x,x^{\prime})=\exp({-tL})
\mathcal{P}(x,x^{\prime})\delta(x,x^{\prime})\;,
\end{equation}
where $\delta(x,x')$ is the delta-function (in the density form).
By utilizing the Fourier integral representation for the covariant delta function (\ref{612}),
we obtain
\begin{equation}
\label{613}
U(t|x,x^{\prime})=
\Delta^{\frac{1}{2}}(x,x^{\prime})\mathcal{P}(x,x^{\prime})
\int\limits_{\mathbb{R}^{n}}\frac{d\xi}{(2\pi)^{n}}\exp\{{\imath
\xi_{\mu^{\prime}}\sigma^{\mu^{\prime}}(x,x^{\prime})}\}\Phi(t|k,x,x^{\prime})\;,
\end{equation}
where
\begin{equation}\label{614}
\Phi(t|k,x,x^{\prime})=\exp(-tA)\cdot\mathbb{I}\;,
\end{equation}
\begin{equation}\label{615}
A=e^{-\imath
\xi_{\mu^{\prime}}\sigma^{\mu^{\prime}}}\mathcal{P}^{-1}\Delta^{-\frac{1}{2}}L\Delta^{\frac{1}{2}}\mathcal{P}e^{\imath
\xi_{\mu^{\prime}}\sigma^{\mu^{\prime}}}\;.
\end{equation}
By using the coincidence limits of the two-point functions, in section 2.2.4,
we obtain the heat kernel diagonal
\begin{equation}
\label{617}
U(t|x,x)=\int\limits_{\mathbb{R}^{n}}\frac{d\xi}{(2\pi)^{n}}\;
\Phi(t|k,x,x)\;.
\end{equation}

 By substituting the
operator (\ref{618}) in equation (\ref{615}), we get
\begin{equation}
\label{619}
A=-e^{-\imath
\xi_{\mu^{\prime}}\sigma^{\mu^{\prime}}}\mathcal{P}^{-1}\Delta^{-\frac{1}{2}}\rho^{-1}\nabla_{\mu}\rho
a^{\mu\nu}\rho\nabla_{\nu}\rho^{-1}\Delta^{\frac{1}{2}}\mathcal{P}e^{\imath
\xi_{\mu^{\prime}}\sigma^{\mu^{\prime}}}+Q\;,
\end{equation}
where $Q=\mathcal{P}^{-1}\bar{Q}\mathcal{P}$.
We rewrite this operator in a more
convenient form
\begin{equation}\label{6i}
A=-\bar{X}_{\mu}a^{\mu\nu}X_{\nu}+Q\;,
\end{equation}
where
\begin{eqnarray}\label{6l}
X_{\nu}&=&e^{-\imath
\xi_{\mu^{\prime}}\sigma^{\mu^{\prime}}}\mathcal{P}^{-1}\Delta^{-\frac{1}{2}}\rho\nabla_{\nu}\rho^{-1}\Delta^{\frac{1}{2}}\mathcal{P}e^{\imath
\xi_{\mu^{\prime}}\sigma^{\mu^{\prime}}}\;,\nonumber\\
\bar{X}_{\mu}&=&e^{\imath
\xi_{\mu^{\prime}}\sigma^{\mu^{\prime}}}\mathcal{P}\Delta^{\frac{1}{2}}\rho^{-1}\nabla_{\mu}\rho\Delta^{-\frac{1}{2}}\mathcal{P}^{-1}e^{-\imath
\xi_{\mu^{\prime}}\sigma^{\mu^{\prime}}}\;.
\end{eqnarray}
It is useful to introduce, now, two quantities
\begin{equation}\label{623}
C_{\nu}=-\rho_{;\;\nu}\rho^{-1}\qquad \textrm{and} \qquad
\bar{C_{\nu}}=-\rho^{-1}\rho_{;\;\nu}\;.
\end{equation}
Then we get
\begin{equation}\label{621}
X_{\nu}=\nabla_{\nu}+C_{\nu}+\zeta_{;\;\nu}+E_{\nu}+\imath
\xi_{\mu^{\prime}}\eta^{\mu^{\prime}}{}_{\nu}\;,
\end{equation}
\begin{equation}\label{621a}
\bar{X}_{\mu}=\nabla_{\mu}-\bar{C}_{\mu}+\zeta_{;\;\mu}+E_{\mu}+\imath
\xi_{\nu^{\prime}}\eta^{\nu^{\prime}}{}_{\mu}\;,
\end{equation}
and
\begin{equation}
\label{622}
A=-(\nabla_{\mu}-\bar{C}_{\mu}+\zeta_{;\;\mu}+E_{\mu}+\imath
\xi_{\rho^{\prime}}\eta^{\rho^{\prime}}{}_{\mu})a^{\mu\nu}(\nabla_{\nu}+C_{\nu}+\zeta_{;\;\nu}+E_{\nu}+\imath
\xi_{\rho^{\prime}}\eta^{\rho^{\prime}}{}_{\nu})+Q\;.
\end{equation}
Finally, a straightforward calculation gives
\begin{equation}
\label{629}
A=H+K+\mathcal{L}\;.
\end{equation}
Here
\bea
\label{630}
H&=&\xi_{\alpha^{\prime}}\xi_{\beta^{\prime}}\eta^{\alpha^{\prime}}{}_{\mu}\eta^{\beta^{\prime}}{}_{\nu}a^{\mu\nu}\;,\\
\label{661a}
K&=&-\imath\xi_{\rho^{\prime}}(B^{\rho^{\prime}\nu}\nabla_{\nu}+G^{\rho^{\prime}})\;,\\
\label{660}
\mathcal{L}&=&-\bar{\mathcal{D}}_{\mu}a^{\mu\nu}\mathcal{D}_{\nu}+Q\;,
\eea
where
\begin{eqnarray}
\label{661b}
B^{\rho^{\prime}\nu}&=&2\eta^{\rho^{\prime}}{}_{\mu}a^{\mu\nu}\;,\nonumber\\
G^{\rho^{\prime}}&=&a^{\mu\nu}{}_{;\;\mu}\eta^{\rho^{\prime}}{}_{\nu}+a^{\mu\nu}\eta^{\rho^{\prime}}{}_{\nu;\;\mu}-\bar{C}_{\mu}a^{\mu\nu}\eta^{\rho^{\prime}}{}_{\nu}+\eta^{\rho^{\prime}}{}_{\nu}a^{\mu\nu}C_{\mu}\nonumber\\
&+&E_{\mu}a^{\mu\nu}\eta^{\rho^{\prime}}{}_{\nu}+\eta^{\rho^{\prime}}{}_{\nu}a^{\mu\nu}E_{\mu}+2\zeta_{;\;\mu}a^{\mu\nu}\eta^{\rho^{\prime}}{}_{\nu}\;,\\
\label{661}
\bar{\mathcal{D}}_{\mu}&=&\nabla_{\mu}+\bar{\mathcal{A}}_{\mu}=\nabla_{\mu}-\bar{C}_{\mu}+\zeta_{;\;\mu}+E_{\mu}\;,\nonumber\\
\mathcal{D}_{\nu}&=&\nabla_{\nu}+\mathcal{A}_{\nu}=\nabla_{\nu}+C_{\nu}+\zeta_{;\;\nu}+E_{\nu}\;,
\end{eqnarray}
with $C_{\nu}$, $\bar{C}_{\mu}$, $\zeta$ defined in
(\ref{623}), (\ref{36b}) and $E_{\mu}$
\be
E_{\mu}=\mathcal{P}^{-1}\nabla_{\mu}\mathcal{P}\;.
\ee

More explicitly we can also
write that
\begin{equation}\label{630a}
\mathcal{L}=-a^{\mu\nu}\nabla_{\mu}\nabla_{\nu}+\mathcal{Y}^{\mu}\nabla_{\mu}+\mathcal{Z}\;,
\end{equation}
where
\bea
\label{630b}
\mathcal{Y}^{\mu}
&=&-a^{\mu\nu}{}_{;\;\nu}+\bar{C}_{\nu}a^{\mu\nu}-a^{\mu\nu}C_{\nu}-2a^{\mu\nu}\zeta_{;\;\nu}-a^{\mu\nu}E_{\nu}-E_{\nu}a^{\mu\nu}\;,\\
\label{632}
\mathcal{Z}&=&-a^{\mu\nu}{}_{;\;\mu}C_{\nu}-a^{\mu\nu}C_{\nu;\;\mu}+\bar{C}_{\mu}a^{\mu\nu}C_{\nu}-a^{\mu\nu}{}_{;\;\mu}\zeta_{;\;\nu}-a^{\mu\nu}\zeta_{;\;\mu\nu}+\bar{C}_{\mu}a^{\mu\nu}\zeta_{;\;\nu}\nonumber\\
&-&\zeta_{;\;\mu}a^{\mu\nu}C_{\nu}-\zeta_{;\;\mu}a^{\mu\nu}\zeta_{;\;\nu}-a^{\mu\nu}{}_{;\;\mu}E_{\nu}-a^{\mu\nu}E_{\nu;\;\mu}+\bar{C}_{\mu}a^{\mu\nu}E_{\nu}-E_{\mu}a^{\mu\nu}C_{\nu}\nonumber\\
&-&E_{\mu}a^{\mu\nu}E_{\nu}-\zeta_{;\;\nu}a^{\mu\nu}E_{\mu}-\zeta_{;\;\nu}E_{\mu}a^{\mu\nu}+Q\;.
\eea

Thus, by using the eq. (\ref{629}) we obtain
\begin{equation}
\label{633}
U(t|x,x)=\int\limits_{\RR^n}\frac{d\xi}{(2\pi)^{n}}\;
e^{-t(H+K+\mathcal{L})}\cdot\mathbb{I}\Big|_{x=x^{\prime}}\;,
\end{equation}
which, by scaling the integration variable $\xi\rightarrow
t^{-\frac{1}{2}}\xi$, takes the form
\begin{equation}\label{634}
U(t|x,x)=(4\pi
t)^{-\frac{n}{2}}\int\limits_{\mathbb{R}^{n}}\frac{d\xi}{\pi^{\frac{n}{2}}}\;
\exp(-H-\sqrt{t}K-t\mathcal{L})\cdot\mathbb{I}\Big|_{x=x^{\prime}}\;.
\end{equation}
It is convenient, to rewrite this equation as
\begin{equation}\label{634a}
U(t|x,x)=(4\pi
t)^{-\frac{n}{2}}\int\limits_{\mathbb{R}^{n}}\frac{d\xi}{\pi^{\frac{n}{2}}}\;
\;e^{-|\xi|^{2}}\exp(-\tilde{H}-\sqrt{t}K-t\mathcal{L})\cdot\mathbb{I}\Big|_{x=x^{\prime}}\;,
\end{equation}
where $|\xi|^{2}=g^{\mu\nu}\xi_{\mu}\xi_{\nu}$ and
\begin{equation}\label{634b}
\tilde{H}=H-|\xi|^{2}\;.
\end{equation}

In order to evaluate the first three coefficients of the
asymptotic expansion of (\ref{634a}) as $t\rightarrow 0$ we use the
Volterra series for the exponent of a sum of two non-commuting operators in (\ref{635}) to obtain
\begin{equation}\label{635b}
\exp(-\tilde{H}-\sqrt{t}K-t\mathcal{L})=e^{-\tilde{H}}-\sqrt{t}\;\Omega+t\Psi+O(t^{\frac{3}{2}})\;,
\end{equation}
where
\begin{equation}\label{635c}
\Omega=\int\limits_{0}^{1}d\tau_{1}e^{-(1-\tau_{1})\tilde{H}}Ke^{-\tau_{1}\tilde{H}}\;,
\end{equation}
and
\begin{equation}\label{635d}
\Psi=\int\limits_{0}^{1}d\tau_{2}\int\limits_{0}^{\tau_{2}}d\tau_{1}e^{-(1-\tau_{2})\tilde{H}}Ke^{-(\tau_{2}-\tau_{1})\tilde{H}}Ke^{-\tau_{1}\tilde{H}}-\int\limits_{0}^{1}d\tau_{1}e^{-(1-\tau_{1})\tilde{H}}\mathcal{L}e^{-\tau_{1}\tilde{H}}\;.
\end{equation}

We are only interested in the terms $a_{0}$ and $a_{1}$ of the
heat kernel expansion, namely the terms of zero order and linear
in the parameter $t$. These terms can be written, respectively, as
\begin{eqnarray}\label{635e}
a_{0}&=&g^{\frac{1}{2}}\tilde{a}_{0}\;,\\
a_{1}&=&g^{\frac{1}{2}}\tilde{a}_{1}\;,
\end{eqnarray}
where
\begin{eqnarray}\label{636}
\tilde{a}_{0}&=&\int\limits_{\mathbb{R}^{n}}\frac{d\xi}{\pi^{\frac{n}{2}}}\;g^{-\frac{1}{2}}\;e^{-|\xi|^{2}}\exp\{-\tilde{H}\}\cdot\mathbb{I}\Big|_{x=x^{\prime}}\;,\\
\label{6360}
\tilde{a}_{1}&=&\int\limits_{\mathbb{R}^{n}}\frac{d\xi}{\pi^{\frac{n}{2}}}\;g^{-\frac{1}{2}}\;e^{-|\xi|^{2}}\Psi\cdot\mathbb{I}\Big|_{x=x^{\prime}}\;.
\end{eqnarray}

The term $t^{\frac{1}{2}}$ of the heat kernel expansion vanishes
identically. This happens because the heat kernel coefficients are
defined as $\xi$-integrals over the whole $\mathbb{R}^{n}$ and the
term $\Omega$ in (\ref{635c}) is an odd function of $\xi$.

\section{Evaluation of the Heat Kernel Coefficients}

\subsection{Local Coefficient $\tilde{a}_{0}$}

We will evaluate the heat kernel coefficients $A_{0}$
and $A_{1}$ using the perturbation theory.
The main idea is to introduce a small deformation parameter
$\lambda$ and evaluate the non-commutative corrections to the
action of Spectral Matrix Gravity. For this purpose we write the
matrix $a_{\mu\nu}$ as
\begin{equation}\label{638}
a^{\mu\nu}=g^{\mu\nu}\,\mathbb{I}+\lambda h^{\mu\nu}\;,
\end{equation}
where $h^{\mu\nu}$ is a traceless
matrix-valued tensor field (a non-commutative
perturbation of the Riemannian metric), satisfying
\begin{equation}\label{638a}
\textrm{tr}_{V} h^{\mu\nu}=0\;.
\end{equation}
Furthermore, we parameterize the matrix-valued density $\rho$
introduced in (\ref{618}) as
\begin{equation}\label{639}
\rho=g^{\frac{1}{4}}e^{\phi}e^{\lambda\sigma}\;.
\end{equation}
Here $\sigma$ is a
traceless matrix-valued scalar field and $\phi$ is a scalar field.
({\it Do not confuse it with the world function introduced in the previous
sections!})
Finally, we also decompose the
endomorphism $Q$,
\begin{equation}\label{6re}
Q=q\cdot\mathbb{I}+\lambda \Theta\;,
\end{equation}
where $\Theta$ is a traceless matrix-valued scalar field.

Now we expand all the quantities in powers of $\lambda$.
On doing so the matrix $\rho$ and its
inverse read
\begin{eqnarray}\label{640}
\rho&=&g^{\frac{1}{4}}\;e^{\phi}\left(1+\lambda\sigma+\frac{\lambda^{2}}{2}\sigma^{2}\right)+O(\lambda^{3})\;,\nonumber\\
\rho^{-1}&=&g^{-\frac{1}{4}}\;e^{-\phi}\left(1-\lambda\sigma+\frac{\lambda^{2}}{2}\sigma^{2}\right)+O(\lambda^{3})\;,
\end{eqnarray}
and its derivative is
\begin{equation}\label{641}
g^{-\frac{1}{4}}\rho_{;\;\nu}=e^{\phi}\left[\lambda\sigma_{;\;\nu}+\frac{\lambda^{2}}{2}(\sigma_{;\;\nu}\sigma+\sigma\sigma_{;\;\nu})\right]+e^{\phi}\phi_{;\;\nu}\left(1+\lambda\sigma+\frac{\lambda^{2}}{2}\sigma^{2}\right)+O(\lambda^{3})\;.
\end{equation}
>From the last two expressions one can easily evaluate the
operators $C_{\nu}$ and $\bar{C}_{\nu}$ obtaining explicitly
\begin{eqnarray}\label{642}
C_{\nu}=-\rho_{;\;\nu}\rho^{-1}=-\phi_{;\;\nu}-\lambda\sigma_{;\;\nu}+\frac{\lambda^{2}}{2}[\sigma_{;\;\nu},\sigma]+O(\lambda^{3})\;,
\end{eqnarray}
\begin{eqnarray}\label{643}
\bar{C}_{\nu}=-\rho^{-1}\rho_{;\;\nu}=-\phi_{;\;\nu}-\lambda\sigma_{;\;\nu}-\frac{\lambda^{2}}{2}[\sigma_{;\;\nu},\sigma]+O(\lambda^{3})\;.
\end{eqnarray}

The operators $\tilde{H}$, $K$ and $\mathcal{L}$ introduced above
in (\ref{630}), (\ref{661a}) and (\ref{660}) depend on the
deformation parameter $\lambda$ as well. By expanding them in
terms of the deformation parameter we get
\begin{eqnarray}\label{644}
\tilde{H}&=&H_{0}+\lambda H_{1}\;,\nonumber\\
K&=&K_{0}+\lambda K_{1}+\lambda^{2}K_{2}+O(\lambda^{2})\;,\nonumber\\
\mathcal{L}&=&\mathcal{L}_{0}+\lambda
\mathcal{L}_{1}+\lambda^{2}\mathcal{L}_{2}+O(\lambda^{2})\;,
\end{eqnarray}
where
\begin{eqnarray}\label{645}
H_{0}&=&\xi_{\alpha^{\prime}}\xi_{\beta^{\prime}}(\eta^{\alpha^{\prime}}{}_{\mu}\eta^{\beta^{\prime}}{}_{\nu}g^{\mu\nu}-g^{\alpha^{\prime}\beta^{\prime}})\;,\\[3pt]
\label{6450}
H_{1}&=&\xi_{\alpha^{\prime}}\xi_{\beta^{\prime}}\eta^{\alpha^{\prime}}{}_{\mu}\eta^{\beta^{\prime}}{}_{\nu}h^{\mu\nu}\;,\\[12pt]
K_{0}&=&-\imath\xi_{\alpha^{\prime}}(2\eta^{\alpha^{\prime}}{}_{\mu}g^{\mu\nu}\nabla_{\nu}+2\eta^{\alpha^{\prime}}{}_{\mu}g^{\mu\nu}\zeta_{;\;\nu}+\eta^{\alpha^{\prime}}{}_{\;\;\mu}{}^{;\;\mu}+E^{\mu}\eta^{\alpha^{\prime}}{}_{\mu}+\eta^{\alpha^{\prime}}{}_{\mu}E^{\mu})\;,\\[3pt]
K_{1}&=&-\imath\xi_{\alpha^{\prime}}(2\eta^{\alpha^{\prime}}{}_{\mu}h^{\mu\nu}\nabla_{\nu}+\eta^{\alpha^{\prime}}{}_{\mu}h^{\mu\nu}{}_{;\nu}+2\eta^{\alpha^{\prime}}{}_{\mu}h^{\mu\nu}\zeta_{;\;\nu}+h^{\mu\nu}\eta^{\alpha^{\prime}}{}_{\mu;\;\nu}+\eta^{\alpha^{\prime}}{}_{\mu}h^{\mu\nu}E_{\nu}\nonumber\\
&+&\eta^{\alpha^{\prime}}{}_{\mu}E_{\nu}h^{\mu\nu})\;,\\[3pt]
K_{2}&=&-\imath\xi_{\alpha^{\prime}}[\eta^{\alpha^{\prime}}_{\;\;\mu}([\sigma_{;\;\nu},h^{\mu\nu}]+[\sigma^{;\;\mu},\sigma])]\;,
\end{eqnarray}
\begin{eqnarray}\label{650}
\mathcal{L}_{0}&=&-\nabla^{2}-(2\zeta^{;\;\mu}+2E^{\mu})\nabla_{\mu}+\phi_{;\;\mu}{}^{;\;\mu}+\phi_{;\;\mu}\phi^{;\;\mu}-\zeta_{;\;\mu}{}^{;\;\mu}\nonumber\\
&-&\zeta^{;\;\mu}\zeta_{;\;\mu}-E_{\mu}{}^{;\;\mu}-E^{\mu}E_{\mu}-2\zeta_{;\;\mu}E^{\mu}+q\;,\\[3pt]
\mathcal{L}_{1}&=&-h^{\mu\nu}\nabla_{\mu}\nabla_{\nu}-[h^{\mu\nu}{}_{;\;\nu}+2\zeta_{;\;\nu}h^{\mu\nu}+(h^{\mu\nu}E_{\nu}+E_{\nu}h^{\mu\nu})]\nabla_{\mu}+\sigma_{;\;\mu}{}^{;\;\mu}\nonumber\\
&+&h^{\mu\nu}{}_{;\;\mu}\phi_{;\;\nu}+h^{\mu\nu}\phi_{;\;\mu\nu}+2\phi^{;\;\nu}\sigma_{;\;\nu}+\phi_{;\;\mu}h^{\mu\nu}\phi_{;\;\nu}-h^{\mu\nu}{}_{;\;\mu}\zeta_{;\;\nu}\nonumber\\
&-&h^{\mu\nu}\zeta_{;\;\mu\nu}-\zeta_{;\;\mu}h^{\mu\nu}\zeta_{;\;\nu}-h^{\mu\nu}{}_{;\;\mu}E_{\nu}-h^{\mu\nu}E_{\nu;\;\mu}+[E_{\mu},\sigma^{;\;\mu}]\nonumber\\
&+&\phi_{;\;\mu}[E_{\nu},h^{\mu\nu}]-E_{\mu}h^{\mu\nu}E_{\nu}-E_{\mu}h^{\mu\nu}\zeta_{;\;\nu}-\zeta_{;\;\mu}h^{\mu\nu}E_{\nu}+\Theta\;,\\[3pt]
\mathcal{L}_{2}&=&([h^{\mu\nu},\sigma_{;\;\nu}]-[\sigma^{;\;\mu},\sigma])\nabla_{\mu}-\frac{1}{2}[\sigma_{;\;\mu}{}^{;\;\mu},\sigma]+h^{\mu\nu}{}_{;\;\mu}\sigma_{;\;\nu}+h^{\mu\nu}\sigma_{;\;\mu\nu}\nonumber\\
&+&\sigma^{;\;\nu}\sigma_{;\;\nu}+\sigma_{;\;\nu}h^{\mu\nu}\phi_{;\;\mu}+\phi_{;\;\nu}h^{\mu\nu}\sigma_{;\;\mu}-\frac{1}{2}[\sigma^{;\;\mu},\sigma]\zeta_{;\;\mu}-\frac{1}{2}\zeta_{;\;\mu}[\sigma^{;\;\mu},\sigma]\nonumber\\
&+&\zeta_{;\;\mu}h^{\mu\nu}\sigma_{;\;\nu}-\sigma_{;\;\nu}h^{\mu\nu}\zeta_{;\;\mu}-\frac{1}{2}([\sigma_{;\;\mu},\sigma]E^{\mu}+E^{\mu}[\sigma_{;\;\mu},\sigma])\nonumber\\
&+&E_{\mu}h^{\mu\nu}\sigma_{;\;\nu}-\sigma_{;\;\mu}h^{\mu\nu}E_{\nu}\;.
\label{6500}
\end{eqnarray}

In the framework of perturbation theory we write, then, the
coefficients $\tilde{a}_{0}$ and $\tilde{a}_{1}$ of the heat
kernel expansion in (\ref{636}) and (\ref{6360}) in terms of the
deformation parameter $\lambda$, namely
\begin{eqnarray}\label{653}
\tilde{a}_{0}(\lambda)&=&a_{0}^{(0)}+\lambda
a_{0}^{(1)}+\lambda^{2}a_{0}^{(2)}+O(\lambda^{3})\nonumber\\
\tilde{a}_{1}(\lambda)&=&a_{1}^{(0)}+\lambda
a_{1}^{(1)}+\lambda^{2}a_{1}^{(2)}+O(\lambda^{3})\;.
\end{eqnarray}
By using the explicit formulas obtained in (\ref{645}) through
(\ref{6500}), we will be able to evaluate all the coefficients of
the Taylor expansions in (\ref{653}).

Next, we introduce a notation that will be useful in the following
calculations. Let $f$ be a function of $\xi$. We define the
Gaussian average of the function $f$ as
\begin{equation}\label{654}
\bra{f}\ket{}=\int\limits_{\mathbb{R}^{n}}\frac{d\xi}{\pi^{\frac{n}{2}}}\;g^{-\frac{1}{2}}\;e^{-|\xi|^{2}}f(\xi)\;.
\end{equation}
The Gaussian averages of the polynomials are well known
\begin{eqnarray}\label{655}
\bra{\xi_{\mu_{1}}\cdots\xi_{\mu_{2n+1}}}\ket{}&=&0\;,\nonumber\\
\bra{\xi_{\mu_{1}}\cdots\xi_{\mu_{2n}}}\ket{}&=&\frac{(2n)!}{2^{2n}n!}g_{(\mu_{1}\mu_{2}}\cdots
g_{\mu_{2n-1}\mu_{2n})}\;,
\end{eqnarray}
where the parentheses $(\;)$ denote the symmetrization over all the
included indices.

For the coefficient of order zero of the heat kernel expansion we
consider the first equation in (\ref{653}). From the formula
(\ref{636}), it is easy to see that the only non-vanishing
contribution to $\tilde{a}_{0}$ is
\begin{equation}\label{656}
\tilde{a}_{0}=\left<\left(1-\lambda
N+\frac{\lambda^{2}}{2}N^{2}\right)\cdot\mathbb{I}\right>\Bigg|_{x=x^{\prime}}+O(\lambda^{3})\;.
\end{equation}
By using the equations (\ref{645}), (\ref{6450}) and after taking
the coincidence limit we obtain the expression
\begin{equation}\label{657}
\tilde{a}_{0}=\left<\left(\mathbb{I}-\lambda
h^{\mu\nu}\xi_{\mu}\xi_{\nu}+\frac{1}{2}\lambda^{2}h^{\mu\nu}h^{\rho\sigma}\xi_{\mu}\xi_{\nu}\xi_{\rho}\xi_{\sigma}\right)\right>+O(\lambda^{3})\;,
\end{equation}
and then, by performing the Gaussian averages, we get
\begin{equation}\label{658}
\tilde{a}_{0}=1-\frac{\lambda}{2}h+\frac{\lambda^{2}}{8}(h^{2}+2h^{\mu\nu}h_{\mu\nu})+O(\lambda^{3})\;,
\end{equation}
where $h=g_{\mu\nu}h^{\mu\nu}$. In order to evaluate the global
coefficient $A_{0}$, we need the trace of
(\ref{658}). Since $h^{\mu\nu}$ is traceless we immediately obtain
\begin{equation}\label{659}
\textrm{tr}_{V}\tilde{a}_{0}
=\textrm{tr}_{V}\left(1+\frac{\lambda^{2}}{8}h^{2}+\frac{\lambda^{2}}{4}h^{\mu\nu}h_{\mu\nu}\right)+O(\lambda^{3})\;.
\end{equation}

\subsection{Coincidence Limits}

In this section we will list the various coincidence limits that we will use
during the calculations performed in this chapter.

Through all this section the subscripts $0$,$1$ and $2$ will be
used to denote terms of different order in the deformation
parameter $\lambda$. More precisely for any quantity $\mathcal{X}$
which contains different orders of $\lambda$ we write
\begin{displaymath}
\mathcal{X}=\mathcal{X}_{0}+\lambda\mathcal{X}_{1}
+\lambda^{2}\mathcal{X}_{2}+O(\lambda^{3})\;,
\end{displaymath}
where $\mathcal{X}_{0}$, $\mathcal{X}_{1}$ and $\mathcal{X}_{2}$
denote, respectively, the zeroth, first and second order in
$\lambda$.

We start with the coincidence limit of the operator $\tilde{H}$ in
(\ref{630}) and its derivatives. More precisely we have
\begin{equation}\label{6A1}
[\tilde{H}]=\lambda\xi_{\mu}\xi_{\nu}h^{\mu\nu}\;.
\end{equation}
For the first derivative we obtain
\begin{equation}\label{6A2}
[\tilde{H}_{;\;\mu}]=\lambda\xi_{\alpha}\xi_{\beta}h^{\alpha\beta}{}_{;\;\mu}\;.
\end{equation}
For the second derivative we get the following formula
\begin{equation}\label{6A3}
[\tilde{H}_{;\;\mu\nu}]=-\frac{2}{3}
\xi_{\alpha}\xi^{\rho}R^{\alpha}{}_{\mu\nu\rho}
-\frac{2}{3}\lambda\xi_{\alpha}\xi_{\sigma}h^{\rho\sigma}
(R^{\alpha}{}_{\mu\nu\rho}+R^{\alpha}{}_{\rho\nu\mu})
+\lambda\xi_{\alpha}\xi_{\beta}h^{\alpha\beta}{}_{;\;\mu\nu}\;.
\end{equation}

Recall, now, the definition (\ref{661a}) for the operator $K$. The
coincidence limits of the terms in $K$ are
\begin{eqnarray}\label{6A4}
\left[B^{\rho^{\prime}\nu}_{0}\right]&=&-2g^{\rho\nu}\;,\\
\left[B^{\rho^{\prime}\nu}_{1}\right]&=&-2h^{\rho\nu}\;.
\end{eqnarray}
For the derivatives of these quantities we have
\begin{eqnarray}\label{6A5}
\left[(\nabla_{\mu}B^{\rho^{\prime}\nu})_{0}\right]&=&0\;,\\
\left[(\nabla_{\mu}B^{\rho^{\prime}\nu})_{1}\right]&=&
-2h^{\rho\nu}{}_{;\;\mu}\;.
\end{eqnarray}
For the other terms we have
\begin{eqnarray}\label{6A6}
\left[G^{\rho^{\prime}}_{0}\right]&=&0\;,\\
\left[G^{\rho^{\prime}}_{1}\right]&=&-h^{\rho\nu}{}_{;\;\nu}\;.
\end{eqnarray}
for the derivatives of (\ref{6A6}) we get
\begin{eqnarray}\label{6A7}
\left[(\nabla_{\nu}G^{\rho^{\prime}})_{0}\right]&=&
\frac{1}{3}R^{\rho}{}_{\nu}+\mathcal{R}^{\rho}{}_{\nu}\;,\\
\left[(\nabla_{\nu}G^{\rho^{\prime}})_{1}\right]&=&
-h^{\rho\alpha}{}_{;\;\alpha\nu}-\frac{1}{3}h^{\rho\alpha}
R_{\alpha\nu}+\frac{2}{3}h^{\mu\alpha}R^{\rho}{}_{\alpha\nu\mu}
+h^{\rho\alpha}\mathcal{R}_{\alpha\nu}\;,\\
\left[(\nabla_{\nu}G^{\rho^{\prime}})_{2}\right] &=&
-[\sigma_{;\;\alpha\nu},h^{\rho\alpha}]
-[\sigma_{;\;\alpha},h^{\rho\alpha}{}_{;\;\nu}]
-[\sigma^{;\;\rho}{}_{;\;\nu},\sigma]-[\sigma^{;\;\rho},\sigma_{;\;\nu}]\;.
\end{eqnarray}

For the operator $\mathcal{L}$ in (\ref{660}) we need the following
coincidence limits
\begin{eqnarray}\label{6A8}
\left[(\mathcal{A}_{\nu})_{0}\right]&=&-[(\bar{\mathcal{A}}_{\nu})_{0}]
=-\phi_{;\;\nu}\;,\\
\left[(\mathcal{A}_{\nu})_{0}\right]&=&-[(\bar{\mathcal{A}}_{\nu})_{0}]
=-\sigma_{;\;\nu}\;,\\
\left[(\mathcal{A}_{\nu})_{0}\right]&=&[(\bar{\mathcal{A}}_{\nu})_{0}]
=\frac{1}{2}[\sigma_{;\;\nu},\sigma]\;.
\end{eqnarray}
We also used, during the calculation, the coincidence limits for
the derivatives of $\mathcal{A}_{\nu}$, namely
\begin{eqnarray}\label{6A9}
\left[(\nabla_{\mu}\mathcal{A}_{\nu})_{0}\right]
&=&-\phi_{;\;\nu\mu}+\frac{1}{6}R_{\nu\mu}-\frac{1}{2}\mathcal{R}_{\nu\mu}\;,\\
\left[(\nabla_{\mu}\mathcal{A}_{\nu})_{1}\right]
&=&-\sigma_{\nu\mu}\;,\\
\left[(\nabla_{\mu}\mathcal{A}_{\nu})_{2}\right]
&=&\frac{1}{2}([\sigma_{;\;\nu\mu},\sigma]
+[\sigma_{;\;\nu},\sigma_{;\;\mu}])\;.
\end{eqnarray}

\subsection{Local Coefficient $\tilde{a}_{1}$}

Now we evaluate the coefficient $\tilde{a}_{1}$. By using the
expressions (\ref{6360}), (\ref{635d}) and (\ref{660}) we have
\begin{eqnarray}\label{662}
\textrm{tr}_{V} \tilde{a}_{1}
&=&\Biggl\{\left<\textrm{tr}_{V}\;[-e^{-\tilde{H}}Q]\right>
+\left<\textrm{tr}_{V}\;\Bigg[\int\limits_{0}^{1}d\tau_{2}\int\limits_{0}^{\tau_{2}}d\tau_{1}e^{-(1-\tau_{2})\tilde{H}}Ke^{-(\tau_{2}-\tau_{1})\tilde{H}}Ke^{-\tau_{1}\tilde{H}}\Bigg]\right>\nonumber\\
&-&\left<\textrm{tr}_{V}\;\Bigg[\int\limits_{0}^{1}d\tau_{1}e^{-(1-\tau_{1})\tilde{H}}\bar{\mathcal{D}}_{\mu}a^{\mu\nu}\mathcal{D}_{\nu}e^{-\tau_{1}\tilde{H}}\Bigg]\right>\Biggr\}\Bigg|_{x=x^{\prime}}\;,
\end{eqnarray}
where the first term of the expression (\ref{662}) has been
obtained by simply using the cyclic property of the trace.

In the following we will evaluate the terms in (\ref{662})
separately. We start with the simplest of them, namely the one
involving the endomorphism $Q$. By using the Taylor expansion in
$\lambda$ of $\tilde{H}$ in (\ref{644}) and the coincidence limits
(\ref{6ca}) we obtain
\begin{equation}\label{663}
\left<\textrm{tr}_{V}\;[-e^{-\tilde{H}}Q]\right>\Big|_{x=x^{\prime}}=\left<\textrm{tr}_{V}\;\left(-Q+\lambda\xi_{\mu}\xi_{\nu}h^{\mu\nu}Q-\frac{\lambda^{2}}{2}\xi_{\mu}\xi_{\nu}\xi_{\rho}\xi_{\sigma}h^{\mu\nu}h^{\rho\sigma}Q\right)\right>\Bigg|_{x=x^{\prime}}+O(\lambda^{3})\;.
\end{equation}
By expanding $Q$ as in (\ref{6re}) and by performing the Gaussian
averages we obtain
\begin{equation}\label{664}
\left<\textrm{tr}_V\;[-e^{-\tilde{H}}Q]\right>\Big|_{x=x^{\prime}}=-Nq+\lambda^{2}\textrm{tr}_{V}\left(\frac{1}{2}h\Theta-\frac{1}{8}h^{2}q-\frac{1}{4}h^{\mu\nu}h_{\mu\nu}q\right)+O(\lambda^{3})\;,
\end{equation}
where we used the property (\ref{638a}).

For the second term in equation (\ref{662}) we get, by using the
definition (\ref{661a}),
\begin{eqnarray}\label{665}
\lefteqn{\left<\textrm{tr}_{V}\;\Bigg[\int\limits_{0}^{1}d\tau_{2}\int\limits_{0}^{\tau_{2}}d\tau_{1}e^{-(1-\tau_{2})\tilde{H}}Ke^{-(\tau_{2}-\tau_{1})\tilde{H}}Ke^{-\tau_{1}\tilde{H}}\Bigg]\right>\Bigg|_{x=x^{\prime}}=}\nonumber\\
&&-\Bigg<\textrm{tr}_{V}\;\Bigg[\xi_{\rho^{\prime}}\xi_{\sigma^{\prime}}\int\limits_{0}^{1}d\tau_{2}\int\limits_{0}^{\tau_{2}}d\tau_{1}e^{-(1-\tau_{2})\tilde{H}}\Big(B^{\rho^{\prime}\nu}\nabla_{\nu}e^{-(\tau_{2}-\tau_{1})\tilde{H}}B^{\sigma^{\prime}\mu}\nabla_{\mu}e^{-\tau_{1}\tilde{H}}\nonumber\\
&&+B^{\rho^{\prime}\nu}\nabla_{\nu}e^{-(\tau_{2}-\tau_{1})\tilde{H}}G^{\sigma^{\prime}}e^{-\tau_{1}\tilde{H}}+G^{\rho^{\prime}}e^{-(\tau_{2}-\tau_{1})\tilde{H}}B^{\sigma^{\prime}\mu}\nabla_{\mu}e^{-\tau_{1}\tilde{H}}\nonumber\\
&&+G^{\rho^{\prime}}e^{-(\tau_{2}-\tau_{1})\tilde{H}}G^{\sigma^{\prime}}e^{-\tau_{1}\tilde{H}}\Big)\Bigg]\Bigg>\Bigg|_{x=x^{\prime}}\;.
\end{eqnarray}
It is straightforward to notice that in the last expression we
need to compute first and second derivatives of the exponentials
containing the operator $\tilde{H}$. These derivatives are
computed by using integral representations, i.e. for the first
derivative we have \cite{avramidi04b}
\begin{equation}\label{666}
\nabla_{\mu}e^{-\tau
\tilde{H}}=-\beta_{\mu}(\tau)e^{-\tau\tilde{H}}\;,
\end{equation}
where
\begin{equation}\label{667}
\beta_{\mu}(\tau)=\int\limits_{0}^{\tau}ds\;e^{-s\tilde{H}}\tilde{H}_{;\;\mu}e^{s\tilde{H}}\;.
\end{equation}
This last integral can be evaluated by referring to the following
formula and by integrating over $s$ \cite{avramidi04b}
\begin{equation}\label{667b}
e^{-\tilde{H}}\tilde{H}_{;\;\mu}e^{\tilde{H}}=\sum_{k=0}^{\infty}\frac{(-1)^{k}}{k!}\underbrace{[\tilde{H},\cdots[\tilde{H}}_{k},\tilde{H}_{;\;\mu}]\cdots]\;.
\end{equation}
By expanding (\ref{667}) in $s$, up to the second order in
$\lambda$, we obtain
\begin{equation}\label{667a}
\nabla_{\mu}e^{-\tau
\tilde{H}}=-\left(\tau\tilde{H}_{;\;\mu}+\frac{1}{2}\tau^{2}[\tilde{H}_{;\;\mu},\tilde{H}]\right)e^{-\tau\tilde{H}}+O(\lambda^{3})\;.
\end{equation}
This last expression can be obtained by recalling that the
coincidence limit for $\tilde{H}$ and its derivatives is of order
$\lambda$ without the zeroth order term.

For the second derivative we write \cite{avramidi04b}
\begin{eqnarray}\label{668}
\nabla_{\mu}\nabla_{\nu}e^{-\tau\tilde{H}}&=&-\int\limits_{0}^{\tau}ds_{1}e^{-(\tau-s_{1})\tilde{H}}\tilde{H}_{;\;\mu\nu}e^{-s_{1}\tilde{H}}+\nonumber\\
&+&\int\limits_{0}^{\tau}ds_{2}\int\limits_{0}^{s_{2}}ds_{1}\Big(e^{-(s_{2}-s_{1})\tilde{H}}\tilde{H}_{;\;\nu}e^{-s_{1}\tilde{H}}\tilde{H}_{;\;\mu}e^{-(\tau-s_{2})\tilde{H}}+\nonumber\\
&+&e^{-(\tau-s_{2})\tilde{H}}\tilde{H}_{;\;\mu}e^{-(s_{2}-s_{1})\tilde{H}}\tilde{H}_{;\;\nu}e^{-s_{1}\tilde{H}}\Big)\;.
\end{eqnarray}
We can express this formula in the same form as (\ref{667a}),
i.e.
\begin{equation}\label{669}
\nabla_{\mu}\nabla_{\nu}e^{-\tau\tilde{H}}=-\left(\tau\tilde{H}_{;\;\mu\nu}+\frac{1}{2}[\tilde{H}_{;\;\mu\nu},\tilde{H}]-\frac{1}{2}\tau^{2}\{\tilde{H}_{;\;\nu},\tilde{H}_{;\;\mu}\}\right)e^{-\tau\tilde{H}}+O(\lambda^{3})\;.
\end{equation}

Now that we have the expressions (\ref{666}) through (\ref{669}), we
can substitute them in (\ref{665}) and we can expand the remaining
exponentials in $\tau$ up to orders $\lambda^{2}$. After the
expansion of the exponentials and after taking the coincidence
limit, we have to evaluate the double integrals of polynomials in
$\tau_{1}$ and $\tau_{2}$ which will yield the numerical
coefficients for the various terms in (\ref{665}). The most general
double integral that we need to evaluate is the following
\begin{equation}\label{670}
I_{1}(\alpha,\beta,\gamma,\delta)=\int\limits_{0}^{1}d\tau_{2}\int\limits_{0}^{\tau_{2}}d\tau_{1}\tau_{1}^{\alpha}(1-\tau_{2})^{\beta}(\tau_{2}-\tau_{1})^{\gamma}(1-\tau_{2}+\tau_{1})^{\delta}\;,
\end{equation}
with $(\alpha,\beta,\gamma,\delta)$ being positive integers so
that the integral is well defined. The solution in closed form of
(\ref{670}) can be found as follows.

This integral is well defined for $\textrm{Re}(\alpha)>-1$,
$\textrm{Re}(\gamma)>-1$ and $0<\tau_{2}<1$. By using the integral
representation of the hypergeometric function we can evaluate the
integral in $\tau_{1}$, i.e.
\begin{equation}\label{6A11}
I_{1}=\frac{\Gamma(1+\alpha)\Gamma(1+\gamma)}{\Gamma(2+\alpha
+\gamma)}\int\limits_{0}^{1}d\tau_{2}\tau_{2}^{1+\alpha
+\gamma}(1-\tau_{2})^{\beta+\delta}{}_{2}F_{1}
\left(1+\alpha,-\delta,2+\alpha+\gamma\;;\frac{\tau_{2}}{1-\tau_{2}}\right)\;.
\end{equation}
Now, by using the linear transformation formula for the
hypergeometric function we get \cite{abramowitz72}
\begin{eqnarray}\label{6A12}
\lefteqn{{}_{2}F_{1}\left(1+\alpha,-\delta,2
+\alpha+\gamma\;;\frac{\tau_{2}}{1-\tau_{2}}\right)=}
\nonumber\\
&&=(1-\tau_{2})^{1+\alpha}{}_{2}F_{1}\left(1+\alpha,2
+\alpha+\gamma+\delta,2+\alpha+\gamma\;;\tau_{2}\right)\;.
\end{eqnarray}
By substituting the last expression in the integral (\ref{6A11}) we
obtain
\begin{eqnarray}\label{6A13}
\lefteqn{I_{1}=\frac{\Gamma(1+\alpha)\Gamma(1+\gamma)}{\Gamma(2
+\alpha+\gamma)}\times}\nonumber\\
&&\int\limits_{0}^{1}d\tau_{2}\tau_{2}^{1+\alpha
+\gamma}(1-\tau_{2})^{1+\alpha+\beta+\delta}{}_{2}F_{1}\left(1
+\alpha,2+\alpha+\gamma+\delta,2+\alpha+\gamma\;;\tau_{2}\right)\;.
\nonumber\\
&&\phantom{2}
\end{eqnarray}
This integral over $\tau_{2}$ is, now, of the following general
form
\begin{equation}\label{6A14}
\mathcal{I}=\int\limits_{0}^{1}dx\;x^{c-1}(1-x)^{d-1}{}_{2}F_{1}
\left(a,b,c\;;x\right)\;,
\end{equation}
which is well define for $\textrm{Re}(c)>0$ and $\textrm{Re}(d)>0$
and has a solution in a closed form \cite{gradshtein07}, namely
\begin{equation}\label{6A15}
\mathcal{I}=\frac{\Gamma(c)\Gamma(d)\Gamma(c+d-a-b)}{\Gamma(c+d-a)
\Gamma(c+d-b)}\;.
\end{equation}
By using the results (\ref{6A15}) in the integral (\ref{6A13}), we
get the solution (\ref{671}), i.e.
\begin{equation}\label{671}
I_{1}(\alpha,\beta,\gamma,\delta)=\frac{\Gamma(1+\alpha)\Gamma(1+\beta)\Gamma(1+\gamma)\Gamma(2+\alpha+\beta+\delta)}{\Gamma(3+\alpha+\beta+\gamma+\delta)\Gamma(2+\alpha+\beta)}\;,
\end{equation}
where $\Gamma(x)$ is the Euler gamma function.

By using the technical details described above and the coincidence
limits for the various terms in (\ref{665}) (see Section 6.3.2), we
obtain
\begin{eqnarray}\label{672}
\left<\textrm{tr}_{V}\;\Bigg[\int\limits_{0}^{1}d\tau_{2}\int\limits_{0}^{\tau_{2}}d\tau_{1}e^{-(1-\tau_{2})\tilde{H}}Ke^{-(\tau_{2}-\tau_{1})\tilde{H}}Ke^{-\tau_{1}\tilde{H}}\Bigg]\right>\Bigg|_{x=x^{\prime}}=\nonumber\\
=\textrm{tr}_{V}\left(\Omega_{0}\right)+\lambda\textrm{tr}_{V}\left(\Omega_{1}\right)+\lambda^{2}\textrm{tr}_{V}\left(\Omega_{2}\right)+O(\lambda^{3})\;,
\end{eqnarray}
where
\begin{eqnarray}\label{673}
\Omega_{0}&=&\frac{1}{6}R\;,\\[3pt]
\label{6730}
\Omega_{1}&=&\frac{1}{6}h^{;\alpha}{}_{;\alpha}
-\frac{1}{6}h^{\mu\nu}{}_{;\;\mu\nu}-\frac{1}{12}hR+\frac{1}{6}h^{\mu\nu}R_{\mu\nu}\;,\\[3pt]
\label{6731}
\Omega_{2}&=&-\frac{1}{16}h_{;\;\mu}h^{;\;\mu}+\frac{1}{6}h^{\mu\nu}{}_{;\;\mu}h_{;\;\nu}-\frac{1}{8}h^{\mu\nu}{}_{;\;\rho}h_{\mu\nu}{}^{;\;\rho}-\frac{1}{12}h
h^{;\alpha}{}_{;\alpha}
+\frac{1}{12}hh^{\mu\nu}{}_{;\;\mu\nu}\nonumber\\
&+&\frac{1}{6}h^{\mu\nu}h_{;\;\mu\nu}
-\frac{1}{6}h^{\mu\nu}h_{\mu\nu}{}^{;\alpha}{}_{;\alpha}
+\frac{1}{12}h^{\mu\nu;\;\rho}h_{\mu\rho;\;\nu}-\frac{1}{6}h^{\mu}{}_{\rho}h^{\nu\rho}R_{\mu\nu}+\frac{1}{48}h^{2}R\nonumber\\
&-&\frac{1}{12}hh^{\mu\nu}R_{\mu\nu}+\frac{1}{24}h^{\mu\nu}h_{\mu\nu}R\;.
\end{eqnarray}

We can finally evaluate the last term in equation (\ref{662}). By
using the definitions (\ref{660}) and (\ref{661}) we can write that
\begin{eqnarray}\label{676}
\lefteqn{\left<\textrm{tr}_{V}\;\Bigg[\int\limits_{0}^{1}d\tau_{1}e^{-(1-\tau_{1})\tilde{H}}\bar{\mathcal{D}}_{\mu}a^{\mu\nu}\mathcal{D}_{\nu}e^{-\tau_{1}\tilde{H}}\Bigg]\right>\Bigg|_{x=x^{\prime}}}\nonumber\\
&&=\Bigg<\textrm{tr}_{V}\;\Bigg[\int\limits_{0}^{1}d\tau_{1}e^{-(1-\tau_{1})\tilde{H}}\Big(a^{\mu\nu}{}_{;\;\mu}\nabla_{\nu}+a^{\mu\nu}\nabla_{\mu}\nabla_{\nu}+a^{\mu\nu}{}_{;\;\mu}\mathcal{A}_{\nu}\nonumber\\
&&+a^{\mu\nu}\mathcal{A}_{\nu;\;\mu}+a^{\mu\nu}\mathcal{A}_{\nu}\nabla_{\mu}+\bar{\mathcal{A}}_{\mu}a^{\mu\nu}\nabla_{\nu}+\bar{\mathcal{A}}_{\mu}a^{\mu\nu}\mathcal{A}_{\nu}\Big)\Bigg]\Bigg>\Bigg|_{x=x^{\prime}}\;.
\end{eqnarray}

In order to evaluate this term we use the derivatives in
(\ref{667a}) and (\ref{669}) and we expand the remaining
exponentials of $\tilde{H}$ in $\tau$ up to terms in
$\lambda^{2}$. During the calculation the numerical coefficients
of the various terms can be evaluated by referring to the
following general integral
\begin{equation}\label{677}
I_{2}(\alpha,\beta)=\int\limits_{0}^{1}d\tau_{1}\tau_{1}^{\alpha}(1-\tau_{1})^{\beta}\;,
\end{equation}
where $(\alpha,\beta)$ are positive integers.
The solution to this integral is easily find by recalling the
integral representation of the hypergeometric function
\cite{abramowitz72}
\begin{equation}\label{6A18}
{}_{2}F_{1}\left(a,b,c\;;z\right)=\frac{\Gamma(c)}{\Gamma(b)
\Gamma(c-b)}\int\limits_{0}^{1}dt\;t^{b-1}(1-t)^{c-b-1}(1-tz)^{-a}\;,
\end{equation}
where $\textrm{Re}(c)>0$ and $\textrm{Re}(b)>0$. From this last
general expression we obtain the integral (\ref{677}) by setting
$z=0$, $\alpha=b-1$ and $\beta=c-b-1$. By recalling that
${}_{2}F_{1}\left(a,b,c\;;0\right)=1$, we finally get
\begin{equation}\label{6A19}
I_{2}(\alpha,\beta)=\frac{\Gamma(1+\alpha)\Gamma(1+\beta)}{\Gamma(2
+\alpha+\beta)}=B(1+\alpha,1+\beta)\;,
\end{equation}
where $B(a,b)$ denotes the Euler beta function.


The explicit form of (\ref{676}) can be obtained with the help of
(\ref{6A19}) and the coincidence limits in Section 6.3.2. After a
straightforward calculation one gets
\begin{eqnarray}\label{679}
\lefteqn{-\left<\textrm{tr}_{V}\;\Bigg[\int\limits_{0}^{1}d\tau_{1}e^{-(1-\tau_{1})\tilde{H}}\bar{\mathcal{D}}_{\mu}a^{\mu\nu}\mathcal{D}_{\nu}e^{-\tau_{1}\tilde{H}}\Bigg]\right>\Bigg|_{x=x^{\prime}}=}\nonumber\\
&&=\textrm{tr}_{V}\left(\Xi_{0}\right)+\lambda\textrm{tr}_{V}\left(\Xi_{1}\right)+\lambda^{2}\textrm{tr}_{V}\left(\Xi_{2}\right)+O(\lambda^{3})\;,
\end{eqnarray}
where
\begin{eqnarray}\label{680}
\Xi_{0}&=&-\phi_{;\;\mu}{}^{;\;\mu}-\phi_{;\;\mu}\phi^{;\;\mu}\;,\\[3pt]
\label{6800}
\Xi_{1}&=&-\sigma_{;\;\mu}{}^{;\;\mu}-h^{\mu\nu}{}_{;\;\mu}\phi_{;\;\nu}-h^{\mu\nu}\phi_{;\;\mu\nu}-2\phi^{;\;\nu}\sigma_{;\;\nu}-\phi_{;\;\mu}h^{\mu\nu}\phi_{;\;\nu}\nonumber\\
&+&\frac{1}{2}h(\phi_{;\;\mu}{}^{;\;\mu}+\phi_{;\;\mu}\phi^{;\;\mu})
-\frac{1}{4} h^{;\alpha}{}_{;\alpha}\;,\\[3pt]
\label{6801}
\Xi_{2}&=& \frac{1}{8}h h^{;\alpha}{}_{;\alpha}
+\frac{1}{4}h^{\mu\nu}h_{\mu\nu}^{;\alpha}{}_{;\alpha}
+\frac{1}{12}h_{;\;\mu}h^{;\;\mu}+\frac{1}{6}h_{\mu\nu;\;\rho}h^{\mu\nu;\;\rho}-\frac{1}{4}h^{\mu\nu}h_{;\;\mu\nu}\nonumber\\
&-&\frac{1}{4}h^{\mu\nu}{}_{;\;\mu}h_{;\;\nu}+\frac{1}{2}hh^{\mu\nu}{}_{;\;\mu}\phi_{;\;\nu}-h^{\mu\nu}{}_{;\;\mu}\sigma_{;\;\nu}-\frac{1}{8}h^{2}\phi_{;\;\mu}{}^{;\;\mu}-\frac{1}{4}h_{\mu\nu}h^{\mu\nu}\phi_{;\;\rho}{}^{;\;\rho}\nonumber\\
&+&\frac{1}{2}h\sigma_{;\;\mu}{}^{;\;\mu}+\frac{1}{2}hh^{\mu\nu}\phi_{;\;\mu\nu}-h^{\mu\nu}\sigma_{;\;\mu\nu}-\frac{1}{8}h^{2}\phi_{;\;\mu}\phi^{;\;\mu}-\frac{1}{4}h^{\mu\nu}h_{\mu\nu}\phi_{;\;\rho}\phi^{;\;\rho}\nonumber\\
&+&h\sigma_{;\;\mu}\phi^{;\;\mu}+\frac{1}{2}hh^{\mu\nu}\phi_{;\;\mu}\phi_{;\;\nu}-\sigma_{;\;\mu}\sigma^{;\;\mu}-2h^{\mu\nu}\sigma_{;\;\mu}\phi_{;\;\nu}\;.
\end{eqnarray}

In the notation of equation (\ref{653}) we can write, now, the
different contributions, in increasing order of $\lambda$, to the
coefficient $\tilde{a}_{1}$. In more details, by using the results
(\ref{664}), (\ref{673}), (\ref{680}) and recalling that
\begin{displaymath}
\tilde{a}_{1}(\lambda)=a_{1}^{(0)}+\lambda
a_{1}^{(1)}+\lambda^{2}a_{1}^{(2)}+O(\lambda^{3})\;,
\end{displaymath}
we get
\begin{equation}\label{683}
a_{1}^{(0)}=\frac{1}{6}R-\phi_{;\;\mu}{}^{;\;\mu}-\phi_{;\;\mu}\phi^{;\;\mu}-q\;.
\end{equation}
Moreover, by using (\ref{6730}) and (\ref{6800}) we obtain
\begin{eqnarray}\label{684}
a_{1}^{(1)}&=&-\sigma_{;\;\mu}{}^{;\;\mu}-h^{\mu\nu}{}_{;\;\mu}\phi_{;\;\nu}-h^{\mu\nu}\phi_{;\;\mu\nu}-2\phi^{;\;\nu}\sigma_{;\;\nu}-\phi_{;\;\mu}h^{\mu\nu}\phi_{;\;\nu}-\frac{1}{6}h^{\mu\nu}{}_{;\;\mu\nu}\nonumber\\
&+&\frac{1}{2}h(\phi_{;\;\mu}{}^{;\;\mu}+\phi_{;\;\mu}\phi^{;\;\mu})
-\frac{1}{12} h^{;\alpha}{}_{;\alpha}
+\frac{1}{6}h^{\mu\nu}R_{\mu\nu}-\frac{1}{12}hR\;.
\end{eqnarray}
Finally, by combining the results in (\ref{664}), (\ref{6731}) and
(\ref{6801}), we have the following expression for the term of
order $\lambda^{2}$ in $\tilde{a}_{1}$, i.e.
\begin{eqnarray}\label{685}
a_{1}^{(2)}&=&-h^{\mu\nu}{}_{;\;\mu}\sigma_{;\;\nu}-h^{\mu\nu}\sigma_{;\;\mu\nu}-\sigma^{;\;\nu}\sigma_{;\;\nu}-\sigma_{;\;\mu}h^{\mu\nu}\phi_{;\;\nu}-\phi_{;\;\nu}h^{\mu\nu}\sigma_{;\;\mu}-\frac{1}{12}h^{\mu\nu}h_{;\;\mu\nu}\nonumber\\
&+&\frac{1}{2}\sigma_{;\;\mu}{}^{;\;\mu}h+\frac{1}{2}hh^{\mu\nu}{}_{;\;\mu}\phi_{;\;\nu}+h\phi_{;\;\mu}\sigma^{;\;\mu}+\frac{1}{2}hh^{\mu\nu}\phi_{;\;\mu\nu}+\frac{1}{2}hh^{\mu\nu}\phi_{;\;\mu}\phi_{;\;\nu}-\frac{1}{12}h^{\mu\nu}{}_{;\;\mu}h_{;\;\nu}\nonumber\\
&+&\frac{1}{12}h^{\mu\nu}{}_{;\;\mu\nu}h
+\frac{1}{12}h^{\mu\nu;\;\rho}h_{\mu\rho;\;\nu}
+\frac{1}{24}h h^{;\alpha}{}_{;\alpha}
+\frac{1}{12}h^{\mu\nu} h_{\mu\nu}{}^{;\alpha}{}_{;\alpha}
+\frac{1}{24}h^{\mu\nu;\;\rho}h_{\mu\nu;\;\rho}\nonumber\\
&-&\frac{1}{8}h^{2}q-\frac{1}{8}h^{2}\phi_{;\;\mu}{}^{;\;\mu}-\frac{1}{8}h^{2}\phi_{;\;\mu}\phi^{;\;\mu}-\frac{1}{4}h^{\mu\nu}h_{\mu\nu}q-\frac{1}{4}h^{\mu\nu}h_{\mu\nu}\phi_{;\;\rho}{}^{;\;\rho}\nonumber\\
&-&\frac{1}{4}h^{\mu\nu}h_{\mu\nu}\phi_{;\;\rho}\phi^{;\;\rho}+\frac{1}{48}h_{;\;\mu}h^{;\;\mu}-\frac{1}{12}hh^{\mu\nu}R_{\mu\nu}-\frac{1}{6}h^{\mu}{}_{\rho}h^{\nu\rho}R_{\mu\nu}+\frac{1}{48}h^{2}R\nonumber\\
&+&\frac{1}{24}h^{\mu\nu}h_{\mu\nu}R+\frac{1}{2}h\Theta\;.
\end{eqnarray}

\section{Construction of the Action}

In order to write the action of Spectral Matrix Gravity, we need
to evaluate the global heat kernel coefficients $A_{0}$ and
$A_{1}$. As we already mentioned above, the coefficients $A_{k}$
are expressed in terms of integrals of the local heat kernel
coefficients $a_{k}$ (which are densities) or the coefficients
$\tilde{a}_{k}$ (which are scalars). 
By using the equation (\ref{659}), we get
\begin{equation}\label{687}
A_{0}=\int\limits_{M}dx\;g^{\frac{1}{2}}\;\textrm{tr}_{V}\left[\mathbb{I}+\frac{\lambda^{2}}{8}(h^{2}+2h_{\mu\nu}h^{\mu\nu})\right]+O(\lambda^{3})\;.
\end{equation}

Now we use the equations (\ref{653}) and (\ref{683})-(\ref{685}) to
compute the coefficient $A_{1}$. By integrating by parts and by
noticing that the trace of a commutator of any two matrices
vanishes, up to terms of order $\lambda^{2}$, we obtain
\begin{eqnarray}\label{688}
A_{1}&=&\int\limits_{M}dx\;g^{\frac{1}{2}}\;\textrm{tr}_{V}\bigg\{-q-\phi^{;\;\mu}\phi_{;\;\mu}+\frac{1}{6}R+\lambda^{2}\bigg(-\sigma^{;\;\mu}\sigma_{;\;\mu}+\frac{1}{2}h\Theta\nonumber\\
&-&\frac{1}{2}h^{;\;\mu}\sigma_{;\;\mu}+\frac{1}{2}hh^{\mu\nu}\phi_{;\;\mu}\phi_{;\;\nu}+h\sigma^{;\;\nu}\phi_{;\;\nu}-\frac{1}{2}h_{;\;\mu}h^{\mu\nu}\phi_{;\;\nu}-2\sigma_{;\;\mu}h^{\mu\nu}\phi_{;\;\nu}\nonumber\\
&-&\frac{1}{12}hh^{\mu\nu}R_{\mu\nu}-\frac{1}{6}h^{\nu}{}_{\rho}h^{\rho\mu}R_{\mu\nu}-\frac{1}{12}h^{\mu\nu}{}_{;\;\mu}h_{;\;\nu}+\frac{1}{12}h^{\mu\nu;\;\rho}h_{\mu\rho;\;\nu}+\frac{1}{48}h^{2}R\nonumber\\
&+&\frac{1}{24}h^{\mu\nu}h_{\mu\nu}R-\frac{1}{48}h_{;\;\mu}h^{;\;\mu}-\frac{1}{24}h^{\mu\nu;\;\rho}h_{\mu\nu;\;\rho}-\frac{1}{8}h^{2}q-\frac{1}{8}h^{2}\phi_{;\;\mu}{}^{;\;\mu}\nonumber\\
&-&\frac{1}{8}h^{2}\phi_{;\;\mu}\phi^{;\;\mu}-\frac{1}{4}h^{\mu\nu}h_{\mu\nu}q-\frac{1}{4}h^{\mu\nu}h_{\mu\nu}\phi_{;\;\rho}{}^{;\;\rho}-\frac{1}{4}h^{\mu\nu}h_{\mu\nu}\phi_{;\;\rho}\phi^{;\;\rho}\bigg)\bigg\}\nonumber\\
&+&O(\lambda^{3})\;.
\end{eqnarray}

The invariant action functional is written as linear combination
of the coefficients $A_{0}$ and $A_{1}$ as shown in (\ref{62})
\begin{eqnarray}\label{689}
S&=&\frac{1}{16\pi
G}\int\limits_{M}dx\;g^{\frac{1}{2}}\bigg\{-6q-6\phi^{;\;\mu}\phi_{;\;\mu}+R-2\Lambda\nonumber\\
&+&\frac{\lambda^{2}}{N}\textrm{tr}_V
\bigg(-6\sigma^{;\;\mu}\sigma_{;\;\mu}-3h^{;\;\mu}\sigma_{;\;\mu}+3h\Theta+3hh^{\mu\nu}\phi_{;\;\mu}\phi_{;\;\nu}+6h\sigma^{;\;\nu}\phi_{;\;\nu}\nonumber\\
&-&3h^{\mu\nu}h_{;\;\mu}\phi_{;\;\nu}-12\sigma_{;\;\nu}h^{\mu\nu}\phi_{;\;\mu}-\frac{1}{2}hh^{\mu\nu}R_{\mu\nu}+\frac{1}{2}h^{\mu\nu}h^{\rho\sigma}R_{\sigma\mu\rho\nu}-\frac{1}{2}h^{\nu}{}_{\rho}h^{\rho\mu}R_{\mu\nu}\nonumber\\[3pt]
&-&\frac{1}{2}h^{\mu\nu}{}_{;\;\mu}h_{;\;\nu}+\frac{1}{2}h^{\mu\nu}{}_{;\;\nu}h_{\mu\rho}{}^{;\;\rho}+\frac{1}{8}h^{2}R+\frac{1}{4}h^{\mu\nu}h_{\mu\nu}R-\frac{1}{8}h_{;\;\mu}h^{;\;\mu}-\frac{1}{4}h^{\mu\nu;\;\rho}h_{\mu\nu;\;\rho}\nonumber\\[3pt]
&-&\frac{3}{4}h^{2}q-\frac{3}{4}h^{2}\phi_{;\;\mu}{}^{;\;\mu}-\frac{3}{4}h^{2}\phi_{;\;\mu}\phi^{;\;\mu}-\frac{3}{2}h^{\mu\nu}h_{\mu\nu}q-\frac{3}{2}h^{\mu\nu}h_{\mu\nu}\phi_{;\;\rho}{}^{;\;\rho}\nonumber\\[3pt]
&-&\frac{3}{2}h^{\mu\nu}h_{\mu\nu}\phi_{;\;\rho}\phi^{;\;\rho}-\frac{\Lambda}{4}h^{2}-\frac{\Lambda}{2}h_{\mu\nu}h^{\mu\nu}\bigg)\bigg\}\nonumber\\
&+&O(\lambda^{3})\;.
\end{eqnarray}
Obviously, the action functional that we
obtained is invariant under the diffeomorphisms
and the gauge transformation $h^{\mu\nu}\rightarrow
Uh^{\mu\nu}U^{-1}$.

The next task is to find the equations of motion for the fields
$\sigma$, $h_{\mu\nu}$ and $\phi$ by varying the action
functional. In this way we will explicitly find the non-commutative
corrections to Einstein's equations.

\section{The Equations of Motion}

By performing the variation with respect to the field $\sigma$, we
obtain the equation
\begin{equation}\label{690}
4\Delta\sigma+\Delta
h-2h\Delta\phi-2h^{;\;\nu}\phi_{;\;\nu}+4h^{\mu\nu}{}_{;\;\nu}\phi_{;\;\mu}+4h^{\mu\nu}\phi_{;\;\mu\nu}+O(\lambda^{3})=0\;.
\end{equation}
Here $\Delta$ is the Laplacian in the Euclidean case and the D'Alambertian in the
pseudo-Euclidean case.
For the matrix-valued field $h^{\mu\nu}$ we obtain the
equation
\begin{eqnarray}\label{692}
\lefteqn{g^{\mu\nu}\Delta\sigma+g^{\mu\nu}\Theta+h\phi^{;\;(\mu}\phi^{;\;\nu)}+g^{\mu\nu}h^{\rho\sigma}\phi_{;\;\rho}\phi_{;\;\sigma}+2g^{\mu\nu}\sigma^{;\;\rho}\phi_{;\;\rho}}\\
&-&h^{;(\mu}\phi^{;\;\nu)}-4\sigma^{;(\mu}\phi^{;\;\nu)}+g^{\mu\nu}h^{\rho\sigma}{}_{;\;\rho}\phi_{;\;\sigma}+g^{\mu\nu}h^{\rho\sigma}\phi_{;\;\rho\sigma}-\frac{1}{6}g^{\mu\nu}h^{\rho\sigma}R_{\rho\sigma}\nonumber\\
&-&\frac{1}{6}hR^{\mu\nu}+\frac{1}{3}h_{\rho\sigma}R^{\sigma\mu\rho\nu}+\frac{1}{3}h^{\rho(\mu}R^{\nu)}{}_{\rho}+\frac{1}{6}h^{;\;(\mu\nu)}+\frac{1}{6}g^{\mu\nu}h^{\rho\sigma}{}_{;\;\rho\sigma}-\frac{1}{3}h^{(\mu}{}_{\rho}{}^{;\;|\rho|\nu)}\nonumber\\
&+&\frac{1}{12}g^{\mu\nu}hR+\frac{1}{6}h^{\mu\nu}R+\frac{1}{12}g^{\mu\nu}\Delta
h+\frac{1}{6}\Delta h^{\mu\nu}-\frac{1}{2}g^{\mu\nu}hq-\frac{1}{2}g^{\mu\nu}h\Delta\phi\nonumber\\
&-&\frac{1}{2}g^{\mu\nu}h\phi_{;\;\rho}\phi^{;\;\rho}-h^{\mu\nu}q-h^{\mu\nu}\Delta\phi-h^{\mu\nu}\phi_{;\;\rho}\phi^{;\;\rho}-\frac{\Lambda}{6}g^{\mu\nu}h-\frac{\Lambda}{3}h^{\mu\nu}+O(\lambda^{3})=0\;.\nonumber
\end{eqnarray}
The variation of the action with respect to the scalar field
$\phi$ yields
\begin{eqnarray}\label{691}
4\Delta\phi&=&-\frac{\lambda^{2}}{N}\textrm{tr}_{V}\bigg(-2hh^{\mu\nu}\phi_{;\;\mu\nu}-2h_{;\;\mu}h^{\mu\nu}\phi_{;\;\nu}-2hh^{\mu\nu}{}_{;\;\mu}\phi_{;\;\nu}\nonumber\\
&-&2h_{;\;\nu}\sigma^{;\;\nu}-2h\Delta\sigma+h^{\mu\nu}{}_{;\;\nu}h_{;\;\mu}+h^{\mu\nu}h_{;\;\mu\nu}+4\sigma_{;\;\mu\nu}h^{\mu\nu}+4\sigma_{;\;\nu}h^{\mu\nu}{}_{;\;\mu}\nonumber\\
&-&\frac{1}{2}h\Delta h-\frac{1}{2}h_{;\;\mu}h^{;\;\mu}+hh_{;\;\mu}\phi^{;\;\mu}+\frac{1}{2}h^{2}\Delta\phi-h^{\mu\nu}\Delta h_{\mu\nu}-h_{\mu\nu;\;\rho}h^{\mu\nu;\;\rho}\nonumber\\
&+&2h_{\mu\nu}h^{\mu\nu;\;\rho}\phi_{;\;\rho}+h^{\mu\nu}h_{\mu\nu}\Delta\phi\bigg)+O(\lambda^{3})\;.
\end{eqnarray}
The equation of motion for the field $g^{\mu\nu}$ can be written
in the following form
\begin{equation}\label{693}
R^{\mu\nu}-\frac{1}{2}g^{\mu\nu}R+\Lambda
g^{\mu\nu}=T^{\mu\nu}+\frac{\lambda^{2}}{N}\textrm{tr}_{V}\mathscr{A}^{\mu\nu}\;.
\end{equation}
Here the tensor $T^{\mu\nu}$ is
\begin{equation}\label{694}
T^{\mu\nu}=6\phi^{;\;(\mu}\phi^{;\;\nu)}-3g^{\mu\nu}\phi_{;\;\rho}\phi^{;\;\rho}-3qg^{\mu\nu}\;,
\end{equation}
which represents the stress-energy tensor for a massless scalar
field. The tensor $\mathscr{A}^{\mu\nu}$ represents, instead, the
stress-energy tensor for the fields $h^{\mu\nu}$ and $\sigma$.

The equation (\ref{693}) is the main result of this chapter. As we can
see, the new fields of our model, $h^{\mu\nu}$ and $\sigma$,
contribute to modify the standard Einstein equations. More
precisely they contribute to an additional term in the
stress-energy tensor.

The tensor $\mathscr{A}^{\mu\nu}$ can be written as the sum of six
terms:
\begin{equation}
\mathscr{A}^{\mu\nu}=\mathscr{A}^{\mu\nu}_{(1)}+\mathscr{A}^{\mu\nu}_{(2)}+\mathscr{A}^{\mu\nu}_{(3)}+\mathscr{A}^{\mu\nu}_{(4)}+\mathscr{A}^{\mu\nu}_{(5)}+\mathscr{A}^{\mu\nu}_{(6)}\;.
\end{equation}
In the first term we have only derivatives of the field $\sigma$
\begin{eqnarray}\label{695}
\mathscr{A}^{\mu\nu}_{(1)}&=&6\sigma^{;\;(\mu}\sigma^{;\;\nu)}-3g^{\mu\nu}\sigma_{;\;\rho}\sigma^{;\;\rho}-6h\sigma^{;\;(\mu}\phi^{;\;\nu)}+3\sigma^{;\;(\mu}h^{;\;\nu)}+6h^{\mu\nu}\sigma^{;\;\rho}\phi_{;\;\rho}\nonumber\\
&+&3\sigma_{;\;\rho}{}^{;\;\rho}h^{\mu\nu}+3g^{\mu\nu}\left(h\sigma^{;\;\rho}\phi_{;\;\rho}-\frac{1}{2}\sigma^{;\;\rho}h_{;\;\rho}-2\sigma_{;\;\rho}\phi_{;\;\sigma}h^{\rho\sigma}\right)\;.
\end{eqnarray}
The second term only contains derivatives of the scalar field
$\phi$, namely
\begin{eqnarray}\label{696}
\mathscr{A}^{\mu\nu}_{(2)}&=&3h^{\rho\tau}\phi_{;\;\rho}\phi_{;\;\tau}\left(h^{\mu\nu}+\frac{1}{2}g^{\mu\nu}h\right)+3\phi_{;\;\tau}\left(h^{\rho\tau}{}_{;\;\rho}h^{\mu\nu}-\frac{1}{2}g^{\mu\nu}h^{\rho\tau}h_{;\;\rho}\right)\nonumber\\
&-&\frac{3}{2}(\phi_{;\;\rho}\phi^{;\;\rho}+\phi_{;\;\rho}{}^{;\;\rho})\left(2h^{(\mu}{}_{\sigma}h^{\nu)\sigma}+hh^{\mu\nu}+\frac{1}{4}h^{2}g^{\mu\nu}+\frac{1}{2}g^{\mu\nu}h^{\alpha\beta}h_{\alpha\beta}\right)\nonumber\\
&+&3h^{\rho\sigma}h^{\mu\nu}\phi_{;\;\rho\sigma}+\frac{3}{2}(\phi^{;\;(\mu\nu)}+\phi^{;\;(\mu}\phi^{;\;\nu)})\left(h_{\rho\sigma}h^{\rho\sigma}+\frac{1}{2}h^{2}\right)\;.
\end{eqnarray}
The third term only contains second derivatives of the
matrix-valued tensor field $h^{\mu\nu}$,
\begin{eqnarray}\label{696a}
\mathscr{A}^{\mu\nu}_{(3)}&=&h^{\alpha(\mu}{}_{;\;(\alpha\sigma)}h^{\nu)\sigma}+\frac{1}{2}h^{\alpha\sigma}{}_{;\;\alpha\sigma}h^{\mu\nu}-\frac{1}{2}hh^{\alpha(\mu;\;\nu)}{}_{;\;\alpha}+\frac{1}{4}g^{\mu\nu}hh^{\rho\sigma}{}_{;\;\rho\sigma}\nonumber\\
&+&\frac{1}{2}h^{\sigma(\mu}h^{\nu)\rho}{}_{;\;\rho\sigma}-\frac{1}{2}h^{\rho\sigma}h^{\mu\nu}{}_{;\;\rho\sigma}-h_{\rho\sigma}{}^{;\;(\nu|\rho|}h^{\mu)\sigma}+h^{(\nu}{}_{\rho}h^{\mu)\rho;\;\sigma}{}_{;\;\sigma}\nonumber\\
&+&\frac{1}{4}hh^{\mu\nu;\;\sigma}{}_{;\;\sigma}-\frac{1}{4}g^{\mu\nu}h^{\sigma}{}_{\rho}h^{\rho\alpha}{}_{;\;[\alpha\sigma]}+\frac{1}{2}h^{\rho\sigma}h_{\rho\sigma}{}^{;\;(\mu\nu)}-\frac{1}{2}g^{\mu\nu}h^{\rho\sigma}h_{\rho\sigma;\;\alpha}{}^{;\;\alpha}\;.\nonumber\\
\phantom{2}
\end{eqnarray}
The fourth term contains only find first derivatives of
$h^{\mu\nu}$, namely
\begin{eqnarray}\label{696b}
\mathscr{A}^{\mu\nu}_{(4)}&=&-\frac{1}{2}h^{\rho\sigma}{}_{;\;\rho}h^{\mu\nu}{}_{;\;\sigma}+\frac{1}{2}h^{;\;\rho}h^{\mu\nu}{}_{;\;\rho}-\frac{1}{2}h^{\rho(\mu}{}_{;\;(\rho}h^{\nu)\lambda}{}_{;\;\lambda)}+\frac{1}{2}g^{\mu\nu}h_{;\;\sigma}h^{\rho\sigma}{}_{;\;\rho}\nonumber\\
&+&\frac{1}{2}h^{\sigma(\mu}{}_{;\;\rho}h^{\nu)\rho}{}_{;\;\sigma}+\frac{1}{4}h^{\rho\sigma;\;(\mu}h_{\rho\sigma}{}^{;\;\nu)}-h_{\sigma\rho}{}^{;\;(\nu}h^{\mu)\sigma;\;\rho}+h^{\rho(\nu;\;|\tau|}h_{\rho}{}^{\mu)}{}_{;\;\tau}\nonumber\\
&+&\frac{1}{2}h^{;\;(\nu}h^{\mu)\rho}{}_{;\;\rho}+\frac{1}{4}g^{\mu\nu}h_{\rho\sigma;\;\alpha}h^{\rho\alpha;\;\sigma}-\frac{1}{2}h_{;\;\rho}h^{\rho(\mu;\;\nu)}-\frac{5}{8}g^{\mu\nu}h^{\rho\sigma;\;\alpha}h_{\rho\sigma;\;\alpha}\;.\nonumber\\
\phantom{}
\end{eqnarray}
The fifth coefficient contains only first and second derivatives
of $h$
\begin{eqnarray}\label{697}
\mathscr{A}^{\mu\nu}_{(5)}&=&\frac{1}{4}h^{;\;\tau}{}_{;\;\tau}h^{\mu\nu}+\frac{1}{4}h^{;\;(\mu\nu)}h+\frac{3}{8}h^{;\;(\mu}h^{;\;\nu)}\nonumber\\
&-&\frac{1}{2}g^{\mu\nu}\left(\frac{1}{2}hh_{;\;\rho}{}^{;\;\rho}-h_{;\;\rho\sigma}h^{\rho\sigma}+\frac{5}{8}h^{;\;\rho}h_{;\;\rho}\right)\;.
\end{eqnarray}
The last term, $\mathcal{A}^{\mu\nu}_{(6)}$, does not contain any
derivative of $h^{\mu\nu}$, namely
\begin{eqnarray}\label{698}
\mathscr{A}^{\mu\nu}_{(6)}&=&3\Theta\left(h^{\mu\nu}+\frac{1}{2}g^{\mu\nu}h\right)-\frac{\Lambda}{2}\left[\left(h^{\mu\nu}+\frac{1}{4}g^{\mu\nu}h\right)h+2h^{(\nu}{}_{\rho}h^{\mu)\rho}+\frac{1}{2}g^{\mu\nu}h^{\rho\sigma}h_{\rho\sigma}\right]\nonumber\\
&-&\frac{1}{2}R_{\alpha\beta}\Bigg[h^{\alpha(\nu}h^{\mu)\beta}+h^{\mu\nu}h^{\alpha\beta}+g^{\mu\nu}\left(hh^{\alpha\beta}+h^{\alpha}{}_{\rho}h^{\beta\rho}\right)+h^{2}g^{\alpha(\mu}g^{\nu)\beta}\nonumber\\
&+&h^{\rho\sigma}h_{\rho\sigma}g^{\alpha(\mu}g^{\nu)\beta}\Bigg]+\frac{1}{2}h^{\alpha\sigma}h^{\rho(\mu}R^{\nu)}{}_{\alpha\rho\sigma}+\frac{1}{4}g^{\mu\nu}h^{\rho\sigma}h^{\alpha\beta}R_{\beta\rho\alpha\sigma}\nonumber\\
&+&\frac{1}{8}(R-6q)\left(g^{\mu\nu}h_{\rho\sigma}h^{\rho\sigma}+4h_{\rho}{}^{(\mu}h^{\nu)\rho}\right)+\frac{1}{16}(R-6q)\left(g^{\mu\nu}h^{2}+4hh^{\mu\nu}\right)\;.\nonumber\\
\phantom{}
\end{eqnarray}

The dynamics described by the equations (\ref{690}), (\ref{692}),
(\ref{691}) and (\ref{693}) can be studied by using an iterative
method. Let us write the solution for the background fields $\phi$
and $g^{\mu\nu}$ as Taylor expansion in the deformation parameter
$\lambda$ as follows
\begin{eqnarray}\label{6willy}
\phi&=&\phi_{0}+\lambda\phi_{1}+\lambda^{2}\phi_{2}+O(\lambda^{3})\;,\nonumber\\
g^{\mu\nu}&=&g^{\mu\nu}_{0}+\lambda
g^{\mu\nu}_{1}+\lambda^{2}g^{\mu\nu}_{2}+O(\lambda^{3})\;.
\end{eqnarray}
By substituting these expressions in equations (\ref{691}) and
(\ref{693}) we obtain, for the terms of order $\lambda^{0}$, the
dynamical equations
\begin{eqnarray}\label{6willy1}
\Delta\phi&=&0\;,\nonumber\\
R^{\mu\nu}-\frac{1}{2}g^{\mu\nu}R+\Lambda g^{\mu\nu}&=&0\;.
\end{eqnarray}
As we can see from the last equations the term $g^{\mu\nu}_{0}$ is
nothing but the solution of the ordinary Einstein equation in
vacuum with cosmological constant. By substituting the solutions
to (\ref{6willy1}) back into the equations of motion for the fields
$\sigma$ and $h^{\mu\nu}$ we get equations of the form
\begin{eqnarray}\label{6willy2}
\Phi_{1}(g^{\mu\nu}_{0},\phi_{0})\sigma&=&O(\lambda^{2})\;,\nonumber\\
\Phi_{2}(g^{\mu\nu}_{0},\phi_{0})h^{\mu\nu}&=&O(\lambda^{2})\;,
\end{eqnarray}
where $\Phi_{1}(g^{\mu\nu}_{0},\phi_{0})$ and
$\Phi_{2}(g^{\mu\nu}_{0},\phi_{0})$ are linear second order
partial differential operators.
By iterating this process we can, in principle, find the solution
to our dynamical equations in form of a Taylor series in
$\lambda$.

\section{Spectrum of Matrix Gravity on De Sitter Space}

The action for Matrix Gravity obtained in the previous section is
a functional of the fields $\phi$, $\sigma$, $h^{\mu\nu}$ and
$g^{\mu\nu}$. The dynamics is described by a system of non-linear
partial differential equations coupled with each other. We
analyze, now, the dynamics of the theory. For simplicity we will
set, from now on, $Q=0$. This particular value for the
matrix-valued scalar $Q$ will not affect our analysis.

As already mentioned above, from the equation of motion (\ref{691})
for the field $\phi$, we can see that to the zeroth order in the
deformation parameter $\lambda$ the field $\phi$ satisfies the
following equation
\begin{equation}\label{699}
\Delta\phi=0\;.
\end{equation}
As it is well known, the solution of the last equation represents
a wave propagating in the whole space. Since we require that
$\phi$ vanishes at infinity, the only solution is
$\phi=O(\lambda^{2})$ in the whole space.

With this solution for the field $\phi$, the matrix-valued
function $\rho$ defined in (\ref{639}) becomes
\begin{equation}\label{6100}
\rho=e^{\lambda\sigma}\;.
\end{equation}
A deeper analysis shows that the matrix-valued scalar field
$\sigma$ is not an independent field. Following
\cite{avramidi04,avramidi04b} the general form of $\rho$ can be
written as
\begin{equation}\label{6101}
\rho=\omega^{-\frac{1}{4}}\;,
\end{equation}
where
\begin{equation}\label{6102}
\omega=-\frac{1}{m!}\varepsilon_{\mu_{1}\ldots\mu_{m}}\varepsilon_{\nu_{1}\ldots\nu_{m}}a^{\mu_{1}\nu_{1}}\cdots
a^{\mu_{m}\nu_{m}}\;.
\end{equation}
By using the decomposition (\ref{638}) of $h^{\mu\nu}$ in equation
(\ref{6102}) we get the following formula, up to the term linear in
the deformation parameter $\lambda$,
\begin{equation}\label{6103}
\sigma=-\frac{1}{4}h+\frac{\lambda}{8}h_{\mu\nu}h^{\mu\nu}+O(\lambda^{2})\;.
\end{equation}
We can write down, now, the action by imposing the constraints
$\phi=O(\lambda^{2})$ and (\ref{6103}). The final result is the
following
\begin{eqnarray}\label{6104}
S&=&\frac{1}{16\pi
G}\int\limits_{M}dx\;g^{\frac{1}{2}}\bigg\{-6\phi^{;\;\mu}\phi_{;\;\mu}+R-2\Lambda+\frac{\lambda^{2}}{N}\textrm{tr}_{V}\bigg[\frac{1}{4}g^{\mu\nu}h_{;\;\mu}h_{;\;\nu}\nonumber\\
&-&\frac{1}{2}hh^{\mu\nu}R_{\mu\nu}+\frac{1}{2}h^{\mu\nu}h^{\rho\sigma}R_{\sigma\mu\rho\nu}-\frac{1}{2}h^{\nu}{}_{\rho}h^{\rho\mu}R_{\mu\nu}-\frac{1}{2}h^{\mu\nu}{}_{;\;\mu}h_{;\;\nu}\nonumber\\
&+&\frac{1}{2}h^{\mu\nu}{}_{;\;\nu}h_{\mu\rho}{}^{;\;\rho}+\frac{1}{8}h^{2}R+\frac{1}{4}h^{\mu\nu}h_{\mu\nu}R-\frac{1}{4}h^{\mu\nu;\;\rho}h_{\mu\nu;\;\rho}\nonumber\\
&-&\frac{\Lambda}{4}h^{2}-\frac{\Lambda}{2}h_{\mu\nu}h^{\mu\nu}\bigg)\bigg\}+O(\lambda^{3})\;.
\end{eqnarray}

The action depends, now, only on the independent tensor fields
$g^{\mu\nu}$, $h^{\mu\nu}$ and the scalar field $\phi$. Therefore,
we will have only two equations that describe the dynamics of the
theory. These dynamical equations can be easily derived from the
ones given in the previous section by imposing the conditions
(\ref{6103}) and $\phi=O(\lambda^{2})$.

The action (\ref{6104}) and the equations of motions for the fields
evaluated in the previous section, assume a simple form on
maximally symmetric background geometries. As we mentioned in the
previous section, the $\lambda^{0}$ term of the background field
$g^{\mu\nu}_{0}$ is solution of the Einstein equations in vacuum
with cosmological constant (\ref{6willy1}). In this section we
consider the De Sitter solution to the equation (\ref{6willy1}). In
this maximally symmetric case the Ricci and Riemann tensors take
the following form
\begin{equation}\label{6110}
R^{\mu}{}_{\nu\alpha\beta}=\frac{1}{n(n-1)}(\delta^{\mu}{}_{\alpha}g_{\nu\beta}-\delta^{\mu}{}_{\beta}g_{\nu\alpha})R\qquad\textrm{and}\qquad
R_{\mu\nu}=\frac{1}{n}g_{\mu\nu}R\;.
\end{equation}
The De Sitter metric gives a solution of the classical equations
provided
\begin{equation}\label{6110a}
R=\frac{2n}{n-2}\Lambda\;.
\end{equation}

Here, and below, we restrict ourselves to the case $n>2$.
By substituting the expressions in equation (\ref{6110}) in the
action (\ref{6104}), we find a form of the action functional valid
in De Sitter geometry, namely
\begin{eqnarray}\label{6111}
&\phantom{2}&S=\frac{1}{16\pi
G}\int\limits_{M}dx\;g^{\frac{1}{2}}\bigg\{-6\phi^{;\;\mu}\phi_{;\;\mu}+R-2\Lambda\nonumber\\
&+&\frac{\lambda^{2}}{N}\textrm{tr}_{V}\bigg[\frac{1}{4}h(-\Delta+\mu_{1})h
-\frac{1}{4}h_{\mu\nu}(-\Delta+\mu_{2})h^{\mu\nu}+\frac{1}{2}h^{\mu\nu}{}_{;\;\nu}h_{\rho\mu}{}^{;\;\rho}-\frac{1}{2}h^{\mu\nu}{}_{;\;\mu}h_{;\;\nu}\bigg]\bigg\}\nonumber\\
&+&O(\lambda^{3})\;,
\end{eqnarray}
where the terms $\mu_{1}$ and $\mu_{2}$ are defined as follows
\begin{eqnarray}\label{6112}
\mu_{1}&=&\frac{n^{2}-5n+8}{2n(n-1)}R-\Lambda\;,
\\[10pt]
\mu_{2}&=&-\frac{n-3}{n-1}R+2\Lambda\;.
\end{eqnarray}

It is interesting, at this point of the discussion, to derive
explicitly the spectrum of the theory. In order to achieve this
result we need to decompose the field $h_{\mu\nu}$ in its
irreducible modes: traceless transverse tensor mode, transverse
vector mode, scalar mode and trace part. In other words we can
write $h_{\mu\nu}$ as
\begin{equation}\label{6112b}
h_{\mu\nu}=\bar{h}_{\mu\nu}^{\bot}+\frac{1}{n}g_{\mu\nu}\varphi+2\zeta_{(\mu;\;\nu)}^{\bot}+\psi_{;\;\mu\nu}\;,
\end{equation}
where the scalar field $\varphi$ is defined as follows
\begin{displaymath}
\varphi=h-\Delta\psi\;,
\end{displaymath}
and the fields $\bar{h}_{\mu\nu}^{\bot}$ and $\zeta_{\mu}^{\bot}$
satisfy the conditions
\begin{equation}\label{6112a}
\nabla^{\mu}\bar{h}_{\mu\nu}^{\bot}=0\;,\qquad
g^{\mu\nu}\bar{h}_{\mu\nu}^{\bot}=0\;,\qquad\nabla^{\mu}\zeta_{\mu}^{\bot}=0\;.
\end{equation}
We can now substitute the expression (\ref{6112}) in the action
(\ref{6111}), and evaluate the terms separately. Explicitly we
obtain
\begin{eqnarray}\label{6113}
\int\limits_{M}dx\;g^{\frac{1}{2}}h^{\mu\nu}{}_{;\;\nu}h_{\rho\mu}{}^{;\;\rho}&=&\int\limits_{M}dx\;g^{\frac{1}{2}}\Bigg[-\frac{1}{n^{2}}\varphi\Delta\varphi-\frac{2}{n}\varphi\left(\Delta+\frac{R}{n}\right)\Delta\psi\nonumber\\
&+&\zeta^{\bot}_{\mu}\left(\Delta+\frac{R}{n}\right)^{2}\zeta^{\bot\mu}-\psi\left(\Delta+\frac{R}{n}\right)^{2}\Delta\psi\Bigg]\;,
\end{eqnarray}
\begin{eqnarray}\label{6114}
\int\limits_{M}dx\;g^{\frac{1}{2}}h^{\mu\nu}{}_{;\;\mu}h_{;\;\nu}&=&\int\limits_{M}dx\;g^{\frac{1}{2}}\Bigg[-\frac{1}{n}\varphi\Delta\varphi-\left(\frac{n+1}{n}\right)\varphi\left(\Delta+\frac{R}{n+1}\right)\Delta\psi\nonumber\\
&-&\psi\left(\Delta+\frac{R}{n}\right)\Delta^{2}\psi\Bigg]\;,
\end{eqnarray}
\begin{eqnarray}\label{6115}
\lefteqn{\int\limits_{M}dx\;g^{\frac{1}{2}}\;h_{\mu\nu}(-\Delta+\mu_{2})h^{\mu\nu}}\nonumber\\
&=&\int\limits_{M}dx\;g^{\frac{1}{2}}\Bigg\{\bar{h}_{\mu\nu}^{\bot}(-\Delta+\mu_{2})\bar{h}^{\bot\mu\nu}+\frac{1}{n}\varphi(-\Delta+\mu_{2})\varphi\nonumber\\
&+&\frac{2}{n}\varphi(-\Delta+\mu_{2})\Delta\psi+2\zeta^{\bot}_{\mu}\left(\Delta+\frac{R}{n}\right)\left(\Delta-\mu_{2}+\frac{n+1}{n(n-1)}R\right)\zeta^{\bot\mu}\nonumber\\
&-&\psi\left[\Delta^{2}+\left(\frac{3R}{n}-\mu_{2}\right)\Delta+\frac{R}{n}\left(\frac{2R}{n-1}-\mu_{2}\right)\right]\Delta\psi\Bigg\}\;,
\end{eqnarray}
and finally
\begin{eqnarray}\label{6116}
&\phantom{2}&\int\limits_{M}dx\;g^{\frac{1}{2}}\;h(-\Delta+\mu_{1})h\nonumber\\
&=&\int\limits_{M}dx\;g^{\frac{1}{2}}\Big[\varphi(-\Delta+\mu_{1})\varphi+2\varphi(-\Delta+\mu_{1})\Delta\psi+\psi(-\Delta+\mu_{1})\Delta^{2}\psi\Big]\;.
\end{eqnarray}

By using the decompositions (\ref{6113}) through (\ref{6116}) we
rewrite the action (\ref{6111}) in terms of the irreducible modes
of $h^{\mu\nu}$, namely
\begin{eqnarray}\label{6117}
S&=&\frac{1}{16\pi
G}\int\limits_{M}dx\;g^{\frac{1}{2}}\Bigg\{-6\phi^{;\;\mu}\phi_{;\;\mu}
+R-2\Lambda+\frac{\lambda^{2}}{N}\textrm{tr}_{V}\Bigg[
-\frac{1}{4}\bar{h}_{\mu\nu}^{\bot}(-\Delta+\mu_{2})\bar{h}^{\bot\mu\nu}
\nonumber\\
&+&\frac{(n-1)(n-2)}{4n^{2}}\;\varphi\left(-\Delta+\frac{n}{2(n-1)}R
-\frac{n(n+2)}{(n-1)(n-2)}\Lambda\right)\varphi\nonumber
\\[10pt]
&-&\frac{1}{2}\zeta^{\bot}_{\mu}\left(\Delta+\frac{R}{n}\right)
\left(\frac{n-2}{n}R-2\Lambda\right)\zeta^{\bot\mu}+\frac{n+2}{4n}
\varphi\left(\frac{n-2}{n}R-2\Lambda\right)\Delta\psi\nonumber
\\[10pt]
&+&\frac{3}{8}\psi\left(\Delta+\frac{2R}{3n}\right)\left(\frac{n-2}{n}R
-2\Lambda\right)\Delta\psi\Bigg]\Bigg\}+O(\lambda^{3})\;.
\end{eqnarray}

It is straightforward to show now that on the mass shell,
(\ref{6110a}), the terms containing the fields $\zeta^{\bot}_{\mu}$
and $\psi$ vanish identically.
More precisely we obtain the following form for the on-shell
action functional
\begin{eqnarray}\label{6118}
S\Big|_{{\rm on-shell}}&=&\frac{1}{16\pi
G}\int\limits_{M}dx\;g^{\frac{1}{2}}\Bigg\{
\frac{4\Lambda}{n-2}
+\frac{\lambda^{2}}{N}\textrm{tr}_{V}\Bigg[-\frac{1}{4}
\bar{h}_{\mu\nu}^{\bot}\left(-\Delta+\frac{4\Lambda}{(n-1)(n-2)}\right)
\bar{h}^{\bot\mu\nu}\nonumber\\
&+&\frac{(n-1)(n-2)}{4n^{2}}\;\varphi
\left(-\Delta-\frac{2n\Lambda}{(n-1)(n-2)}\right)\varphi\Bigg]\Bigg\}
+O(\lambda^{3})\;.
\end{eqnarray}
Thus on the mass shell the only remaining fields are the (traceless)
matrix-valued traceless transverse tensor  $\bar{h}_{\mu\nu}^{\bot}$ and
the (traceless) matrix-valued scalar field $\varphi$. This action looks
exactly the same as in General Relativity, the only difference being
that the fields are matrix-valued and traceless. Therefore, it describes
$(N-1)$ spin-2 particles and $(N-1)$ spin-0 particles. Note also, that
exactly as in General Relativity, the scalar conformal mode is
unstable if the cosmological constant $\Lambda$ is assumed to be positive.

\section{Concluding Remarks}

In this chapter we studied a non-commutative deformation of General
Relativity (called Spectral Matrix Gravity) proposed in
\cite{avramidi04b} where the non-commutative limit has been
explicitly evaluated. The approach of the paper \cite{avramidi04b}
to construct the action for Matrix Gravity differs from the one proposed
in
\cite{avramidi04,avramidi03,fucci08}. In the latter the action of
our model was a straightforward generalization of the
Hilbert-Einstein action in which the measure and the scalar
curvature were matrix-valued quantities. This last approach seems
to have some intrinsic arbitrariness due to the freedom of
choosing the particular form of the matrix-valued measure (for a
discussion see \cite{fucci08}). In order to avoid these issues, in
\cite{avramidi04b}  the action of Matrix Gravity was defined as a
linear combination of the first two global heat kernel
coefficients (\ref{62}) of a non-Laplace type partial differential
operator.

By using the covariant Fourier transform method we were able to
evaluate the coefficients $A_{0}$ and $A_{1}$, and as a result,
the action functional within the perturbation theory  in the
deformation parameter $\lambda$. The main result of this chapter is
the derivation of the modified Einstein equations in (\ref{693}) in
the weak deformation limit. In this case the pure non-commutative
fields, namely $h^{\mu\nu}$ and $\sigma$, contribute to the
right-hand side of the Einstein equation, that is, the
stress-energy tensor. The explicit form of these non-commutative
correction terms has been derived in (\ref{695}) through (\ref{698})
for the first time.


Some of the physical implications of Matrix Gravity have been
extensively discussed in \cite{avramidi03,avramidi04,avramidi04b}.
This theory exhibits non-geodesic motion which can be related to
a violation of the equivalence principle, moreover, because of the new
gauge symmetry, there are new physical conserved charges. At last, this
theory represents a consistent model of interacting spin-2 particles
on curved space which usually was a problem. An interesting question is
the limit as $N\rightarrow\infty$ of our model, this might be related to
matrix models and string theory.

As it is outlined in the introduction Matrix Gravity can be considered
as a Gravitational Chromodynamics describing the gravitational interaction
of a new degree of freedom that we call {\it gravitational color}.
Whether or not
it is related to the color of QCD is an open question. Let's suppose
for simplicity that it is the same, and that the gauge group of Matrix Gravity
is nothing but $SU(3)$. If one pushes this
analogy with QCD
to its logical limit then this would mean that the theory predicts
that the {\it gravitational interaction of quarks depends on their colors}.
Exactly as in QCD the strong interaction between quark of color $i$ and
a quark of color $j$ is transmitted by gluon of type $(ij)$, the
gravitational interaction between quark of color $i$ and
a quark of color $j$ is transmitted by the graviton of type $(ij)$.
In this case, all particles in the electro-weak sector, including photon,
do not feel the gravitational color. In that  sense it is `{\it dark}'.
Notice that in the non-relativistic limit the Newtonian potential will
also become `matrix-valued'.
One can go even further. Since the usual (white) mass is determined by
the sum of the color masses, one can assume that the color masses can be even
negative. Then the gravitational interaction of such particles would include
non only attractive forces but also repellent forces (antigravity?).
This feature could then solve the mystery of singularities in General
Relativity.

The consequences of our model in the ambit of cosmology are easily seen
by inspecting equations (\ref{693}) and (\ref{694}). The deformation of the
energy-momentum tensor in (\ref{694}) is written only in terms of the
non-commutative part $h^{\mu\nu}$. It would be interesting to study
whether or not $h^{\mu\nu}$ could account for a field of negative pressure.
If so, our model could describe the dynamics of dark energy.
Furthermore, the distortion of the gravitation expansion of the universe
due to the non-commutative degrees of freedom of the gravitational field
will certainly have some
effect on the anisotropy of the cosmic background
radiation, nucleosynthesis and structure formation. Of course, a detailed
analysis of these effects requires a careful study of the fluctuations
in the early universe.

Of course the validity
of these statements requires further investigations.
The ultimate goal of this theory is to construct a consistent theory of
the gravitational field which is compatible with the Standard Model and able
to solve the current open issues which afflict General Relativity, i.e. the problems
of the origin of dark matter and dark energy, the recent anomalies found in
the solar system (Pioneer Anomaly, flyby anomaly, etc.) and last, but not least,
the problem of quantization of the gravitational field.

In summary, we would like to stress that our model makes it possible to make
a number of {\it very specific predictions}
that can serve as experimental tests of the theory.





\chapter[KINEMATICS IN MATRIX GRAVITY]
{KINEMATICS IN MATRIX GRAVITY\footnotemark[5]}

\footnotetext[5]{The material in this chapter has been published in \emph{General Relativity and Gravitation}: I. G. Avramidi and G. Fucci, Kinematics in Matrix Gravity, \emph{Gen. Rel. Grav.} (2008) DOI 10.1007/s10714-008-0713-6}

\begin{chapabstract}

We develop the kinematics in Matrix Gravity, which is a modified theory
of gravity obtained by a non-commutative deformation of General
Relativity. In this model the usual interpretation of gravity as
Riemannian  geometry is replaced by a new kind of geometry, which is
equivalent  to a collection of Finsler geometries with several Finsler
metrics depending  both on the position and on the velocity. As a result
the Riemannian geodesic  flow is replaced by a collection of Finsler
flows. This naturally leads to a model in which  a particle is described
by several mass parameters.
If these mass parameters are different then the
equivalence principle is violated. In the non-relativistic limit this
also leads to corrections to the Newton's  gravitational potential. We
find the first and second order corrections to the usual Riemannian
geodesic flow and evaluate the anomalous nongeodesic  acceleration in a
particular case of static spherically symmetric background.
\end{chapabstract}

\section{Introduction}

In this chapter we investigate the motion of test particles in an extended
theory of gravity, called Matrix Gravity, proposed in a series of
recent papers
\cite{avramidi03,avramidi04,avramidi04b} and presented in Chapter 5.
The main goal of the present chapter is to investigate the motion of
test particles in a simple model of Matrix Gravity and study the
non-geodesic corrections to General Relativity.

The outline is as follows. In Sect. 2.  we develop the
kinematics in Matrix Gravity. In Sect. 3. we compute the first and
second order  non-commutative corrections to the usual Riemannian
geodesic flow. In Sect. 4 we find a static spherically  symmetric
solution of the dynamical equations of Matrix Gravity in a  particular
case of commutative $2\times 2$ matrices. In Sect. 5 we  evaluate the
anomalous acceleration of test particles in this background. In Sect. 6
we discuss our results.

\section{Kinematics in Matrix Gravity}

\subsection{Riemannian Geometry}

Let us recall how the geodesic motion appears in General
Relativity, that is, in Riemannian  geometry (for more details,
see \cite{avramidi04}).  First of all, let
\be
F(x,\xi)=\sqrt{-|\xi|^2}\,,
\ee
where
$\xi_\mu$ is a non-vanishing cotangent vector at the point $x$,
and $|\xi|^2=g^{\mu\nu}(x)\xi_\mu\xi_\nu$
(recall that the signature of our metric is $(-+\dots+)$).
Obviously, this is a homogeneous function of $\xi$ of degree $1$,
that is,
\be
F(x,\lambda\xi)=\lambda F(x,\xi)\,.
\ee
Let
\be
H(x,\xi)=-\frac{1}{2}F^2(x,\xi)
=\frac{1}{2}|\xi|^2\,.
\ee
This is, of course,
a homogeneous polynomial of $\xi_\mu$ of order $2$, and,
therefore,  the Riemannian metric can be recovered by
\be
g^{\mu\nu}(x)=
\frac{\partial^2}{\partial\xi_\mu\partial\xi_\nu}H(x,\xi)\,.
\ee
Now, let us consider a Hamiltonian system with the Hamiltonian
$H(x,\xi)$
\bea
\frac{dx^\mu}{dt}&=& \frac{\partial
H(x,\xi)}{\partial \xi_\mu}
=g^{\mu\nu}(x)\xi_\nu\,,\\[12pt]
\frac{d\xi_\mu}{dt}&=&-\frac{\partial H(x,\xi)}{\partial x^\mu} =
-\frac{1}{2}\partial_\mu g^{\alpha\beta}(x) \xi_\alpha\xi_\beta\,.
\eea
The trajectories of this Hamiltonian system are, then,
nothing but the geodesics of the metric $g_{\mu\nu}$. Of course,
the Hamiltonian is conserved, that is,
\be
g^{\mu\nu}(x(t))\xi_\mu(t)\xi_\nu(t) =-E\,,
\ee
where $E$ is a constant
parameter.

\subsection{Finsler Geometry}

As it is explained in \cite{avramidi04,avramidi04b} Matrix
Gravity is closely related to {\it Finsler geometry} \cite{rund59}
rather than Riemannian geometry. In this section we
follow the description of Finsler geometry outlined in
\cite{rund59}.
To avoid confusion we should note that we present
it in a slightly modified equivalent form, namely, we start with the
Finsler function in the cotangent bundle rather than in the
tangent bundle.

Finsler geometry is defined by a
Finsler function $F(x,\xi)$ which is a homogeneous function of
$\xi_\mu$ of degree $1$ and the Hamiltonian
\be\label{73}
H(x,\xi)=-\frac{1}{2}F^2(x,\xi)\,.
\ee
Such Hamiltonian is still
a homogeneous function of $\xi_\mu$ of degree $2$, that is,
\be
\xi_\mu\frac{\partial}{\partial \xi_\mu}H(x,\xi)=2H(x,\xi)\,,
\ee
but {\it not necessarily a polynomial} in $\xi_\mu$!

Now, we define a tangent vector $u$ by
\be
u^\mu=\frac{\partial}{\partial\xi_\mu}H(x,\xi)\,,
\ee
and the \emph{Finsler metric}
\be
\label{76}
G^{\mu\nu}(x,\xi)=\frac{\partial^2}
{\partial\xi_\mu\partial\xi_\nu }H(x,\xi)\,.
\ee

The difference with the Riemannian metric is, obviously, that the
Finsler metric does depend on $\xi_\mu$, more precisely, it is a
homogeneous function of $\xi_\mu$ of degree $0$, i.e.
\be
G^{\mu\nu}(x,\lambda\xi)=G^{\mu\nu}(x,\xi)\,,
\ee
so that it
depends only on the direction of the covector $\xi$ but not on its
magnitude. This leads to a number of useful identities, in
particular,
\be
H(x,\xi)=\frac{1}{2}
G^{\mu\nu}(x,\xi)\xi_\mu\xi_\nu\,,
\ee
and
\be
u^\mu=G^{\mu\nu}(x,\xi)\xi_\nu\,.
\ee

Now, we can solve this equation for $\xi_\mu$ treating $u^\nu$ as
independent variables to get
\be
\xi_\mu=G_{\mu\nu}(x,u)u^\nu\,,
\ee
where $G_{\mu\nu}$ is the inverse Finsler metric defined by
\be
G_{\mu\nu}(x,u) G^{\nu\alpha}(x,\xi)=\delta^\alpha_\mu\,.
\ee
By using the results obtained above we can express the Hamiltonian
$H$ in terms of the vector $u^{\mu}$, more precisely we have
\be
H(x,\xi(x,u))=\frac{1}{2} G_{\mu\nu}(x,u)u^\mu u^\nu\,.
\ee

The derivatives of the Finsler metric obviously satisfy the
identities
\be
\frac{\partial}{\partial
\xi_\alpha}G^{\beta\gamma}(x,\xi) =\frac{\partial}{\partial
\xi_\beta}G^{\gamma\alpha}(x,\xi) =\frac{\partial}{\partial
\xi_\gamma}G^{\alpha\beta}(x,\xi)\,,
\ee
\be
\xi_\mu\frac{\partial}{\partial \xi_\mu}G^{\nu\alpha}(x,\xi) =
\xi_\mu\frac{\partial}{\partial \xi_\nu}G^{\mu\alpha}(x,\xi)= 0\,,
\ee
and, more generally,
\be
\xi_\mu \frac{\partial^k}{\partial
\xi_{\nu_1} \dots\partial \xi_{\nu_k}}G^{\mu\alpha}(x,\xi) =0\,.
\ee
This means, in particular, that the following relations hold
\be
\frac{\partial u^\mu}{\partial
\xi_\alpha}=G^{\mu\alpha}(x,\xi)\,, \qquad \frac{\partial
\xi_\alpha}{\partial u^\mu}=G_{\mu\alpha}(x,u)\,.
\ee

It is easy to see that the metric $G_{\mu\nu}(x,u)$ is a
homogeneous function of $u$ of degree $0$, that is,
\be
u^\mu\frac{\partial }{\partial u^\mu}G_{\nu\alpha}(x,u)=0\,,
\ee
and, therefore, $H(x,\xi(x,u))$ is a homogeneous function of $u$
of degree $2$. This leads to the identities
\be
\xi_\mu=\frac{1}{2}
{\partial \over\partial u^\mu}H(x,\xi(x,u))\,,
\ee
\be
G_{\mu\nu}(x,u)={1\over 2} {\partial^2 \over\partial u^\mu\partial
u^\nu }H(x,\xi(x,u))\,.
\ee

Finally, this enables one to define the Finsler interval
\be
ds^2=G_{\mu\nu}(x,\dot x)dx^\mu dx^\nu\,,
\ee so that
\be
d\tau=\sqrt{-ds^2}= \sqrt{-G_{\mu\nu}(x,\dot x)\dot x^\mu\dot
x^\nu}\; dt =F(x,\xi(x,\dot x))dt\,, \label{733f}
\ee
where
\be
\dot x^\mu=\frac{dx^\mu}{dt}\,,\qquad \xi_\mu=G_{\mu\nu}(x,\dot
x)\dot x^\nu\,.
\ee
By treating $H(x,\xi)$ as a Hamiltonian we
obtain a system of first order ordinary differential equations
\bea
\frac{dx^\mu}{dt}&=& \frac{\partial H(x,\xi)}{\partial
\xi_\mu}
\,,\\[12pt]
\frac{d\xi_\mu}{dt}&=&- \frac{\partial H(x,\xi)}{\partial
x^\mu}\,.
\eea
The trajectories of this Hamiltonian system
naturally replace the geodesics in Riemannian geometry.  Again, as
in the Riemannian case, the Hamiltonian is conserved along the
integral trajectories
\be
H(x(t),\xi(t))=-E\,.
\ee
Of course, in
the particular case, when the Hamiltonian is equal to
$H(x,\xi)=\frac{1}{2}|\xi|^2$,  all the
constructions derived above reduce to the standard structure of Riemannian
geometry.

\subsection{Induced Finsler Geometry in Matrix Gravity}

The kinematics in Matrix Gravity is defined as follows. In
complete analogy with the above discussion we consider the matrix
\be
A(x,\xi)=a^{\mu\nu}(x)\xi_\mu\xi_\nu\,,
\ee
where $a^{\mu\nu}$ is the matrix-valued metric (\ref{638})
As we mentioned in the introduction this expression has been already
encountered in physics, in particular, in \cite{barcelo02} it is shown
that it is the most general structure describing ``analog models'' for gravity.

This is a
Hermitian matrix, so it has real eigenvalues $h_{i}(x,\xi)$,
$i=1,2,\dots, N$. We consider a generic case when the eigenvalues
are simple. We note that the eigenvalues $h_{i}(x,\xi)$ are
homogeneous functions (but not polynomials!) of $\xi$ of degree
$2$. Thus, each one of them, more precisely $\sqrt{-h_i(x,\xi)}$,
can serve as a Finsler function. In other words, we obtain $N$
different Finsler functions, and, therefore, $N$ different Finsler
metrics. Thus, quite naturally, instead of a single Riemannian
metric and a unique Riemannian geodesic flow there appears $N$
Finsler metrics and $N$ corresponding flows. In some sense, the
noncommutativity leads to a ``splitting'' of a single geodesic to
a system of close trajectories.

Now, to define a unique Finsler metric
we need to define a unique Hamiltonian, which is a homogeneous
function of the momenta of degree $2$. It is defined in terms of the
Finsler function as in (\ref{73})
which is a homogeneous function of the momenta of degree
$1$.
To define a unique Finsler function we can proceed as follows. Let
$\mu_i$, $i=1,\dots, N$, be some dimensionless real parameters
such that
\be\label{714}
\sum_{i=1}^N \mu_i=1\,,
\ee
so that there
are $(N-1)$ independent parameters. Then we can define the Finsler
function by
\be
F(x,\xi)=\sum_{i=1}^N \mu_i \sqrt{-h_i(x,\xi)}\,.
\label{7233xx}
\ee
Notice that,  in the commutative limit, as $\varkappa\to 0$
and $a^{\mu\nu}=g^{\mu\nu}\II$,  all eigenvalues of the matrix
$A(x,\xi)$ degenerate to the same  value,
$h_i(x,\xi)=|\xi|^2$, and, hence, the Finsler
function becomes $F(x,\xi)=\sqrt{-|\xi|^2}$.
In this case the Finsler flow degenerates to the usual Riemannian
geodesic flow.

Next, we define the Hamiltonian according to eq. (\ref{73})
\bea
H(x,\xi)&=&
-\frac{1}{2}
\left(\sum_{i=1}^N \mu_i \sqrt{-h_i(x,\xi)}\right)^2
\nonumber\\
&=&
\frac{1}{2}\sum_{i=1}^N \mu_i^2 h_i(x,\xi)
-\sum_{1\le i<j\le N}\mu_i\mu_j \sqrt{h_i(x,\xi)h_j(x,\xi)}\,.
\label{7234xx}
\eea
In a particular case, when all parameters $\mu_i$ are equal, i.e.
$\mu_i=1/N$, the Finsler function reduces to
\be
F(x,\xi)=\frac{1}{N}\sum_{i=1}^N \sqrt{-h_i(x,\xi)}
=\frac{1}{N}\;\tr \sqrt{-A(x,\xi)}\,.
\label{7235xx}
\ee
By using the
decomposition of the matrix-valued metric $a^{\mu\nu}$ 
(\ref{638})
one can see that
\be
\frac{1}{N}\tr
A(x,\xi)=|\xi|^2\,,
\ee
and, therefore,
\be
\frac{1}{N}\sum_{i=1}^N h_i(x,\xi)=|\xi|^2\,.
\ee
Thus, we conclude that in this particular case
\be H(x,\xi)
=\frac{1}{N}\left( \frac{1}{2} |\xi|^2
-\frac{1}{N} \sum_{1\le i<j\le N}
\sqrt{h_i(x,\xi)h_j(x,\xi)}\right)\,.
\ee

It is difficult to give a general physical picture of these models since
the Hamiltonian is non-polynomial in the momenta. Hamiltonian systems with
homogeneous Hamiltonians have not been studied as thoroughly as the
usual systems with quadratic Hamiltonians and a potential.

\subsection{Kinematics}

The problem is, now, how to use these mathematical tools to
describe the motion of physical massive test particles in Matrix
Gravity. The motion of a massive
particle in the gravitational field is
determined
in General Relativity by the action which is proportional to the
interval, that, is,
\be
S_{\rm particle} =-\int\limits_{P_1}^{P_2}
m\sqrt{-g_{\mu\nu}(x)dx^\mu dx^\nu} =-\int\limits_{t_1}^{t_2} m
\sqrt{-|\dot x|^2}dt\,,
\ee
where
$m$ is the mass of the particle,
$P_1$ and $P_2$ are the initial and the final position of the
particle in the spacetime,
$t$ is a parameter, $t_1$ and $t_2$ are the initial and the final
values, $\dot x^\mu=\frac{dx^\mu}{dt}$ and
$|\dot x|^2=g_{\mu\nu}(x)\dot
x^\mu\dot x^\nu\,.$
This action is, of course,
reparametrization-invariant. So, as always, there is a freedom of
choosing the parameter $t$. We can always choose the parameter to be the
{\it affine parameter}  such that $|\dot x|^2$ is constant, for example,
if the parameter is the proper time $t=\tau$, then $|\dot x|^2=-1$.
The Euler-Lagrange equations for this functional are, of course,
\be
\frac{D \dot x^\nu}{dt}
=\frac{d^2 \dot x^\nu}{dt^2}
+\Gamma^\nu{}_{\alpha\beta}(x)\dot x^\alpha\dot x^\beta
=0\,,
\ee
where $\Gamma^\mu{}_{\alpha\beta}$ are the standard
Christoffel symbols of the metric
$g_{\mu\nu}$.
Of course, the equivalence principle holds since these equations do not
depend on the mass.

In Matrix Gravity a particle is described
instead of one mass
parameter $m$ by   $N$ different mass
parameters
\be
m_i=m\mu_i\,,
\ee
where
\be
m=\sum_{i=1}^N m_i\,.
\ee
The parameters
$m_i$
describe the ``tendency'' for a particle to move along
the trajectory determined by the corresponding Hamiltonian
$h_i(x,\xi)$.
 In the commutative limit we only observe the total mass $m$.

We define the Finsler function $F(x,\xi)$ and the Hamiltonian
$H(x,\xi)$ as in eqs. (\ref{7233xx}) and (\ref{7234xx}).
Then the action for a
particle in the gravitational field has the form
\be
S_{\rm
particle} =-\int\limits_{t_1}^{t_2} m F(x,\xi(x,\dot x))\;dt \,.
\ee
Thus, the Finsler function $F(x,\xi(x,\dot x))$ (with the
covector $\xi_\mu$ expressed in terms of the tangent vector $\dot
x^\mu$) plays the role of the Lagrangian. To study the role of
non-commutative corrections, it is convenient to rewrite this
action in the form that resembles the action in General
Relativity.
\be
S_{\rm particle}
=-\int\limits_{t_1}^{t_2} m_{\rm eff}(x,\dot x) \sqrt{-|\dot x|^2}dt\,,
\ee
with some
{\it ``effective mass'' $m_{\rm eff}(x,\dot x)$ that depends
on the location and on the velocity of the particle}
\be
m_{\rm eff}(x,\dot x)
=\sum_{i=1}^N m_{i}
\sqrt{\frac{h_i(x,\xi(\dot x))}{|\dot x|^2}}\,.
\ee

This action is again reparametrization-invariant. Therefore, we
can choose the natural arc-length parameter so that
$F(x,\xi(x,\dot x))=1$. Then the equations of motion determined by
the Euler-Lagrange equations have the same form
\be
\frac{d^2
x^\mu}{dt^2}+\gamma^\mu{}_{\alpha\beta}(x,\dot x)\dot x^\alpha \dot
x^\beta=0\,,
\ee
where $\gamma^\mu{}_{\alpha\beta}(x,\dot x)$ are
the Finsler Christoffel coefficients defined by the equations that
look identical to the usual equations but with the Finsler metric
instead of the Riemannian metric, that is,
\be\label{79}
\gamma^\mu{}_{\alpha\beta}(x,\dot x)
=\frac{1}{2}G^{\mu\nu}(x,\xi(x,\dot x))
\left(\frac{\partial}{\partial x^\alpha} G_{\nu\beta}(x,\dot x)
+\frac{\partial}{\partial x^\beta} G_{\nu\alpha}(x,\dot x)
-\frac{\partial}{\partial x^\nu} G_{\alpha\beta}(x,\dot x)
\right)\,.
\ee

To study the role of non-commutative corrections it is convenient
to rewrite these equations in a covariant form in the Riemannian
language.  In
the commutative limit, as $\varkappa\to 0$, we can expand all our
constructions in power series in $\varkappa$ so that the
non-perturbed quantities are the Riemannian ones. In particular,
we have
\be
\gamma^\mu{}_{\alpha\beta}(x,\dot x)=\Gamma^\mu{}_{\alpha\beta}(x)
+\theta^\mu{}_{\alpha\beta}(x,\dot x)\,,
\ee
where
$\theta^\mu{}_{\alpha\beta}$ are some tensors of order $\varkappa$.
Then the equations of motion can be written in the form
\be
\frac{D \dot x^\nu}{dt}=A^\nu_{\rm anom}(x,\dot x)\,,
\ee
where
\be
\frac{D \dot x^\nu}{dt} =\frac{d^2
x^\mu}{dt^2}+\Gamma^\mu{}_{\alpha\beta}(x)\dot x^\alpha \dot x^\beta
\ee
and
\be\label{712}
A^\nu_{\rm anom}(x,\dot
x)=-\theta^\nu{}_{\alpha\beta}(x,\dot x) \dot x^\alpha \dot
x^\beta\,,
\ee
is the {\it anomalous nongeodesic acceleration}.

\section{Perturbation Theory}

We see that the motion of test particles in matrix Gravity is quite
different from that of General Relativity. The most important difference
is that particles exhibit a {\it non-geodesic motion}. In other words,
there is no Riemannian metric such that particles move along the
geodesics of that metric. It is this anomalous acceleration that we are
going to study in this chapter.

In the commutative limit the action of a particle in Matrix Gravity
reduces to the action of a particle in General  Relativity with the mass
$m$ determined by the sum of all masses $m_i$.  In
this chapter we consider  two different cases. In the first case, that we
call the {\it nonuniform model}, we assume that all mass parameters are
different, and in the second case, that we call the {\it uniform model},
we discuss what happens if they are equal to each other.

\subsection{Nonuniform Model: First Order in $\varkappa$}

So, in this section we study the generic case when the parameters
$\mu_i$ are different.
As we already mentioned above, in this case the Finsler function
$F(x,\xi)$ is given by (\ref{7233xx}).
By using the decomposition $a^{\mu\nu}=g^{\mu\nu}\mathbb{I}+\varkappa h^{\mu\nu}$ of the matrix-valued metric
$a^{\mu\nu}$ we have
\be
A(x,\xi)=a^{\mu\nu}(x)\xi_\mu\xi_\nu
=|\xi|^2\II+\varkappa h^{\mu\nu}(x)\xi_\mu\xi_\nu\,.
\ee
Therefore, the
eigenvalues of the matrix $A(x,\xi)$ are
\be
h_i(x,\xi)=|\xi|^2+\varkappa\lambda_i(x,\xi)\,,
\ee
where
$\lambda_i(x,\xi)$ are the eigenvalues of the matrix
$h^{\mu\nu}(x)\xi_\mu\xi_\nu$. In the first order in $\varkappa$
we get the Finsler function
\be F(x,\xi)=\sqrt{-|\xi|^2}\left(1
+\varkappa\frac{1}{2}\frac{P(x,\xi)}{|\xi|^2}\right)
+O(\varkappa^2)\,,
\ee
and the Hamiltonian
\bea
H(x,\xi)&=&\frac{1}{2}|\xi|^2 +\varkappa\frac{1}{2}P(x,\xi)
+O(\varkappa^2)\,,
\nonumber\\
\eea
where
\be
P(x,\xi)=\sum_{i=1}^N\mu_i\lambda_i(x,\xi)\,.
\ee
By using the fact
that $P(x,\xi)$ is a homogeneous function of $\xi$ of order
$2$, we find the Finsler metric
\be
G^{\mu\nu}(x,\xi)=g^{\mu\nu}(x)
+\varkappa q^{\mu\nu}(x,\xi)
+O(\varkappa^2)\,,
\ee
and its inverse
\be
G_{\mu\nu}(x,u)=g_{\mu\nu}(x)
-\varkappa q_{\mu\nu}(x,\xi(x,u))
+O(\varkappa^2)\,,
\ee
where
\be
q^{\mu\nu}(x,\xi)
=\frac{1}{2}\frac{\partial^2}{\partial \xi_\mu\partial\xi_\nu}
P(x,\xi)\,.
\label{7275xx}
\ee
Here the indices are raised and lowered with the Riemannian
metric, and
\be
u^\mu(x,\xi)=G^{\mu\nu}(x,\xi)\xi_\nu\,,\qquad
\xi_\mu(x,u)=G_{\mu\nu}(x,u)u^\nu\,.
\ee
Since $P(x,\xi)$ is a homogeneous function of $\xi$ of order $2$
we have
\be
P(x,\xi)=q^{\mu\nu}(x,\xi)\xi_\mu\xi_\nu\,.
\ee
Note that since $\tr h^{\mu\nu}=0$ the matrix $h^{\mu\nu}\xi_\mu\xi_\nu$
is traceless, which implies that the sum of its eigenvalues is equal to
zero. Thus, in the uniform case, when all mass parameters $\mu_i$ are
the same, the function $P(x,\xi)$ vanishes. In this case the effects of
non-commutativity are of the second order in $\varkappa$; we study this
case in the next section.

We also note that
\be
|\xi|^2=|u|^2-2\varkappa P(x,\xi(x,u))
+O(\varkappa^2)\,.
\ee
Thus, our Lagrangian is
\be
F(x,\xi(x,\dot x))=
\sqrt{-|\dot x|^2}\left(1
-\varkappa \frac{1}{2}
\frac{P(x,\xi(x,\dot x))}{|\dot x|^2}
\right)
+O(\varkappa^2)\,,
\ee
Finally, we compute the Christoffel symbols
to obtain
\be
\theta^\mu{}_{\alpha\beta}(x,\dot x)
=-\frac{1}{2}\varkappa g^{\mu\nu}\left(
\nabla_\alpha q_{\beta\nu}(x,\dot x)
+\nabla_\beta q_{\alpha\nu}(x,\dot x)
-\nabla_\nu q_{\alpha\beta}(x,\dot x)
\right)
+O(\varkappa^2)\,,
\ee
and the covariant derivatives are defined with the Riemannian metric.

Thus, the anomalous acceleration is
\be
A^\mu{}_{\rm anom}=
\frac{\varkappa}{2}
g^{\mu\nu}\left(
2\nabla_\alpha q_{\beta\nu}(x,\dot x)
-\nabla_\nu q_{\alpha\beta}(x,\dot x)
\right)\dot x^\alpha \dot x^\beta
+O(\varkappa^2)\,,
\ee

\subsection{Uniform Model: Second Order in $\varkappa$}

So, in this section we will simply assume that all mass parameters
are equal, that
is,
\be
m_i=\frac{m}{N}\,.
\ee
In this case the Finsler function
$F(x,\xi)$ is given by (\ref{7235xx}).
By using the decomposition of the matrix-valued metric and
the fact that $\tr h^{\mu\nu}=0$ we get
the Finsler function
 \be
F(x,\xi)=\sqrt{-|\xi|^2}\left(1
-\varkappa^2\frac{1}{8}S^{\mu\nu\alpha\beta}(x)
\frac{\xi_\mu\xi_\nu\xi_\alpha\xi_\beta}{|\xi|^4}\right)
+O(\varkappa^3)\,,
\ee
and the Hamiltonian
\bea
H(x,\xi)&=&\frac{1}{2}|\xi|^2\left(1
-\varkappa^2\frac{1}{4}S^{\mu\nu\alpha\beta}(x)
\frac{\xi_\mu\xi_\nu\xi_\alpha\xi_\beta}{|\xi|^4}\right)
+O(\varkappa^3)\,,
\eea
where
\be
S^{\mu\nu\alpha\beta}=\frac{1}{N}\tr
(h^{\mu\nu}h^{\alpha\beta})\,.
\ee
By using the above, we compute the
Finsler metric
\be
G^{\mu\nu}(x,\xi)=g^{\mu\nu}(x)
-\varkappa^2\frac{1}{4}S^{\mu\nu\alpha\beta}(x)
\frac{\xi_\alpha\xi_\beta}{|\xi|^2}
+O(\varkappa^3)\,,
\ee
and its inverse
\be
G_{\mu\nu}(x,u)=g_{\mu\nu}(x)
+\varkappa^2\frac{1}{4}S_{\mu\nu\alpha\beta}(x)
\frac{u^\alpha u^\beta}{|u|^2}
+O(\varkappa^3)\,.
\ee
We also note that
\be
|\xi|^2=|u|^2+\varkappa^2\frac{1}{2}S_{\mu\nu\alpha\beta}(x)
\frac{u^\mu u^\nu u^\alpha u^\beta}{|u|^2}
+O(\varkappa^3)\,.
\ee
Thus, our Lagrangian is
\be
F(x,\xi(x,\dot x))=
\sqrt{-|\dot x|^2}\left(1
+\varkappa^2\frac{1}{8}S_{\mu\nu\alpha\beta}(x)
\frac{\dot x^\mu \dot x^\nu \dot x^\alpha \dot x^\beta}{|\dot x|^4}\right)
+O(\varkappa^3)\,,
\ee
Finally, we
compute the Christoffel symbols to obtain
\be
\theta^\mu{}_{\alpha\beta}(x,\dot x)
=\varkappa^2\frac{1}{8}g^{\mu\nu}\left(
\nabla_\alpha S_{\beta\nu\rho\sigma}
+\nabla_\beta S_{\alpha\nu\rho\sigma}
-\nabla_\nu S_{\alpha\beta\rho\sigma}
\right)\frac{\dot x^\rho\dot x^\sigma}{|\dot x|^2}
+O(\varkappa^3)\,.
\ee

Thus, the anomalous acceleration is
\be
A^\mu{}_{\rm anom}=-
\frac{\varkappa^2}{8}g^{\mu\nu}\left(
2\nabla_\alpha S_{\beta\nu\rho\sigma}
-\nabla_\nu S_{\alpha\beta\rho\sigma}
\right)
\frac{\dot x^\rho\dot x^\sigma\dot x^\alpha\dot x^\beta}{|\dot x|^2}
+O(\varkappa^3)\,,
\ee
Notice that with our choice of the parameter $t$ we have
$F(x,\xi(x,\xi))=1$,
and, therefore, in the equations of motion we can substitute with
the same accuracy
\be
|\xi|^2=-1+O(\varkappa^2)\,,\qquad
|\dot x|^2=-1+O(\varkappa^2)\,.
\ee
Therefore, we obtain finally
\be
A^\mu{}_{\rm anom}=-
\frac{\varkappa^2}{8}g^{\mu\nu}\left(
2\nabla_\alpha S_{\beta\nu\rho\sigma}
-\nabla_\nu S_{\alpha\beta\rho\sigma}
\right)
\dot x^\rho\dot x^\sigma\dot x^\alpha\dot x^\beta
+O(\varkappa^3)\,.
\label{7283xx}
\ee

\subsection{Non-commutative Corrections to Newton's Law}

Now, we will derive the non-commutative corrections to the Newton's Law.
We label the coordinates as
\be
x^0=t, \qquad x^1=r, \qquad x^2=\theta,\quad x^3=\varphi\,,
\ee
and consider the static spherically symmetric (Schwarzschild)
metric
\be
ds^2=-U(r)dt^2+U^{-1}(r)dr^2+r^2(d\theta^2+\sin^2\theta\,d\varphi^2)\,,
\ee
where
\be
U(r)=1-\frac{r_g}{r}\,,\qquad
r_g=2GM\,,
\ee
and $M$ is the mass of the central body.
It is worth recalling that here $t$ is the coordinate time. In the
previous sections we used $t$ to denote an affine parameter of the
trajectory that we agreed to choose to be the proper time. In the
present section we use $\tau$ to denote the proper time and $t$ to
denote the coordinate time.

The motion of test particles
in Schwarzschild geometry is
very well studied in General Relativity, see, for example
\cite{weinberg72}.
Assuming that the particle moves in the
equatorial plane $\theta=\pi/2$ away from the center, that is,
$dr/d\tau>0$,
the equations of motion have the following integrals
\cite{weinberg72}
\bea
\dot x^0=\frac{ dt}{d\tau}&=&\frac{E}{m}\frac{1}{U(r)}\,,\\
\dot x^1=\frac{dr}{d\tau}&=&\sqrt{\frac{E^2}{m^2}
-\left(1+\frac{L^2}{m^2}\frac{1}{r^2}\right)U(r)}\,,\\
\dot x^2=\frac{d\theta}{d\tau}&=&0\,,\qquad
\theta=\frac{\pi}{2}\,,\\
\dot x^3=\frac{d\varphi}{d\tau}&=&\frac{L}{m}\frac{1}{r^2}\,,\\
\eea
where $m$, $L$, and $E$ are the mass of the particle, its orbital
momentum and the energy.

In the non-relativistic limit for weak gravitational fields,
assuming
\be
E=m+E'\,,
\ee
with $E'<<m$, and $r>>r_g$
one can identify the coordinate time with the proper time,
so that
\be
\dot x^0=\frac{dt}{d\tau}=1\,.
\ee
Further, for the non-relativistic motion we have
$\dot r$, $r\dot \theta$, $r\dot\varphi<<1$,
and the radial velocity reduces, of course, to the standard Newtonian
expression
\bea
\dot x^1=\frac{dr}{d\tau}&=&\sqrt{\frac{2E'}{m}
-\frac{L^2}{m^2}\frac{1}{r^2}
+\frac{r_g}{r}}
\,,
\eea
which for $L=0$ becomes
\bea
\dot x^1=\frac{dr}{d\tau}&=&\sqrt{\frac{2E'}{m}
+\frac{r_g}{r}}
\,,
\eea

It is worth stressing that the anomalous acceleration due to
non-commutativity in the non-relativistic limit
can be interpreted as a correction to the Newton's Law. Assuming that
a particle is moving in the equatorial plane, $\theta=\pi/2$,
with zero orbital momentum, $\varphi={\rm const}$, the equation of motion
is
\bea
\frac{d^2 r}{dt^2}&=&-\frac{\partial}{\partial r}V_{\rm eff}(r)
\nonumber\\
&=&-\frac{GM}{r^2}+A^r{}_{\rm anom}\,,
\eea
where in the uniform model
\be
A^{r}{}_{\rm anom}= \frac{\varkappa^{2}}{8} \partial_r S^{0000}
+O(\varkappa^3)
\,,
\label{7314xx}
\ee
with $S^{0000}=\frac{1}{N}\tr h^{00}h^{00}$,
and
in the non-uniform model
\be
A^{r}{}_{\rm anom}= -\frac{\varkappa}{2} \partial_r
q^{00}
+O(\varkappa^{2})
\label{7315xx}
\,,
\ee
with $q^{00}$ being the component of the tensor $q^{\mu\nu}$
defined by (\ref{7275xx}).
This gives the non-commutative corrections to Newton's Law:
in the uniform model,
\bea
V_{\rm eff}(r)=-\frac{GM}{r}-\frac{\varkappa^{2}}{8} S^{0000}(r)
+O(\varkappa^3)\,,
\eea
and, in the nonuniform model,
\bea
V_{\rm eff}(r)=-\frac{GM}{r}+\frac{\varkappa}{2} q^{00}(r)
+O(\varkappa^2)
\,.
\eea
Here, of course, the tensor components $S^{0000}$ and $q^{00}$
should be obtained by the solution of the non-commutative
Einstein field equations (in the perturbation theory).

\subsection{Static Spherically Symmetric Solutions}

In the present chapter we study the effects of Matrix Gravity
in the simplest possible case restricting ourselves
to a {\it commutative algebra}.
The commutativity assumption enormously simplifies the dynamical
equations. By recalling equation (\ref{561}) it is easy to show that  in this case the dynamical equations
look exactly as the Einstein equations in the vacuum
\begin{equation}
\label{744}
\mathcal{R}_{\mu\nu}=\Lambda b_{\mu\nu}\;,
\end{equation}
where ${\cal R}_{\mu\nu}$ is the matrix-valued Ricci tensor defined by ${\cal R}_{\mu\nu}={\cal R}^\alpha{}_{\mu\alpha\nu}$.

In this section we are going to study, in particular, a static spherically symmetric solution of the equation
(\ref{744}).
We present the
matrix-valued metric $a^{\mu\nu}$ by writing the ``matrix-valued
Hamiltonian''
\begin{equation}\label{719xx}
a^{\mu\nu}\xi_\mu\xi_\nu=
A(r)(\xi_0)^2
+B(r)(\xi_1)^2
+\mathbb{I}\frac{1}{r^2}
\left[(\xi_2)^2
+\frac{1}{\sin^{2}\theta}(\xi_3)^2\right]\;,
\end{equation}
or the ``matrix-valued interval''
\begin{equation}\label{719}
b_{\mu\nu}dx^\mu dx^\nu=
A^{-1}(r)dt^{2}+B^{-1}(r)dr^{2}
+\mathbb{I}\;r^{2}\left(d\theta^{2}
+\sin^{2}\theta\, d\varphi^{2}\right)\;,
\end{equation}
where the coefficients $A(r)$ and $B(r)$
are commuting matrices that depend only on the
radial coordinate $r$.
This simply means that we choose the following ansatz
\begin{eqnarray}\label{729xxx}
a^{00}&=& A\;,\qquad
a^{11}=B\;, \nonumber\\[10pt]
a^{22}&=&\frac{1}{r^2}\;\mathbb{I}\;, \qquad
a^{33}=\frac{1}{r^{2}\sin^{2}\theta} \;\mathbb{I}\;.
\end{eqnarray}

Next, by computing the connection coefficients ${\cal A}^\alpha{}_{\mu\nu}$ and
the matrix-valued Ricci tensor we obtain
the equations
of motion
\begin{eqnarray}
\mathcal{R}_{00}&=&
A^{-1}B\left[
\frac{1}{2}A^{-1}A^{\prime\prime}
-\frac{3}{4}A^{-2}(A^{\prime})^2
+\frac{1}{4}A^{-1}A^{\prime}B^{-1}B^{\prime}
+\frac{1}{r}A^{-1}A^{\prime}
\right]
=\Lambda A^{-1}\;,
\nonumber
\\
\label{740}\\
\mathcal{R}_{11}&=&
\frac{1}{2}A^{-1}A^{\prime\prime}
-\frac{3}{4}A^{-2}(A^{\prime})^2
+\frac{1}{4}A^{-1}A^{\prime}B^{-1}B^{\prime}
-\frac{1}{r}B^{-1}B^{\prime}
=\Lambda B^{-1}\;,
\label{741}\\
\mathcal{R}_{22}&=&
-\frac{r}{2}B'
-B
+\frac{r}{2}BA^{-1}A^{\prime}
+\mathbb{I}
=\Lambda
r^2\cdot\II\;,
\label{742}\\
\mathcal{R}_{33}&=&\sin^{2}\theta\;\mathcal{R}_{22} =\Lambda
r^2\sin^2\theta\cdot\II \;. \label{743}
\end{eqnarray}
where the prime
denotes differentiation with respect to $r$.

By using the equations (\ref{740}) and (\ref{741}) we find
\begin{equation}\label{746}
A^{-1}A^{\prime}+B^{-1}B^{\prime}=0\;;
\end{equation}
the general solution of this equation is
\begin{equation}\label{747}
A(r)B(r)=C_1\;,
\end{equation}
where $C_1$ is an arbitrary constant matrix from our algebra.
We require that at the spatial infinity as $r\to \infty$ the matrices
$A$ and $B$ and, therefore, the matrix $C$ as well,
are non-degenerate.

By using
this relation we obtain further  from eqs. (\ref{741})
and (\ref{742}) two
compatible equations for the matrix $B$
\be
B''+\frac{2}{r}B'+2\Lambda=0\,,
\label{740xxx}
\ee
and
\be
rB'+B=(1-\Lambda r^2)\II\,.
\label{742xxx}
\ee
The general solution of the eq. (\ref{742xxx}) is
\be
B(r)=\left(1-\frac{1}{3}\Lambda r^2\right)\II+\frac{1}{r}C_2\,,
\ee
where $C_2$ is another arbitrary constant matrix  from our algebra. It
is not difficult to see that this form of the matrix $B$ also satisfies
the eq. (\ref{740xxx}). The matrix $A$ is now obtained from the equation
(\ref{747})
\be
A(r)=C_1\left[\left(1-\frac{1}{3}\Lambda r^2\right)\II
+\frac{1}{r}C_2\right]^{-1}\,.
\ee

We will also
require that in the limit
$\varkappa\rightarrow 0$ we
should get the standard
Schwarzschild solution with the cosmological constant
\be
B(r)=-A^{-1}(r)=\left(1-\frac{1}{3}\Lambda r^2
-\frac{r_g}{r}\right)\II\,,
\ee
where $r_g$ is the gravitational radius of the central body of mass
$M$,
\be
r_g=2GM\,,
\label{7330xx}
\ee
that is, in that limit the matrices $E$ and $C$ should be
\be
C_1=-\II\,, \qquad C_2=-r_g\II\,.
\ee

\subsection{$2\times 2$ Matrices}

To be specific, we restrict ourselves further
to real symmetric $2\times 2$ matrices
generated by
\begin{equation}\label{713a}
\mathbb{I}=
\left(%
\begin{array}{cc}
  1 & 0 \\
  0 & 1 \\
\end{array}%
\right)
\,,\qquad
\textrm{and}\qquad\tau=\left(%
\begin{array}{cc}
  0 & 1 \\
  1 & 0 \\
\end{array}%
\right)\;.
\end{equation}
In this case the
constant matrices $C_1$ and $C_2$ can be expressed in terms of four
real parameters
\be
C_1=\alpha\II+\theta\tau\,,\qquad
C_2=\mu\II+L\tau\,,
\ee
where $\theta=\varkappa\bar\theta$ and $L=\varkappa\bar L$ are the
parameters of first order in the deformation parameter $\varkappa$.
Here the parameters $\alpha$ and $\theta$ are dimensionless and
the parameters $\mu$ and $L$ have the dimension of length.

Then the matrix
$B(r)$ has the form
\begin{equation}\label{720}
B(r)=\left(1-\frac{1}{3}\Lambda r^2+\frac{\mu}{r}\right)\II
+\frac{L}{r}\tau\,.
\end{equation}
Next, noting that $\tau^2=\II$, and by using the relation
\be
(a\II+b\tau)^{-1}=\frac{1}{a^2-b^2}\left(a\II-b\tau\right)\,,
\ee
we obtain the matrix $A(r)$
\be
A(r)=\varphi(r)\II+\psi(r)\tau\,,
\ee
where
\be
\varphi(r)=\frac{\alpha\left(1-\frac{1}{3}\Lambda r^2\right)
+\frac{\alpha\mu-\theta L}{r}}
{\left(1-\frac{1}{3}\Lambda r^2+\frac{\mu}{r}\right)^2
-\frac{L^2}{r^2}}\,,
\ee

\be
\psi(r)=\frac{\theta\left(1-\frac{1}{3}\Lambda r^2\right)
+\frac{\theta \mu-\alpha L}{r}}{\left(1-\frac{1}{3}\Lambda r^2
+\frac{\mu}{r}\right)^2
-\frac{L^2}{r^2}}\,.
\ee

The parameters $\alpha, \theta, \mu$ and $L$ should be determined
by the boundary conditions at spatial infinity. The question of
boundary conditions is a subtle point since we do not know the
physical nature of the additional degrees of freedom.
We will simply require that the diagonal part of the metric
is asymptotically De Sitter. This
immediately gives
\be
\alpha=-1\,.
\ee
Now, we introduce a new parameter
\be
r_0=|\Lambda|^{-1/2}\,,
\ee
and require that for $r_g<<r<<r_0$, the diagonal part of the
metric, more precisely, the function $\varphi(r)$ is
asymptotically Schwarzschild, that is,
\be
\varphi(r)=-1-\frac{r_g}{r}+O\left(\frac{r_g^2}{r^2}\right)
+O\left(\frac{r^2}{r_0^2}\right)\,.
\ee
This fixes the parameter $\mu$
\be
\mu=-r_g+\theta L\,.
\ee
The parameters $\theta$ and $L$ remain undetermined.

Finally, by introducing new parameters
\be
\rho=(1+\theta^2) L -\theta r_g
\ee
\be
r_{\pm}=r_g-(\theta\pm 1)L\,
\ee
we can rewrite our solution in the form
\be
\varphi(r)=\frac{-r\left(r-\frac{1}{3}\Lambda r^3
-r_g+2\theta L\right)}
{\left[r-\frac{1}{3}\Lambda r^3-r_-\right]
\left[r-\frac{1}{3}\Lambda r^3-r_+\right]}\,,
\ee

\be
\psi(r)=\frac{r\left[\theta\left(r-\frac{1}{3}\Lambda r^3\right)
+\rho\right]}
{\left[r-\frac{1}{3}\Lambda r^3-r_-\right]
\left[r-\frac{1}{3}\Lambda r^3-r_+\right]}\,
\label{73125xx}
\,.
\ee
Of course, as $\varkappa\rightarrow 0$
both parameters $L=\varkappa\bar L$ and
$\theta=\varkappa\bar\theta$
vanish and we
get the standard
Schwarzschild solution with the cosmological constant.

Notice that the matrix-valued metric $a^{\mu\nu}$
becomes singular when
the matrices $A$ and $B$ are not invertible, that is,
when
\be
\det A(r)=0\,.
\ee
The solutions of this equation are the roots of the cubic
polynomials
\be
r-\frac{1}{3}\Lambda r^3-r_-=0
\qquad\mbox{and}\qquad
r-\frac{1}{3}\Lambda r^3-r_+=0
\label{74146}
\ee

Recall that the standard Schwarzschild coordinate singularity,  which
determines the position of the event horizon, is located at $r=r_g$. The
presence of singularities depends on the values of the parameters. We
analyze, now, the first eq. in (\ref{74146}). In the case $\Lambda\le 0$
the polynomial has one root if $r_->0$ and does not have any roots if
$r_-<0$. In the case $\Lambda>0$ it is easy to see that: i) if $r_-
>(2/3) r_0$, then there are no roots, ii) if $0< r_- <(2/3) r_0$, then
the polynomial has two roots, and ii) if $r_-<0$, then the polynomial
has one root. The same applies to the second eq. in (\ref{74146}).

We emphasize that there are two cases {\it without any singularities at
any finite value of $r$}. This happens if either: a) $\Lambda\le 0$ and
$r_\pm <0$, or b) $\Lambda>0$ and $r_\pm> (2/3)r_0$. This can certainly
happen for large values of $|\theta|$ and $|L|$.
In particular, if
$\theta$ and $L$ have the same signs and
\be
|\theta|>1+\frac{r_g}{|L|}\,,
\ee
then both $r_\pm$ are negative, $r_\pm <0$, and
if $\theta$ and $L$ have opposite
signs and
\be
|\theta|>1+\frac{\frac{2}{3}r_0-r_g}{|L|}\,,
\ee
then $r_\pm> (2/3)r_0$. This is a very interesting phenomenon which is
entirely new and due to the additional degrees of freedom.

We would like to clarify some points.
The parameters $\mu_{i}$ introduced in
the previous sections describe the properties of
the test particle, that is, the matter.
The parameters
$\theta$ and $\rho$ introduced
in the static and spherically symmetric solution of non-commutative Einstein
equations describe the properties of the gravitational field, that is,
the properties of the source of the gravitational field, that is, the
central body.
The parameters $\theta$ and $\rho$ are not related to the parameters $\mu_i$.

\section{Anomalous Acceleration}

In
this section we are going to evaluate the anomalous acceleration
of non-relativistic test particles in the static spherically symmetric
gravitational field of a massive central body.

All we have to do is to evaluate the
components of the anomalous acceleration (\ref{7283xx}).
As we will see the only essential component of the anomalous
acceleration is the radial one $A^r{}_{\rm anom}$.
All other components of
the anomalous acceleration are negligible in this limit.
As we will see below, the anomalous acceleration is caused by
the radial gradient of the
component $h^{00}$ of the matrix-valued metric, which is
\be
\varkappa h^{00}
=\psi(r)\tau\,,
\label{75158}
\ee
where $\psi(r)$ is given by
(\ref{73125xx}).
Our analysis is restricted
to the perturbation theory in the deformation parameter
$\varkappa$ (first order in $\varkappa$ in the non-uniform model
and second order in $\varkappa$ in the uniform model).
That is, we should expand our result in powers of $\rho$ and $\theta$ and
keep only linear terms in the non-uniform model and quadratic
terms in the uniform model.

For future use we write the function $\psi(r)$
in the first order in the parameter $\varkappa$
\be
\psi(r)=\frac{r\left[\theta\left(r-\frac{1}{3}\Lambda r^3\right)
+\rho\right]}
{\left(r-\frac{1}{3}\Lambda r^3-r_g\right)^2}
+O(\varkappa^2)\,
\,,
\ee
and for $r<<r_0$
\be
\psi(r)=\frac{r(\theta r
+\rho)}{(r-r_g)^2}
+O(\varkappa^2)
\,,
\label{73125xxz}
\ee
and, finally, for $r_g<<r<<r_0$,
\be
\psi(r)=\theta+\frac{\rho}{r}
+O(\varkappa^2)
\,\,.
\label{73125xxzt}
\ee

We would like to emphasize at this point
that the perturbation theory we are going to perform is only
valid for small corrections. When the corrections become large
we need to consider the exact equations of motion
(\ref{712}).

\subsection{Uniform Model}

In the non-relativistic limit the formula
for the anomalous radial acceleration
(\ref{7314xx}) gives
\bea
\label{7333zz}
 A^{r}{}_{\rm anom}
&=& \frac{1}{4}\psi(r)\psi'(r)
+O(\varkappa^3)\,.
\eea
The derivative of the function $\psi(r)$ is easily computed
\be
\psi'(r)=\omega(r)\psi(r)\,,
\ee
where
\be
\omega(r)=
\frac{1}{r}
+\frac{\theta(1-\Lambda r^2)}
{\theta\left(r-\frac{1}{3}\Lambda r^3\right)+\rho}
-\frac{1-\Lambda r^2}
{r-\frac{1}{3}\Lambda r^3-r_-}
-\frac{1-\Lambda r^2}
{r-\frac{1}{3}\Lambda r^3-r_+}\,.
\ee
Thus, we obtain finally
\bea
\label{7333yy}
 A^{r}{}_{\rm anom}
&=& \frac{1}{4}\psi^2(r)\omega(r)
+O(\varkappa^3)\,.
\eea

Recall that the parameters $\rho$ and $\theta$ are of first order
in $\varkappa$.
Strictly speaking we should expand this formula
in $\rho$ and $\theta$ keeping only quadratic terms;
we get
\bea
A^{r}{}_{\rm anom}
&=&\frac{1}{4}
\frac{\left[\theta \left(r-\frac{1}{3}\Lambda r^3\right)+\rho\right]r}
{\left(r-\frac{1}{3}\Lambda r^3-r_g\right)^5}
\Biggl\{
\left(r-\frac{1}{3}\Lambda r^3-r_g\right)
\left[\theta \left(2r-\frac{4}{3}\Lambda r^3\right)+\rho\right]
\nonumber\\
&&
-2r(1-\Lambda r^2)
\left[\theta \left(r-\frac{1}{3}\Lambda r^3\right)+\rho\right]
\Biggr\}
+O(\varkappa^3)\,.
\eea
For $r<<r_0$ (that is, $|\Lambda| r^2<<1$) this becomes
\bea
A^{r}{}_{\rm anom}
&=&-\frac{1}{4}
\frac{r\left(\theta r+\rho\right)
\left[(\rho+2\theta r_g)r+\rho r_g-\frac{2}{3}\theta\Lambda r^4\right]}
{(r-r_g)^5}
+O(\varkappa^3)\,.
\eea
We need to keep the term linear in $\Lambda$ since we do not know
the values of the parameters $\theta$ and $\rho$.
Finally, for $r_g<<r<<r_0$ we obtain
\bea
A^{r}{}_{\rm anom}
&=&-\frac{1}{4}
\left(\theta + \frac{\rho}{r}\right)
\left(\frac{\rho+2\theta r_g}{r^2}
-\frac{2}{3}\theta\Lambda r\right)
+O(\varkappa^3)\,.
\eea

\subsection{Non-uniform Model}

Similarly, in the non-uniform model the anomalous acceleration is
given by
eq. (\ref{7315xx}).
In the $2\times 2$ matrix case considered above
the eigenvalues of the matrix
$
h^{\mu\nu}\xi_\mu\xi_\nu
$
are
\be
\lambda_{1,2}=\pm \frac{1}{2}\tr( h^{\mu\nu}\tau)\xi_\mu\xi_\nu\,.
\ee
Therefore,
\be
P(x,\xi)=\mu_1\lambda_1+\mu_2\lambda_2
=\gamma \frac{1}{2}\tr( h^{\mu\nu}\tau)\xi_\mu\xi_\nu\,,
\ee
where
\be
\gamma=\mu_1-\mu_2\,.
\ee
Thus
\be
q^{\mu\nu}=\frac{\gamma}{2}\tr( h^{\mu\nu}\tau)\,.
\ee
So, we
obtain
\be
\varkappa q^{00}=\gamma\psi(r)\,.
\ee
Thus
\bea
\label{7222zz}
A^{r}_{\rm anom}
&=&-\frac{1}{2}\gamma\psi'(r)+O(\varkappa^2)
\nonumber\\
&=&-\frac{1\gamma}{2}\psi(r)\omega(r)
+O(\varkappa^2)\,.
\eea

Now, we recall that $\rho$ and $\theta$ are of first order in $\varkappa$
and expand in powers of $\rho$ and $\theta$
keeping only linear terms
\bea
\label{7222xzz}
A^{r}_{\rm anom}
&=&-\frac{1}{2}\frac{\gamma}
{\left(r-\frac{1}{3}\Lambda r^3-r_g\right)^3}
\Biggl\{\left(r-\frac{1}{3}\Lambda r^3-r_g\right)
\left[\theta\left(2r-\frac{4}{3}\Lambda r^3\right)+\rho\right]
\nonumber\\
&&
-2r(1-\Lambda r^2)\left[
\theta\left(r-\frac{1}{3}\Lambda r^3\right)+\rho
\right]\Biggr\}
+O(\varkappa^2)\,.
\eea
In the case $r<<r_0$ (when $|\Lambda| r^2<<1$) this takes the form
\bea
\label{7222xzza}
A^{r}_{\rm anom}
&=&\frac{1}{2}\gamma
\frac{\left[(\rho+2\theta r_g)r+\rho r_g
-\frac{2}{3}\theta\Lambda r^4\right]}
{(r-r_g)^3}
+O(\varkappa^2)\,.
\eea
Finally, for $r_g<<r<<r_0$ we obtain
\bea
\label{7222xzzb}
A^{r}_{\rm anom}
&=&\frac{1}{2}\gamma
\left[\frac{(\rho+2\theta r_g)}{r^2}
-\frac{2}{3}\theta\Lambda r
\right]
+O(\varkappa^2)\,.
\eea

\section{Conclusions}

In this chapter we described the kinematics of test particles in the
framework of a recently developed modified theory of gravitation,
called Matrix Gravity \cite{avramidi03,avramidi04,avramidi04b}.
We outlined the motivation for this theory, which is a non-commutative
deformation of General Relativity. Matrix Gravity can be interpreted in terms
of a collection of Finsler geometries on the spacetime manifold rather than
in terms of Riemannian geometry.
This leads, in particular, to a
new phenomenon of \emph{splitting} of Riemannian geodesics into a
system of trajectories (Finsler geodesics) close to the Riemannian
geodesic. More precisely, instead of one
Riemannian metric we have several Finsler metrics and different
mass parameters which describe the tendency to follow a particular
Finsler geodesics determined by a particular Finsler metric.
As a result the test particles
exhibit a \emph{non-geodesic motion} which can be interpreted in terms of
an anomalous acceleration.

By using a commutative algebra we found
a static spherically symmetric solution of the
modified Einstein equations.
In this case a
completely new feature appears due to the presence of
additional degrees of freedom. The coordinate singularities of our
model depend on additional parameters (constants of integration).
Interestingly,
there is a  range of values for these free parameters in which
\emph{no singularity occurs}. This is just one of the intriguing
differences between Matrix Gravity and General Relativity.

The description of matter in Matrix Gravity needs additional study.
In this chapter we studied just the behavior of classical test particles.
We propose to describe a gravitating particle by several mass parameters rather
than one parameter as in General Relativity.
We considered two models of matter: a uniform one, in which all mass parameters are
equal, and a non-uniform one, in which the mass parameters are different.
The choice of one model over the other should be
dictated by physical reasons.
It is worth emphasizing that in the generic non-uniform model
the {\it equivalence principle is violated}.

The interesting question
whether the matter is described by only one mass
parameter or more than one mass parameters as well as the more general question
of the physical origin of multiple mass
parameters requires further study.
Since we do not know much about the physical  origin of the color
masses, we do not assume that they are positive. We do not exclude the
possibility that some of the mass parameters can be negative or zero.
This would imply, of course, that in this theory there is also
gravitational repulsion (antigravity). This could help solve the problem
of the gravitational collapse in General Relativity, which is caused by
the  infinite gravitational attraction.



\chapter[A MODEL FOR THE PIONEER ANOMALY]
{A MODEL FOR THE PIONEER ANOMALY\footnotemark[6]}

\footnotetext[6]{The material in this chapter has been published as a preprint:
I. G. Avramidi and G. Fucci, A Model for the Pioneer Anomaly, arXiv: 0811.1573 [gr-qc]}

\begin{chapabstract}

In a previous work we showed that massive test particles exhibit a
non-geodesic acceleration in a modified theory of gravity obtained by a
non-commutative deformation of General Relativity (so-called Matrix
Gravity). We propose that this non-geodesic acceleration might be the
origin of the anomalous acceleration experienced by the Pioneer 10 and
Pioneer 11 spacecrafts.
\end{chapabstract}

\section{Introduction}

The Pioneer anomaly has been studied by many authors (see
\cite{anderson98,anderson02,moffat04,reynaud06,jaekel06,tangen07}
and the references in these papers) and it has a pretty strong
experimental status \cite{laemmerzahl06}. It exhibits itself in an
anomalous acceleration of the Pioneer 10 and 11 spacecrafts in the
range of distances between $20 {\rm AU}$ and $50 {\rm AU}$ ($\sim
10^{14}{\rm cm}$) from the Sun. The acceleration is directed
toward the Sun and has a magnitude of \cite{anderson98,anderson02}
\be A^r_{\rm anom}\approx (8.74\pm1.33)\times  10^{-8} {\rm cm/s}^2 \,.
\ee
In the last
years there have been many attempts to explain the Pioneer anomaly
by modifying General Relativity (see, for example,
\cite{moffat04} and the references therein). However, there is
also some evidence \cite{tangen07} that it could not be explained
within standard General Relativity since it exhibits a {\it
non-geodesic motion}. That is, it cannot be explained by just
perturbing the Schwarzschild metric of the Solar system. It seems,  from
the analysis of the trajectories, that the spacecrafts do not  move
along the geodesics of any metric. Another puzzling fact is  that there
is no measurable anomaly in the motion of the planets  themselves, which
violates the equivalence principle. In other  words, the heavy objects
like the planets, with masses greater  than $\sim 10^{27}{\rm g}$, do
not feel any anomaly while the smaller objects, like the Pioneer
spacecrafts, with masses of  order $\sim 10^{5}{\rm g}$, do experience
it.

There are also some interesting numerical coincidences regarding the
Pioneer  anomaly (noticed in \cite{Makela07} as well). Recall that the
cosmological distance, which can be defined either by the Hubble
constant $H$ or by the cosmological constant $\Lambda$, is of  order
\be\label{81}
r_0\sim \frac{c}{H} \sim
\frac{1}{\sqrt{\Lambda}} \sim 10^{28} {\rm cm}
\ee
and the Compton
wavelength of the proton is of order
\be
r_1\sim \frac{\hbar}{m_p
c}\sim 10^{-13} {\rm cm}\,.
\ee
Now, we easily see, first of all,
that there is the following numerical relation
\be
\left(\frac{r_1}{r_{\rm anom}}\right) \sim \left(\frac{r_{\rm
anom}}{r_0}\right)^2\,,
\ee
where $r_{\rm anom}\sim 10^{14}{\rm
cm}$ is the distance at which the anomaly is observed. This means
that
\be
r_{\rm anom}\sim
\left(\frac{\hbar}{m_pc\Lambda}\right)^{1/3} \sim
\left(\frac{\hbar c}{m_p H^2}\right)^{1/3}\;.
\label{814xxx}
\ee
Secondly, the
characteristic distance determined by the value of the anomalous
acceleration, $A_{\rm anom}\sim 10^{-8} {\rm cm/sec}^2$, is of the
same order as the cosmological distance
\be
r_2\sim
\frac{c^2}{A_{\rm anom}}\sim 10^{28} {\rm cm}\,,
\ee
which simply
means that
\be
A_{\rm anom}\sim Hc\sim c^2\sqrt{\Lambda}\,.
\ee
It is very intriguing to speculate that the {\it Pioneer effect is
the result of some kind of interplay between the microscopic and
cosmological effects at the macroscopic scales}.

In this chapter we apply the investigation of  motion of test particles in
an extended  theory of gravity, called Matrix Gravity, initiated in
\cite{avramidi08d} to study the anomalous acceleration of Pioneer 10 and
Pioneer 11 spacecrafts.



We would like to stress that this study is just a first
attempt to analyze the phenomenological  effects of Matrix Gravity. We
do not claim that this simple model definitely solves the mystery of the
anomaly. Our aim is just to  propose another candidate for its origin.
Only future tests and  more detailed models can describe
the Pioneer anomaly in full  capacity. This work does not represent the
final answer, but  just a first attempt of studying this phenomenon
within the  framework of Matrix Gravity.

\section{Anomalous Acceleration in Matrix Gravity}

The anomalous non-geodesic
acceleration was derived within perturbation theory in the
deformation parameter in \cite{avramidi08d} and presented in the previous Chapter. We study the two cases
mentioned above.

We consider a simple model of $2\times 2$ real symmetric
commutative matrices. The static spherically symmetric
solution of the matrix
Einstein equations for this model
was obtained in \cite{avramidi08d}. By using the results on the
motion of test particles obtained in Chapter 7,
and by recalling that
in the non-relativistic limit the only essential component
of the anomalous  acceleration is the radial one $A^r{}_{\rm anom}$, we obtain for the uniform model
\bea
\label{8333zz}
A^{r}_{\rm anom}
&=&
\frac{1}{2}\frac{\partial}{\partial r}\left[
-\frac{1}{1-\frac{r_g}{r}}
+f_1(r)-2\theta f_2(r)+\left(\frac{\theta}{2}f_1(r)
+(1+\theta^2)f_2(r)\right)^{2}
\right]\,,
\;\;\;\;\;\;\;
\eea
and for the non-uniform model,
\bea
\label{8222zz}
A^{r}_{\rm anom}
&=&
-\frac{1}{2}\frac{\partial}{\partial r}\left[
\frac{1}{1-\frac{r_g}{r}}
+(\gamma\theta-1)f_1(r)+\gamma(1+\theta)^2f_2(r)
\right]
\,.
\eea
In this last formulas we have introduced the functions
\be
u(r)=1-\frac{1}{3}\Lambda r^2-\frac{r_g}{r}\,,
\ee
and
\be
f_1(r)=\frac{u(r)}{\left[u(r)+(\theta+1)\frac{L}{r}\right]
\left[u(r)+(\theta-1)\frac{L}{r}\right]}\;,
\ee
\be
f_2(r)=\frac{\frac{L}{r}}{\left[u(r)+(\theta+1)\frac{L}{r}\right]
\left[u(r)+(\theta-1)\frac{L}{r}\right]}\;.
\ee

We would like to emphasize at this point  that the perturbation theory
is only  valid for small corrections. Obviously, when the corrections become large
one needs  to consider the exact equations of motion.

\section{Pioneer Anomaly}

We have two free parameters in our model, ${\theta}$ and $L$
(and $\gamma$ in the non-uniform model).
We estimate these parameters to match the value of the observed
anomalous acceleration of the Pioneer spacecrafts.

First of all, we recall the
observed value of the cosmological constant
$\Lambda \approx 2.5 \cdot 10^{-56} {\rm cm}^{-2}$;
therefore,
$
r_0 \approx |\Lambda|^{-1/2}= 6.3 \cdot 10^{27} {\rm cm},
$
and the gravitational radius of the Sun
$
r_g\approx 1.5\cdot 10^5 {\rm cm}\,.
$
The relevant scale of the Pioneer anomaly is
$r_{\rm anom}\sim 10^{14}- 10^{15} {\rm cm}$,
therefore, we can restrict our analysis to the range
$r_g<<r<<r_0$. The values of the dimensionless parameters are
$\frac{r_g}{r}\sim 10^{-8}$,
$\frac{r}{r_0}\sim 10^{-15}$, and
$\frac{r_g}{r_0} \sim 10^{-23}$.
We also recall that the value of the anomalous acceleration is
$
A^r_{\rm anom}\approx 8.7 \cdot 10^{-8} {\rm cm/s}^2 \,.
$
We should stress that
our  analysis only applies to the range of distances
relevant   for the study of the Pioneer anomaly. Therefore, strictly
speaking,  from a formal point of view, one cannot extrapolate our
equations beyond this interval.
Since the parameters  $\frac{r_g}{r}$, $\frac{r}{r_0}$ and
$\frac{r_g}{r_0}$  are negligibly small (compared to $1$)
they can be omitted.

By using the eqs. (\ref{8333zz}) and (\ref{8222zz}), and by defining
$\rho=(1+\theta^{2})L-\theta r_{g}$, we obtain \cite{avramidi08d}
(in the usual units, $c$ being the speed of light)
for $r_g<<r<<r_0$: in the uniform model,
\bea
A^{r}{}_{\rm anom}
&=&-\frac{c^2}{4}
\left(\theta + \frac{\rho}{r}\right)
\left(\frac{\rho+2\theta r_g}{r^2}
-\frac{2}{3}\theta\Lambda r\right)\,,
\label{8219xx}
\eea
and in the non-uniform model,
\bea
\label{8222xzzb}
A^{r}_{\rm anom}
&=&\frac{c^2}{2}\gamma
\left(\frac{\rho+2\theta r_g}{r^2}
-\frac{2}{3}\theta\Lambda r
\right)\,.
\label{8220xx}
\eea

{\it Uniform Model.}
First, we restrict to the case of vanishing cosmological constant.
Then the function
(\ref{8219xx})
takes the form
\bea
A^{r}{}_{\rm anom}(r)
&=&-\frac{c^2}{4}
\left({\theta}+\frac{{\rho}}{r}\right)
\frac{{\rho+2\theta r_g}}{r^2}\,.
\eea
It has an extremum if the signs of
${\theta}$ and ${\rho}$ are different, which occurs at
$
r_*=-\frac{3}{2}\frac{{\rho}}{{\theta}}
$
and
is equal to
\be
A^r_{\rm anom}(r_*)=-\frac{c^2 {\theta}^3}{27}
\frac{(\rho+2\theta r_g)}{\rho^2}
\;.
\ee
Now, we assume that $r_*\sim r_{\rm anom}
\sim 10^{14}\,{\rm cm}$ and
$A^r_{\rm anom}(r_*)\sim
-10^{-8}{\rm cm}/{\rm sec}^2$
to
estimate the parameters
\be
{\rho}\sim 10^7 {\rm cm} \,,\qquad
{\theta}\sim -10^{-7}\,.
\ee

If we leave the cosmological constant there is another range of
parameters that should be investigated. Namely, when the term
$\frac{2{\theta}}{3 r_0^2}r$ becomes comparable with the term
$\frac{{\rho}}{r^2}$.
In this case the anomalous acceleration can be written, by
dropping negligible terms, as
\begin{equation}
A^{r}{}_{\rm anom}(r)=
-\frac{c^{2}}{4r_{0}}\left(\frac{{\theta}{\rho}
r_{0}}{r^{2}}
+\frac{2{\theta}^2}{3r_{0}} r\right)\;.
\end{equation}


We note that the term $\frac{c^{2}}{4r_{0}}$ gives the right magnitude of
the anomalous acceleration. If
we assume that the two terms in the parentheses
are comparable at the characteristic
length $r_{\rm anom}$ and are of order $1$, then we get
an estimate
\begin{equation}
{\rho}\sim
\frac{r_{\rm anom}^3}{r_{0}^{2}}{\theta}
\qquad\textrm{and}\qquad
{\theta} \sim
\left(\frac{r_{0}}{r_{\rm anom}}\right)^{\frac{1}{2}}\;,
\end{equation}
and, therefore,
\begin{equation}
{\rho}\sim 10^{-7}{\rm cm}\qquad\textrm{and}\qquad
{\theta}\sim
10^{7}\;.
\end{equation}

{\it Nonuniform Model.}
In the non-uniform model
we have an additional parameter $\gamma$. The function
has an extremum at
\be
r_*=\left(\frac{3{\rho} r_0^2}{{\theta}}\right)^{1/3}.
\ee
Now, we assume that $r_*\sim r_{\rm anom}\sim 10^{14}{\rm cm}$; then
\begin{equation}\label{810}
\frac{{\rho}}{{\theta}}
=
\frac{r_*^{3}}{3r_{0}^{2}}\sim
10^{-13} {\rm cm}\;.
\end{equation}
Further, by assuming $A^r{}_{\rm anom}(r_*)\sim
-10^{-8}{\rm cm}/{\rm sec}^2$ and using the eq.
we estimate the parameter $\gamma$
\begin{equation}
\gamma \sim 10^{13}\;.
\end{equation}

It is interesting to notice that, in this case, ${\rho}/\theta$ has the
same order of magnitude of the Compton wavelength of the proton.
Moreover, by using (\ref{810}), we confirm
the coincidence (\ref{814xxx})
mentioned in the
introduction.
This is very intriguing; it allows one to speculate
that the anomalous acceleration could be a result of an interplay
between the microscopic and macroscopic worlds; in other words,
the Pioneer anomaly could be a quantum effect.

\section{Conclusions}

In this chapter we applied the kinematics of test particles
\cite{avramidi08d} in Matrix Gravity \cite{avramidi04,avramidi04b} to
the study of the Pioneer anomaly.  

We considered two models: a uniform one, in which a particle
is described by a single mass parameter, and a non-uniform one, in which
a particle is described by multiple mass parameters.  The choice of one
model over the other should be dictated by physical reasons.  The
interesting question of whether the matter is described by only one mass
parameter or more than one mass parameters  requires further study. If
the Pioneer  anomaly is a new physical phenomenon we have to accept the
fact that  the equivalence principle does not hold. If this  is the
case, a model with different mass parameters  (violating the equivalence
principle) would be more  appropriate to describe the motion of test
particles in the Solar system.



\chapter{CONCLUSIONS}	

In this work we carried out a detailed study of two major subjects. The first part of
this Dissertation is devoted to the study of non-perturbative aspects of quantum electrodynamics
on Riemannian manifolds by mainly utilizing heat kernel asymptotic expansion techniques.
The second part focuses on the analysis of low energy aspects of a newly developed theory
of the gravitational field called Matrix Gravity.

In the following we will present a summary of the main results obtained in this
work and we will also describe some ideas for future directions of research.

\section{Summary of the main results}

It is needless to say that non-perturbative results are of fundamental importance
in Physics. On one hand, the techniques used to obtain such results
are very interesting from the mathematical point of view and, on the other hand,
completely new physical phenomena, which are not predicted in perturbation theory,
can be discovered.
Here, we will present a list of the main results obtained in this Dissertation.


\begin{enumerate}

\item{We established the existence of a {\it new type} of {\it non-perturbative} asymptotic
expansion for the heat kernel for the Laplacian, and its trace, on homogeneous Abelian bundles.}

\item{We developed a \emph{new} perturbation theory for the Laplacian, the heat semigroup and
the heat kernel for the case in which the $U(1)$ curvature (electromagnetic field) is
much stronger than the Riemannian curvature.}

\item{We explicitly evaluated the {\it universal tensor functions} in the heat kernel
coefficients. They are analytic functions of the dimensionless quantity $tF$. These universal
functions were not known in the literature and they have been evaluated here for the first time.}

\item{We evaluated, for the first time, the first three non-diagonal heat kernel asymptotic coefficients,
in  (\ref{363}), (\ref{363a})
and (\ref{364}) in powers of the Riemannian curvature and in terms of the new universal tensor functions. }

\item{We found the first three diagonal heat kernel asymptotic
coefficients in (\ref{370}), (\ref{372f}) and (\ref{372}) in powers of the Riemannian curvature and in terms of the new universal tensor functions. This result is completely new and it has been presented here
for the first time in the literature.}

\item{We proved, here, that {\it all} the off-diagonal odd-order heat kernel asymptotic
coefficients are odd order polynomials in the normal coordinate $u$, and, therefore,
they vanish on the diagonal.}

\item{We developed an algebraic framework for the calculation of the heat kernel
asymptotic coefficients which {\it only} relies on the algebra of the commutators
of the relevant operators.}

\item{We proved, in this work, that the new non-perturbative heat kernel
asymptotic coefficients can be written in terms of polynomials that
we called {\it generalized Hermite polynomials}. For these polynomials
we found the generating function and evaluated explicitly the first six polynomials.}

\item{We proved a formula that gives the $n$-th power of the sum
of two operators which satisfy the Heisenberg algebra, in terms of a finite sum
involving powers of the operators and powers of their commutator. }

\item{We evaluated the imaginary part of the effective action
for both scalar and spinor fields in a $n$-dimensional curved spacetime under the influence of a strong
electromagnetic field.}

\item{We explicitly evaluated the coefficient
linear in the Riemannian curvature, $b_{2}(t)$, of the heat kernel asymptotic expansion.
By using this coefficient, we found the imaginary part of the effective action,
of zeroth and first order in the Riemannian curvature, non-perturbative in the electromagnetic field, in $n$-dimensions.}

\item{We generalized the classical result obtained by Schwinger for the creation of pairs
in the electromagnetic field:  For the first time in literature, we found an expression for the creation
of scalar and spinor particles in curved spacetime induced by the gravitational field.
This essentially non-perturbative effect in curved spacetime was completely unknown and it
has been found here for the first time.}

\item{We explicitly evaluated the imaginary part of the effective action
for both scalar and spinor fields in a number of different interesting limiting cases:
\begin{itemize}

\item{The physical dimension: $n=4$.}
\item{Massless limit (Supercritical Electric Field): $m^{2}\ll E, B$.}
\item{Pure electric field: $B\ll m^{2}, E$, in $n$ dimensions and $n=4$.}

\end{itemize}}

\item{We have discovered {\it new} infrared divergences in the imaginary part of the
effective action for massless spinor fields in four dimensions (or supercritical electric field),
which is induced purely by the gravitational correction.}

\item{In this work, we obtained, for the first time in literature, the dynamical
equations for Matrix General Relativity in absence of matter by varying a generalization
of the Hilbert-Einstein action.}

\item{We proposed an action to describe non-commutative matter and we derived,
for the first time, the non-commutative Einstein equations in presence of matter. }

\item{We evaluated the action of Spectral Matrix Gravity in the
weak deformation limit up to the second order in the deformation parameter. Namely, up
to quadratic terms in the Riemannian curvature.}

\item{We evaluated, for the first time, the dynamical equations for
Spectral Matrix Gravity in the weak deformation limit, and we found the spectrum
of the theory on a DeSitter background.}

\item{We proposed a model for kinematics of test particles in Matrix Gravity.
In particular we proposed that test particles are described by several mass parameters and that they
do not move along the geodesics predicted by General Relativity. That is, we predict
the violation of the equivalence principle.}

\item{We evaluated,
in perturbation theory, the non-geodesic part of the motion interpreted
as an anomalous acceleration. We found the anomalous acceleration, up to the first
order in the perturbation parameter, when all the mass parameters are the same (uniform
model) and we also found the anomalous acceleration, up to the second
order in the perturbation parameter, when they are all different (non-uniform model). }

\item{We found a static spherically symmetric solution of Matrix General Relativity in a
simple model of Abelian $2\times 2$ matrices. }

\item{
In particular, we found that
there exists a range for the free parameters in the theory for which there is no horizon in the static and spherically symmetric solution.
This is a completely new feature, which is absent in General Relativity.
 In this particular Abelian case we computed explicitly the anomalous acceleration for both the
uniform and non-uniform models.}

\item{We applied our results
to the analysis of a recently found anomaly in the trajectories of the Pioneer $10$ and $11$
spacecrafts.  We found that there is a range for the free parameters of our theory
which can be adjusted to give the right order of magnitude for the Pioneer anomaly .}

\end{enumerate}

\section{Future directions of research}

The results in this Dissertation open a variety of different opportunities for
future research. In the following we present a list of some of the ideas involving
the results we obtained for the non-perturbative heat kernel asymptotic expansion on
homogeneous Abelian bundles.

\begin{itemize}

\item{The study of complex manifolds is very important in mathematics. An Hermitian
manifold is a even-dimensional complex manifold on which a smooth Hermitian metric is defined.
It can be shown that on this manifold there always exist an antisymmetric $2$-form, called the
K\"{a}hler form, which is obtained from the Hermitian metric. It is well known, then,
that an Hermitian manifold is a K\"{a}hler manifold if the K\"{a}hler $2$-form is closed.
If we assume that the curvature on the K\"{a}hler manifold is small, we satisfy all the
conditions for the analysis done in Chapter $3$ by just replacing the electromagnetic $2$-form
with the K\"{a}hler $2$-form. Such analysis would give the heat kernel asymptotic expansion
coefficients for the Laplacian defined on K\"{a}hler manifolds. This would be an important
study especially in connection with String Theory in which one often uses complex manifolds.}

\item{Two dimensional quantum field theory has become very important over the past years. In fact,
in two dimensions, one can obtain exact results for many cases of interest, which one can use to
get a better understanding of how a particular quantum system would behave in four dimensions.
Our results, when restricted to dimension two, would dramatically simplify. The heat kernel
asymptotic coefficients that we found can be used to evaluate the effective action for quantum fields
in two dimensional curved space under the influence of a strong
electromagnetic field. In particular, the coefficient $b_{2}$ (linear in the Riemannian curvature) found here is responsible for
the divergences in two dimensions and can be used to regularize the theory. Moreover, if one considers
conformal fields, the same coefficients would give the conformal anomaly.}

\item{During the evaluation of the non-perturbative heat kernel asymptotic coefficients, we introduced the
generalized Hermite polynomials to represent the derivatives of the Schwinger heat kernel.
These are polynomials in normal coordinates with
coefficients given by analytic functions of the electromagnetic $2$-form.
It will be very interesting, from the mathematical point of view, to systematically study
the most important properties of these polynomials. In particular, by finding
the differential equations they satisfy one could attempt to prove some orthogonality
relations between them. Moreover, it would be interesting to see if we can find some different
representations, other than a Rodrigues-type formula find here, for the generalized Hermite
polynomials. Lastly, it would be very interesting to find a recurrence relation especially
for their derivatives.}

\item{The Dirac operator has been intensively studied both in mathematics and theoretical physics.
It is globally defined on a manifold $M$ provided that $M$ is orientable and
has a spin structure. It can be proved that if $M$ is compact, the Dirac operator is
elliptic, self-adjoint, and it has a discrete spectrum. Of particular relevance is
the index of the Dirac operator because it gives a measure of the difference between
left and right spinors due to the topology of the manifold $M$. This gives origin to
the chiral anomaly \cite{esposito98,rennie90}. It can be proved that the index of an elliptic
operator, and in particular of the Dirac operator, can be expressed in terms of the heat kernel.
The results obtained in this work could be used in order to study
the index of the Dirac operator in four dimensions under the condition of
large parallel $U(1)$ curvature. It would be very interesting to analyze in what
extent a strong electromagnetic field influences the number of left and right spinors
and the chiral anomaly.}

\item{The non-perturbative
results that we obtained earlier, are valid for an arbitrary Riemannian metric.
Considering, in particular, Einstein manifolds, for which the metric satisfies Einstein's
equations, would be very interesting in astrophysical situations in which the electromagnetic
field is stronger than the gravitational field. In fact, by specifying our result for the imaginary part of the
effective action in the Schwarzschild metric, we could describe particle creation
in the surrounding areas of objects like pulsars and magnetars. 
In particular, once the imaginary part of the effective action for scalar and spinor fields is known, one can
evaluate the current of created particles in the neighborhood of such astrophysical objects.}

\end{itemize}

Matrix Gravity is obviously of recent discovery, and much more research is needed to be done in
order to unveil all of its features. In the following we list some ideas for future research.

\begin{itemize}

\item{It is well known that General Relativity cannot be quantized in a consistent way,
because Einstein's theory is non-renormalizable. A satisfactory
quantized theory of the gravitational field is the most important unsolved problem
in present day theoretical physics. It would be interesting to study in detail
whether or not Matrix Gravity can be successfully quantized. In fact, Matrix Gravity
is nothing but a generalized $\sigma$-model, and therefore the problems of quantization
are the same as a field theoretical $\sigma$-model.}

\item{An interesting study would be the analysis of some simple non-commutative,
static and symmetric
solutions of the dynamical equations, like a non-commut-ative Schwarzschild metric and
a non-commutative DeSitter metric. The first would describe a non-commutative black hole
while the second would represent a non-commutative cosmological model. The singularities in
General Relativity appear when the geodesics in the spacetime cannot be prolonged to arbitrary
values of their affine parameter, this means that all the geodesics converge to the singular
point at a certain time.
Since in our model the trajectory of particles is described by a bundle of trajectories,
there is the possibility that the non-commutative black hole would be free of singularities.
This would be an important feature of our model which is not present in General Relativity.
By analyzing a non-commutative cosmological model, like DeSitter, one could try to
understand if our model predicts the accelerated expansion of the Universe without relying
on the concept of Dark Energy. Having a theory that does not rely on the mysterious Dark Energy to
explain the accelerated expansion of the Universe would be of fundamental importance.}

\item{Lastly, an important problem afflicting General Relativity is the flat rotational
curves of galaxies. It is well known that the radial velocity of a non-relativistic fluid in rotation decreases,
following a specific law, with the distance from the rotational axis. However, observations
show that the radial velocity stays approximately constant regardless of the distance from
the center. In order to explain this discrepancy with the theory, the idea of Dark Matter has
been introduced. By looking at the rotational curves of galaxies one is able to
derive the distribution of Dark Matter (which is unobservable) within the galaxy and in
its neighborhoods. Obviously this idea is not satisfactory from a theoretical point of view.
It would be interesting, then, to understand if a non-commutative model for galaxies could
could explain the flat rotational curves without relying on the concept of Dark Matter.}

\end{itemize}


\thispagestyle{thesheadings}
\begin{Bibliographyno}

\bibitem{abramowitz72}
Abramowitz M. and  Stegun I. A.
Handbook of Mathematical Functions with Formulas, Graphs and
Mathematical Tables, (New York: Wiley) (1972)

\bibitem{anderson98}
Anderson J. D., Laing P. A., Lau E. L., Liu A. S., Nieto M. M. and  Turyshev S. G.,
Indication, from Pioneer 10/11, Galileo, and Ulysses data, of an
apparent anomalous, weak, long-range  acceleration, \emph{Phys. Rev.
Lett.} {\bf 81} (1998) 2858-2861

\bibitem{anderson02}
Anderson J. D., Laing P. A., Lau E. L., Liu A. S., Nieto M. M. and  Turyshev S. G.,
Study of the anomalous acceleration of Pioneer 10 and  11, \emph{Phys.
Rev.} D {\bf 65} (2002) 082004

\bibitem{avramidi90a} Avramidi I. G.,
The covariant technique for calculation of the heat kernel
asymptotic expansion,
\emph{Phys. Lett.}  B {\bf 238} (1990) 92-97

\bibitem{avramidi91} Avramidi I. G.,
The covariant technique for calculation of one-loop effective
action, \emph{Nucl. Phys. B} {\bf 355} (1991) 712-754
Erratum:  \emph{Nucl. Phys. B} {\bf 509} (1998) 557-558

\bibitem{avramidi93} Avramidi I. G.,
A new algebraic approach for calculating the heat
kernel in gauge theories, \emph{Phys. Lett.} B {\bf 305} (1993) 27-34

\bibitem{avramidi94} Avramidi I. G.,
Covariant methods for calculating the low-energy
effective action in quantum field theory and quantum gravity,
arXiv:gr-qc/9403036, 48 pp (1994)

\bibitem{avramidi94a}
Avramidi I. G.,
The heat kernel on symmetric spaces via
integrating over the group  of isometries, \emph{Phys. Lett.} B {\bf 336} (1994)
171-177

\bibitem{avramidi95} Avramidi I. G.,
Covariant algebraic calculation of the one-loop effective potential
in non-Abelian gauge theories and a new approach to stability problem,
\emph{J. Math. Phys.} {\bf 36} (1995) 1557-1571

\bibitem{avramidi95a} Avramidi I. G.,
Covariant algebraic method for calculation of the
low-energy heat kernel, \emph{J. Math. Phys.} {\bf 36} (1995) 5055-5070 , Erratum: \emph{J. Math. Phys.}
{\bf 39} (1998) 1720

\bibitem{avramidi96} Avramidi I. G.,
A new algebraic approach for calculating the heat kernel in quantum gravity,
\emph{J. Math. Phys.} {\bf 37} (1996) 374-394

\bibitem{avramidi97} Avramidi I. G.,
Covariant approximation schemes for calculation of the heat kernel in quantum
field theory, in: Quantum Gravity, Eds. V. A. Berezin, V. A. Rubakov and D. V.
Semikoz (Singapore: World Scientific), pp. 61-78 (1997)

\bibitem{avramidi99}
Avramidi I. G.,
Covariant techniques for computation of the heat
kernel, \emph{Rev. Math. Phys.} {\bf 11} (1999) 947-980

\bibitem{avramidi00} Avramidi I. G.,
Heat Kernel and Quantum Gravity, Lecture Notes in Physics, Series
Monographs, LNP: m64 (Berlin: Springer-Verlag) (2000)

\bibitem{avramidi01}
Avramidi I. G. and Branson T.,
Heat kernel asymptotics of
operators with non-Laplace principal part, {\it Rev. Math. Phys.}
{\bf 13} (2001) 847-890

\bibitem{avramidi01a} Avramidi  I. G.,
Effective Action Approach to Quantum Field Theory,
Lectures for the course in quantum field theory, University of Naples ``Federico II" (2001)

\bibitem{avramidi02}
Avramidi I. G.,
Heat kernel approach in quantum field theory,
\emph{Nucl. Phys. Proc. Suppl.} {\bf 104} (2002) 3-32

\bibitem{avramidi02a}
Avramidi I. G. and Branson T.,
A discrete leading symbol and spectral
asymptotics for natural differential operators, {\it J. Functional
Analysis} {\bf 190} (2002) 292-337

\bibitem{avramidi03} Avramidi I. G.,
A noncommutative deformation of general relativity, \emph{Phys.
Lett. B} {\bf 576} (2003) 195-198

\bibitem{avramidi04} Avramidi I. G.,
Matrix General Relativity: a new look at old problems, \emph{Class.
 Quantum Grav.} {\bf 21} (2004)  103-120

\bibitem{avramidi04b} Avramidi I. G.,
Gauged gravity via spectral asymptotics of non-Laplace type
operators, \emph{J. High Energy Phys.} 07 030 (2004)

\bibitem{avramidi07} Avramidi I. G.,
Lectures on Analytic and Geometric Methods for Heat Kernel Applications in Finance,
NATIXIS Corporate and Investment Bank, Paris, France (2007)

\bibitem{avramidi08}
Avramidi I. G.,
Heat kernel on homogeneous bundles,
\emph{Int. J. Geom. Meth. Mod. Phys.} {\bf 5} (2008) 1-23

\bibitem{avramidi08a} Avramidi I. G.,
Heat kernel on homogeneous bundles over symmetric spaces,
\emph{Comm. Math. Phys.} doi: 10.1007/s00220-008-0639-6 (2008)

\bibitem{avramidi08b}
Avramidi I. G.,
Heat kernel asymptotics on symmetric spaces,
Proc. Midwest Geometry Conference,
\emph{Comm. Math. Anal. Conf.} {\bf 01} (2008) 1-10

\bibitem{avramidi08d} Avramidi I. G. and Fucci G.,
Kinematics in matrix gravity,
\emph{Gen. Rel. Grav.} DOI: 10.1007/s10714-008-0713-6 (2008)

\bibitem{avramidi08e}
Avramidi I. G. and Fucci G.,
Nonperturbative heat kernel asymptotics on homogeneous Abelian bundles,
\emph{Comm. Math. Phys.} (2009) DOI: 10.1007/s00220-009-0804-6

\bibitem{avramidi08c} Avramidi I. G.,
Mathematical tools for calculation of the effective action in quantum gravity,
in: Quantum Gravity, Ed. B. Booss-Bavnbek, G. Esposito and M. Lesch,
(Berlin, Springer, 2009)

\bibitem{avramidi08f} Avramidi I. G. and Fucci G.,
A model for the Pioneer anomaly, arXiv:0811.1573 [gr-qc], 12pp

\bibitem{avramidi09a}
Avramidi I. G. and Fucci G.,
Low-energy effective action in non-perturbative electrodynamics in curved spacetime,
arXiv:0902.1541 [hep-th], 41pp (submitted to Journal of Mathematical Physics)

\bibitem{aschieri05}
Aschieri P., Blohmann C., Dimitrijevi\'c M., Meyer F., Schupp P. and
Wess J.,
A gravity theory on noncommutative spaces,
\emph{Class. Quantum Grav.} {\bf 22} (2005) 3511-3532

\bibitem{barcelo02}
Barcel\'o C., Liberati S. and Visser M.,
Refringence, field theory and normal modes,
\emph{Class. Quantum Grav.} {\bf 19} (2002) 2961-2982

\bibitem{barvin85} Barvinsky  A. O. and  Vilkovisky G. A.,
The generalized Schwinger-DeWitt
technique in gauge theories and quantum gravity, \emph{Phys. Rep.} {\bf 119} (1985) 1-74

\bibitem{bateman53}
Bateman H. and Erdeyi A.,
Higher Transcendental Functions, (New-York: McGraw-Hill), vol. 2 (1953)

\bibitem{camporesi90}
Camporesi R.,
Harmonic analysis and propagators on homogeneous
spaces, \emph{Phys. Rep.} {\bf 196} (1990) 1-134

\bibitem{carroll01}
Carroll S. M., Harvey J. A., Kosteleck\'y V. A., Lane C. D. and Okamoto T.,
Noncommutative field theory and Lorentz violation,
\emph{Phys. Rev. Lett.} {\bf 87} (2001) 141601

\bibitem{chamseddine93}
Chamseddine H. A., Felder G. and Fr\"{o}hlich J.,
Gravity in non-commutative geometry, \emph{Commun. Math. Phys.} {\bf 155} (1993)
205-218

\bibitem{connes} Chamseddine H. A. and Connes A.,
The spectral action principle, \emph{Commun. Math. Phys.} {\bf 186} (1997) 731-750

\bibitem{dewitt65} DeWitt B. S.,
Dynamical Theory of Groups and Fields,
(Gordon and Breach Science Publishers) (1965)

\bibitem{dewitt67a} DeWitt B. S.,
Quantum theory of gravity II: The manifestly covariant theory,
\emph{Phys. Rev} {\bf 162} (1967) 1195-1238

\bibitem{dewitt67b} DeWitt B. S.,
Quantum theory of gravity III: The application of the covariant theory,
\emph{Phys. Rev} {\bf 162} (1967) 1239-1256

\bibitem{dewitt75} DeWitt B. S.,
Quantum field theory in curved spacetime,
\emph{Phys. Rep.} C {\bf 19} (1975) 295-357

\bibitem{dewitt03} DeWitt B. S.,
The Global Approach to Quantum Field Theory,
(Oxford University Press, Oxford) (2003)

\bibitem{dewitt08} DeWitt B. S. and Esposito G.,
An introduction to quantum gravity,
\emph{Int. J. Geom. Meth. Mod. Phys.} Vol. 5 (2008) 101-156

\bibitem{dirac27} Dirac P. A. M.,
The quantum theory of the emission and absorption of radiation,
\emph{Proc. Royal Soc. London} Series A, 114 (1927) 243-265

\bibitem{esposito97} Esposito G., Kamenshchik  A. Y. and Pollifrone G.,
Euclidean Quantum Gravity on Manifolds with Boundaries,
(Kluwer Academic Publishers, Netherlands) (1997)

\bibitem{esposito98} Esposito G.,
Dirac Operators and Spectral Geometry,
(Cambridge University Press, Cambridge) (1998)

\bibitem{feynman42} Feynman  R. P.,
The Principle of Least Action in Quantum Mechanics,
Ph.D. Dissertation, Princeton University (1942)

\bibitem{fucci08} Fucci G. and Avramidi I. G.,
Noncommutative Einstein equations \emph{Class. Quantum Grav.}
{\bf 25} (2008) 025005

\bibitem{fucci09} Fucci G. and Avramidi I. G.,
Non-commutative corrections in spectral matrix gravity,
\emph{Class. Quantum Grav.} {\bf 26} (2009) 045019

\bibitem{fulling89} Fulling S. A.,
Aspects of Quantum Field Theory in Curved Space-Time,
(Cambridge: Cambridge University Press) (1989)

\bibitem{fock37}
Fock V. A.,
The proper time in classical and quantum
mechanics, {\it Izv. USSR Acad. Sci. Phys.}
{\bf 4-5} (1937) 551-568

\bibitem{gilkey95}
Gilkey P. B.,
Invariance Theory, the Heat Equation and the
Atiyah-Singer Index Theorem, (Boca Raton: CRC Press) (1995)

\bibitem{gradshtein07}  Gradshtein I. S. and Ryzhik I. M.,
Table of Integrals, Series and Products, Eds. A. Jeffrey and D.
Zwillinger (Oxford: Academic) (2007)

\bibitem{harikumar06} Harikumar E. and Rivelles V. O.,
Noncommutative gravity, \emph{Class. Quant. Grav.} {\bf 23} (2006) 7551-7560

\bibitem{hurt83}
Hurt N. E.,
Geometric Quantization in Action: Applications of
Harmonic Analysis in Quantum Statistical Mechanics and Quantum Field
Theory, (Dordrecht: Reidel Publishing) (1983)

\bibitem{jaekel06}
Jaekel M-Th. and Reynaud S.,
Gravity tests and the Pioneer anomaly,
in: Laser, clocks and drag-free: exploration of relativistic
gravity in space, Eds. H. Dittus, C.Lammerzahl and S. Turyshev,
(Berlin: Springer) 193 (2006)

\bibitem{kirsten01}
Kirsten K.,
Spectral Functions in Mathematics and Physics,
(Boca Raton: CRC Press) (2001)

\bibitem{konechny02} Konechny A. and Schwarz A.,
Introduction to M(atrix) theory and noncommutative geometry,
\emph{Phys. Rep.} {\bf 360} (2002) 353-465

\bibitem{laemmerzahl06}
L\"ammerzahl C., Preuss O. and Dittus H.,
Is the physics within the solar system really understood?, arXiv:gr-qc/060452 (2006)

\bibitem{madore94} Madore J. and Mourad J.,
A noncommutative extension of gravity, \emph{Int. J. Mod. Phys.} D
{\bf 3} (1994) 221-224

\bibitem{madore99} Madore J.,
An Introduction to Noncommutative
Differential Geometry and its Physical Applications (Cambridge:
Cambridge University Press) (1999)

\bibitem{Makela07} M\"{a}kel\"{a} J.,
Pioneer anomaly: an interesting numerical coincidence,
arXiv:0710.5460v1 [gr-qc] (2007)

\bibitem{mann84} Mann R. B.,
Five theories of gravity, \emph{Class. Quantum Grav.} {\bf 1} (1984)
561-572

\bibitem{moffat04}
Moffat J. W.,
Modified gravitational theory and the Pioneer 10 and 11
spacecraft anomalous acceleration, arXiv:gr-qc/0405076 (2004)

\bibitem{rennie90} Rennie R.,
Geometry and topology of chiral anomalies in gauge theories, \emph{Adv. Phys.} {\bf 39} (1990) 617-779

\bibitem{reynaud06}
Reynaud S. and Jaekel M-Th.,
Long range gravity tests and the Pioneer
anomaly, arXiv:gr-qc/0610160 (2006)

\bibitem{rund59}
Rund H.,
The Differential Geometry of Finsler Spaces,
(Berlin: Springer-Verlag) (1959)

\bibitem{ryder96}
Ryder L. H.,
Quantum Field Theory - 2nd Edition,
(Cambridge University Press: Cambridge) (1996)

\bibitem{sasakura04}
Sasakura N.,
Heat kernel coefficients for compact fuzzy spaces, \emph{J. High Energy
Phys.} 0412 009 (2004)

\bibitem{sasakura05} Sasakura N.,
Effective local geometric quantities in fuzzy spaces from heat kernel
expansions,
\emph{J. High Energy Phys.} 0503 015 (2005)

\bibitem{schwinger51} Schwinger  J. S.,
On gauge invariance and vacuum polarization,
\emph{Phys. Rev.} {\bf 82} (1951) 664-679

\bibitem{schwinger54} Schwinger  J. S.,
The theory of quantized fields V.,
\emph{Phys. Rev.} {\bf 93} (1954) 615-628

\bibitem{synge60}  Synge J. L.,
Relativity: The General Theory,
(Amsterdam: North Holland) (1960)

\bibitem{szabo06}
Szabo R. J.,
Symmetry, gravity and noncommutativity,
\emph{Class. Quantum Grav} {\bf 23} (2006) R199-R242

\bibitem{taylor96} Taylor M. E.,
Partial Differential Equations: The Basic Theory,
Springer-Verlag (New York) 1996

\bibitem{tangen07} Tangen K.,
Could the Pioneer anomaly have a
gravitational origin?,
\emph{Phys. Rev.} {\bf D76} (2007) 042005

\bibitem{vandeven98}
Van de Ven A. E. M.,
Index free heat kernel coefficients, \emph{Class. Quant. Grav.} {\bf 15} (1998)
2311-2344

\bibitem{vassile03} Vassilevich  D. V.,
Heat kernel expansion: User's manual,
\emph{Phys. Rep.} {\bf 388} (2003) 279-360

\bibitem{vassilevich} Vassilevich D. V.,
Quantum noncommutative gravity in two dimensions, \emph{Nucl.Phys.} {\bf B715} (2005)
695-712

\bibitem{wald87} Wald R. M.,
A new type of gauge invariance for a collection of massless spin-2
fields: II. Geometrical interpretation, \emph{Class. Quantum
Grav.} {\bf 4} (1987) 1279-1316

\bibitem{weinberg72} Weinberg S.,
Gravitation and Cosmology: Principles and Applications of the General Theory of Relativity,
(John Wiley and Sons) (1972)

\end{Bibliographyno}	
\pagestyle{plain}
\end{document}